%% file: main.tex
\documentclass[12pt,a4paper,twoside]{book} 

\usepackage{thesis_style}

\begin{document}

\frontmatter
\include{latex/00_frontmatter}

\singlespacing
\include{latex/001_acknowledgements}
\include{latex/01_abstract}

\include{latex/02_works}
\setcounter{tocdepth}{1}
\tableofcontents

\onehalfspacing

\mainmatter

\include{latex/10_overview}

\part{Background and technical introduction}

\include{latex/11_introduction}
\include{latex/12_apparatus}

\part{Experiments with ultracold metastable helium atoms}
\include{latex/21_transitions}

\include{latex/22_tuneout}

\include{latex/23_depletion}

\include{latex/24_conclusion}

\cleardoublepage
\bookmarksetup{startatroot}
\singlespacing
\printbibliography
\addcontentsline{toc}{chapter}{Bibliography}

\appendix
\include{latex/A1_lattice_build}

\end{document}

%% file: latex/00_frontmatter.tex
\begin{titlepage}
\begin{center}

\vspace{0.45cm}
\LARGE{\textbf{Metrology and Many-Body Physics with Ultracold Metastable Helium\\}}
\vspace{0.5cm}
\large{Jacob Alexander Ross\\}
\mbox{}
\vfill
\vspace{0.15cm}
\normalsize{\emph{A thesis submitted to \\}}
\large{The Australian National University\\}
\normalsize{\emph{for the degree of\\}}
\large{Doctor of Philosophy \\}
\vspace{0.5cm}
\normalsize{March 2016 - October 2021}\\
\normalsize{Revised September 2022}\\


\end{center}
\end{titlepage}

\pagebreak
\hspace{0pt}
\vfill
\begin{centering}
\noindent\emph{``We shall not cease from exploration\\
And the end of all our exploring \\
Will be to arrive where we started \\
And know the place for the first time."\\
\emph{T. S. Eliot}}\\ 
\end{centering}
\vfill
\hspace{0pt}
\pagebreak

\newpage
\begin{flushleft}
\Large{Metrology and Many-Body Physics with \\Ultracold Metastable Helium}
\vspace{0.1cm}
\hrule
\vspace{1cm}
\large{\textbf{Jacob Alexander Ross\\}}
He* BEC group\\
Quantum Science \& Technology department\\
Research School of Physics\\
Joint Colleges of Science\\
Australian National University\\
Canberra, Australia

\vspace{1cm}
\begin{table}[h]
\begin{tabular}{c l}
\emph{\large{Supervisory committee}} & \large{Dr Sean S. Hodgman}\\
                              & \large{Professor Andrew G. Truscott}\\
                              & \large{Professor Kenneth G. H. Baldwin}
\end{tabular}
\end{table}

\vspace{2cm}
\large{\textbf{Declaration\\}}
Except where acknowledged in the customary manner, the material presented in this thesis is, to the best of my knowledge, original and has not been submitted in whole or part for a degree in any university.\\
\end{flushleft}
\begin{flushright}
\vspace{0.5cm}
\rule{3cm}{1pt}\\
Jacob A. Ross
\end{flushright}

\vfill
\begin{center}
\small{$\copyright$ Jacob Alexander Ross 2021\\
All rights reserved}
\end{center}

\mbox{}
\begin{figure*}[b]
\begin{center}
\includegraphics[width=0.7\textwidth]{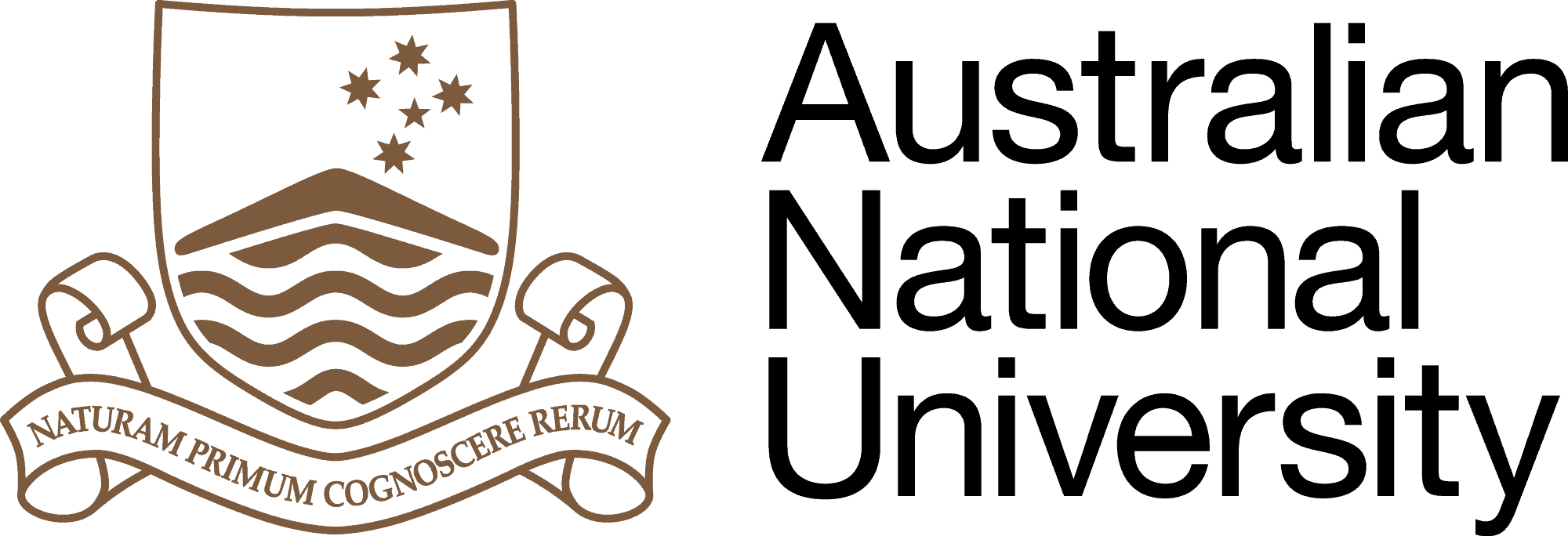}
\end{center}
\end{figure*}

\newpage
\noindent\large{For my parents - finally you can see me in robes.}\newline
\mbox{}
\vfill
\begin{flushright}
\large{\emph{``We who cut mere stones must always be envisioning cathedrals"\\} 
- Quarry workers' creed}
\end{flushright}

\cleardoublepage

%% file: latex/001_acknowledgements.tex
\section*{Acknowledgements}

My first gratitude is to Sean and Andrew for taking a chance on my younger self. 
An ageless adage reads that research is unpredictable, but even so I doubt that anyone could have guessed how this process would unfold.
It is a substantial credit to you that the obstacles along the way, both personal and technical, did not derail the enterprise altogether.
The ink of this compendium is coloured with a trace of the fortitude that I could only have distilled through persisting under your unwavering belief that I could see this through.
And if you ever did tire of guiding my sight to the next turn of the track, it is a further credit to you that the alacrity of your advice never faltered.

To my entire panel, Kenneth Baldwin, Andrew Truscott, and Sean Hodgman, I owe a litany of gratitudes for shaping my understanding of what it is to be a scientist.
For one, your guidance of my writing instilled in me the metrologist's aspiration for ever more precise language, accurate claims, and the exact elucidation of implications (even if I never did master brevity).
For another, I would be professionally impoverished without your frank counsel about the stategems of academia, your readiness to find time for a confused student among many and commendable responsibilities, your clear-sighted technical intuition, and your tutelage in the art of prioritization.
Finally, for imparting the dual lessons that asking for help is almost always better done sooner, and that one's character grows in measure of the time spent fighting through the thickets of ignorance.

Elsewhere, the seemingly interminable humour and optimism of Piotr Deuar was invaluable.
Fittingly, some real understanding started to form only after I figured out how to get into the IFPAN campus after dark in the pouring rain without speaking a word of Polish... 
If that's a metaphor for this whole venture, then your contribution would be that of a small hand-held candle lantern, distant but always a source of cheer.
My journey through Europe en route to your graciously-hosted visit turned out to be more influential than you appreciate. 

At the abuse of yet another aphorism, it takes a village to raise a student. I am grateful to many mavericks and role-models for enlivening my life in the school, teaching me to see matters physical and academic from different angles. With particular acknowledgement to Rose Ahlefeldt, Geoff Campbell, Nanda Dasgupta, and to my fellow graduate students especially Alex Bennet, Jessica Eastman, Patrick Everitt, Abbas Hussein, Ruvi Lecamwasam, Matthias Wurdack, Richard Taylor, Kieran Thomas, and Alexis Tuen, thanks for your camaraderie in both celebration or commiseration at the vicissitudes of science.
More than that, it takes all types to make a village. In short survey, I must thank Ross Tranter and Colin Dedman for the indispensable technical support (if I came away with any measurable fraction of your wizardry and know-how, I am all the richer for it), and also Karen Nulty, Liudmila Mangos, Sonia Padrun, and Nikki Azzopardie for making sense of the administrative apparatus of the university (and always forgiving the overdue progress reports).

There is a unique gratitude to those who, wittingly or not, became adoptive `pastoral types' through the years (in equal senses of herding confused mammals and of doling out spiritual guidance). To Inger and Victoria, you'll know that my brief salute here is one of silent respect and not for lack of inspiration. In short, thank you for your incisive wisdom, elevating the soul, and for helping me remember to believe in myself. To Tim, thank you for encouraging me down the unbeaten track. Designing and executing the research education and development retreat was a great privilege and I hope in earnest to pay forward the investment you made in me. 

Glancing back some distance, I cannot overlook the central figures of my formative years as a proto-physicist. 
I will forever and fondly remember the influence of Ian MacArthur and Darren Grasso in fostering my curiosity, guiding me deeper into formalism and up ladders of abstraction.
It may surprise you to see this journey culminate in a thesis not on bits and manifolds, but on `this crude matter'. Trust me, you're not alone.
But I did learn from Darren that our creativity dwells alongside the diversity of our knowledge, and so our conversations were just as defining as the present matters.
If it is clich\'{e} to claim one `would not be here if not for...' it is only so because of the inescapable contingency of our success on those who walk beside us.
And thus I owe a foundational gratitude to David Thomson, and Lorraine Donaldson before that, for opening my eyes to the joys of physics.

The challenges of living through a PhD surely pale in comparison to those of living with a PhD student.
To that, I raise a glass especially to Charlotte, Lukas, Kendal, and Ryan for your good humour and compassionate ears after those countless late homecomings on days when nothing seemed to work.
Special mention must go to Alice and Morgan for showing me that not only was it normal to feel out of one's depth, alone in the woods, out of rope, or out of patience, one could also expect to get through nonetheless, and seeing you cut your paths gave me greater confidence in mine.
Whatever good fortune let me count Zachary Brown, Sarah Jackson, Katie Jameson, Stephanie Jones, and Brianna Sage among my scientist peers is beyond me. Your impression may well last a lifetime, and I am grateful for the serious conversations as much as the serious fun. Singular recognition must be made of Josh Izaac, my physics pal and `older brother' from the high school days, for the kinship and for reminding me that my various grievances are usually sensible, and always surmountable. This reflection feels incomplete without wondrously many others who've shared their light along the way - you shall remain nameless here, but if for some strange reason you find yourself reading this, know I write with you in mind.

I must make a penultimate nod to my peers at the centre of the circle,  whose dear friendship may be the gift I most cherish from the past half-decade. It has been an honour to be found worthy of your company, and I hope to enjoy it for years to come.
To Aqeel Akber, for always striving for the human aspect and the extension of the self; to Geoff Bonning, for always thinking bigger; To Prithvi Reddy, Lauren Bezzina, and Bryce Henson, for your trust and your time, for sharing in both the elated peaks and dismal the pits of our journeys, and especially for everything other than the organizing and the physics. 
To my brothers-in-arms and fellow helionauts, Dongki Shin and Bryce (again) in turn: David, your painstaking meticulousness set a standard to which I can only aspire, and the contrast of your rigorous creativity with your humility and luminous curiosity will always be a wonder to me. Bryce, I seem to learn something new from you each day.  Your attention to detail is maddening and compelling, and maybe sometime I'll think of something you haven't already. Until then I look forward to trying to keep up. 

To my family, I will eternally be grateful for your encouragement and support, even if you didn't quite know what I was doing or why, and for your patience during my digressions when trying to explain. Finally, to Hannah: My gratitude for your arrival can scarcely fit within these pages. I lost count long ago of times you talked me back on track or down from outright renunciation. For this, and boundless things beyond, I wonder from which well this good fortune springs. At the eve of our next chapter together, I celebrate the unread pages and look ahead with joy.

\newpage

%% file: latex/02_works.tex
\section*{Works discussed in this thesis}

\begin{itemize}

	\item \fullpaper{Frequency measurements of transitions from the $2\triplet P_2$ state to the $5\singlet D_2$, $5\triplet S_1$, and $5\triplet D$ states in ultracold helium}{\underline{J. A. Ross}, K. F. Thomas, B. M. Henson, D. Cocks, K. G. H. Baldwin, S. S. Hodgman, A. Truscott}{Physical Review A}{2020}{https://journals.aps.org/pra/abstract/10.1103/PhysRevA.102.042804}

	\item \fullpaper{Measurement of a helium tune-out frequency: an independent test of quantum electrodynamics}{B. M. Henson, \underline{J. A. Ross}, K. F. Thomas, C. N. Kuhn, D. K. Shin, S. S. Hodgman, Y. H. Zhang, L. Y. Tang, G. W. F. Drake, A. T. Bondy, A. G. Truscott, K. G. H. Baldwin}{Science}{2022}{https://www.science.org/doi/10.1126/science.abk2502}
	
	\item \fullpaper{Survival of the quantum depletion of a condensate after release from a harmonic trap in theory and experiment}{\underline{J. A. Ross}, P. Deuar, D. K. Shin, K. F. Thomas, B. M. Henson, S. S. Hodgman, A. G. Truscott}{Nature Scientific Reports}{2022}{https://www.nature.com/articles/s41598-022-16477-9}
	
	\item \fullpaper{Rapid generation of metastable helium Bose-Einstein condensates}{A. H. Abbas, X. Meng, R. S. Patil, \underline{J. A. Ross}, A. G. Truscott, S. S. Hodgman}{Physical Review A}{2021}{https://journals.aps.org/pra/abstract/10.1103/PhysRevA.103.053317}

\end{itemize}	

\section*{Other publications from the course of study}
\begin{itemize}
	\item \fullpaper{Trap frequency measurement with a pulsed atom laser}{B. M. Henson, K. F. Thomas, Z. Mehdi, \underline{J. A. Ross}, S. S. Hodgman, A. G Truscott}{Optics Express}{2022}{https://doi.org/10.1364/OE.455009}

	\item \fullpaper{Direct measurement of the forbidden $2\triplet S_1 - 3\triplet S_1$ atomic transition in helium}{K. F. Thomas, \underline{J. A. Ross}, B. M. Henson, D. K. Shin, K. G. H. Baldwin, S. S. Hodgman, A. G. Truscott}{Physical Review Letters}{2020}{https://journals.aps.org/prl/abstract/10.1103/PhysRevLett.125.013002}

	\item \fullpaper{Entanglement-based 3D magnetic gradiometry with an ultracold atomic scattering halo}{D. K. Shin, \underline{J. A. Ross}, B. M. Henson, S. S. Hodgman, A. G. Truscott}{New Journal of Physics}{2020}{https://iopscience.iop.org/article/10.1088/1367-2630/ab66de/meta}

	\item \fullpaper{Approaching the adiabatic timescale with machine learning}{B. M. Henson, D. K. Shin, K. F. Thomas, \underline{J. A. Ross}, M. R. Hush, S. S. Hodgman, A. G. Truscott}{Proceedings of the National Academy of Science}{2018}{https://www.pnas.org/content/115/52/13216.short}

	\item \fullpaper{Widely tunable, narrow linewidth external-cavity gain chip laser for spectroscopy between 1.0-1.1 $\mu$m}{D. K. Shin, B. M. Henson, R. I. Khakimov, \underline{J. A. Ross}, C. J. Dedman, S. S. Hodgman, K. G. H. Baldwin, A. G. Truscott}{Optics Express}{2016}{https://www.osapublishing.org/abstract.cfm?uri=oe-24-24-27403}
\end{itemize}

\newpage

%% file: latex/10_overview.tex
\chapter{Overview}
\markboth{OVERVIEW}{}

\begin{adjustwidth}{3cm}{0cm}
\begin{flushright}
\singlespacing
{\emph{``In the weeks that had just passed, Commander Norton\\
had often wondered what he would say at this moment.\\
But now that it was upon him, history chose his words,\\
and he spoke almost	automatically, 	barely aware \\
of the echo from the past: `Rama Base.\\
 \emph{Endeavour} has landed.'"}\\ 
- Arthur C Clarke\footnote{\emph{Rendezvous with Rama}, Harcourt Brace Jovanovich (1973)}}
\end{flushright}
\end{adjustwidth}
\onehalfspacing
\vspace{1cm}

\section*{Prologue}\label{sec:prologue}
\addcontentsline{toc}{section}{Prologue}

	\dropcap{Splitting} a ray of light from the solar chromosphere during the total eclipse of 1868, Pierre Jules C\'{e}sar Jansen resolved a bright yellow line through a spectroscope.
	As no element known on earth emitted this colour, a new element was identified and named helium after the Greek sun titan, Helios. Helium is now understood to comprise some 24 per cent of the ordinary matter in the universe, outweighing the sum of all heavier elements, and to consist primarily of a primordial nuclear $\alpha$ particle neutralized by two electrons. To this day, helium remains a nucleation point of cosmological knowledge. For example, spectrometry of the atmosphere of WASP-107b revealed absorption of light from its parent star at 1083.331 nm, intrinsic to helium, and led to the ascertainment of the exoplanet's atmospheric erosion rate of some $10^{10}-3\times10^{11}$ grams per second \cite{Spake18}. On earth, the same absorption line is employed in a handful of laboratories around the world to drive helium towards a new extremum in the cosmos. While helium fuses into carbon at some 10$^8$ kelvin in the furnaces at the centre of giant stars, and the near-vacuum conditions in the Boomerang nebula reach a single degree kelvin, the helium studied in this thesis momentarily sustains temperatures as low as $10^{-8}$ kelvin.

	Deep in the ultracold regime, dilute gases take on the unfamiliar character of quantum degeneracy, departing from the familiar ideal gas in the sense that the spin-symmetry of the constituent atoms now determines the statistical features of the ensemble. Atoms with integer spin $n$ are Bosons and are not bound by the Pauli exclusion principle as Fermions, with half-integer spin $\frac{1}{2}(2n+1)$, are. The quantum-degenerate behaviour of dilute bosonic gases has the character of a collection of atoms residing in a common quantum state, behaving as waves with a length scale larger than the space between the atoms, and with an emergent order distilled by evaporating away the chaos of thermal motion. Highly ordered, quiescent, and exquisitely isolated, ultracold dilute gases present scientists with almost perfectly idealized conditions to study the structure of matter, its interaction with light, and the emergence of collective phenomena from constituents. This thesis touches on each of these themes in turn: First, by extending the proud history of optical spectroscopy in helium; second, by measuring the frequency of the tune-out point near 413 nm with sufficienct accuracy to check the veracity of state-of-the-art calculations in quantum electrodynamics; and finally, tracing the tiny effects of weak interactions in ultra-dilute superfluids by counting individual atoms. Weaving through these themes is a common thread - precise quantification of subtle processes through careful attention to weak signals.

\subsection*{Indivisible and unattainable}

	Although his writings are lost to history, the greek philosopher Democritus is remembered for his hypothesis that there was a smallest thing: That one could break mountains into boulders, boulders into stones, stones to sand ... to something irreducible. He called these \emph{atomos}, for indivisible.	This was astoundingly prescient: The atomic theory, as it came to be known, would not find empirical validation for another two millenia. And, like all theories that prove to be correct, it too reached its point of failure a few hundred years thereafter. 	The framework that would subsume the atomic theory would also synthesize the resolution of the `two clouds obscuring the sky of physics' described by William Thomson, 1$^{st}$ Baron Kelvin, in an address to the Royal Institution of Great Britain: The experimental finding by Michelson and Morley that the speed of light was isotropic, and the poor predictions of the Maxwell-Boltzmann statistical mechanics at low temperatures.

	The early validation of atomic theory (of indivisibles, as opposed to the modern theory of atomic \emph{structure}) came from the success of the kinetic theory of gases in explaining the empirical laws of Avogadro, Boyle, and Gay-Lussac, and their synthesis in the ideal gas law. Although Boyle himself raised the prospect of a minimum absolute temperature, the first estimation of it value in celsius was made by Guillaume Amontons by extrapolating the contraction of a cooling air column to the point where its volume would vanish: -240 $^\circ$C. This was improved by Johann Lambert to the value -270 $^\circ$C, close to the present value of $-273.15~^\circ$C, as determined by William Thomson and hence defined as zero degrees Kelvin. Amontons was right about one thing, though: The absolute zero of temperature is an unattainable asymptote, as codified in the third law of thermodynamics \cite{Masanes17}. At these extreme conditions, one of Kelvin's clouds presented itself in the divergence of the predicted specific heat capacity of gases from experimental measurements, worsening at low temperatures. 
	The classical atomic theory was further challenged when the \emph{indivisibles} were found to divide. 
	Certain elements emit varieties of radiated particles and transform into other elements. 
	This is now understood in light of Ernest Rutherford's thesis that all atoms contain positively-charged nuclei, considering evidence from the scattering experiments conducted by Hans Geiger and Ernest Marsden. 
	This would eventually be married with the understanding of the distinct discovery that each element emits of light with specific wavelengths, called \emph{Fraunhofer lines}.
	The first cloud was thus clarified by Einstein's synthesis of the photoelectric effect and Max Planck's postulate that photon energies came in discrete units\footnote{An experimentalist through and through, Planck made this postulate not out of some theoretical inspiration: It just made the theory fit the data.}. 
	The new proposal, that matter and light both carried energy in quantized units, was the seed crystal around which a revolution in physics would soon nucleate.

	The empirical Rydberg constant relating radiated photon wavelengths to the series of (integer) quantum numbers $n$, via $E=R(n_{1}^{-2}-n_{2}^{-2})$, was derived by Niels Bohr by considering the consequences of quantizing the angular momentum of electrons `orbiting' the nucleus in units of Planck's constant (and thus energy, as spin was not yet understood) \cite{Bohr1913}
	\footnote{The Planck constant has the same units as angular momentum, but it comes from the phase-space integral $\int p~dq$ in the calculation of the \emph{action}.}.
	Thus the structure of the Hydrogen spectrum was grounded in an understanding of the structure of the supposedly indivisible atoms. 
	The picture was not yet complete: The so-called `fine structure' lines, Pieter Zeeman's observation that magnetic fields can alter spectral lines \cite{Zeeman1897}, and the presence of doublet lines at very similar wavelengths remained unexplained until the resolution of Kelvin's second cloud.
	The experiment of Albert Michelson and Edward Morley became the empirical grounding of Einstein's special theory of relativity 
	\footnote{Ironically, Michelson claimed (at the inauguration of the Ryerson Physics Laboratory at the University of Chicago) that the `great principles [had] already been discovered,' and that physics would `henceforth be limited to finding truths in the sixth decimal place'. This is often misattributed to Kelvin, confused with his comment about the two `clouds'. While Michelson's experiment \emph{did} disprove Kelvin's hypothesis of a luminiferous ether, it disproved Michelson all the more spectacularly.} \cite{Michelson1887}.
	Among the triumphs of relativistic quantum mechanics was the \emph{prediction} of the existence of antimatter in essentially the same swing as explaining the doublet lines, Zeeman effect, and fine structure in terms of the spin angular momentum of the electron \cite{DiracH}.
	And yet certain details were still unresolved. 
	Of central imporance was the observation by Willis Lamb and Robert Retherford that two of Hydrogen's energy levels, predicted to be identical by Dirac's relativistic quantum theory, were in fact distinct \cite{Lamb47,Lamb50}.
	The explanation was found by Hans Bethe by renormalization of the proton and electron masses \cite{Bethe47}, laying the foundation for the first relativistic quantum field theory, \emph{quantum electrodynamics}, the `jewel of physics' that crystallized following the dissolution of the two clouds.

\subsection*{The foundation stone}

	Quantum electrodynamics (or \emph{QED}) describes the interaction of charged particles with the electromagnetic field, whose fundamental excitations are identified with photons - particles of light.
	QED therefore describes the physics that governs all we see with our eyes and indeed the enormous variety of condensed matter from metals to neurons.
	After Bethe's successful prediction of the Lamb shift, some astoundingly fast progress was made within just a couple of years, building upon the construction of QED by Richard Feynman, Nobuo Tomonaga, and Julian Schwinger \cite{FeynmanNobel}.
	In the theory of QED any observable can be expressed as a sum over constituent processes of increasing complexity.
	Lower approximations account for the ingoing and outgoing particles, more complex ones for interactions between them mediated by force-carrying bosons, and yet more complicated ones by the fleeting influence of `virtual' pairs of matter-antimatter twins, which exert some influence on the outcome before annihilating away (or not).
	The more complex processes, by measure of the number of interactions in the corresponding Feynman diagram, are weighted by increasing powers of the fine structure constant $\alpha=\frac{e^2}{4\pi\epsilon_0\hbar c}\approx1/137$.
	This picture of the basic substance of the visible universe makes accurate predictions of measurable effects, beginning with the Lamb shift, extending to the anomalous magnetic moment of the electron (now known to some parts per billion \cite{Aoyama15,Hanneke08}), and currently compares very well with the measurements from state of the art of atomic spectroscopy and particle accelerators.
	As the first synthesis of special relativity and quantum mechanics, QED laid foundations for the standard model of particle physics and still stands as the most accurate quantitative description of the world to date. 

	The concurrent advance of experimental and theoretical precision has yielded ever more accurate determinations of basic quantities such as $\alpha$, the Rydberg constant, and the sizes of the nuclei of light elements \cite{NIST_Constants}. 
	Currently, these fundamental constants are so called by simple empirical fact, not by derivation from some physical principle.
	There is considerable work, too broad to review here, to compare values determined on earth with astronomical observations to search for variations of these values across space or time.
	Precise determinations in atomic systems, including helium, can contribute to this search by providing references on earth for comparison with radiation of cosmological origin.

	Hints towards extensions of our current understanding may already be visible in other known discrepancies between theory and experiment.
	One prominent anomaly is the disagreement between determinations of the proton radius using different measurement techniques.
	The latest update to the CODATA recommended values of the physical constants \cite{CODATA21} note that the uncertainty in the proton radius has been reduced in comparison to the previous value by recent measurements in hydrogen spectroscopy \cite{Beyer17,Bezginov19}.
	However the update also notes that the tension `has not been fully resolved,' concluding:  `Further experiments are needed'.
	Further outstanding anomalies include a very recent 4.2$\sigma$ difference between calculated and measured values of the muon magnetic moment  (considering two combined experiments \cite{Abi21}) and hints of broken lepton symmetry in b-quark decay \cite{LHCb21}.
	The space of possible theories that could explain this data is vast, and so further measurements are required to constrain or discard competing theories.
	Precision measurements in atomic systems can provide such information, as in the ongoing quest to determine the nuclear charge radii of the $^3$He and $^4$He isotopes.
	This mission is timely:  The recent measurement of the alpha particle radius in muonic helium \cite{Krauth21} is the counterpart of electronic $^4$He in a valuable complement to the analogous experiments in hydrogen.
	The state of these ongoing campaigns, including outstanding disagreements between predicted and measured energy levels in helium, are also discussed in chapter \ref{chap:transitions}.
	
	The first major work in this thesis contributes to this effort by providing the first measurements of a transition from the low-lying $n=2$ manifold to the higher $n=5$ manifold in $^4$He.
	The energy of higher-lying levels can be computed with greater accuracy \cite{Drake07}, and so transitions between low- and high-lying states can serve as constraints for the ionization energy of the lower states. While this measurement does not have the precision required to resolve QED effects, the method could be employed with a more accurate frequency reference and obtain frequency measurements competitive with state-of-the-art QED.
	The second major work provides a test of QED through the measurement of a tune-out wavelength which is a stringent test of the QED predictions of oscillator strengths, a complementary scheme to energy level measurements, and is discussed in chapter \ref{chap:tuneout}.
	
\subsection*{Approaching the unapproachable} 

	Simultaneous with the high-profile crusade in high-energy physics during the late 20$^\textrm{th}$ century was an overturning of our understanding of \emph{low}-energy phenomena.
	The inoculation of quantum theory into statistical mechanics improved the predictions of material properties such as conductivities and specific heats, especially in the low-temperature regime where the Maxwell-Boltzmann statistics start to diverge from the actual behaviour of materials and one enters the domain of quantum statistical mechanics.
	In this regime, spin, the quantity originally drawn from relativistic quantum mechanics, was soon to be found to have pivotal importance in explaining the structure and dynamics of solid objects at room temperature and below, and eventually fertilized the burgeoning field of ultracold atomic physics.

	The first piece came into play with the successful liquefaction of helium by Heike Kammerlingh Onnes in 1908, which could be called the ground-breaking moment that began the era of low-temperature physics\footnote{Onnes was awarded the Nobel prize for his work, as were many of the persons named in this chapter.}.
	A mere three years later, Onnes used liquid helium to cool solid mercury and documented a vanishing of the resistance of the metal below 4.2 kelvin that he called superconductivity.
	Nearly three decades later, Pyotr Kapitza \cite{Kapitza38} and the duo of John Allen and Don Misener \cite{Allen38} made near-simultaneous observations  of super\emph{fluidity} in liquid helium (published in the same issue of \emph{Nature})\footnote{Later examination of Onnes' notebooks would reveal he also observed the effect in his experiments with mercury, but apparently did not recognize the significance.},  putting into play another piece that would eventually be connected to the first by the thread of quantum theory. 

	While working in Dhaka, a young admirer of Einstein named Satyendra Nath Bose derived the black-body spectrum starting from the assumption that photons, being indistinguishable, had fewer macrostates than an otherwise-identical ensemble of distinguishable particles \cite{Bose24}.

	In 1925, extending Bose's work to atoms (in particular, attending to the conservation of particle number), Einstein predicted  that bosonic atoms would \emph{condense} into a new state of matter, diverging from the predictions of the Maxwell-Boltzmann statistical mechanics \cite{Einstein25}. 	
	It would take until the turn of the millenium before the predicted Bose-Einstein condensation would be realized in the laboratory.
	However, the observed superfluidity in helium was almost immediately postulated by Fritz London to be connected to this condensation effect \cite{London38} (published in the same issue as Allen, Kapitza, and Misener's reports on superfluidity).
	Three years later Lev Landau formalized a model of superfluidity in terms of the qualities of the excitation spectrum \cite{Landau41}, which eventually led to the construction of the two-fluid model that was first proposed by Laszlo Tisza in 1940 \cite{Tisza38}.
	In the two-fluid model a superfluid is approximated by a coexistence of a normal component with a frictionless component comprised of elementary excitations.
	In 1947, Nikolay Bogoliubov provided the microscopic theory underlying the model, showing that it was not the bosonic atoms themselves, but excitations in collective degrees of freedom that underwent condensation and gave rise to the superfluid part \cite{Bogoliubov47}.
	In this picture the thermal depletion of the condensate was thus distinguished from the quantum depletion which was induced by interactions and persisted in the limit of zero temperature.
	The depletion of condensates is now appreciated to be relevant to a broader range of systems than just bosonic gases.
	For instance, both the BEC and BCS regimes of superconductivity are characterized by the condensation of dimerized fermions and Cooper pairs, respectively, regardless of the substrate in which they are produced.
	In a rapid succession of works initiated by Landau, Lifshitz, Lars Onsager, and Oliver Penrose \cite{Yang62}  the concept of \emph{off-diagonal long-range order} was established which entails the occurrence of superfluidity in both bosons and fermions \cite{PitaevskiiStringari}.

	The study of superfluid helium and of condensation in general is again connected to cosmological scales by evidence of superfluidity in the crust of neutron stars \cite{Baym69,Martin16,Page11}, and at the frontier of physics in models where condensed primordial axions are presented as dark matter candidates \cite{Mielke09}.
	Bose's postulate reaches deeper into fundamental particle physics via the spin-statistics theorem.
	With spin itself now being understood in terms of the symmetry groups of fundamental particles, Bose's work on distinguishability plays a foundational role in the statistical mechanics, and thus thermodynamics, of the fundamental fields in the standard model. 
	For this seminal contribution, the particles of integer spin are now generically called bosons.
	The general statistics of indistinguishable particles now bears both the name Bose-Einstein statistics in honour of these pioneering physicists. 
	A more well-known consequence of Bose-Einstein statistics is the now-ubiquitous laser, which is distinguished from Bose-Einstein condensation by the fact that the chemical potential of a photon gas is zero \cite{Klaers10,Schmidt16}.
	Among innumerable other applications, the laser would prove instrumental in the controlled realization of atomic condensates in the laboratory.

	Following the development of the central techniques of laser cooling \cite{Phillips82,Chu85}, magneto-optical \cite{Raab87} and purely optical trapping \cite{Chu86}, magnetic trapping \cite{Migdall85},	evaporative cooling \cite{Petrich95},	and sub-doppler polarization gradient cooling \cite{Lett88}, BEC was finally realized in three labs  utilizing all of these techniques to produce magnetically trapped condensates of alkali atoms \cite{Anderson95,Davis95,Bradley95,Cornell02}.	
	The optical trapping of condensates followed shortly thereafter \cite{StamperKurn98}, as did the achievement of degenerate Fermi gases \cite{DeMarco99,Truscott01}	and Bose-Fermi mixtures \cite{Schreck01}.
	In the two intervening decades over a hundred quantum-gas labs have come into operation around the world using at least 19 elements for various purposes\footnote{See \url{https://everycoldatom.com/}{everycoldatom.com}.}.

	Ultracold atomic systems hold the record for the lowest kinetic temperatures on earth in \emph{pico}-kelvin regime \cite{Kastberg95, Manning14}.
	Naturally, the coldest object in the universe will tend to heat up by virtue of being surrounded by a universe, but even a perfectly isolated condensate has a coherence time intrinsically limited by interaction with the ever-present thermal modes \cite{Sinatra09}.
	
	Aside from the study of the basic physics of degenerate matter, ultracold atom experiments have found a dizzying range of applications including foundational tests of quantum mechanics \cite{Lopes15,Manning15},	matter-wave interferometry \cite{Cronin09},	studies of light-matter interactions and atomic structure (e.g. photoassociation\cite{Jones06} and precision spectroscopy \cite{Campbell17,Marti18}). Numerous advanced techniques have been developed to confine and control ultracold atoms in boxes \cite{Meyrath05}, rings \cite{Gupta05}, and shells \cite{Gaunt13} in one \cite{Kinoshita04} or two \cite{Rychatrik04} dimensions. 
	Recently the condensation of molecules \cite{Zwirlein03} has led to the study of state-resolved chemistry and controllable reactions \cite{Balakrishnan16}, paving the way towards new foundational studies in quantum chemistry and nano-assembly \cite{Reynolds20}. 
	The subfield of ultracold atoms in optical lattices \cite{LewensteinLattices,Bloch05,Bloch08,Bloch12,Gross17} has blossomed in recent years, and some work towards the realization of an optical lattice for metastable helium is reported in appendix \ref{chap:lattice}.
	
	As the means of control become more sophisticated, interrogation techniques have developed apace. Absorption imaging is the most popular and well known, and led to the famous images of condensates emerging from a thermal gas \cite{Nobel01Note}. Other optical methods like phase contrast imaging \cite{MakingProbingUnderstanding} and sideband imaging \cite{Lye99} have found utility as non-destructive imaging modalities. Atomic fluorescence is also used for accurate determinations of trap populations \cite{VassenReview} and, with the advent of high-numerical-aperture optics in vacuum, has progressed so far as site-resolved imaging in optical lattices. 
	All these readout methods have found use in combination with interrogation techniques like multi-photon techniques, in particular Bragg  \cite{Stenger99} and Raman \cite{Hagley99,Cola04} spectroscopy. Atom lasers \cite{Mewes97,Bloch99} have also been widely deployed in combination with imaging methods.
	
	The applications of such techniques have included studies of basic characteristics of degenerate matter, such as the Bogoliubov transformation and quasiparticle excitation spectrum  \cite{Steinhauer02,Vogels02}, vortex formation in rotating condensates \cite{Madison00}, fluctuations in the condensate population in accordance with the canonical ensemble picture, \cite{Kristensen19}, quantification of the quantum depletion \cite{Xu06,Lopes17_depletion}, and direct measurement of the equation of state \cite{Mordini20}. 
	This thesis also describes contributions to the study of quantum depletion, in chapter \ref{chap:QD}.

	If one thing is obvious, it is that the field of ultracold atoms is impossible to thoroughly review and summarize within the scope of this dissertation. The survey above just provides some bearings by which to orient the following chapters with respect to the ongoing work in the field. Below I present a short overview of work done with metastable helium, the focal element of this thesis, and then lay out the structure of this dissertation.

\subsubsection*{Coming into focus}

	Among the zoo of atomic species cooled to degeneracy, helium has two particular characteristics that distinguish it as a candidate for condensation. The first is the structural simplicity that renders its energy levels and transition rates tractable to highly accurate calculations using quantum-electrodynamic atomic structure theory. While hydrogen has reached degeneracy \cite{Fried98}, the apparatus is even more complicated than helium beamlines and thus there is an advantage in the relative ease of working with helium\footnote{This is certainly not to say that working with helium is easy, as testified in chapter \ref{chap:apparatus} and appendix \ref{chap:lattice}.}.
	The second is the peculiar singly-excited $2\triplet S_1$ state, which has its own distinguished notation - \mhe. 
	This state possesses two superlative properties: On one hand, it has an extraordinarily large energy (for an atomic transition) of 19.8 eV relative to the ground state. 
	On the other hand, this state can only decay to the ground state and this transition has a lifetime of 7870(510) seconds \cite{Hodgman09_mhe}. 
	The latter fact means the former is rendered experimentally relevant, and this conjunction is exploited in cold atom labs to detect individual helium atoms either directly, by measuring small pulses of current on solid detectors after atoms impact their surface, or indirectly by monitoring the production of ions from interatomic collisions that release the stored potential energy, disintegrating one of the colliding atoms. 

	Helium-4 was initially condensed by two labs in France \cite{Robert01,Santos01}, followed by the Netherlands \cite{Tychkov06}, and the ANU \cite{Dall07} where the works in this thesis were undertaken. Since the first realization in Canberra, helium condensates have been produced in the USA \cite{Doret09} and Austria \cite{Keller14}, and labs in France \cite{Bouton15}, the Netherlands \cite{Flores15}, and Canberra \cite{Abbas21} have brought additional \mhe~BEC machines online.
	Fermionic $^3$He$^*$ has been cooled to degeneracy also but less often due to the extra experimental complexity of additional lasers and gas recyclers to recollect the rare and expensive $^3$He gas.
	A more detailed survey of the scientific works conducted with degenerate helium is presented in chapter \ref{chap:theory}.

\section*{Pr\'{e}cis}\label{sec:precis}
\addcontentsline{toc}{section}{Pr\'{e}cis}  


	This thesis documents three experiments that were conducted in the ANU helium BEC laboratory over the period of 2018-2021 and one construction project that I worked on through 2016-2018. The structure of the dissertation is as follows.
	In chapter \ref{chap:theory} I present the relevant theoretical background in order to introduce the core concepts of this thesis. This includes a short survey of atomic structure and atom-light interactions, the defining properties of BEC, and the particular affordances and interest of helium.
	In chapter \ref{chap:apparatus} I describe the experimental apparatus used to perform the major works in chapters \ref{chap:transitions}, \ref{chap:tuneout}, and \ref{chap:QD}, including details about the laser systems, atom lasers, and implementations of the cooling and trapping sequences.
	 
	Chapters \ref{chap:transitions} and \ref{chap:tuneout} concentrate on two of the laser-spectroscopic works conducted by our group over 2018-2021.
	Chapter \ref{chap:transitions} contains an account of the measurement of a handful of lines from the $2\triplet P_2$ level to states in the $n=5$ manifold. 
	The measurements were made by disturbing an early stage of the laser cooling sequence by driving transitions from the excited state which is populated via the cooling transition. 
	The perturbation was transduced into a reduced trap population by the evaporative cooling ramp. 
	Measuring the atom loss resolves a number of absorption lines whose center frequencies we determine with an order of magnitude greater precision than the previous measurement. We also make a first direct spectroscopic observation of the spin-forbidden $2\triplet P_2 - 5\singlet D_2$ transition. The results are published in \cite{Ross20} and laid the groundwork for a subsequent measurement of the strongly forbidden $2\triplet S_1 - 3 \triplet S_1$ transition \cite{Thomas20}.
	Chapter \ref{chap:tuneout} touches the cutting edge of laser spectroscopy in the form of a measurement of a tune-out frequency in \mhe~near 413 nm which is able to discern the predicted contributions of QED effects.

	Chapter \ref{chap:QD} deviates from the theme of spectroscopy and focuses instead on the study of basic BEC physics. The quantum depletion is a feature of any interacting condensate and has been subject to much attention in recent years, particularly in light of its direct connection to a new thermodynamic quantity called the \emph{contact}. Studies of the contact using far-field techniques have so far yielded results inconsistent with theoretical frameworks that are otherwise uncontroversial. In this context I revisit an observation of the depletion in the far-field and discern more carefully what can be inferred about the ultra-dilute tails observed in the experiment. 

	In appendix \ref{chap:lattice} I discuss the contributions I made to the refurbishment of a retired cold-helium beamline and the subsequent upgrade including major extensions to the vacuum system, construction of  an optical dipole trap loaded from an evaporatively-cooled magnetic trap, and installation of a resonant absorption-image acquisition and processing system. In late 2018 I changed the focus of my work to laser spectroscopy of helium (leading to the works \cite{Henson22,Thomas20,Ross20} and chapters \ref{chap:transitions} and \ref{chap:tuneout}), and while we were waiting for the new laser system to arrive I commenced the work that comprises chapter \ref{chap:QD}. Since my departure from the lattice lab, A number of	other graduate students have been in residence and subsequently achieved condensation, as reported in the publication \cite{Abbas21}.
	
	Chapter \ref{chap:conclusion} summarizes the contributions of the works in this thesis. Whereas each chapter contains a discussion of the near-term outlook in terms of building upon these works, chapter \ref{chap:conclusion} concludes with a brief look toward the horizon.


\vfill

\begin{adjustwidth}{6cm}{0cm}
\begin{flushright}
\singlespacing
\emph{
``Who are we? And more than that: \\
I consider this not only one of the tasks,\\
 but \emph{the} task, of science, \\
the only one that really counts.''}\\
- Erwin Schr\"{o}dinger \cite{SchrodingerQuote}
\end{flushright}
\end{adjustwidth}
\onehalfspacing

%% file: latex/11_introduction.tex
\chapter{Theoretical background}
\markboth{\thechapter. THEORETICAL BACKGROUND}{}
\label{chap:theory}
	\begin{adjustwidth}{0cm}{0cm}
	\begin{flushright}
	\singlespacing
	\emph{``No one's mouth is big enough to utter the whole thing."\\} 
	- Alan Watts
	\end{flushright}
	\end{adjustwidth}
	\onehalfspacing
	\vspace{1cm}

	\noindent{A} cold atom experimentalist draws on a panoply of tools (both conceptual and instrumental) which are all intricate and absorbing in their own right.
	This chapter presents the essential ideas needed to give form and context to the content of the major works reported in this thesis.
	We will glance at the hydrogen atom to establish a language in which to elucidate the deceptively simple structure of helium.
	 We must be equipped with some study of the coupling between electromagnetic fields and light, given the ubiquitous use of laser sources and magnetic trapping in this dissertation, and the focus on laser spectroscopy\footnote{The \emph{focus} of a lens or mirror is the point of maximum concentration of light that is refracted or reflected from a distant or uniform source.
	Originally, the term referred to the fireplace at the centre of traditional single-room dwellings; still the brightest point of light, but also the source itself.}.
	Of course, the fun doesn't end when the lights turn off: Even in the dark, helium exhibits explosive two-body interactions, which impose limits on the size and lifetime of atomic condensates.
	Finally, we will review the basic features of the emergence of macroscopic coherence in the form of a Bose-Einstein condensate.
	

\section{Atoms and light}
\label{sec:atoms_and_light}
	The discussion in this section is a short tour of the relevant atomic physics.
	Many high-quality textbooks go into much greater detail than is required for this thesis and the reader is referred to, for example \cite{FootAtomic,BinneyBook} for deeper coverage.
	In this section, only the essential points are presented, omitting lengthy calculations, with references provided for the complete working.
	Our picture of atoms, their structure, and the interaction with electromagnetic fields, will be made in terms of their modern description in the language of quantum mechanics.

	Quantum mechanics is the study of systems whose state at any time $t$ is completely specified by a wavefunction $\ket{\psi(t)}$ and whose dynamics are determined by the time-dependent Schr\"{o}dinger equation,
	\begin{equation}
		i\hbar\frac{\partial}{\partial t}\ket{\psi(t)} = \hat{H}\ket{\psi(t)}.
		\label{eqn:TDSE}
	\end{equation}
	The wavefunction $\ket{\psi}$ is represented by a \emph{state vector} that is an element of the complex Hilbert space $\mathcal{H}$.
	The \emph{Born rule} postulates that the probability of a quantum system being observed in the state $\ket{\psi}$ given that it is known to be in the state $\ket{\phi}$ is given by squared inner product $|\braket{\psi}{\phi}|^2$.
	States are orthogonal if the inner product is zero, but the system may evolve naturally from $\ket{\phi}$ to have a nonzero projection onto $\ket{\psi}$ under the action of $\hat{H}$.
	The Hamiltonian $\hat{H}$ is a linear operator with the Hermitian property $\hat{H}^\dagger=H$, where the dagger denotes the conjugate transpose operation\footnote{Formally, $\mathcal{H}$ is a vector space $\mathbb{C}^D$ of dimension $D$ which is complete with respect to the $L^2$ norm $||x||=\sqrt{\braket{x}{x}}$ induced by the inner product $\braket{x}{y}\rightarrow\mathbb{C}$, and $\hat{H}\in\mathcal{B}$, the Banach space of bounded linear operators $\hat{O}:\mathcal{H}\rightarrow\mathcal{H}$.
	$\mathcal{B}$ is also a vector space with a norm (the trace norm) but not an inner product.
	The states themselves are defined up to scalar multiplication, and hence are actually rays in $\mathcal{H}$ better thought of as points in projective space; we will simply assume they are normalized as $\bra{\psi}\psi\rangle=1$ for brevity.
	We will say no more of the consternations stemming from the Born rule.}.	
	The Hermitian property guarantees $\hat{H}$ is \emph{normal}, $[\hat{H}^\dagger, \hat{H}]=0$ and therefore the time-independent Schr\"{o}dinger equation
	\begin{equation}
		\hat{H}\ket{\psi} = E\ket{\psi}
		\label{eqn:TISE}
	\end{equation}
	specifies the eigenvectors $\ket{e_n}$ of $\hat{H}$ which provide a complete orthonormal basis for $\mathcal{H}$.
	This allows any (pure) quantum state to be written in the form $\ket{\psi} = \sum_n a_n\ket{e_n}$, with complex coefficients $a_n$ called \emph{amplitudes}.
	An interpretation of this fact in the light of the Born rule is that the energy eigenstates $\ket{e_i}$ correspond to distinguishable  and mutually exclusive  states of a system, which can be discriminated if their energy eigenvalues $E_i$ are different.

	The energy eigenbasis for a single charged particle bound in a central potential\footnote{Better known as the hydrogen atom.} have the form,
	\begin{equation}
	\psi_{nlm}(r,\theta,\phi) = 
	\sqrt{\left(\frac{2}{na_0 ^*}\right)^3\frac{(n-l-1)!}{2n(n+l)!}}e^{-\rho/2}\rho^l L_{n-l-1}^{2l+1}(\rho) Y^{m}_{l}(\theta,\phi)
	\end{equation}
	when written in spherical coordinates $(r,\theta,\phi)$, where $\rho = 2r/na_0 ^*$ and $a_0 ^* = \frac{4\pi\epsilon_0 \hbar^2}{\mu e^2}$ is the reduced Bohr radius.
	The radial Laguerre polynomials $L_{n-l-1}^{2l+1}(\rho)$ and the spherical harmonics $Y_{l}^{m}(\theta,\phi)$ are labeled by the angular momentum $l$, the magnetic quantum number $m$, and the energy is fixed by the principal quantum number $n$ as $E_n = -hcR_\infty/n^2$, where $h$ is the Planck constant, $c$ is the speed of light, and $R_\infty$ is the Rydberg constant.
	The bound-state energy $E_n$ is negative - one must do work to ionize the atom and produce the free ion-electron pair whose energy is defined to be zero.
	The quantum numbers $l$ and $m$ distinguish states that are otherwise degnerate in $n$, and the latter serve to lift the degeneracy via the Zeeman shift as I discuss in a later section.
	Here,  $h=2\pi\hbar$ is the Planck constant, $\varepsilon_0$ is the electric permittivity of free space, and $\mu$ is the Bohr magneton.

	Let us consider an atom immersed in an electric field oscillating with frequency $\omega = 2 \pi f$ (with $f$ in Hz), and write the Hamiltonian in the form
	\begin{equation}
		\hat{H}(t) = \hat{H}_0 + \hat{H}_{I}(t)
	\end{equation}
	where the bare atomic Hamiltonian $H_0$ sets the energy scale of the system, and the monochromatic time-dependent perturbation takes the form $\hat{H}_I(t) = \Lambda \cos(\omega t)$.
	We will give physical meaning to $\Lambda$ in a moment.
	The time-dependent state can be written in terms of the eigenbasis $\ket{\psi_n}$ of $\hat{H}_0$,
	\begin{equation}
		\ket{\Psi(t)} = \sum_n c_n(t)e^{-i\omega_n t}\ket{\psi_n},
	\end{equation}
	where $\omega_n= E_n/\hbar$.
	Substitution into Eqn \ref{eqn:TDSE} reduces to the coupled set of differential equations 
	\begin{equation}
		i\hbar\dot{c}_n(t) = \sum_m e^{-i(E_m-E_n)t/\hbar}\bra{\psi_m}\hat{H}_I\ket{\psi_n},
	\end{equation} 
	which can be solved succinctly after making some simplifications.
	Let us first assume (without loss of generality) that the atom is initially in the state $\psi_1$, where $c_1(0)=1$ and $c_i(0)=0~\forall i\neq1$.
	Second, that the light field is oscillating close to resonance with the transition to a state $\psi_2$, and that all other resonances are far enough away that we can approximate the atom with a two-level system with a resonant frequency $\omega_0 =(E_2-E_1)/\hbar$.
	Finally, suppose that the wavelength of the light is much larger than the atom. 
	This is the \emph{dipole approximation}, wherein the electric field has a constant value throughout space, but oscillates in time as $\textbf{E}(t) = {E}_0 \textrm{Re}(e^{-i\omega t}\hat{\mb{u}})$, where $\hat{\mb{u}}$ is the unit polarization vector, and $\textbf{r} = r\hat{\textbf{r}}$.
	The interaction energy is then given by retaining only the dipole operator $-e\textbf{r}$ from the multipole expansion of the electronic charge distribution\footnote{This is a good approximation when the wavelength $\lambda = c/f$ is much larger than the atom. This is universally applicable for our purposes as the size of the helium atom is $\approx 30$ pm, some 0.1\% of the shortest wavelength of light used in this thesis}.
	The interaction Hamiltonian can then be written as
	\begin{equation}
		\hat{H}_I = e\textbf{r}\cdot\textbf{E}(t),
	\end{equation}
	and the excitation probability can be written in the form \cite{FootAtomic,BinneyBook}
	\begin{equation}
		P_2(t) = \Omega^2 \frac{\sin^2(\frac{1}{2}(\omega_0-\omega)t)}{(\omega_0-\omega)^2},
		\label{eqn:transition_prob}
	\end{equation}
	in terms of the Rabi frequency
	\begin{equation}
		\Omega = \frac{\bra{\psi_1}e\textbf{r}\cdot\textbf{E}\ket{\psi_2}}{\hbar},
	\end{equation}
	which is, in general, the frequency of oscillation between pairs of states in response to some external driving function.
	We are now ready to impart a physical meaning to the coupling term: An atom exposed to an oscillating field will respond by oscillating between energy eigenstates, each of which having their own charge distribution through space, and so the dipole moment of the atom also oscillates in time.
	
	The dipole operator $-e\textbf{r}$ only couples to the radial part of the wavefunction, which allows a separation of the expectation value in the preceding equation into radial and angular parts, $\bra{2}\textbf{r}\cdot\hat{\mb{u}}\ket{1} = \bra{2}R\ket{1}\mathcal{I}$.
	Setting aside the radial part $R$, the angular integral $\mathcal{I}$ can be written as a contraction over the spherical harmonic basis functions, 
	\begin{equation}
		\mathcal{I}=\int_{0}^{2\pi}\int_{0}^{\pi} Y^{*}_{l_2,m_2}(\theta,\phi)\hat{r}\cdot\hat{\mb{u}}Y_{l_1,m_1}(\theta,\phi)\sin\theta d\theta d\phi
	\end{equation} 
	which is zero unless some constraints, known as \emph{selection rules}, are satisfied.
	To proceed, we assume that the atom is immersed in a magnetic field and define the $z$ axis to be the direction of the magnetic field vector $\textbf{B}$\footnote{This is true for almost all the contexts we encounter in this thesis, but where a magnetic axis is not present, one should perform an average over all angles as required.}.
	The dipole operator can then be written as the superposition of the linear and circular oscillating field components,  
	\begin{equation}
		\hat{r}\cdot\hat{\mb{u}}\propto A_{\sigma^-}Y_{1,-1} + A_z Y_{1,0} + A_{\sigma^+}Y_{1,+1},
		\label{polz_decomp}
	\end{equation}
	where the $A_{\sigma^\pm}$ are the amplitudes of the clockwise- and anti-clockwise circular polarization, and $A_z$ the amplitude of linear polarization in the atomic reference frame.
	When this expression is inserted into the integral $\mathcal{I}$, the algebra of spherical harmonics ensures \cite{FootAtomic}
	\begin{equation}
		\Omega\propto a\delta_{l_2,l_1+1}\delta_{m_2,m_1+m} + b\delta_{l_2,l_1-1}\delta_{m_2,m_1+m}
	\end{equation}
	where $m=\pm1,0$ as in the expansion of the dipole operator, and $a$, $b$ are constants whose exact values are not required here.
	Thus $\mathcal{I}=0$ unless $\Delta l=\pm1$ and either $\Delta m_l=0$ or $\Delta m_l=\pm1$.
	Transitions not satisfying these conditions are said to be \emph{forbidden}, but in reality they still occur, and can be accounted for by including higher terms in the series expansion of the interaction Hamiltonian.
	Emission or absorption events with $\Delta m_l=0$ are called $\pi$-transitions, and correspond to the dipole moment induced by light with linear polarization along the quantization axis, whose amplitude is captured by the $A_z$ term in Eqn.
	\ref{polz_decomp}.
	The $\sigma$ transitions couple to oscillations in the plane normal to the quantization axis, and correspond to transitions where $\Delta m_l=\pm 1$, driven by the $A_{\sigma^\pm}$ terms.

	This description of the coupling between atomic states and the electric field is sufficient, with some further work, to derive the Einstein rate equations for absorption and stimulated emission \cite{FootAtomic}.
	However, the more familiar phenomenon of spontaneous emission remains out of reach and requires the full-blown theory of quantum electrodynamics for an explanation, so we shall not proceed to derive it here.
	Instead we will have to be satisfied with a classical model which captures some of the essential features so far left unmentioned.

	\subsection*{Classical oscillator model}

	In the \emph{Lorentz oscillator} picture, one approximates an atom by a positive charge fixed at the origin with a harmonically-coupled negative charge with a single degree of freedom $x(t)$ whose potential is zero at $x(t)=0$.
	The electron's equation of motion is $\ddot{x} + \Gamma\dot{x} + \omega_0^2x = -e E/m_e$, where
	\begin{equation}
		\Gamma = \frac{e^2\omega^2}{6\pi \varepsilon_0 m_e c^3}
	\end{equation}
	is the damping rate corresponding to radiative decay.
	The spectrum of an exponentially decaying oscillator has the familiar Lorentzian lineshape,

	\begin{equation}
		\alpha(\omega) = \frac{6\pi\varepsilon_0c^3}{\omega_0^2}\frac{\Gamma}{\omega_0^2-\omega^2-\textrm{i}(\omega^3/\omega_0^2)\Gamma},
		\label{eqn:lorentzian}
	\end{equation}

	where the polarizability $\alpha(\omega)$, whose quantum-mechanical origin is the central interest of chapter \ref{chap:tuneout}, determines the amplitude of the dipole response via $\pvec(t) = \alpha(\omega)E(t)$.
	This response function features the characteristic full-width at half-maximum (FWHM) scale of each transition, which is the inverse of the lifetime $\tau=1/\Gamma$.

	The interaction energy of the dipole and the electric field is $U_{dip} = -\frac{1}{2}\langle\textbf{p}\cdot\textbf{E}\rangle = -\frac{1}{2\varepsilon_0 c}Re(\alpha)I(\rvec)$, which inherits a spatial structure from the intensity $I(\rvec)$.
	When the polarizability is positive (that is, when the light is red-detuned from $\omega_0$ and the denominator has a positive real part) then the dipole oscillates in-phase with the field and so the interaction energy is minimized at the intensity maxima.
	This is the operational principle of optical dipole traps, wherein the \emph{dipole force} $F_{dip} = -\nabla U_{dip} \propto \nabla I(\rvec)$ therefore confines atoms to the focus of a red-detuned Gaussian beam\footnote{The dipole force can also be generated with blue-detuned beams to create repulsive potential barriers, or confine atoms atoms in the dark core of a blue-detuned Bessel beam.}.
	Ashkin made the first demonstration of the dipole force by trapping micron-sized particles in 1970 \cite{Ashkin70}.
	Ashkin suggested 3D trapping of atoms in 1978 \cite{Ashkin78}, and in 1986 Chu \emph{et al.} accomplished the first optical trap of neutral atoms \cite{Chu86}. 
	In 1987 Ashkin demonstrated optical trapping of single cells, viruses, and bacteria \cite{Ashkin87cell, Ashkin87virus}, adding to the list of techniques propagated from physics labs into precision biology.
	The first BEC produced exclusively using optical trapping was achieved in 2001 \cite{Barrett01}.
	Ashkin was awarded the Nobel prize in 2018 for his development of optical trapping.
	The dipole force is most pertinent to chapter \ref{chap:lattice} as it is the basic principle underpinning optical lattice traps, and to chapter \ref{chap:tuneout} wherein the frequency at which Re$(\alpha(\omega))=0$ is measured as a test of quantum electrodynamics.

	In the case where the detuning $\Delta$ is small in comparison to $\omega_0$ (as will be the case throughout this dissertation), the dipole potential can be written in the form \cite{Grimm00}
	\begin{equation}
		U_{dip}(\textbf{r}) = \frac{3\pi c^2}{2\omega_0^3}\frac{\Gamma}{\Delta}I(\textbf{r}),
	\end{equation}
	and absorption of light is captured by the imaginary part of the polarizability \cite{FootAtomic}, which is related to the scattering rate as
	\begin{equation}
		\Gamma_{sc} = \frac{\textrm{Im}(\alpha)I(\rvec)}{\hbar\varepsilon_0 c}.
	\end{equation}
	Light scattering competes with the dipole force because repeated absorption of photons with momentum $\hbar k$ at a rate $\Gamma_{sc}$ gives rise to an effective force of $F_{sc}=\Gamma_{sc}\hbar k$, not to mention the deleterious effects of heating by repeated absorption events.
	Fortunately, in all situations relevant to our concerns here, the scattering rate can be written 
	\begin{equation}
		\Gamma_{sc}(\textbf{r}) = \frac{3\pi c^2}{2\hbar\omega_0^3}\left(\frac{\Gamma}{\Delta}\right)^2 I(\textbf{r}),
	\end{equation}
	from which it can be seen that $\Gamma_{sc}/U_{dip} \propto \Gamma/\Delta$ - that is, for large enough detuning, the scattering rate is dominated by the dipole force.

\section{Helium} 

	The idealized model of the hydrogen atom is fine for illustrating some important features of atomic physics, but in heavier atoms than hydrogen\footnote{Or as astronomers tend to call them, `metals'.}, interactions between electrons also play an important role, which is true even for the humble helium atom.
	Although the structure of helium is simple enough that theoretical calculations can confidently accrue many significant figures, with precision rivalling similar calculations for hydrogen, the presence of a second electron does considerably complicate the physics\footnote{As the old joke goes, atomic physicists count `\emph{one, two, many...}' \cite{FootAtomic}}.
	The helium Hamiltonian 
	$$
	\left(\frac{-\hbar^2}{2m}\nabla_{1}^2+\frac{-\hbar^2}{2m}\nabla_{2}^2 + \frac{e^2}{4\pi\epsilon_0}\left(-\frac{Z}{r_1}-\frac{Z}{r_2}+\frac{1}{r_{12}}\right) \right)\ket{\psi} = E \ket{\psi}
	$$

	\noindent for the two-electron wavefunction $\ket{\psi}$ includes kinetic terms $\propto\nabla_i^2\ket{\psi}$ and a central potential $\propto 1/r$ for each electron, plus a repulsive interaction inversely proportional to the electron sepration $r_{12}$.
	The presence of a second electron also introduces another defining feature of quantum mechanics: spin.
	The electron wavefunctions must be antisymmetric under exchange of particle labels because all fermions obey the Pauli exclusion principle.
	On the other hand, the Hamiltonian is invariant under exchange of the electrons.
	If we denote the exchange operator by $\hat{X}$, then we have $[\hat{H},\hat{X}] = 0$, implying $\hat{X}$ and $\hat{H}$ have the same eigenstates.
	Therefore the energy eigenbasis satisfies $\hat{X}\ket{e} = -\ket{e}$.

\begin{figure}
		\includegraphics[width=\textwidth]{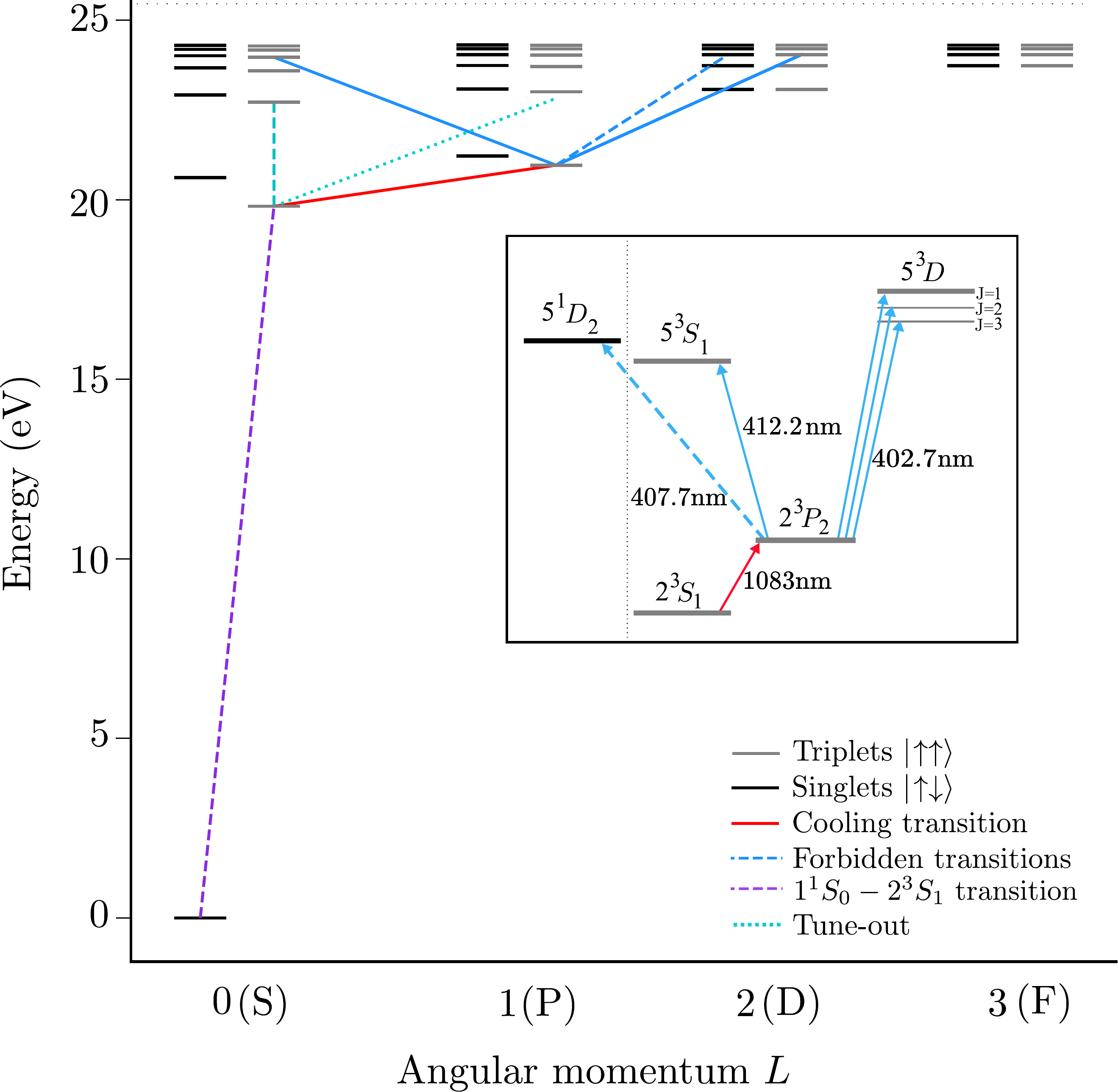}
		\caption{Energy levels of the singly-excited $^4$He atom, grouped by orbital angular momentum quantum number $L$. The $n=1$ manifold contains the unique $1^{1\!}S_0$ ground state - the Pauli exclusion principle precludes the existence of any $1^{3\!}S$ state. All singlet states (with total electron spin $S=0$) are marked with black lines, and the triplet states ($S=1$) in black. Levels up to $n=10$ and $L=3$ are shown along with notable optical transitions. The 1083.331nm cooling and trapping $2^{3\!}S_1\rightarrow2^{3\!}P_2$ transition is shown in solid red. The $63$nm transition between the ground and metastable excited $\MetastableState$ state is indicated in purple, and indicated as \emph{forbidden} by a dashed line. The transitions discussed in Chapter \ref{chap:transitions} and \cite{Ross20} are detailed in the inset (not to scale) , illustrating the fine structure splitting between states with different azimuthal quantum number $J$ which is too small to resolve at the scale of the main diagram. These were measured in the same campaign as the tune-out wavelength (dotted line, Chapter \ref{chap:tuneout} and Ref. \cite{Henson22}) and the forbidden $2^{3\!}S_1\rightarrow3^{3\!}S_1$ transition (reported in \cite{Thomas20}, not discussed in this thesis). The dotted line shows the 24.6 eV ionization threshold \cite{Drake07}.}
		\label{fig:full_level_diagram}
	\end{figure}

	It must be, then, that a given eigenstate must be separable into odd and even parts as either $\ket{\psi} =  \psi^{S}_{space}\psi^{A}_{spin}$ or $\ket{\psi}=\psi^{A}_{space}\psi^{S}_{spin}$.
	We can enumerate the possibilities for the symmetric (spin) wavefunctions, 
	
	\begin{align}
	\psi^S_\text{spin} =& \ket{\uparrow\uparrow},\\
		&\ket{\downarrow\downarrow},\\
		&(\ket{\uparrow\downarrow}+\ket{\downarrow\uparrow})/\sqrt{2}
	\end{align}
	 and the antisymmetric term
	\begin{equation}
	\psi^A_\text{spin} = (\ket{\uparrow\downarrow}-\ket{\downarrow\uparrow})/\sqrt{2}
	\end{equation}

	\noindent which are all degenerate in energy because the atomic Hamiltonian does not couple to the spin part of the wavefunction.
	For a singly-excited helium atom, as will always be the case in this thesis, the interaction term splits the spatial part of the wavefunction for a given set of quantum numbers into symmetric and antisymmetric forms
	$$
	\psi^{S}_{space} = \frac{1}{\sqrt{2}}\left(\ket{1\textrm{s},nl} +\ket{nl,1\textrm{s}}\right),
	$$
	$$
	\psi^{A}_{space} = \frac{1}{\sqrt{2}}\left(\ket{1\textrm{s},nl}  - \ket{nl,1\textrm{s}}\right)
	$$
	in terms of the spatial parts $\ket{n_1,l_1,n_2,l_2}$ of the two-electron wavefunction, with one electron fixed in the 1s state.
	These adjoin the spin wavefunctions to form the $n\triplet L_{J}$ and $n\singlet L_{J}$ states, referred to as (triplet) ortho- and (singlet) para-helium, respectively\footnote{In what follows, we use the spectroscopic convention and label states by the $n^{2S+1}L_J$, where $J=S+L$ because the helium nucleus is spinless, and we drop the $(1s)$ term as the second electron will invariably be in the ground state in all cases we consider. Doubly-excited helium is highly unstable because the first excited state (19.8eV) has about 80\% of the ionization energy of helium (24.6 eV \cite{Drake07}).}.
	It can be shown via degenerate perturbation theory that the exchange antisymmetry produces the \emph{exchange energy} difference between the singlet and triplet states, where the  orthohelium state has a lower energy.
	 The $\metastable$ state, also denoted \mhe, distinguishes helium amongst the zoo of atomic species available to the cold atom physicist by virtue of its 19.8 eV excitation energy and $\approx 7800$ s lifetime \cite{Hodgman09_mhe}.
	 This state is forbidden to decay to the ground state by the $\delta L\neq0$ selection rule and also because the dipole operator does not couple states of different spin, such as the \mhe~state and the ground state.
	Decay to the ground state from the metastable state is  therefore called \emph{doubly forbidden}.
	Forbidden transitions can generally occur by higher-order interactions that are omitted from the dipole approximation, and the \mhe~state in particular can decay via a magnetic dipole transition.
	Nonetheless, the two-hour lifetime is effectively a ground state for the purposes of ultracold helium experiments, which generally last less than a minute.
	Fortuitously, the metastable state is connected to the $2\triplet P_2$ state by a transition with a wavelength of 1083.331nm, which is far more readily accessible with compact laser systems than the $\leq 63$ nm X-ray transitions from the true ground state.
	The $2\triplet P_2$ state also provides an essentially closed transition cycle for laser cooling as its decay is dominated by transitions to the \mhe~state, obviating the need for complicated optical repumping schemes as required for laser cooling of alkali atoms.
	There is also a forbidden transition to the true ground state, but as this decay is some billionfold slower, it is negligible for our purposes (nonetheless, it was measured in the ANU lab \cite{Hodgman09_23P}). The relevant features of the Helium level structure are shown in Fig. \ref{fig:full_level_diagram}.

	The astute reader will notice the absence of presentation of an explicit form for the electron wavefunctions in the helium atom.
	Indeed,	helium is not analytically solvable because its eigenfunctions are not separable into a form $\Psi = \psi_1\otimes\psi_2$.
	A variational approach is required for tractable and accurate calculations, which was developed by Hylleraas \cite{Hylleraas1920,Hylleraas1929,Hylleraas1930}.
	In the intervening century, numerical methods for calculating the energy levels and transition rates in the helium atom have kept pace with precision experiments, and have incorporated the effects of relativity, nuclear recoil, and finite optical wavelength. A more detailed survey of recent progress is given in chapter \ref{chap:transitions}.

	\subsection*{Magnetic fields and the Zeeman effect}

	The inclusion of spin introduces another important feature of atomic spectra, the Zeeman effect, whose discovery heralded a `watershed' moment of modern physics.
	The Zeeman effect refers to the phenomenon of spectral line splitting that occurs when an atom is immersed in a DC magnetic field\footnote{Oscillating magnetic fields have other effects on atoms, for example inducing magnetic dipole (or higher order) transitions involving spin-flips.}.
	The interaction energy of an atom in a magnetic field, 
	
	\begin{equation}
		H = -\text{\boldmath$\mu$}\cdot \textbf{B},
	\end{equation}
	has contributions from both orbital and spin angular momenta ($\textbf{L}$ and $\textbf{S}$, respectively) through the atom's magnetic moment,
	\begin{equation}
		\text{\boldmath$\mu$} = -\mu_B\textbf{L} - g_s\mu_B \textbf{S}.
	\end{equation}
	Working in the $\ket{LSJm_J}$ basis, where $J$ and $m_J$ are the total angular momentum and its z-projection, yields the energies $E_Z = g_J \mu_B B m_J$, where $\mu_B$ is the Bohr magneton..
	The atomic g-factor can be written as
	\begin{equation}
		g_J = \frac{3}{2} + \frac{S(S+1)-L(L+1)}{2J(J+1)},
	\end{equation}
	using the approximate value of the electron g-factor $g_s=2$.
	The eigenstates of the field-free atomic Hamiltonian will be $(2J+1)$-fold degenerate and specified by the $\ket{Lm_L S m_S}$ quantum numbers.
	Adding the magnetic field interaction breaks this degeneracy by introducing off-diagonal terms to the Hamiltonian when expressed in the $LS$ basis,  leading to the Zeeman splitting.
	The field-free and magnetic-interaction terms can be written in a common basis in terms of the Clebsch-Gordan coefficients and then diagonalized, as described in chapter \ref{chap:transitions}.
	The anomalous Zeeman effect arises in triplet states because the inter-level spacing depends on $m$ and $g_J$, which splits spectral lines as well as levels, whereas singlet states have $g_J=1$ and transitions between them do not fan out in the same fashion.

	For all the intricacy of atomic structure, life would be very dull indeed were it not for the interactions between them.
	Indeed, the material reality of the world depends, in a sense, less on the structure of its building blocks and more on how they fit together\footnote{This could be said to underpin the transferability of systems from a natural instantiation to analytic or simulated contexts, because there is truth in the \emph{structure} which is independent of the \emph{substrate} - this theme is elaborated in appendix \ref{chap:lattice}.}.
	The varied and central roles of interactions will be revisited in later chapters, traversing the spectrum from isolated, to weakly-interacting, to distinctly many-body systems.

\section{Interacting atoms}

	Here we briefly review elastic scattering, and then turn to important inelastic scattering processes present in our experiments.
	A detailed primer in atomic scattering physics can be found in the classic texts \cite{PitaevskiiStringari} and \cite{PethickSmith}, with more detail in the latter.
	An exhaustive review of low-temperature scattering studies up to the turn of the millenium can be found in \cite{Weiner99}.
	I focus here on two-body collisions, which are the dominant interactions in the low-density regime of ultracold helium.
	Low densities, which are necessary conditions to minimize the two-body loss processes characteristic to metastable helium, imply that low temperatures are required to achieve high phase space density and reach the degenerate regime.
	Two-body collisions are the crucial enabler for thermalization during the evaporative cooling employed to reach such low temperatures, so long as the relaxation times are shorter than the sample lifetime.

	Neglecting spin-orbit coupling and relativistic effects the two-body scattering problem reduces to the Schr\"{o}dinger equation in the centre-of-momentum frame,
	\begin{equation}
	\left(\frac{\hbar^2}{2m^*}\Delta + V(r) - E\right)\psi(r) = 0,
	\end{equation}
	where $\psi$ is the wavefunction capturing the relative motion of the two atoms	in terms of the separation $r=|\rvec_1-\rvec_2|$ between the particles and the reduced mass $m^*=m_1m_2/(m_1+m_2)$.
	In the asymptotic regime where $r$ is much larger than the scale of the interaction potential $V(r)$, the solution takes the form of a superposition of the initial plane wave and the scattered solution,
	\begin{equation}
	\psi(r) \propto e^{ikz} + f(\theta)\frac{e^{ikr}}{r},
	\end{equation}

	where $k=\sqrt{2m^*E/\hbar}$ is the plane wave-vector of the initial approach and $\theta$ is the angle of scattering from the direction of incidence.
	A general solution can found by expanding $f(\theta)$ into a convenient basis of \emph{partial waves} (spherical harmonics) which are labeled $s,p,d,f,..$ in order of increasing angular momentum.
	In the low-energy limit, $f(\theta)$ is independent of angle and only the spherically symmetric s-wave term contributes, and the limit $f(\theta)\rightarrow-a$ is accordingly called the s-wave scattering length.
	In the millikelvin regime the scattering physics is determined by just a few partial waves \cite{McNamara07}, and in the ultracold (microkelvin or colder) regime only the s-wave scattering channel is significant.
	The total cross section, which is the total probability that a near collision results in particle scattering, approaches $\sigma=8\pi a^2$ for polarized bosons\footnote{For fermion pairs with odd total spin, the cross section tends to zero because of the Pauli exclusion principle, and thus the s-wave scattering length vanishes.} \cite{PitaevskiiStringari,Przybytek05}.

	The s-wave scattering length is also an important determinant of the energetics of degenerate matter such as BEC.
	Because BECs are dominated by long-wavelength behaviour, a theoretical treatment can be considerably simplified by considering only the \emph{effective interactions}.
	By formulating the scattering problem in momentum space, the effective interaction strength for low-energy scattering  $g=4\pi \hbar^2 a/m$, also referred to as the pseudopotential, can be found by integrating out the high-frequency modes (also known as the Born approximation).
	This necessarily washes out extremely short-range correlations but makes fairly accurate calculations much more tractable by reducing the size of the basis set used in a calculation.

	In molecular collisions the scattering process will obviously depend on the relative orientation of the molecules.
	In collisions between single atoms, though, there is a more subtle orientation-dependence which arises from the total spin of the two-particle system.
	The three possible configurations between pairs of \mhe~atoms correspond to the singlet $^1\Sigma_g^+$, triplet $^3\Sigma_u^+$, and quintet $^5\Sigma_g^+$ Born-Oppenheimer molecular potentials\footnote{The subscript \emph{g} and \emph{u} are short for \emph{gerade} and \emph{ungerade} (German for even and odd) label the reflection symmetry of the two-body wavefunction.} with total spin 0, 1, and 2 .
	When the atoms are spin-polarized, as \mhe~atoms are when confined in magnetic traps, then the only scattering that occurs is in the quintet channel.
	In low-energy scattering contexts dominated by s-wave scattering, odd partial waves do not contribute to the interaction potential and hence the triplet $^3\Sigma_u^+$ potential is dominated by the quintet $^5\Sigma_g^+$ term for all interactions with nonzero total spin \cite{Leo01}. 
	As such the inter-species scattering lengths $a_{1,1}$, $a_{-1,-1}$, $a_{0,1}$, and $a_{0,-1}$ are all equal \cite{Leo01,Vassen16}.
	The most accurate determination of the s-wave scattering length in these configurations is 7.512 nm \cite{Moal06}, in agreement with calculations performed the year before the measurements \cite{Przybytek05}.
	When the total spin is zero, the singlet potential contributes and so $a_{1,-1}\approx8.8$ nm and $a_{0,0}\approx3.8$ nm \cite{Leo01,Vassen16}.
	In general this thesis will be concerned with interactions between spin-polarized helium atoms and hence will use the abbreviation $a\equiv a_{1,1}$ unless specified otherwise.

	A powerful tool available in some cold atom experiments are Feshbach resonances.
	A detailed description is found in \cite{Chin10}, but from an operational standpoint they allow control of the scattering length as $a = \tilde{a}(1-\Delta/(B-B_0))$, where B is the strength of an ambient magnetic field, $B_0$ is the resonance value of the field, $\tilde{a}$ is the value when the field is far from a resonance and $\Delta$ sets the resonance width.
	The scattering length can thus be tuned in size and even in sign.
	However, the stability of a BEC with arbitrary population requires a positive s-wave scattering length.
	Attractive interactions leading to instability and collapse of a condensate with number above a critical value $N_\textrm{cr}\sim\mathcal{O}(a_{ho}/|a|)$, where $a_{ho} = \sqrt{\hbar/m\omega}$ is the length scale of the ground state of the harmonic trap housing the atoms \footnote{In some negative-temperature states created in an optical lattice, a BEC can be stable with attractive interactions \cite{Braun13}.} \cite{PitaevskiiStringari}.
	Switching from stable to unstable configurations permits one to examine condensate collapse (as in the spectacular Bosenova experiments \cite{Cornish00}) and also to explore the BEC-BCS crossover \cite{Bourdel04}.
	The spinless nucleus of $^4$He prohibits coupling of bound states within an $m_J$ manifold from crossing an open-channel threshold, precluding this pathway to a Feshbach resonance \cite{Goosen10}.
	Nonetheless, Feshbach resonances induced by spin-spin interactions between helium atoms have been predicted \cite{Venturi99, Goosen10}, but have not observed to date \cite{Borbely12}.
	In a recent work \cite{Hirsch21}, the authors (including members of the ANU He* lab) describe calculations with a newer method predicting, unfortunately, the absence of Feshbach resonances between $^4$He-$^4$He collisions.
	
	Inelastic scattering processes are those which exchange energy between the internal and motional states of either atom.
	They can be represented as a complex scattering potential \cite{Leo01} which permits losses from on-shell scattering channels.
	An important inelastic process characteristic of metastable noble gases is \emph{Penning ionization} \cite{VassenReview}.
	 This can occur through the decay channels
	\begin{equation}
		\textrm{He}^*+\textrm{He}^*\rightarrow 
		\begin{cases}
			\textrm{He}+ \textrm{He}^+ + e^-&\textrm{(PI)}\\
			\textrm{He}_{2}^{+} + e^-&\textrm{(AI)}
		\end{cases}
	\end{equation}
	wherein the first channel is formally called Penning ionization and the second is  Auto-ionization and the rate of the latter is generally very small in comparison to the former \cite{Muller91}.
	Indeed, the energy of the metastable state is sufficient to ionize any neutral atom (except helium or neon) from its ground state, and underpins the single-atom sensitivity of our solid state detector (described in section \ref{sec:DLD}).
	Aside from attracting intensive study in its own right \cite{Partridge10,Stas06,McNamara07}, this explosive potential was a significant hurdle for researchers attempting to achive Bose-Einstein condensation with helium.
	The density achieved in early magneto-optical traps (MOTs) was limited to some hundredfold less than the alkali-metal MOTs of the day \cite{Bardou92,Kumukura92,Mastwijk98}.
	Helium MOT densities were limited by losses through two-body collisions involving atoms in the $\metastable$ and those excited to the $2\triplet P_2$ state by the trapping beams, as opposed to rescattering pressure as in the case of alkali metals.
	Indeed, while the Penning loss rate constant\footnote{The rate constant $\beta_{i,j}$ for losses via collisions between atoms in states $i$ and $j$ yields an absolute loss rate per volume per time through the product with the respective densities, $\Gamma_{i,j} = \beta_{i,j}n_i n_j$ cm$^{-3}$s$^{-1}$.} between pairs of $\metastable$ atoms is of order $2\times10^{-9}$ cm$^3$/s, the loss rate for $\metastable-2\triplet P_2$  collisions is around $10^{-7}$ cm$^3$/s.
	Thus early MOTs had loss rates which were a population-weighted average of these rates, around $7\times10^{-8}$ cm$^3$/s \cite{Weiner99}.
	Such light-assisted losses  limited the population achievable in MOTs until larger beams and detunings were used \cite{Tol99}.
	The Penning ionization rate is some 20-fold lower at large detunings compared to near-resonant light \cite{Mastwijk98}, reducing this loss rate to the order of $5\times10^{-9}$ cm$^3$/s at large detunings.
	Fortunately, the inelastic scattering cross-sections depend on the molecular potentials in such a way that condensation becomes attainable: When all the atoms are polarized in the either of the $m_J=\pm1$ states, the incoming state has a total spin of 2, whereas the reaction products have a total spin of 1, and so this process is forbidden.
	In reality, it does occur through a weak virtual spin-dipole transition \cite{Shlyapnikov94}, but slowly enough that spin-polarized \mhe~exhibits a $10^4$-fold reduction in the Penning ionization rate.
	At field strengths above 50 G, however, the suppression weakens \cite{Shlyapnikov94,Borbely12}.
	Other noble gases also exhibit highly energetic metastable states, but the lifetime and suppression of Penning ionization decreases with increasing mass \cite{Orzel99, Spoden05}.
	Thus helium may be the only noble gas ever to cooled to degeneracy.

\section{Bose-Einstein condensation}
\label{sec:BEC_theory}

	The fifth state of matter\footnote{The familiar first phases, solid, liquid, and gas, are vanishingly rare in cosmological terms.
	The fourth, plasma, is the state of at leats 99\% of the ordinary matter in the universe \cite{Plasmastuff}.
	Helium comprises about 23\%, most of which being primordial baryons formed during the recombination epoch.}  has a long and storied history\cite{Mukundanote}.
	Interest in BEC was amplified back in the middle of the 20th century when Fritz London proposed that Bose-Einstein condensation was connected to the superfluid phenomenon in liquid helium.
	Nikolay Nikolayevich Bogolyubov formalized this connection and so, historically speaking, helium was the element which hosted the earliest experimental realization of Bose-Einstein condensation, albeit with a very small condensed fraction\footnote{Nikolay was a darling of Russian theoretical physics, receiving his PhD-equivalent qualification at 19 and made important contributions to quantum field theory. In his famous paper on the problem of interacting bosons, his name is transliterated as \emph{Bogolubov}. \emph{Bogolyubov} and \emph{Bogoliubov} are also common transliterations.}.
	While liquid helium is a rare thing in cosmological terms, BEC may have existed already for millions of years in the superdense quark matter of neutron stars \cite{Haskell18, Martin16,Baym69,Page11}, wresting the claim of cosmic novelty from human hands. 
	Nonetheless, the essentially pure atomic condensates and the emerging study of molecular condensates in laboratory settings are among the most extreme conditions in the universe, and are not believed to occur naturally elsewhere.
	Following the oft-cited seven decades between the initial theoretical descriptions and the experimental realization of atomic Bose-Einstein condensates (BEC) \cite{Davis95,Bradley95,Anderson95}, the field has become quite industrious at the eve of its centenary.
	As pithily put by a review only five years after the Nobel-winning experiments, `Any attempt to review recent progress is out of date as soon as it is published' \cite{Courteille01}.
	This is no less true today, as the number of ultracold quantum gas experiments worldwide\footnote{See \url{everycoldatom.com}} now number nearly 200 and numerous companies have been founded on the promise of selling better sensors and computers using technologies based on BEC physics.
	There are numerous treatments of the theory of Bose-Einstein condensation, for example the classic textbooks \cite{PitaevskiiStringari,PethickSmith} and review articles \cite{Dalfovo99, Yukalov11_basics,Courteille01}.
	The essential background for discussion here draws on these standard sources unless otherwise cited.

	As for what a BEC \emph{is}, there are several workable operational definitions, but a precise and lab-relevant definition is surprisingly elusive.
	The canonical description of Bose-Einstein condensation is the condition where the de Broglie wavelength associated with thermal kinetic energy
	\begin{equation}
		\lambda_T = \frac{h}{\sqrt{2\pi m k_B T}}
	\end{equation}
	is comparable to the interparticle spacing, coinciding with a macroscopic occupation of the single-particle ground state as the de Broglie waves of many bosons constructively interfere.
	This condition can also be restated as the point when the phase space density
	\begin{equation}
		\aleph = n \lambda_T^3
	\end{equation}
	exceeds the critical value of $\zeta(3)\approx2.612$. 
	This is generally introduced of as the point where the volume over which the atoms are `delocalized'  exceeds the average volume per particle, and so the spin-statistics of the particles dictate the deviation from the statistics of an ideal gas.
	Although it is commonly said that at this point the `wavefunctions overlap', the wavefunctions of the individual particles in a trapped gas overlap wherever the wavefunction is supported - that is, throughout the trap.
	Furthermore, the de Broglie wavelength pertains to the wavelength associated with the plane-wave motion of a free particle rather than the spatial scale of the `wavepacket' of a particle, and as given above is defined as the wavelength corresponding to the \emph{average} particle velocity, rather than fully characterizing the ensemble. 
	Thus even above the critical temperature there are many states with long wavelengths who will be populated, and so the `delocalization' picture is not so sharp.
	A more precise criterion is to compute, in the framework of the grand canonical ensemble, the number of particles in the ground state. For a Bose gas this yields
	\begin{equation}
		N_0 = \frac{1}{\exp((E_0-\mu)/k_B T)-1},
	\end{equation}
	where $E_0$ is the ground state energy of the single-particle Hamiltonian and $\mu$ is the chemical potential of an adjoining reservoir. 
	One can then show that, below the critical temperature $T_c$, the ground-state population diverges, corresponding to Bose-Einstein condensation.
	However, in laboratory realizations of BEC, there is no reservoir attached to the system under study, but rather an isolated gas at some finite entropy which contains both the thermal and condensed fractions (if the latter is present). 
	
	Moreover, in the presence of interactions, this criterion faces another issue: the stationary states of an interacting gas of $N$ atoms cannot be written as a product $\ket{\Psi} = \ket{\psi_i}^{\otimes N}$ of single-particle eigenstates $\ket{\psi_i}$.
	That is, while measurements are confined to observables of single particles, the single-particle states are not eigenstates and so it is not sensible to talk of their `macroscopic occupation'.
	The Penrose-Onsager criterion \cite{Penrose56} provides an alternative in terms of the density matrix $\rho$ for the isolated composite system comprised of all the gas particles.
	The single-particle density matrix is then the expected value of the one-body field operator
	\begin{equation}
		\rho^{(1)} = \Tr\left(\rho\hat{\Psi}^\dagger\hat{\Psi}\right),
	\end{equation}
	whose eigenvalues $p_{l}^{1}$ give the occupation probability of the $l^{\rm th}$ eigenvector of $\rho^{(1)}$.
	The eigenvectors themselves are the single-particle modes (which may be, in general, some superposition of the non-interacting eigenstates).
	If any eigenvalue $p_{l}^{i}$ is proportional to $N$ in the limit $N\rightarrow\infty$, then the system is said to have undergone Bose-Einstein condensation (or, simply, \emph{condensed}) into the $l^\textrm{th}$ mode.
	Of course, real systems are subject to atom losses and heating, violating the assumptions of equilibrium underpinning both the approaches above.
	
	This is all to illustrate the point that the real world is full of intricacies and the `intuitive' pictures of BEC, despite being useful pedagogical tools, can skirt around some important physical features.
	Ultimately though, this is a thesis concerned with experiments, and we shall say little more than remarking on the compelling agreement between abstract and actual condensates irrespective of the preceding issues.
	Indeed, the Penrose-Onsager criterion has been shown to be a good characterization of a non-Hermitian polariton condensate \cite{Manni12} in that off-diagonal long range order (i.e. phase coherence) emerges along with the growh of a single eigenvalue of the one-body density matrix.
	As the saying goes, if it interferes like a condensate \cite{Andrews97} , undergoes number fluctuations like a condensate \cite{Kristensen19},  has HBT correlations like a condensate \cite{Schellekens05,Jeltes07}, Kibble-Zureks like a condensate \cite{Anquez16}, and quacks like a condensate \cite{Duck01}, then it probably \emph{is} a condensate.

	While most atomic condensates, and all of those in this thesis, are trapped in non-uniform potentials, many important features of condensates are easier to state for homogeneous systems.
	One can usually extend calculations to harmonically trapped systems by a local density approximation, wherein one performs a density-weighted average across a condensate, considering each volume element as a homogenous condensate in its own right.
	Thus, for the most part the following discussion will focus on homogeneous systems for simplicity's sake.
	I present some particular results in the case of a harmonically trapped gas at the end of this section.

	\subsection*{Bogoliubov theory}
	The fundamental theoretical object of interest is the Hamiltonian of a bosonic quantum field with two-body interactions $\hat{H} = \hat{K} + \hat{I}$, with kinetic part
	\begin{equation}
		\hat{K} = \int\left(\frac{\hbar^2}{2m}\nabla\hat{\Psi}^\dagger(\textbf{r})\nabla\hat{\Psi}(\textbf{r})\right)d\textbf{r}
		\label{eqn:ham1}
	\end{equation}
	and the interaction term
	\begin{equation}
		\hat{I} = \frac{1}{2}\int\left(\hat{\Psi}^\dagger(\textbf{r}')\hat{\Psi}^\dagger(\textbf{r})V(\textbf{r}'-\textbf{r}) \hat{\Psi}(\textbf{r}')\hat{\Psi}(\textbf{r})\right)d\textbf{r}'d\textbf{r}
		\label{eqn:ham2}
	\end{equation}
	where $\Psi(r)$ are the field operators subject to the bosonic commutation relations
	\begin{align}
		[\Psi(r),\Psi^\dagger(r')] &= \delta(r-r')\\
		 [\Psi^\dagger(r),\Psi^\dagger(r')]&=[\Psi(r),\Psi(r)]=0.
	\end{align}	
	We can then write the field operator in the suggestive form	
	\begin{align}
		\hat{\Psi} &= \psi_0 \hat{a}_0 + \sum_{i\neq0}\psi_i \hat{a}_i
	\end{align}
	in terms of an orthonormal basis of single-particle modes $\psi_i$ and corresponding field operators $\hat{a}_i$.
	In doing so we distinguish $\pvec=0$ as the condensed mode, and say that condensation occurs when $N_0=\langle\hat{a}^\dagger_0\hat{a}_0\rangle =\mathcal{O}(N)$.
	The observation that the condensed mode has a population of order $N$ means that in the thermodynamic limit ($N\rightarrow\infty,~V\rightarrow\infty$), one particle here or there will not really make a measurable difference.
	This argument can be expressed quantitatively as the Bogoliubov approximation wherein the annihilation and creation operators for the condensed mode are replaced with complex numbers as per
	\begin{equation}
		\hat{a}_0 = \sqrt{N_0}e^{i\alpha}, \hat{a}_0^\dagger= \sqrt{N_0}e^{-i\alpha}, 
	\end{equation}
	which permits the condensate wavefunction to take the form
	\begin{align}
		\hat{\Psi} &= \sqrt{N_0}e^{i\alpha} \psi_0 + \delta\hat{\Psi}\\
					&= \Psi_0 + \delta\hat{\Psi}
	\end{align}

	The first term is the condensate wavefunction (the \emph{mean-field} term), and the second corresponds to the population of non-condensed modes thanks to the effect of interactions, which are captured by the quasiparticle picture sketched in the next section.	The emergence of a condensate has many of the hallmarks of a classical phase transition: kinetic effects are necessary to redistribute energy and reach steady-state; a unique critical temperature $T_c$ exists; below $T_c$ an \emph{order parameter} takes on a nonzero value; and condensation is equivalent to the spontaneous breaking of a U(1) gauge symmetry \cite{Yukalov11_symmetry}.
	Above the critical temperature, $|\Psi_0|=0$, and in general it exhibits a discontinuous derivative at the critical temperature.
	Hence, in the Landau-Ginzburg framework, condensation is a second-order phase transition\footnote{In the Ehrenfest picture one is instead concerned with the number of times one must differentiate some state function (e.g.
	specific heat, compressibility, pressure, free energy) before finding a discontinuity at the critical point.
	In this picture, the transition is first-order as one has continuous state functions with discontinuous derivatives.}.
	The other hallmark of Landau-Ginzburg phase transitions is the spontaneous breaking of symmetry as one crosses from the disordered to the ordered phase (as when a solid breaks the translational symmetry of the fluid phase).
	Condensates do exhibit such symmetry breaking: The Hamiltonian has a $U(1)$ gauge symmetry, but the ground state of a condensate spontaneously chooses a fixed but unpredictable phase $\alpha$.
	By interfering two independently prepared condensates, one observes interference fringes \cite{Andrews97}, and indeed, the fringe locations will change with each realization and measurement.
	More directly, one can interfere light leakage from a reservoir-coupled photon condensate against a reference beam, and observe that phase jumps occur in the output when the condensate field drops to zero.
	That is, the re-emergence of the condensate is heralded by the selection of a new, specific, phase, apparently uncorrelated with the phase that existed before it \cite{Schmitt16}.

	Symmetry breaking is a subtle point discussed infrequently in standard textbooks.
	It happens that one can substitute complex numbers for the field operators,  even when $\langle N_0\rangle \rightarrow 0$ and still obtain correct results \cite{Ginibre67}.
	That is, condensation is not necessary for the Bogoliubov approximation to be valid.
	However, it \emph{is} the case that the onset of condensation coincides with the ground state breaking\footnote{This is sometimes called `breaking the gauge symmetry'.
	This is a misnomer, as a gauge symmetry is a property of the theory, not a state, and all consistent theories must be gauge symmetric throughout.
	These subtleties are discussed at length in lucid terms in \cite{Poniatowski19}.} the $U(1)$ symmetry \cite{Suto05}.
	

	\subsection*{Harmonically trapped condensates}

	Turning back toward harmonic gases, we can begin from the full Hamiltonian (Eqns.
	(\ref{eqn:ham1},\ref{eqn:ham2}) and produce an effective Schr\"{o}dinger equation.
	By assuming slow variations in the density, integrating out short-wavelength modes as in the Born approximation, one can derive the Gross-Pitaevskii equation (GPE),
	\begin{equation}
		i\hbar\frac{\partial \Psi_0(r,t)}{\partial t} = \left(-\frac{\hbar^2\nabla^2}{2m} + V(r,t) + g|\Psi_0(r,t)|^2\right)\Psi_0(r,t).
		\label{eqn:GPE}
	\end{equation}
	which is valid for arbitary interactions dominated by the s-wave scattering length.
	 If one further assumes that the condensate density varies on scales larger than the healing length $\xi = \hbar/\sqrt{2mgn}$, one can make the Thomas-Fermi approximation and ignore the kinetic term in the GPE, whereby the condensate density profile can be written as

	\begin{equation}
		n(\textbf{r}) = \frac{\mu-V(\textbf{r})}{g},
	\end{equation}
	where $\mu$ is the chemical potential.
	The chemical potential for the harmonically trapped condensate also fixes the average energy per particle as
	\begin{equation}
		\mu = \frac{7}{5}\frac{E}{N} = \frac{\hbar\bar{\omega}}{2}\left(\frac{15 N a}{a_{ho}}\right)^{2/5} ,
	\end{equation}
	recalling $a_{ho} = \sqrt{\hbar/m\bar{\omega}}$ is the length scale of the ground state of the harmonic oscillator, with $\bar{\omega}$ being the geometric mean of the $x$, $y$, and $z$ trapping frequencies.
	For a Bose gas confined in a harmonic potential of the form
	\begin{equation}
		V = \frac{1}{2} m \omega_x^2 x^2 + \frac{1}{2} m \omega_y^2 y^2 + \frac{1}{2} m \omega_z^2 z^2,
	\end{equation}
	the condition $\mu-V(\textbf{r})=0$ determines the boundary of the condensate, giving the Thomas-Fermi radii $R_i = \sqrt{2\mu/m \omega_i^2}$, where $i\in\{x,y,z\}$.
	This produces the famous inverted-parabola density profile with a peak density of
	\begin{equation}
		n_0 = \frac{1}{8 \pi}\left( (15N_0)^2 \left(\frac{m \bar{\omega}}{\sqrt{a \hbar}}\right)^{6}\right)^{1/5}.
		\label{eqn:n0}
	\end{equation}
	The Thomas-Fermi approximation is valid when $N a/a_{ho}\gg1$. 
	This is when the mean-field energy significantly outweighs the kinetic term.
	The phase space density for condensation is achieved at the (ideal) critical temperature 
	\begin{align}
		T_c^{0} &= \frac{\hbar \bar{\omega}}{k_B}\left(\frac{N}{\zeta(3)}\right)^{1/3}\\
				&\approx0.94\frac{\hbar \bar{\omega} N^{1/3}}{k_B}.
	\end{align}
	where $\bar{\omega}=(\omega_x\omega_y\omega_z)^{1/3}$ is the geometric mean of the trapping frequencies, and the condensed fraction below the critical temperature is 
	\begin{equation}
		\frac{N_0}{N} = 1 - \left(\frac{T}{T_c^{0}}\right)^3.
	\end{equation}
	The rest of the population, occupying excited single-particle states, is called the thermal fraction.
	The population $N_T$ of the thermal fraction saturates as $\frac{N_T}{N} = \left(\frac{k_B T}{\hbar \bar{\omega}}\right)^3$ in a non-interacting Bose gas, wherein any atoms added to the gas fall into the condensate mode, regardless of the density of the gas.
	Real gases generally don't show such a feature, and this bifurcation is corrected by including the effect of interactions on the thermodynamics of the Bose gas.	 
	Interactions between particles reduce the critical temperature in a harmonic trap, which can be understood in terms of repulsive interactions reducing the density at a given temperature and thus requiring a lower temperature to achieve a give $\aleph$.
	The resulting shift in the critical temperature can be supplemented with a correction for the finite population of the condensate and written as
	\begin{equation}
		\frac{\delta T_c}{T_c^{0}} = -1.3 \frac{a}{a_{ho}} N^{1/6} -0.73\frac{ \langle\omega\rangle}{\omega_{ho} N^{1/3}}
	\end{equation}
	where the latter term, the correction for finite atom number, includes the arithmetic mean trapping frequency $\langle\omega\rangle$ and vanishes in the thermodynamic limit $(N\rightarrow\infty,~V\rightarrow\infty,~N/V$ finite).
	These deviations from ideal behaviour have been observed in experiments \cite{Tammuz11,Smith11}.

	As for the practical production of condensates, much quality literature has been written on the theory and techniques of atomic cooling employed to reach degeneracy.
	These days, such techniques are standard across hundreds of laboratories and so space will not be spared for their general consideration here.
	The curious reader is directed to \cite{MakingProbingUnderstanding,Courteille01,MetVdS, TychkovThesis} for detailed discussions.
	Rather, in the next chapter I discuss the specifications of the apparatus I used in the course of this research. 
	
\subsection{Cooling techniques}
	\label{sec:doppler_basics}
	In brief, both BEC machines discussed in this thesis exist to achieve magnetic trapping of metastable helium in ultra-high vacuum (UHV) conditions, and from there apply further operations as required by the science at hand.
	Magnetic trapping is required because the limits of optical cooling are too hot to achieve degeneracy with the densities realizable with optically trapped helium.
	However, optical cooling is a necessary first stage of cooling to reach the cold (millikelvin) regime before evaporatively cooling to the ultracold (microkelvin) regime en route to degeneracy in the nanokelvin regime.
	Doppler cooling is the most-used optical cooling technique in this thesis for two main reasons.
	First, helium's spinless nucleus means it has no hyperfine structure and precludes some other forms of sub-Doppler cooling.
	Second, Doppler cooling is sufficient to reach phase space densities high enough to seed evaporative cooling sequences which achieve condensation.

	Doppler cooling effectively produces a velocity-dependent force by exploiting both radiative pressure and the Doppler effect, which I illustrate here with reference to a two-level atom with resonant frequency $\omega_0$.
	An atom moving with velocity $\vec{v}$ with respect to the source of a laser (approximated as a plane wave with wavevector $\vec{k} = \omega/c$ in the reference frame of the laser emitter) will see a radiation field with a Doppler-shifted frequency $\omega' = (1 - \vec{k}\cdot \vec{v}/c)\omega$.
	If the laser is detuned from the transition one has the resonance condition $\vec{k} \cdot \vec{v}/c = (1 - \omega_0/\omega)$, hence if the detuning $\Delta = \omega-\omega_0$ is negative (i.e. the laser is \emph{red} of the resonance) then the light is Doppler-shifted into resonance in the atomic frame for some $\vec{k} \cdot \vec{v}<0$, i.e. when the atom is moving toward the source of the laser. 
	When the atom absorbs a photon and is excited to the upper state of the transition it acquires the photon recoil momentum $\vec{p} = \hbar \vec{k}$.
	If the laser is red-detuned then this impulse opposes the atomic motion along the $\vec{k}$ direction and slows the atom down.
	The atom eventually decays to the ground state on a timescale $\tau = 1/\Gamma$ where $\Gamma$ is the excited-state linewidth.
	The emission events (at a rate $\Gamma$) are isotropic and so the integrated momentum imparted from the many emission recoils is zero.
	The atom is then excited again at a rate $\Gamma_{sc}$ (the real part of Eqn. \ref{eqn:lorentzian}), which can be written 
	\begin{equation}
		\Gamma_{sc} = \frac{s}{2 }\frac{\Gamma}{1 + s + (2\delta/\Gamma)^2}
	\end{equation}	
	in terms of the (atom-frame) detuning $\delta$ and $s=I/I_\textrm{sat}$, where the saturation intensity $I_\textrm{sat} = \pi h c \Gamma/(3\lambda^3)$ is $0.167$ mW/mm$^2$ for the cooling transition in \mhe~\cite{BaldwinReview}.
	The cycling absorption-emission events culminate in an effective force $\vec{F} \approx \Gamma_\textrm{sc}\hbar\vec{k}$ (when $\Gamma_\textrm{sc}>\Gamma$) called the \emph{radiative force}. 
	The scattering rate $\Gamma_\textrm{sc}$ depends on the detuning, hence the radiative force inherits a velocity-dependence via the Doppler effects and so is occasionally called \emph{optical molasses} because it is a friction-like force that reduces the atomic kinetic energy.

	The fundamental temperature limit for laser cooling is the \emph{recoil limit} set by the relation $k_B T = \hbar^2k^2/2m$, where $k$ is the wavevector of the cooling light.
	This is the energy scale fixed by a single-photon emission, and is about 2 $\mu$K for Helium.
	While this is too hot to achieve condensation in the traps considered in this thesis (whose critical temperature is about 1 $\mu$K), we generally only use laser cooling down to the \emph{Doppler limit}.
	The latter is given by $T = \hbar\Gamma/2 k_B\approx 38~\mu$K and corresponds to the minimum velocity difference which can be distinguished by the cooling laser.
	In practise this is set by the linewidth $\Gamma$ of the cooling transition (1.6MHz for the $2\triplet S_1 - 2\triplet P_2$ transition), as the laser linewidth is smaller by about an order of magnitude \cite{Shin16}.
	These limits of optical cooling mean that laser cooling of helium to the ground state of our traps is not possible.
	Furthermore, helium has a spinless nucleus and thus no hyperfine structure, precluding the polarization-gradient cooling employed in alkali atoms.
	Fortunately, condensation is achievable via evaporative cooling in magnetic traps (and in an optical dipole trap, as discussed in appendix \ref{chap:lattice}).

	The machinery used to cut the pathway from room temperature to degeneracy is described in the next chapter. 

%% file: latex/12_apparatus.tex
\chapter{Experimental infrastructure}
\markboth{\thechapter. EXPERIMENTAL INFRASTRUCTURE}{}
\label{chap:apparatus}
	\begin{adjustwidth}{0cm}{0cm}
	\begin{flushright}
	\singlespacing
	\emph{``Every sensor is a temperature sensor.\\
	 Some sensors measure other things too."\\} 
	- Engineer's aphorism\footnote{Quoted without attribution in Elecia White's \emph{Making Embedded Systems}, O'Reilly media (2011).}
	\end{flushright}
	\end{adjustwidth}
	\onehalfspacing
	\vspace{1cm}
	{While} nature abhors a vacuum, experimentalists abhor a background.
	Ultracold atom experiments require ultra-high vacuum (UHV) conditions because collisions with a background gas can easily overcome the frail forces that hold the atoms in their traps.
	The major works discussed in chapters \ref{chap:transitions},  \ref{chap:tuneout} and \ref{chap:QD} were undertaken on a machine which has been in productive operation for many years\footnote{This machine is described in great detail in past publications \cite{Swansson04,Dall07_BEC} and PhD dissertations \cite{HodgmanThesis,ManningThesis,ShinThesis,DallThesis}.
	This chapter presents a summary description. The motivated reader is referred to the aforementioned sources for more information, and to the texts \cite{FootAtomic,MetVdS} for general information about cooling and trapping techniques.}.
	This machine achieved BEC in 2007 \cite{Dall07_BEC}, utilizing a bright cold beam of helium \cite{Swansson04} and optimized Zeeman slower \cite{Dedman04}.
	Key technical facilitations were the installation of auxiliary field coils for active cancellation of stray magnetic fields \cite{Dedman04}, and ambient air temperature control to the centikelvin level at the BEC chamber by control of the cooling air mixture \cite{Dedman15}.
	The machine has had a storied history since then, hosting a range of experimental themes including testing quantum-electrodynamic predictions of electronic transition rates \cite{Dall08}, momentum correlations between many atoms \cite{Hodgman17,Dall13,Manning13} (especially Hanbury Brown-Twiss correlations \cite{Manning10,Dall11a,Hodgman11,Rugway11,Rugway13}), studies of atom lasers \cite{Dall07_laser,Dall08a,Henson18_BCR,Manning10} and guided matter waves \cite{Dall10, Dall11a,Dall11}, and quantum-optics demonstrations such as single-particle sources \cite{Manning14}, ghost-imaging \cite{Khakimov16,Hodgman19}, and foundational topics such as Wheeler's delayed choice thought experiment \cite{Manning15}, and Bell-type correlations between pairs of freely propagating atoms \cite{Shin19}.
	This machine was also used for proof-of-principle demonstrations of a machine-learning assisted optimization of a quantum transport protocol \cite{Henson18_ML} and 3D magnetic gradiometry with pairs of atoms \cite{Shin20}.
	While I was a collaborator on the latter two works, they are not included in this thesis.
	My main work on this machine include laser spectroscopic measurements of several transition energies \cite{Ross20,Thomas20} (the former constituting chapter \ref{chap:transitions}); a measurement of the 413 nm tune-out wavelength (as first reported in \cite{Henson15} and improved during the course of this thesis, discussed in Chapter \ref{chap:tuneout}); and measurement of the quantum depletion of a condensate as described in chapter \ref{chap:QD}.

	About half the duration of this course of study was devoted to the refurbishment of a retired cold-atom experiment towards loading a \mhe~BEC into an optical lattice.
	The two machines can be distinguished by their intended final trapping method: The first-described machine can therefore be called the \emph{BiQUIC machine}, and the upgraded apparatus referred to as the \emph{lattice machine}.
	Before its erstwhile retirement, the lattice machine was used for the study of electron scattering and  dual-species MOT experiments \cite{Uhlmann05,Byron10,Byron10a} and determination of electronic transition rates in \mhe~\cite{Hodgman09_23P} .
	The renovation of the lattice machine was motivated by the insufficient optical access into the BEC chamber of the BiQUIC machine, preventing a straightforward upgrade to incorporate several additional trapping beams.
	BEC was recently achieved in the lattice machine by the resident graduate students \cite{Abbas21}, and the construction is ongoing.
	As the lattice machine has yet to produce scientific results in its newfound youth, I will discuss the scientific motivation for its construction and my contributions to the project in appendix \ref{chap:lattice}.
	The present chapter is concerned with the anatomy of the BiQUIC machine, which hosted the works described in chapters \ref{chap:transitions}, \ref{chap:tuneout}, and \ref{chap:QD}.

\section{Helium beamline}

\subsection{Vacuum system}
	Our vacuum is maintained by continuous operation of several turbomolecular pumps\footnote{One of the earliest accomplishments of significant vacuum was by Lavoisier, who used pumps from the fire station to evacuate a chamber.
	Since then vacuum has advanced considerably, and at ultrahigh vacuum the exhaust is so dilute that one does not have the viscosity to operate pumps in the true sense.
	Nonetheless, the name remains.}, backed by roughing pumps that vent directly to the atmosphere.
	For the UHV chambers where BEC is created, an additional turbomolecular pump operates between the vacuum-adjacent turbo and the roughing pump.
	
\begin{figure}
		\centering
		\includegraphics[width=\textwidth]{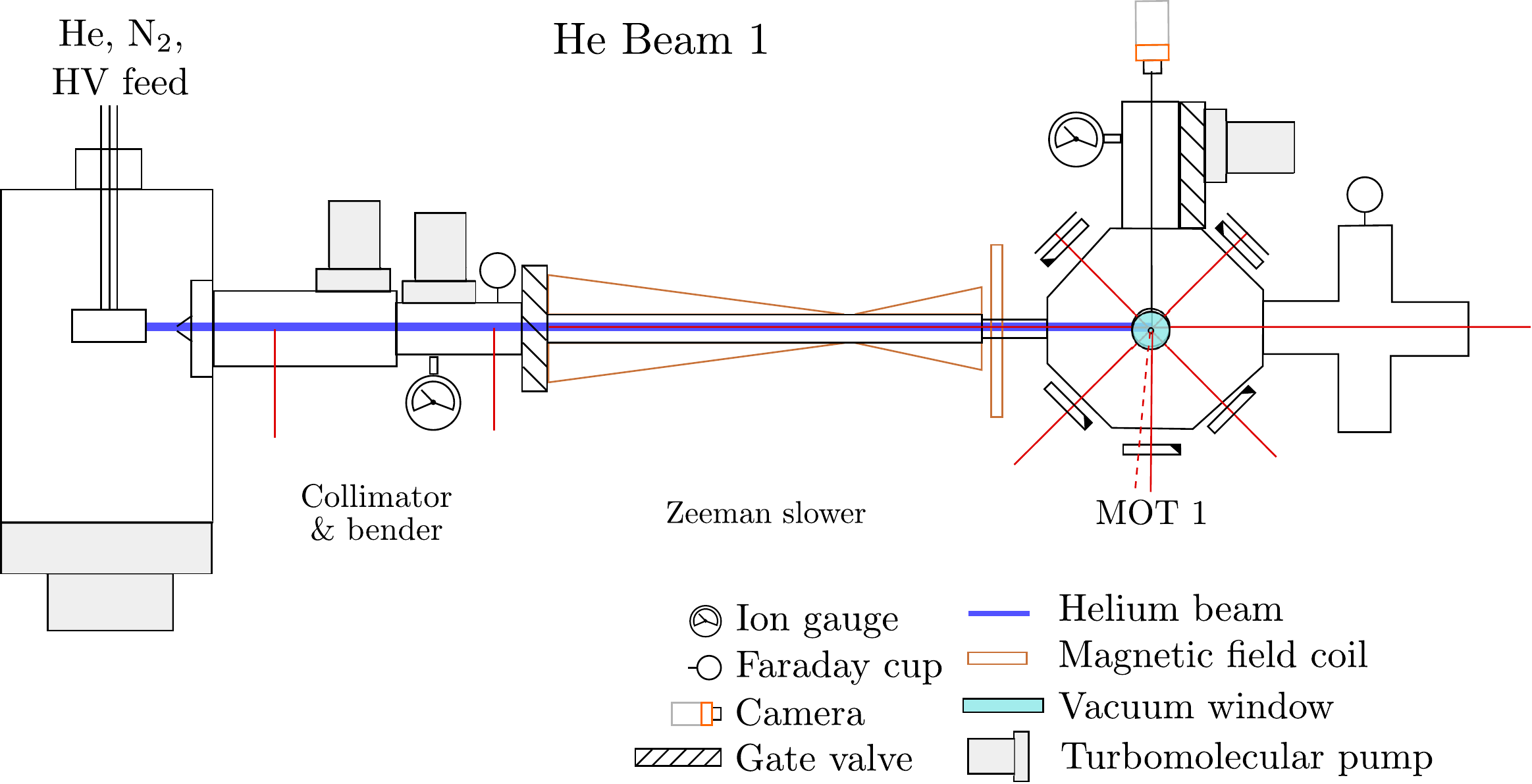}
		\caption{Vacuum system and laser insertion optics for initial cooling of the helium from room temperature to the mK scale in the first MOT. The helium source is cooled by liquid $\textrm{N}_2$ and sustained by a high-voltage (HV) DC discharge.
		Cooling and trapping beams (solid red lines) are transmitted via free-space links from the AOM table and inserted through flange-mounted windows.
		The collimation and bending stage provide intial cooling and background reduction before feeding into the Zeeman slower, which reduces the longitudinal velocity of the atoms so they can be captured by the first MOT at the end of beam 1. 
		The horizontal beam in the first MOT is retro-reflected though a quarter-wave plate mounted on a mirror with a 2 mm hole bored through the centre, which feeds cold helium from MOT 1 into science chamber (see Fig. \ref{fig:apparatus_2})}
		\label{fig:apparatus}
	\end{figure}

\begin{figure}
		\centering
		\includegraphics[width=\textwidth]{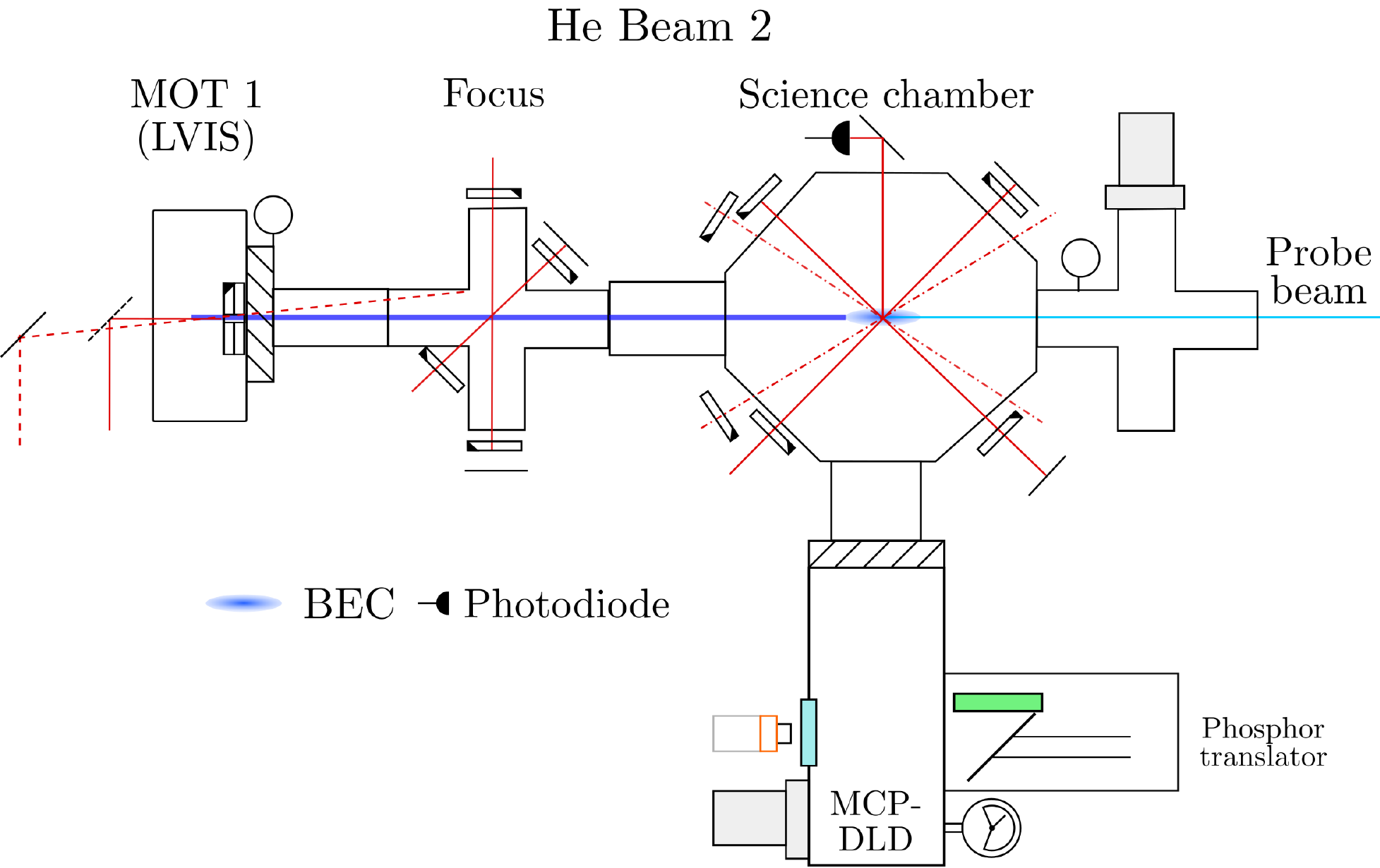}
		\caption{Schematic of the second beamline. The low-velocity intense source (LVIS) of \mhe~ atoms is formed by kicking atoms from the first MOT into the ultra-high vacuum conditions of the science chamber with a slightly blue-detuned `push' beam. Doppler cooling beams (red dot-dash lines) lower the temperature before commencing evaporative cooling. Two other features are present in this diagram, and abesnt in the lattice machine. The spectroscopic laser (blue line) is inserted through a vacuum window along the weak ($x$) axis of the trap. A phosphor-screen-backed MCP (green) is mounted above a 45$^\circ$ mirror permitting imaging of the phosphor pattern with a camera. This is typically used for troubleshooting and alignment rather than scientific purposes, and so is moved out of the fall-line with an in-vacuum translation mount. }
		\label{fig:apparatus_2}
	\end{figure}

	The vacuum systems form the skeleton of both machines, which have a common structure illustrated schematically in Figs. \ref{fig:apparatus} \& \ref{fig:apparatus_2}.

	The placement of some of the turbomolecular pumps, Faraday cups, and pressure gauges differ between the machines, but this makes no functional difference.
	However, the beamline of the lattice machine does not feature a deflection stage or a focusing stage for the \emph{low-velocity intense source} of helium (LVIS).
	The BiQUIC machine has more limited optical access and has featured the addition of a tunable titanium-sapphire laser (section \ref{sec:spec_laser}).
	A major difference between the two setups is the configuration of the chamber that houses the final trapping stage where BEC is achieved (which is still under development in the lattice machine).
	Otherwise, the overall structure of the vacuum systems is the same and so only the BiQUIC machine is shown in its entirety in in Figs. \ref{fig:apparatus} \& \ref{fig:apparatus_2}.
	For comparison, the apparatus around the lattice machine's science chamber is shown and described in appendix \ref{chap:lattice}.

\subsection*{\mhe~atom source}

	The metastable \mhe~state is generally produced by a DC electric discharge as opposed to optical excitation due to the lack of convenient X-ray light sources at 63 nm.
	Electrons accelerated by a strong potential gradient can ionize helium atoms, and thus also have enough energy to excite atoms into the \mhe~state (or achieve it by recombination and/or relaxation along a decay pathway including the $\MetastableState$ state).
	In our experiments, a grounded copper block cooled by liquid nitrogen serves as the anode and reaction vessel.
	A tungsten cathode is held at 2 kV and placed in front of the vessel, separated by electrically insulating boron nitride spacers.
	The breakdown voltage of helium gas in this geometry is lowest at a pressure above the standard operating setting, and even then an additional 2.5 kV pulse is required to achieve ignition before turning the source pressure down into the operating regime.
	Unfortunately, the DC discharge only excites about .01\% of the atoms into the \mhe~state \cite{Stas06}.
	Further, the light mass of helium means that even at temperatures around 70 K (thanks to the liquid nitrogen cooling) the atoms have fairly high velocities necessitating a Zeeman slower stage before the initial trapping stage. 
	The source chamber tends to operate consistently over a year or so, permitting development and execution of a handful of experimental campaigns.
	However, the copper housing degrades with use.
	The first symptom of an ailing source is intermittent extinctions and, if not promptly addressed, an inability to successfully strike.
	This is remedied by extracting the source chamber from the vacuum system and skimming a few microns of material from the interior surface of the block, a procedure which is  straightforward but painstaking.

\subsection*{Beam forming}

	The gas expands out of the forward-facing aperture of the reaction vessel and a small solid angle is selected by a skimming nozzle between the source chamber and the rest of the helium beamline.
	The skimmer also functions as a differential pumping stage by rejecting atoms with large transverse velocities, and keeping the first laser cooling stages at a lower pressure than the source chamber.
	The \mhe beam is collimated by a single laser beam reflected in a figure-4 configuration, forming a a 2D optical molasses \cite{Lett81,Rooijakkers96} and providing cooling in the two transverse degrees of freedom.
	Before insertion into the chamber, a cylindrical lens expands the laser beam into an ellipse with a roughly 5:1 aspect ratio. The long axis is parallel to the helium beamline. 
	This extends the interaction zone to about 10 cm. 
	The beams are offset in the image plane here for clarity, but in reality they are offset along the axis of the helium beam.
	In the schematic, the beam passes through the four-way intersection of the laser beams. 
	The laser detuning (red of resonance) ensures that only atoms moving against the laser propagation are Doppler-shifted towards resonance, hence reducing the beam divergence.

	This is followed by a `bending' stage wherein a single laser beam, picked off from the collimator beam, deflects the metastable atoms by a few degrees through a second stage of differential pumping, which blocks the line of sight from the source to the MOT, into the Zeeman slower chamber.
	This provides better selectivity of \mhe~atoms by reducing the number of unexcited atoms that enter the Zeeman slower and first MOT chamber.

	Zeeman slowers, so named for the exploitation of the Zeeman effect\footnote{In the earier days of cold atom science they were sometimes called \emph{Zeeman-compensated slowers} \cite{Mastwijk98}}, are necessary to slow atoms from thermal characteristic of their sources to below the capture velocity of the magneto-optical traps located at the end of the slower.
	An atom moving with respect to a laser source (as described in section \ref{sec:doppler_basics}) in a magnetic field $\textbf{B} = B\hat{b}$ will have its electronic resonances shifted by the Doppler and Zeeman effects, which can be summarized as
	\begin{equation}
		\omega_{0,\text{lab}} = \omega_{0}\left(1 + \frac{k\cdot v}{c}\right) + \frac{\mu_B B}{\hbar}\left(g_e m_e - g_g m_g\right),
	\end{equation}
	where the $g$-factors and magnetic quantum numbers $m$ of the ground and excited state are denoted by $g$ and $e$, respectively. 
	Hence, a carefully architected magnetic field can compensate for the changing Doppler shift as atoms decelerate.
	The $\sigma^+$ light employed in our Zeeman slower drives the cooling transition with $g_g m_g = 2\times 1$ and $g_e m_e = \frac{3}{2}\times2$, leading to a field-dependent shift of $\approx1.4$ MHz/G.
	
	The magnetic field profile for maximum deceleration depends on the initial velocity, and so the spread of initial velocities according to the Maxwell-Boltzmann distribution necessitates another figure of merit to optimize against.
	In our experiments the Zeeman slower coil windings are built to create a field profile that maximizes the number of atoms slowed below the capture velocity of the first MOT \cite{Dedman04}.
	The Zeeman slower thus reduces the most probable longitudinal velocity of the beam from $\approx700$ m/s to $\approx70$ m/s, which permits loading of the MOT to a steady-state population in about one second.


\subsection*{First Magneto-Optical Trap}
	The magneto-optical trap combines the radiation pressure with a spatially-varying magnetic field to confine atoms in three dimensions.
	Current-carrying coils in an anti-Helmholtz configuration create a locally linear magnetic strength (about the zero in the centre) with a gradient of the form $(B',B',-2B')$ where the stronger gradient is along the axis of symmetry. 
	Because the Zeeman shift of the levels in the helium $2\triplet S_1$ and $2\triplet P_2$ manifold are linear functions of the magnetic field strength (at least in regimes relevant to our traps, where the fine structure splitting of the $2\triplet P$ states is a thousandfold larger than the Zeeman shift), the cooling transition exhibits a Zeeman shift which is proportional to the distance from the centre of the trap.
	We drive the $\sigma^+$ transition between the $2\triplet S_1(m_J=1)$ and $2\triplet P_2(m_J=2)$ states, and red-detune the laser to ensure the light is resonant on an ellipsoidal shell around the trap centre.
	The atoms also exhibit a Doppler broadening from their motion within the trap, but this amounts to a broadening of $\approx100$ kHz at $\approx1$ mK, which is tiny in comparison with the $\approx 100$ MHz ($\approx 70\Gamma$) shift in the $1.6$ MHz-wide transition, a small correction to the picure just outlined\footnote{The Maxwellian distribution of velocities in a thermal cloud leads to a broadening of the absorption line due to the range of Doppler shifts experienced by the atoms. The Doppler-broadened linewidth can be calculated through $\Delta_{D} = \sqrt{8 k_B T\log(2)/(mc^2)}\omega_0$, where $\Delta$ is the broadened linewidth and $\omega_0$ is the resonant frequency \cite{FootAtomic}.}.
	The net result is that as an atom moves away from the origin of the trap, the probability of it absorbing a photon from the beam pointing back towards the origin increases, thus creating a restoring force along each axis.

	This technique was first proposed by Jean Dalibard and eventually realised by Raab \emph{et al} \cite{Raab87}.
	The population of early helium MOTs were limited because the small size of the trap and low detunings led to high densities and thus rapid deterioration by Penning losses.
	Later attempts achieved larger atom numbers by operating with larger beams and detunings (-22$\Gamma$), obtaining lower densities than in other species and mitigating losses due to Penning ionization \cite{Tol99}.
	The multifaceted physics of MOTs has proven rich ground for study in general \cite{Townsend95,Walker90}, and also in the specific case of helium, for which the two-body loss rate has been studied \cite{Tol99} along with the roles of light-assisted collisions \cite{Stas06,McNamara07}. 
	Temperatures in a MOT are, at best, on the order of the steady-state molasses temperature $\approx1$ mK \cite{Lett81}.
	Further evaporative cooling is therefore required to reach quantum degeneracy.
	
	The background pressure in the first MOT chamber is still too high to maintain a long-lived magnetic trap.
	Therefore, the first MOT is used as a source for a secondary helium beam which feeds into the science chamber.
	The MOT chamber features a $\approx$2 mm hole bored through a compound quarter-waveplate-and-mirror mounted inside the vacuum chamber (see Figs. \ref{fig:apparatus} \& \ref{fig:apparatus_2}), which adjoins the transfer conduit supplying the BEC chamber with cold helium.

	A slightly blue-detuned `push' beam is used to kick atoms through the aperture at the back of the first MOT into the second MOT along ballistic trajectories \cite{Swansson04}.
	In practise the beam points slightly upward, and the orientation is optimized by maximizing the saturated fluorescence signal in the second MOT (see section \ref{sec:he_detection}for a description of the technique).
	The atoms pass through a `focus' stage consisting of two crossed, retro-reflected laser beams with a $\approx$1 cm waist.
	Wire coils mounted around the (horizontal) 2$\frac{3}{4}$" window flanges create a quadrupole magnetic field surrounding the beamline such that the circularly-polarized beams address only atoms on the half of the beam proximal to the optical insertion.
	This final stage of transverse cooling increases the efficiency of loading the second MOT from the \mhe~beam.
	The focusing stage typically increases the saturated fluorescence signal in the second MOT by at least 50\%.

\subsection*{The Science Chamber}

	The second trapping chamber is evacuated to a pressure of some $10^{-11}$ mbar, yielding an excellent magnetic trap lifetime (permitting optical interrogation periods of up to 25 seconds after the $\approx$20 second BEC production sequence in one recent measurement \cite{Thomas20}).
	The chamber is surrounded by three pairs of square magnetic coils which actuate active cancellation of stray fields from the earth and nearby electronic equipment\footnote{The `nuller' reduces, but does not eliminate, 50Hz noise from the AC mains supply to the laboratory.} \cite{Dedman07}.
	This assists the creation of a stable magnetic trap with two coil pairs in a Bi-planar quadrupole Ioffe configuration (BiQUIC) \cite{Dall07}.
	These coils also generate the field which maintains the second MOT, wherein the final stage of optical cooling is provided by a pair of low-intensity Doppler cooling beams inserted at 15$^\circ$ relative to the horizontal \cite{Dall07_BEC}.
	The transfer between the two traps is achieved by simply switching off the light.

	Magnetic traps exploit the interaction between the atomic magnetic dipole and a magnetic field, taking the form $E_B = -\mu\cdot \textbf{B} = -\hbar g m_J B$.
	A magnetic field with a local extremum can therefore form a confining potential for the states whose energy is minimized at this point.
	It follows from the Earnshaw's theorem that a magnetic field in free space cannot have any local maxima \cite{Harms00,MakingProbingUnderstanding}, and so one must select an atomic state with $g m_J>0$ for confinement at a local minimum of the magnetic field strength.
	In \mhe~this uniquely selects the $m_J=+1$ state for magnetic trapping, which also ensures complete spin-polarization and the attendant reduction in Penning losses.
	The BiQUIC design is a solution to a critical issue with early traps \cite{Migdall85}: The quadrupole fields generated by coils in an anti-Helmholtz configuration feature a zero crossing in the trap centre.
	Atoms in a magnetic field undergo transitions to untrapped states, known as \emph{Majorana transitions}, at a rate proportional to $\exp(-B/\omega)$, where $\omega$ is the trapping frequency and $B$ is the magnitude of a (uniform) magnetic field \cite{Sukumar97}.
	In harmonic traps the Zeeman splitting at the trap minimum thus sets the upper bound on the Majorana transition rate \cite{Brink06}.
	Many variations on the quadrupole trap exist to mitigate this issue, including the Time-Orbiting Potential (TOP) \cite{Petrich95}, Ioffe-Pritchard \cite{Pritchard83}, cloverleaf \cite{Mewes96}, and QUIC \cite{Esslinger98} traps, while other groups have added a repulsive dipole potential to the trap zero with a focused and blue-detuned laser beam \cite{Davis95}.
	The magnetic trap enables forced evaporative cooling to below the condensation threshold, some three orders of magnitude colder than the limits of optical cooling.

	\begin{figure}
	\begin{minipage}{0.55\textwidth} 
	\vspace{0pt}
		\includegraphics[width=\textwidth]{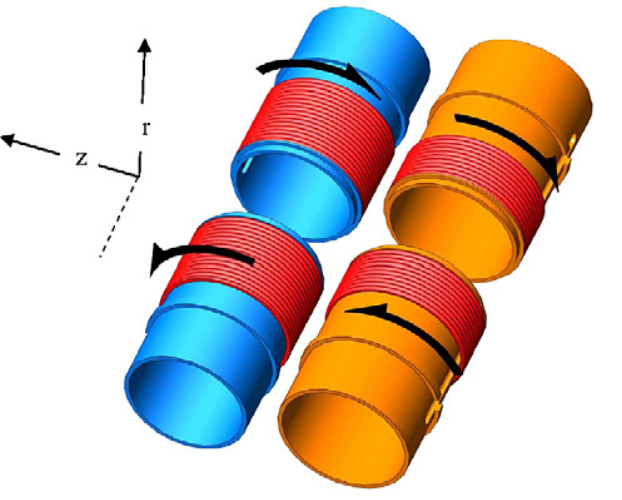}

	\end{minipage}
	\hfill
	\begin{minipage}{0.45\textwidth}
	\vspace{0pt}
	\caption{Schematic diagram of the Bi-QUIC coil design. Black arrows indicate the direction of current flow. The cylindrical coordinate axes indicate the axis of symmetry ($z$), which has a lower trapping frequency than the 'tight' axes. In this thesis, the experiments typically use traps with a `weak' axis with frequency $\omega_z\approx55$ Hz and `tight' axis with frequency $\omega_\perp\approx 430$ Hz. Diagram reproduced with permission from \cite{Dall07}}
	\end{minipage}
	\end{figure}

	A comprehensive introduction to evaporative cooling can be found in \cite{Ketterle96}; see \cite{FootAtomic,PethickSmith} for more compact discussions. 
	Here I present a short, intuitive picture of evaporative cooling in a semiclassical, non-interacting model. 
	A gas of bosons at finite temperature will consist of atoms whose energies are described by Bose-Einstein statistics (see chapter \ref{chap:theory}). 
	At equilibrium, the expected population of single-particle states falls off exponentially with energy $E$. 
	Therefore there will be some atoms which have much greater energy than the average. 
	If these atoms are confined in a magnetic trap, then the more energetic atoms will be more likely to be found further from the trap origin.
	The wavefunction of the high-energy atoms will thus be supported in regions with larger magnetic fields than the lower-energy atoms.
	Therefore it is possible to selectively remove the high-energy atoms by driving the transition between trapped and untrapped states with radio-frequency (RF) radiation tuned to the resonance near the Zeeman splitting seen by only the most energetic atoms in the trap.
	After a timescale on the order of a few times the (inverse) inter-atomic collision rate, the gas rethermalizes (i.e. regains the Bose-Einstein statistics after having its tail cut off) at a lower temperature by virtue of losing its most energetic constituents.
	Repeating this process for lower RF frequencies removes atoms with progressively lower energies, cooling the sample.
	If one ramps the radio frequency down too quickly, then the loss rate of atoms can counteract the loss of temperature and result in a decrease in phase-space density, i.e. moving further from the condensation threshold.
	Conversely, a ramp that is slow will be more efficient but also more subject to other trap loss channels and needlessly prolong the experimental cycle.
	Theoretically optimal evaporative cooling ramps are discussed in \cite{Ketterle96}, and in practise we apply a continuous RF sweep from 20 MHz down to $\approx800$ kHz, with an empirically-optimized and roughly exponential shape over about 20 seconds.
	The endpoint of the ramp depends on the configuration of the BiQUIC trap (specifically the trapping frequency and the bias at the trap centre) and the desired size and temperature of the condensate.

	Two radio antennas provide the RF signals for evaporative cooling and atom lasers, respectively.
	The evaporative cooling ramp is generated by an arbitrary waveform generator pre-loaded with a time-varying frequency ramp created with the LabView control software (see section \ref{sec:DAQ}).
	The secondary coil is driven by a function generator according to a waveform that is pre-loaded at the start of the experimental sequence.
	Both signals are passed through separate dedicated 30W amplifiers before insertion into the chamber.
	A vacuum window on the opposite side to the helium beam provides optical access for a spectroscopic probe beam or the 1550nm beams used for atom-optics techniques (employed in works not relevant to this thesis).
	The final relevant component is a diagnostic photodiode mounted outside the chamber and is used to measure a saturated fluorescence signal (see section \ref{sec:he_detection}), which serves as a diagnostic for the in-trap population.

\section{Light sources}\label{ssec:lasers}
\subsection*{Cooling and trapping light}
	
	\begin{figure}
		\centering
		\includegraphics[width=\textwidth]{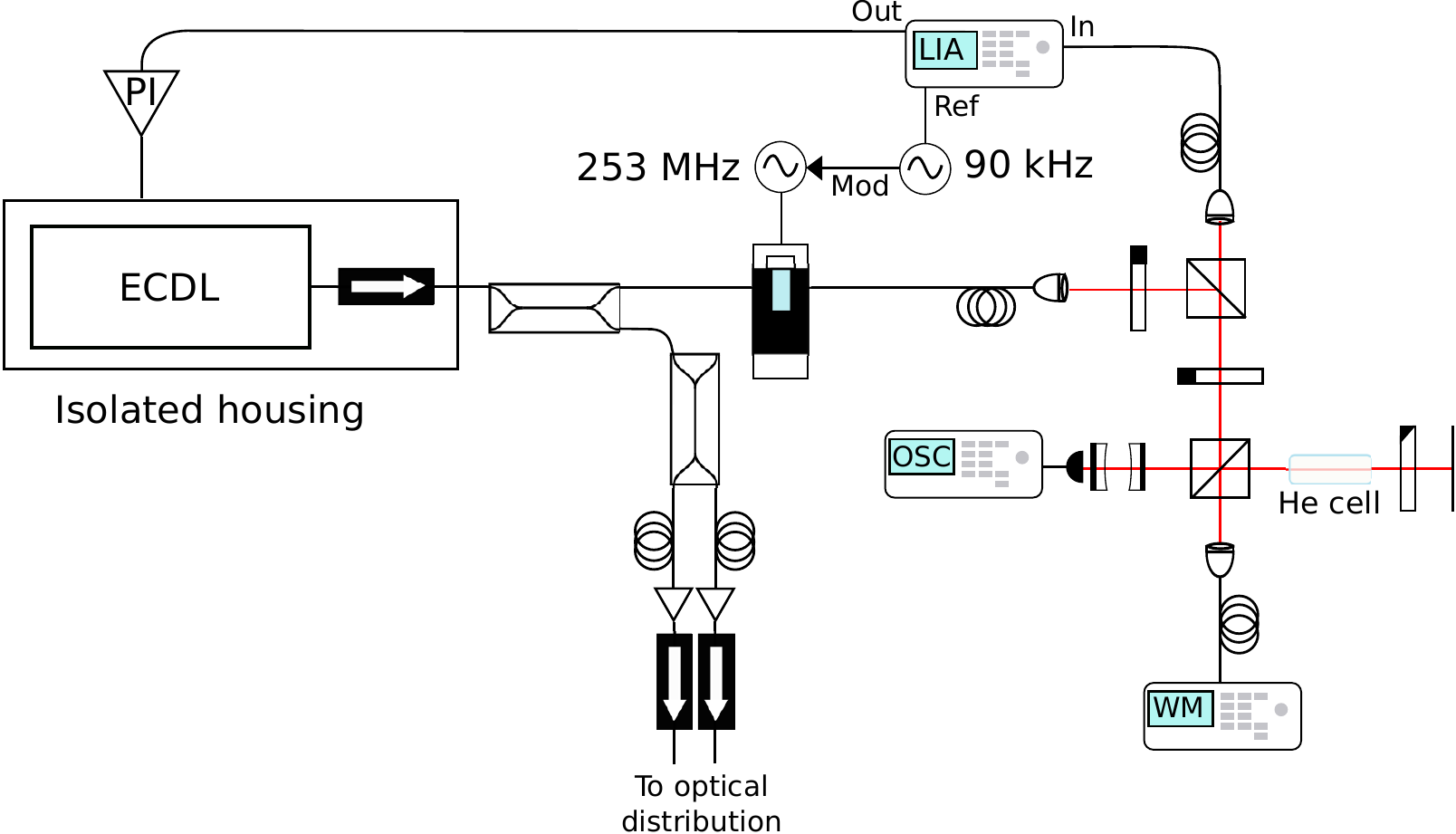}
		\caption{Schematic of the generation and control system for the main cooling laser operating at 1083.331 nm described in the text. Light from the laser diode is filtered by the cavity resonance and internal diffraction grating, which is actuated by a piezoelectric crystal. Part of the light is picked off with an in-fibre beamsplitter and sent to the experiments. The rest passes through an in-fibre AOM which increases the frequency by 253 MHz and outputs light into the saturated-absorption spectroscopy system (lower right). The SAS offset is in turn modulated at 90kHz and the demodulated signal (output of the LIA) is proportional to the gradient of the SAS signal, which is minimized at the absorption peak. Hence the demodulated output serves as the error signal for a PID controller that feeds back to the piezoelectric actuator of the frequency-selective diffraction grating. Further details about the spectroscopy technique can be found in \cite{ShinThesis,FootAtomic}.}
		\label{fig:main_laser}
	\end{figure}

	The principal light source for both machines is the 1083.331 nm cooling and trapping laser.
	The light source, depicted in Figure \ref{fig:main_laser}, is an external-cavity diode laser (ECDL) that operates with a locked linewidth below 100 kHz and is detailed in the publication \cite{Shin16} and in PhD thesis of D. K. Shin \cite{ShinThesis}.
	The output of the ECDL passes through an in-fibre beamsplitter, from which one output is blue-shifted by 253 MHz by an acousto-optical modulator (AOM) and locked to a helium gas cell by saturation absorption spectroscopy.
	The remaining arm is coupled by optical fibres to the fibre amplifiers of both experiments.
	Each experiment features its own fibre amplifier that produces up to 5 Watts of laser power from the $\approx$3 mW supplied by the ECDL and then distributes the light to the laser cooling and trapping chambers via a suite of dedicated AOMs (depicted in section \ref{sec:new_optics}).


	In each machine, a suite of AOMs are used to independently control the power and frequency of each of several specialized beams\footnote{The active component of an AOM is a crystal is driven at radio frequencies by an amplified voltage-controlled oscillator.	An acoustic standing wave, constituted by phonons with frequency matched to the driving frequency, scatters light into discrete diffraction modes width momenta shifted by the phonon-lattice momentum.	The action is analogous to a diffraction grating, and described by the Kapitza-Dirac mechanism.	Thus the AOM transduces electronic signals into optical frequency shifts.	An analogous effect, with the role of atoms and light interchanged, is the operating principle of optical lattices.}.
	The AOM suites are supplied by light picked off from the output of the fibre amplifier by a half-waveplate and beamsplitter pair per AOM.
	The first-order diffraction is usually chosen for its more efficient transmission.
	The pointing of the diffracted beams vary with the AOM frequency, which is a problem when the ultimate destination of the light is separated from the AOM by several meters of optical path length, after which the deflection can become significant.
	To circumvent this, the AOMs are set up in a double-pass ``cat's-eye" configuration so the deflection in one pass is canceled by the second pass in the opposite direction.
	A quarter-wave plate placed in the optical path through the cat's-eye ensures the doubly-diffracted beams are transmitted back through the beamsplitter into the distribution optics.
	
	AOMs offer the advantage of fast switching times, but some light can leak from the AOM suite through the optical paths into the vacuum chamber.
	As such we used electronic shutters and an opaque enclosure to prevent this from occurring.
	The shutter action as well as the detuning and diffraction efficiency of the AOMs are actuated by the control software described later in this chapter.
	
	The supply of light which is 253 MHz red of the field-free resonance reduces the risk of stray light interacting with atoms and disrupting the delicate cooling and trapping processes.
	Furthermore, the fact that AOMs are limited in operation to a maximum frequency on the order of a hundred MHz or so means that the large detunings required for certain laser cooling applications could not be achieved using such double-pass setups seeded by resonant light\footnote{Cascading double-pass systems are certainly possible, but adding more moving parts with efficiencies generally below 80\% is undesirable.}.
	The lattice machine also makes use of 1550nm light for trapping in an optical dipole (and eventually lattice), whose generation and application is discussed in appendix \ref{chap:lattice}.

\subsection*{Spectroscopic laser}
\label{sec:spec_laser}
	\begin{figure}
		\centering
		\includegraphics[width=\textwidth]{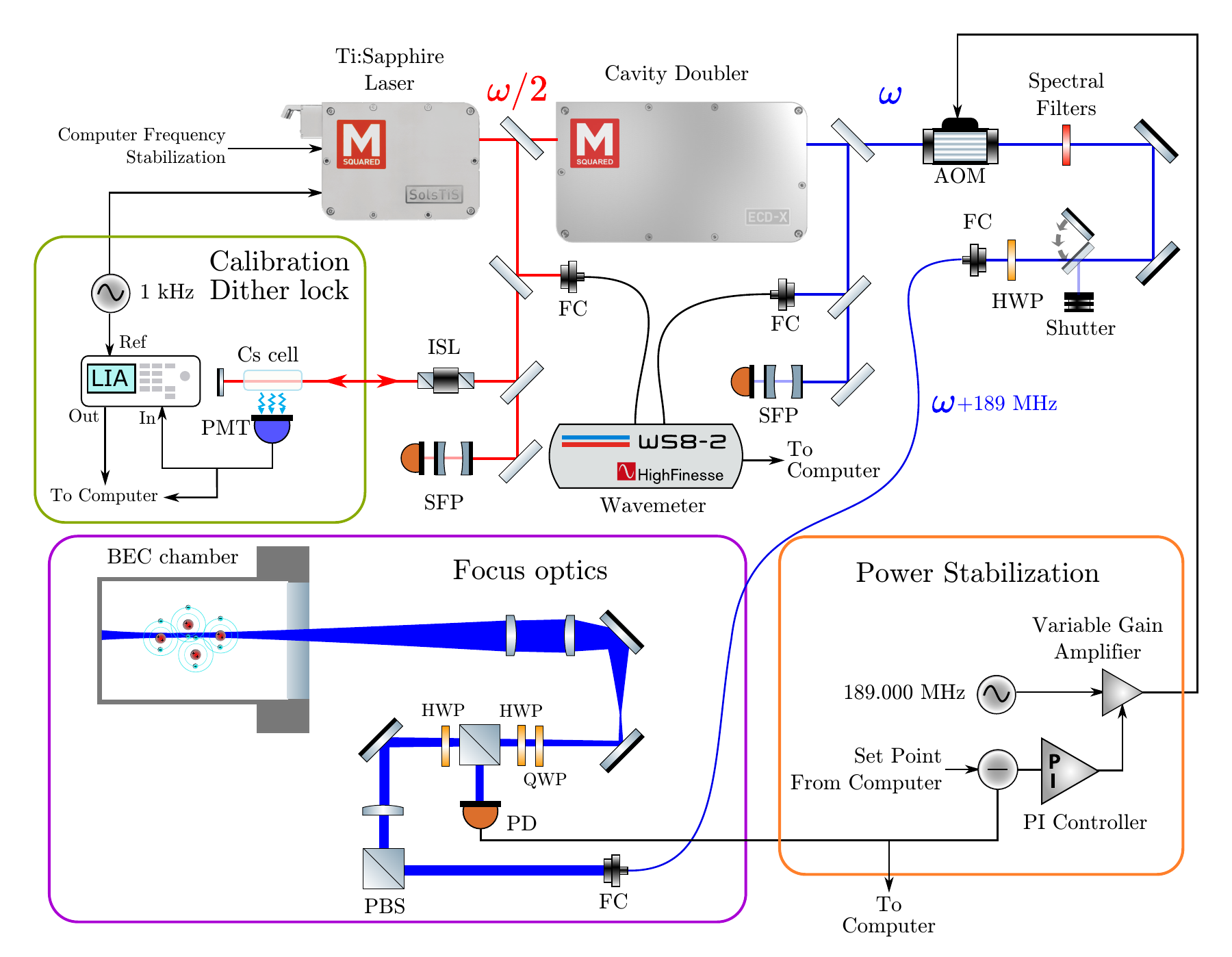}
		\caption{Schematic of the spectroscopic laser system.  The main laser modules supply light through an optical fibre, after three stages of optical filtering, to the insertion optics. The input to the fibre is the first diffraction mode from the AOM that serves as the actuator for an electronic PID loop whose power set point (P set) is compared to the reading of a photodiode (P act) at the insertion optics. The laser frequency is stabilized by a software-based PID loop that actuates the Ti:S cavity length (via inboard piezos) to reduce the between the desired and actual wavelengths ($\lambda$ set and $\lambda$ act, resp.). A pair of scanning Fabry-Perot cavities (SFP)  monitor the light from each laser module to verify both are operating in a single-mode regime. The switches at the input to the Ti:S are shown in the configuration used to calibrate the wavemeter. In this procedure the flipper mirror (coupling red light to the SFP) is removed and light enters a cesium gas cell. The fluorescence signal is measured by a photomultiplier tube backed by a photodiode, and this provides the signal for a dither lock at 1 kHz, demodulated by a lock-in amplifier (LIA).}
		\label{fig:tunable_laser}
	\end{figure}

	A light source unique to the BiQUIC machine was the tunable laser system used to generate light in the $402-430$ nm range for laser spectroscopy (namely the works \cite{Ross20,Henson22} corresponding to chapters \ref{chap:transitions} and \ref{chap:tuneout} respectively, and also Ref. \cite{Thomas20}).
	This laser system is depicted in Figure \ref{fig:tunable_laser}.
	Light at 1064nm from a Lighthouse Photonics \emph{Sprout} module was frequency doubled by second-harmonic generation to pump an M-squared SolsTi:S titanium-sapphire laser which we operated around 800 nm.
	The output from the SolsTi:S was doubled again in an M-squared ECD-X module to the target wavelengths.
	A fraction of the light is fed into a High Finesse WS-8 wavemeter.
	A MATLAB software lock uses the wavemeter output to stabilize the laser to within 100 kHz of the target wavelength, and allows automatic scans across the region of interest by automatically updating the laser set point (details are provided in section \ref{sec:DAQ}).
	We use the wavemeter to lock the tunable laser with respect to the red light, so the instrumental uncertainty is doubled in determinations of the blue frequency.
	High Finesse specifies the absolute accuracy of the WS-8 at 2 MHz within 2 nm of a transition line, and 10 MHz otherwise.
	We calibrated the wavemeter once every day or so with respect to the two-photon crossover transition between the $6^2P_{\frac{3}{2}} (F=4)$ and $6^2P_{\frac{3}{2}} (F=5)$ lines in a cesium vapor cell.
	We used saturated absorption spectroscopy to lock the red light to the Cs transition while feeding light into the wavemeter for the calibration.

	We used the first diffracted mode of an AOM, driven at 189 MHz, to control the beam power.
	The output of the AOM was fed into an optical fibre which coupled the light to the vacuum insertion optics.
	{After exiting the fibre, the light was linearly polarized and a small fraction was directed onto a photodiode, whose voltage served as the input parameter to a PID loop.
	The control loop had a 3dB bandwidth of 170 kHz and was able to stabilize the beam power to within a relative error of 0.3\%.}
	There, we used waveplates to set the polarization and a telescope to magnify the beam up to $\approx$2 cm waist.
	A final lens fixed to a three-axis translation mount was used to focus and align the beam.
	For the measurements of the tune-out and forbidden transition wavelengths, we focused the beam to a $\approx10~\mu$m diffraction-limited waist at the site of the BEC to achieve strong interaction given the weak signals of interest.
	For the $2\triplet P_2\rightarrow 5\triplet D$ transitions described in chapter \ref{chap:transitions}, the beam was collimated and operated at a reduced power to mitigate power broadening and saturation of these much stronger resonances.
	The optical power was regulated with reference to a photodiode that sampled the beam after the fibre via a polarizing beamsplitter.

	The wavemeter logs, photodiode voltage, and transmission of light from both the red and blue beams through two scanning Fabry-Perot cavities (SFPs) were recorded through the DAQ system in order to provide diagnostics in post-processing.
	Part of the data analysis pipeline then automatically discarded shots where either SFP showed multiple peaks within a single free-spectral range (FSR), indicating that one of the lasers was running in a multimode regime, or where the photodiode trace indicated a laser supply failure, or where other anomalous behaviour was detected.

\section{Detection of metastable helium atoms}
\label{sec:he_detection}
	Two optical detection methods are employed in the ANU helium labs.
	Resonant absorption imaging is only used in the lattice machine, and is therefore discussed in appendix \ref{chap:lattice}.
	Both machines make use of saturated fluorescence measurements to measure the number of atoms in a trap.
	In this method, bright resonant light ($I\gg I_\textrm{sat}$) is applied to the atoms such that the population of the excited state saturates at 50\%.
	If the light is applied while the trap remains on, then while some atoms may decay to a trapped state, repeated absorption events eventually drive them out of the trap.
	Therefore, one expects a sharp peak and subsequent decay in the optical power re-emitted from the trap.
	This radiation can be captured by a lens and focused onto a photodiode, producing an analog voltage which can be used to measure the number of atoms confined in the trap.
	The difference between peak and steady-state voltage after the pulse is a direct probe of the atomic population (see appendix \ref{chap:lattice} for further discussion).
	This technique is employed as a diagnostic tool for fast readout of the second MOT population when optimizing the alignment of cooling and trapping beams. 

	A unique diagnostic available in helium experiments is the production of helium ion-electron pairs.
	This process can be monitored by in-vacuum electron multipliers mounted near the trapping region.
	Proximity to the trap is desirable for efficient collection of the resulting particles.
	Aside from investigating Penning ionization itself, ion detection has been applied in photoassociation spectroscopy \cite{Herschbach00,Koelemeij04}, to monitor the onset of Bose-Einstein condensation \cite{Tychkov06}, and for high-precision laser spectroscopy \cite{Rengelink18}.
	Ion detection garners only a brief mention in works relevant to appendix \ref{chap:lattice}, wherein we hoped to detect ions as a diagnostic when aligning an optical dipole trap\footnote{The ion detector in the BiQUIC machine stopped working some years ago and has not been replaced: the utility of this diagnostic is outweighed by the effort required to break vacuum for the first time in a decade or so, disassemble the chamber and surrounding optics, reassemble everything, bake the chamber, and realign all the optics.}.
	In an accident a year or so after I left that lab, a ceramic feed-through connecting the ion detector to the external electronics cracked.
	This broke vacuum, necessitating a rebake of the chamber, and saw the detector retired.

\subsection*{Single-atom detection with the MCP-DLD stack}
	\label{sec:DLD}
	The principal detection scheme in this thesis is single-atom sensing in the far-field regime using a multichannel plate and delay-line detector combination (MCP-DLD)\footnote{In this context far-field means that the density of the cloud is determined by the initial velocity distribution and that effects of the finite initial size are negligible. This is suitable in our experiments because the condensate is smaller than the detector resolution. See chapter 5 of D. K. Shin's PhD thesis \cite{ShinThesis} for a quantitative discussion.}.

	\begin{figure}
		\centering
		\includegraphics[width=\textwidth]{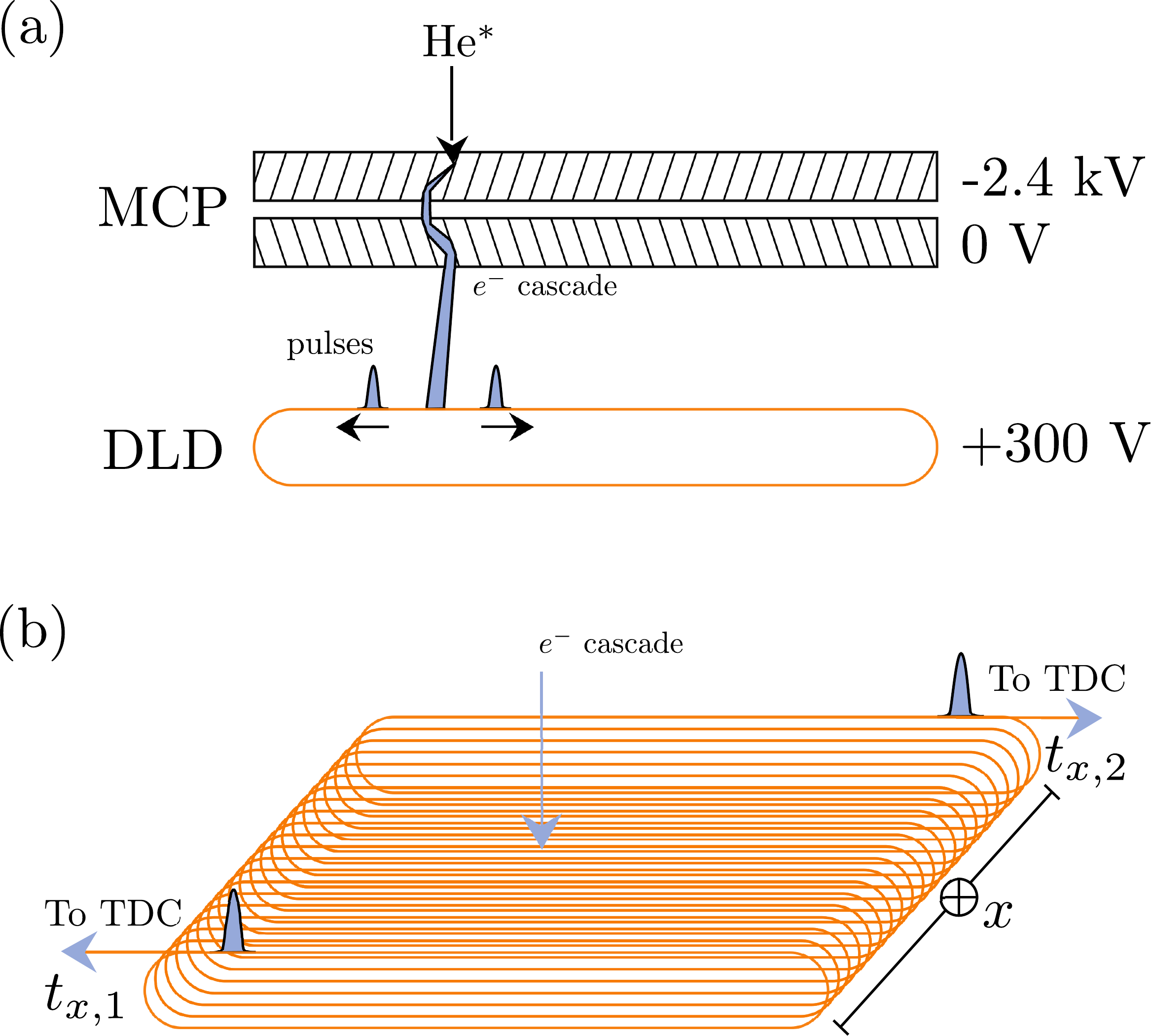}
		\caption{Schematic sketch of the MCP-DLD detector. The twin plates of the multi-channel electron multiplier (MCP) have a 2.4 kV potential between the upper and lower surfaces which accelerates electrons along the pores after they are freed from the surface by the energy released from \mhe~impact events. The resultant shower of electrons deposits a charge excess on a winding of the delay-line detector (DLD), which disperses as oppositely-travelling current pulses along the line. The arrival times ($t_{i,1}$ and $t_{i,2}$) of each pulse along each axis are received by the TDC and used to infer the respective coordinate of the electron cascade (only the winding along $x$ is shown for simplicity, but in the BiQUIC machine the DLD consists of two orthogonal windings).}
		\label{fig:MCP_DLD}
	\end{figure}

	In some configurations, such as the one at the \mhe~group at VU Amsterdam, the MCP itself can be used as an ion- or metastable-atom detector.
	In our machines, the MCP is paired with either a phosphor screen (usually only used for diagnostic purposes) or more typically with a delay line detector to form the MCP-DLD that is the workhorse of most experiments conducted at ANU.
	A summary description is given here, and detailed explanations of the detector stack and signal processing pipeline can be found in \cite{ShinThesis, HodgmanThesis, ManningThesis}.
	The MCP consists of two plates, each of which feature 10 $\micron$-diameter pores arranged in a square grid with 20 $\micron$ between the centres of the pore openings.
	The plates are 80 mm in diameter, with the upper surface 848 mm below the BiQUIC trap centre.
	The freefall time-of-flight of the centre of mass of the cloud is 417 ms, and so the maximum detectable horizontal velocity is about 9.5 m/s, sufficient to capture almost all of the thermal fraction for clouds below the critical temperature.
	The detector plates are grounded for most of the experimental sequence, and then the top plate is ramped to $-2.4$ kV over about 2 seconds before dropping the trap.
	The negative voltage repels electrons from the surface of the plate, protecting the detector surface from degradation and reduces the background count rate.
	The background rate, also called the \emph{dark count} rate, is typically 0.56 Hz/cm$^2$ when operating the plates at $-2.4$ kV, and is negligible for the purposes of experiments in this thesis.

	When an atom strikes the surface of a pore after falling from the trap, a second-order process releases a free electron with most of the 19.8 eV as kinetic energy \cite{Jagutzi02,Hotop96}.
	These electrons are accelerated down the pore by the strong electric field, and themselves impact the pore surface and eject more electrons, triggering an electron avalanche that amplifies each atom impact into over 10$^6$ electrons.
	The electron shower exits the back of the detector and is accelerated by a +300 V potential towards the delay-line detector (DLD).
	The DLD in the BiQUIC machine consists of two coils of wire each wound in a helical pattern and arranged perpendicular to one another.
	 The arrival of an electron cascade causes a current pulse to travel along each wire in both directions from the point of impact.
	The pulses pass through a fast pre-amplifier and then through a constant-fraction discriminator which converts the analog pulses into a digital signal.
	Both of these processes take place within a Roentdek DLATR6 which is located outside the vacuum chamber.
	The digital pulses are transmitted to a Roentdek TDC8HP time-digital converter (TDC) which registers the arrival times, relative to the arrival of the main trigger signal from the LabView control software, and writes these to a \verb|txt| file as \verb|(channel,time)| pairs.
	Finally, a custom C++ script passes over the file and converts the timing data to \verb|(t,x,y)| tuples.
	The conversion from \verb|(channel,time)| to \verb|(t,x,y)| is performed separately, and indeed the latter usually by a different machine, so as to provide maximum CPU availability to the acquisition function.	Further details about the architecture, construction, and calibration of these detectors is found in \cite{HodgmanThesis,ManningThesis}.

	A second MCP, backed by a phosphor screen and a mirror at 45$^\circ$ to the vertical axis, is mounted on an in-vacuum translation stage below the main chamber.
	The mirror directs light from the phosphor screen to a CCD camera mounted outside the vacuum chamber.
	This detection method is not typically used for scientific purposes these days, but is a useful additional diagnostic when the MCP-DLD detector stack appears to be faulty.
	The phosphor screen does have a much larger dynamic range than the DLD, however, which makes it useful for visual spotting of subtle distortions in the BEC profile, such as used in the alignment of the spectroscopic probe beam.


	 \begin{figure}
	 	\centering
	 	\includegraphics[width=\textwidth]{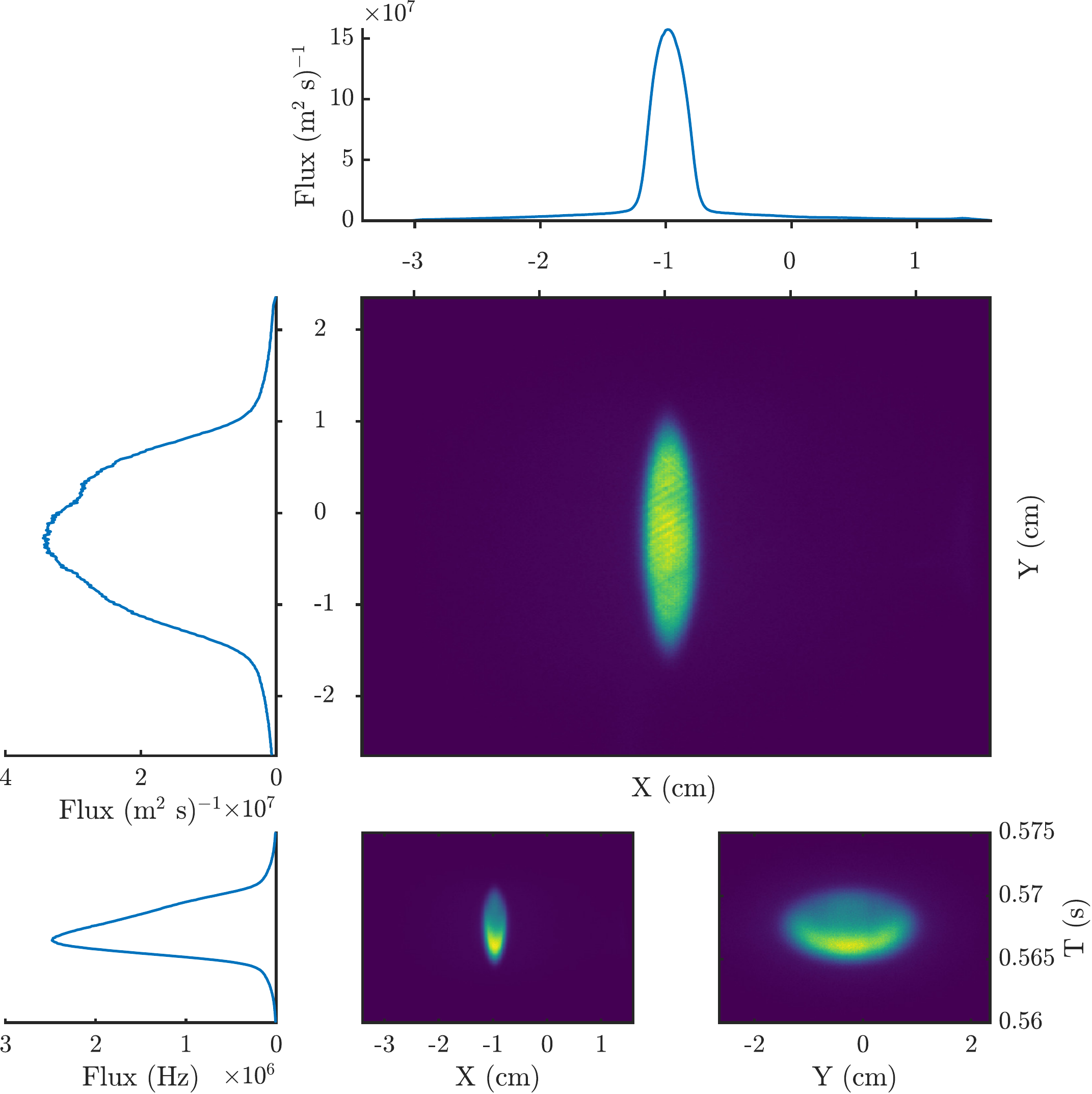}
	 	\caption{MCP readout of a dropped BEC with small thermal fraction, just visible as dilute wings outside the central peak. The detector saturation is evident in the time-of-flight profiles (bottom row, all with common vertical axis) as a sudden downturn in the detected flux, whereas the full BEC has a parabolic profile. The density- and space-dependence of the saturation is visible in the 2D sections (lower middle and lower right). The one-dimensional fluxes along X and Y are averaged over the time interval displayed in the bottom row, and the total count rate (bottom left) is integrated over the entire detector.}
	 	\label{fig:dropped_bec}
	 \end{figure}

	 \newpage
	 \begin{figure}
	 	\centering
	 	\includegraphics[width=\textwidth]{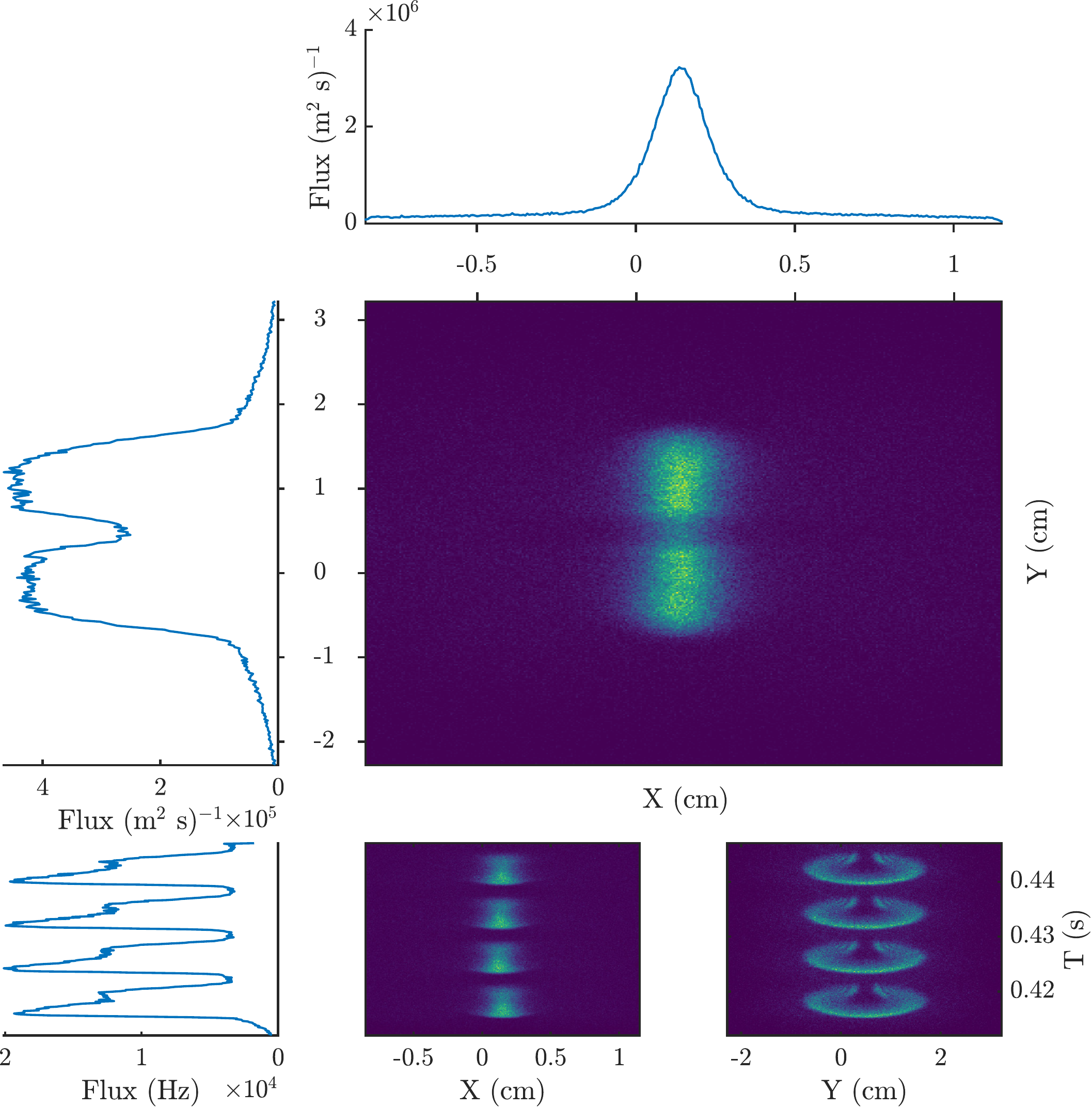}
	 	\caption{MCP readout of the first few pulses from a pulsed atom laser, displayed similarly to Fig. \ref{fig:dropped_bec}. When atoms are transferred into the untrapped state, they are repelled from the trap by the mean-field energy of the remaining condensate, leading to pulse broadening especially in the tight $(y,z)$ axes. In the dilute outcoupled pulse, the speed of sound is low enough that the relative velocity of the pulse and BEC is supersonic in the outcoupled cloud, leading to the complex interference patterns which have been identified as \emph{Bogoliubov-Cerenkov radiation}\cite{Henson18_BCR}. The peak flux is about a thousandth of that for a dropped BEC, and hence saturation is avoided. The flat top of the Y-projection is due to the distortion of the profile during the trap escape, not detector saturation.}
	 	\label{fig:dropped_pal}
	 \end{figure}
	 \newpage
\subsection*{Atom lasers}
	\label{sec:atomlaser}

	The most straightforward way to use an MCP-DLD to interrogate a BEC is simply to drop the atoms onto the detector.
	An example of this straightforward detection scheme is illustrated in Fig.
	\ref{fig:dropped_bec}.
	A drawback of this method is that the high atom fluxes in large BECs can temporarily deplete the available current carriers near the surface of the detector, resulting in a nonlinear reduction in detection efficiency known as \emph{saturation}.
	This means that for even moderately sized condensates, a determination of the condensate population by counting detection events or by straightforward fitting of the time-of-flight profile is impractical, and one cannot compute interesting quantities like correlation functions with meaningful accuracy.
	Another limitation of the trap-release method is that it is a completely destructive measurement - one cannot re-trap the cloud or only drop part of it, which slows down data acquisition for even simple purposes like determining the trapping frequencies.
	Atom lasers can circumvent both of these issues.

	The basic principle of an atom laser is to transfer some portion of the condensed atoms into a state that is no longer confined by the trapping potential.
	This was first achieved (shortly after the first realisation of BEC) by applying pulses of RF radiation to a magnetically trapped condensate \cite{Mewes97}.
	Later, a continuous-wave atom laser was demonstrated and used to perform RF spectroscopy of the cloud, revealing the symmetry-breaking action of gravity which pulls the BEC away from the minimum of the magnetic field \cite{Bloch99}.
	Optical Raman transitions\footnote{In atom optics, \emph{Bragg} transitions, which impart a momentum transfer, are distinguished from  \emph{Raman} transitions which induce a change in the electronic state.} can also be used as the outcoupling mechanism \cite{Hagley99} with the advantage of imparting a controllable momentum to the outcoupled atoms, but are not employed in the works in this thesis.

	Atom lasers are so named because the trapped condensate acts as a reservoir of coherent matter waves, in analogy to lasers as a source of coherent light \cite{Naraschewski99,Glauber63}.
	The first-order coherence of atom lasers is evident from the observation of interference fringes between matter-wave beams \cite{Andrews97} and the higher-order coherence manifests in the many-particle correlations detected in \mhe~atom lasers \cite{Manning10,Dall11a,Rugway11}.
	Atom lasers can also be collimated \cite{Bloch99}, directed with collimated laser beams playing the analogous role of an optical waveguide \cite{Guerin06,Couvert08}, and manipulated with optically-induced `mirrors' and beamsplitters \cite{Bloch01}.
	The potentially continuous operation of atom lasers \cite{Chikkatur02} brings the analogy with conventional lasers even closer.
	Atom laser physics has been studied in detail and widely deployed through the field of atom optics in the two intervening decades since their conception, including all of the experiments described in this thesis.
	

	A simple explanation of the mechanism of an atom laser requires a two-level atom with magnetically trapped and untrapped states labeled $\ket{1}$ and $\ket{0}$.
	An RF pulse tuned to the splitting $\omega_{10}$ induces an oscillation between the two states with Rabi frequency $\Omega$ for a duration $t$.
	For an ensemble of $N$ atoms, the independent single-particle oscillations produce a many-body product state which evolves as \cite{Mewes97}
	\begin{align}
		\ket{\Psi} &= (\cos(\Omega t/2)\ket{0} + \sin(\Omega t/2)\ket{1})^{\otimes N} \\
		& = \sum_{n=0}^{ N}\sqrt{\frac{N!}{n!(N-n)!}}\cos(\Omega t/2)^{N-n}\sin(\Omega t/2)^n\ket{N-n,n},
		\label{eqn:PAL_Rabi}
	\end{align}
	where $\ket{N-n,n}$ denotes the state with $n$ atoms coupled out of the trapped state and $N-n$ atoms remaining.
	The outcoupled fraction is then $\sin^2(\Omega t/2)$.

	In practise, the Zeeman splitting varies across the trap due to the inhomogeneous magnetic field, and so a narrow RF pulse would only be resonant with a section of the condensate.
	While useful for detailed spectroscopy of the condensate, the outcoupling rate would depend on the condensate population.
	This can be circumvented by applying short RF pulses resonant with the minimum Zeeman splitting of the trap.
	If such pulses are sufficiently short, the Fourier broadening of the finite-duration pulse is wider than the RF width of the condensate.
	The typical radio-frequency width of the condensate, i.e.
	the difference between the resonant frequency at the centre and the edge of the condensate, is set by the chemical potential of the condensate and is typically less than 10 kHz in our experiments.
	The outcoupling pulse typically consists of 6-10 cycles of RF tuned to the Zeman splitting of the trap bias, which typically ranges from 0.6-1 MHz depending on the configuration that is most suited to the experiment at hand. 
	This leads to pulses that range from about 5-25 $\mu$s, where the number of cycles is the most direct means to control the outcoupling rate (c.f. Eqn. \ref{eqn:PAL_Rabi}).
	An example of the MCP-DLD readout from a pulsed atom laser is shown in Figure \ref{fig:dropped_pal}.



\section{Data acquisition \& control}
\label{sec:DAQ}
	
	Control of the optical components, coil current supplies, and RF radiation is coordinated by a central LabView program which transmits pulse sequences to the machine via National Instruments pulse generation cards.
	The cards themselves are connected to the CPU by a PCI bus, and synchronize their internal clocks at the beginning of each shot.
	The cards also record analog inputs such as Fabry-Perot cavity scan traces, photodiode traces, and mains voltage traces for post-processing diagnostics.
	The spectroscopic laser lock and tuning operations are also performed on the main control computer, as described in section \ref{sec:spec_laser}.

	The LabView control is supplemented by a command that calls a MATLAB subroutine (referred to as the `interface'), which is customizable to suit the purposes of a given experiment\footnote{The software developed to analyse the data from various experiments is invariably written in MATLAB.	Many core capabilities are stored in the repository \url{https://github.com/HeBECANU/Core_BEC_Analysis}, and generally each project will have a unique public repository as well.}.
	The motivation for this extension was for automatic and more flexible variation of experimental sequences. 
	This subroutine can return the \verb|.xml| file path of a LabView pulse sequence file for use in the next experimental cycle, and even modify the sequences, albeit in a limited way due to the particulars of LabView's sequence specification format.
	This extension allows for automatic optimization of experimental sequences, as was demonstrated in \cite{Henson18_ML}, but obviously is of no use for mechanical tasks like beam alignment\footnote{One can dream, though.}.
	Moreover, running back-to-back measurement and calibration shots would otherwise have to be done manually, as would updates of the spectroscopic laser wavelength during scans across the spectral region of interest.
	Therefore without the automatic update procedures enabled by the interface, the multi-day data collection runs embarked upon in this thesis would have become considerably more time-consuming and error-prone. 
	
	The interface script also writes metadata to a log file, including the POSIX timestamp, type of shot executed (e.g.
	\verb|calibration| or \verb|AtomLaser|), shot number, and other relevant parameters.
	The timestamps in this logfile can be cross-referenced with the \verb|txt| files that are written by the TDC computer and with the LabView logs of analog inputs in order to match the experimental parameters with data files for post-processing purposes.
	The data analysis pipelines for modern experiments can be complex, and the experiments in this thesis generally comprise a handful of different diagnostics processed in parallel.
	The tagging of shots by timestamp cross-referencing allows separation of relevant shots for each processing subroutine.
	
\subsection{Laser lock control protocol}

\begin{figure}
		\centering
		\includegraphics[width=\textwidth]{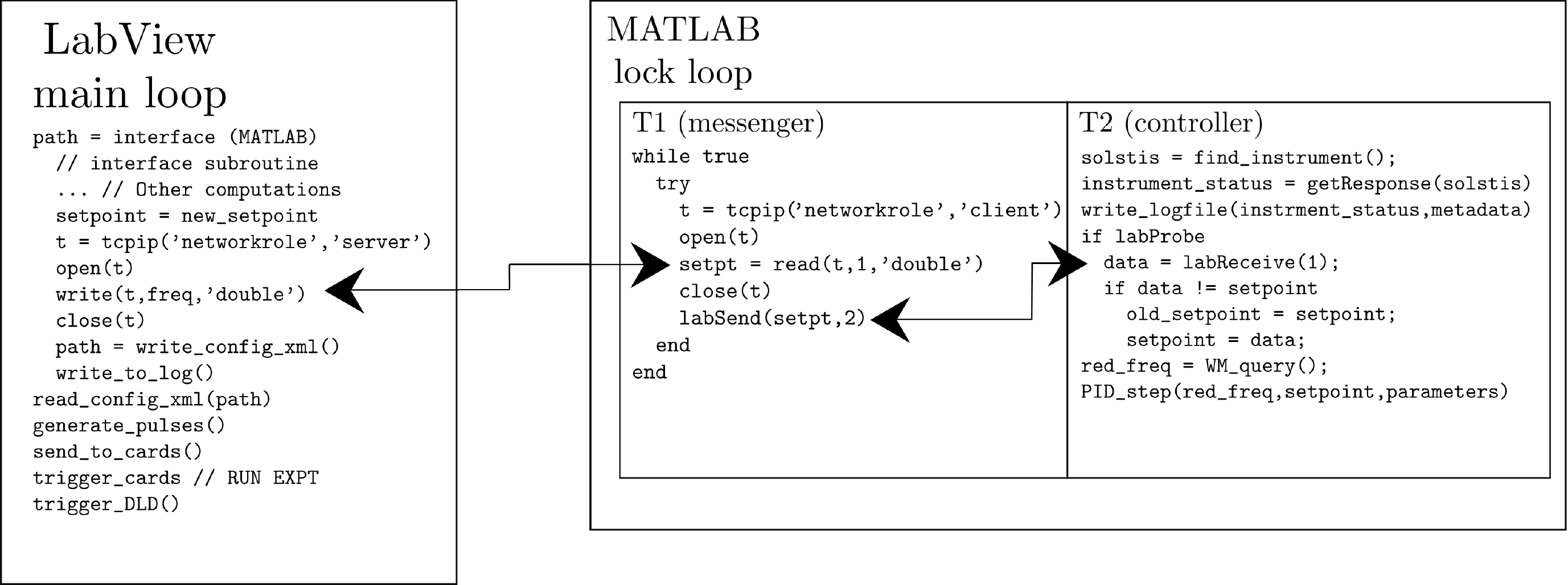}
		\caption{Schematic of the three major components of the lock loop used to scan the laser setpoint during spectroscopic measurements. As detailed in the main text, the main LabView script calls MATLAB subroutine which, among other tasks, communicates with a parallel pair of worker threads T1 and T2 which actuate the SolsTi:S controls.}
		\label{fig:tcp_control}
	\end{figure}

	As mentioned, the interface plays an important role in scanning the spectroscopic laser set point across regions of interest (be they electronic transitions as in chapter \ref{chap:transitions} or the tune-out wavelength in chapter \ref{chap:tuneout}). 
	The relationships between the subroutines involved in the laser control system are illustrated in Fig. \ref{fig:tcp_control}.
	The major features of this system architecture are two parallel processes which communicate with each other via MATLAB's cross-thread communication (\verb|labSend(data,to_process_id)| and \verb|labReceive(from_process_id)|).
	The messenger thread (T1) continuously looks for a TCPIP connection with the interface, which runs at the start of each experimental sequence and opens a TCPIP channel for the thread to connect with and receive the new setpoint.
	The control thread (T2) operates a PID loop by querying the wavemeter for its present reading (via USB connection) and also querying the laser control module for relevant parameters (particularly internal photodiode readings and voltages across piezoelectric actuators for interal optomechanics).
	The control loop then compares the desired set point to the measured set point and updates the actuator voltages according to the output of a discrete PID function.
	The control thread (T2) also contains a subroutine which is triggered when T1 sends the new setpoint through via \verb|labSend()| (which sets the \verb|labSend| bit to 1).
	If this condition is met, then T2 updates the setpoint for the PID loop with the new value obtained from \verb|interface| via T1.
	There is a stall condition here where the TCPIP port is opened by \verb|interface| but T1 is still waiting for the timeout of the last query, hence \verb|interface| will hang until the next loop of T1.
	This is not a major problem as the interface script must terminate before the pulse generation cards are triggered (they are sequential in the LabView program), so rather than interrupting the experimental execution it simply reduces the PID bandwidth briefly at the start of each run.
	The loop bandwidth is speed-limited by the frequent condition checks, device queries, MATLAB's limited speed as an interpreted language, and competition with other processes for CPU time. 
	Nonetheless the lock loop maintains a clock speed on the order of 10-15 Hz.
	The system has sufficient bandwidth to achieve a $\sim900$~ms rise time (10\% to 90\%) for a $\sim300$~MHz frequency step, and during interrogation we measured a typical (in-loop) standard deviation of 170~kHz. 
	The stabilization of the laser frequency uses measurements of the red side of the laser system (before the doubling cavity), so interruptions to the doubling cavity do not impact the lock.

\subsection*{Optimizing machine performance}

	Despite the stabilized environment, optics are prone to wander slightly over time.
	The stable temperature is especially important as the cooling beams are coupled from the AOM tables to the vacuum chamber optics by free-space links.
	Drift in the relative position of the optics tables would therefore result in misalignment of many laser beams, and degrade performance of the machine.
	The conditioned air is supplied through three HEPA filters in the lab ceiling.
	The resulting positive pressure in an area enclosed by PVC strip walls keeps dust from entering the area on air currents.

	If the second MOT is loading and a saturated fluorescence signal is visible, one can usually optimize all of the optical alignments on this signal, which is a factor of ten faster than the full BEC creation sequence.
	If this is not available, logs of diagnostics taken at several points throughout the beamline can facilitate diagnosis of the point of failure.
	The most commonly employed tests are Faraday cups which produce a current when impacted by \mhe~atoms, as read out on a picoammeter.
	These current sensors are placed after the deflection stage, at the end of the Zeeman slower (with a large hole bored through it to allow the slowing beam to pass through), between the first MOT and the focus stage, and behind the second MOT chamber.
	In case the first MOT is not operational, a Xenics Bobcat CCD camera mounted above the chamber provides a view of the trapping region and direct visual feedback about the presence, shape, and density of the MOT.

%% file: latex/21_transitions.tex
\chapter{Frequency measurements of resonances between the second and fifth manifolds in {$^4$}\mhe}
\markboth{\thechapter. TRANSITION FREQUENCY MEASUREMENTS}{}
\label{chap:transitions}

\blankfootnote{\noindent The contents of this chapter relate to the work published in \textbf{Frequency measurements of transitions from the $2\triplet P_2$ state to the $5\singlet D_2$, $5\triplet S_1$, and $5\triplet D$ states in ultracold helium} by J. A. Ross, K. F. Thomas, B. M. Henson, D. Cocks, K. G. H. Baldwin, S. S. Hodgman, A. Truscott, \href{https://journals.aps.org/pra/abstract/10.1103/PhysRevA.102.042804}{\emph{Physical Review A} \textbf{102}} (2020)}

  {The} appearance of ordinary matter arises from interactions between charged particles and light.
	This phenomenon is the domain of the theory of quantum electrodynamics (QED), which provides the most accurate quantitative predictions of any physical theory to date.
	The theory of QED is the workhorse of modern atomic structure calculations, whose only inputs are the CODATA values of three physical constants: the proton-electron mass ratio, the Rydberg constant, and the fine structure constant $\alpha$.
	These constants of nature can be constrained with state-of-the-art atomic spectroscopy, which is accurate enough to match theoretical uncertainties in table-top experiments.
	Thanks to the quality of modern theory and experiment, atomic structure measurements reprise their role in frontier tests of physics.
	
\section{Introduction}

  In 1964, Charles Schwartz proposed the determination of $\alpha$ from the fine structure intervals of the $2\triplet P$ manifold in helium, which are subject to strong QED effects \cite{Schwartz64}.
	The contemporary knowledge of helium's structure greatly exceeds Schwartz's anticipation of parts-per-million accuracy.
	For example, the $2\triplet S_1 - 2\triplet P$ and $2\triplet P - 3\triplet D$ intervals measured by Cancio Pastor \textit{et al.} \cite{Pastor04} and Luo \emph{et al.} \cite{Luo15}, respectively, both have relative uncertainties better than 50 parts per \emph{trillion}, providing Lamb shift measurements accurate to several ppm.
	The measurement of the $\PStateManifold$ fine structure splitting by Smiciklas \textit{et al.} to sub-kilohertz precision determines $\alpha$ to several ppb \cite{Smiciklas10}.
	Measurements of the $2^{3\!}P_1-2^{3\!}P_2$ interval by Kato \emph{et al.}, accurate to 25Hz, would constrain $\alpha$ to less than one ppb given a similarly accurate measurement of the $2^{3\!}P_0 - 2^{3\!}P_1$ transition and QED calculations including terms of order $\alpha^7$ \cite{Kato18}.

  \begin{figure}
  	
  	\begin{minipage}[t]{0.38\textwidth}
			\vspace{0pt}
      \caption{Energy level diagram for $^4$He showing the transitions measured in this work (blue) are driven by a tunable laser referred to in the text as the \emph{probe beam}.
				A laser tuned to the $2\triplet S_1-2\triplet P_2$ transition (red, referred to as \emph{pump beam}) populates the lower state of the target transitions.
				 The doubly forbidden $1^{1\!}S_0 - 2^{3\!}S_1$ transition is excited in a high voltage discharge source.
				Transitions across the dotted line are \emph{forbidden} by the $\Delta S=0$ selection rule.
				Level splittings are not to scale.}
      \label{fig:lvl_diag}
      \end{minipage}
      \hfill
      \begin{minipage}[t]{0.6\textwidth}
      \vspace{0pt}
      \includegraphics[width=\textwidth]{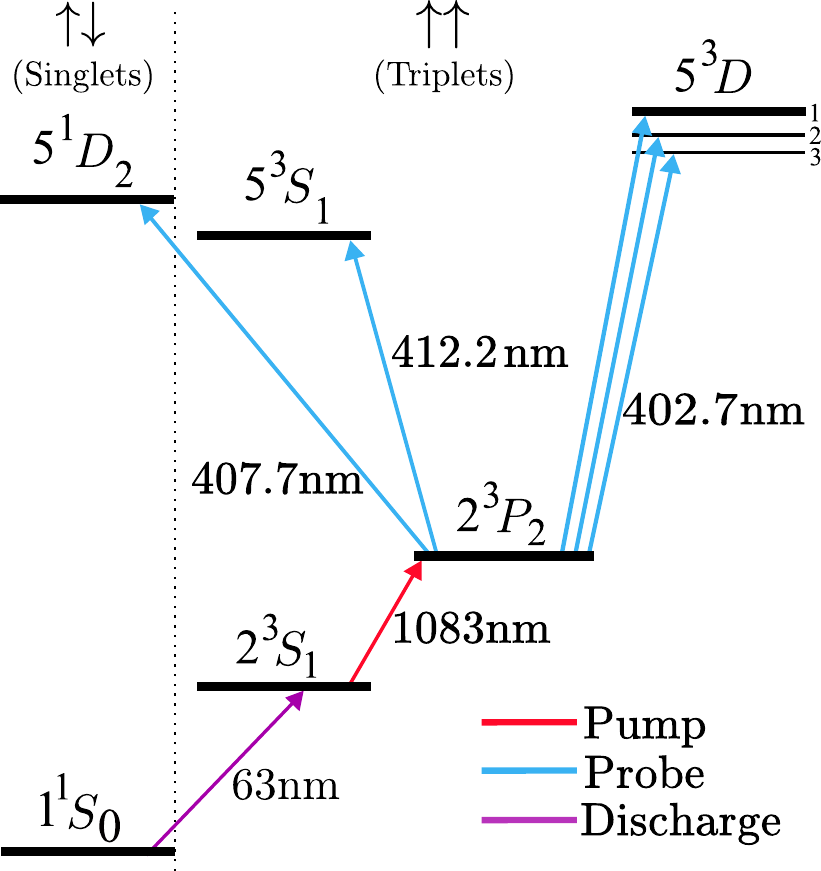}
      \end{minipage}
  \end{figure}

  A concurrent issue is the so-called `proton radius puzzle': Determinations of the proton charge radius from Lamb shift measurements in muonic and electronic hydrogen \cite{Pohl10, Bezginov19}, electron-hydrogen scattering experiments \cite{Beyer17,Xiong19}, and isotope shifts in light muonic atoms \cite{Kalinowski19,Pohl16} disagree significantly with both the CODATA recommended value and with other recent experiments \cite{Fleurbaey18}.
	 Helium is a promising candidate to provide insight into this unresolved issue because its simple structure is tractable to QED calculations.
	Furthermore, ongoing theoretical work \cite{Pachucki15,Pachucki17,Pachucki11,Pachucki10,Morton12,Morton06,Patkos16,Patkos17} and recent high-precision measurements \cite{Rooij11,Notermans14,Notermans16,Rengelink18} find a $4\sigma$ discrepancy between the difference $\delta = r^2(^3\textrm{He}) - r^2(^4\textrm{He})$ of squared nuclear charge radii obtained from the isotope shifts of the $\MetastableState - \PStateManifold$ and ${2^{1\!}S_{0}} - \MetastableState$ transitions \cite{Pachucki15,Patkos17}.
	The completed calculation of QED effects to order $\alpha^7$ will allow determination of the absolute nuclear charge radii accurate to better than 1\% \cite{Pachucki17}.
	Along with these $\alpha^7$ contributions, measurement of the $2\triplet S-2\triplet P$ spacing to within 1.4 kHz would allow a determination of the nuclear charge radius to below 0.1\% accuracy, better than expected from the muonic helium Lamb shift \cite{Wienczek19}.
	The corresponding QED calculations to seventh order have been completed (Ref. \cite{Patkos21}) since the publication of the work in this chapter. 
	The authors note, however, that the predicted energies of the $2\triplet S - 3 \triplet D$ and $2\triplet P - 3 \triplet D$ transitions disagree significantly with experiments, and refrain from computing the nuclear charge radius until these discrepancies are resolved.
	 
  Notable among recent studies of helium's structure are the measurements of \emph{forbidden} transitions between the singlet and triplet manifolds.
	Such transitions are made possible in reality due to relativistic effects and are extremely narrow, therefore precise measurements of their spectral features can provide stringent tests of QED \cite{Lach01}.
	The work in this chapter complements existing measurements of forbidden lines in helium at 1557 nm \cite{Rooij11,Rengelink18}, 887 nm \cite{Notermans14}, and 427 nm \cite{Thomas20}.

	This chapter concerns frequency measurements of the transitions from the $2^3P_2$ state to five states in the $n=5$ manifold of $^4$He, illustrated in Fig.
	\ref{fig:lvl_diag}.
	These measurements improve on the precision of previous measurements by at least an order of magnitude \cite{Martin60}, and resolve the fine structure splitting of the $\PStateManifold_2 - 5\triplet D$ transition for the first time.
	The results also include the first direct measurement of the spin-forbidden $2^{3\!}P_2 - 5^{1\!}D_2$ transition in helium, whose transition rate is four orders of magnitude smaller than the other transitions discussed here.  
	Theoretical transition energies agree with the observed values within our experimental uncertainty.

\section{Measurement technique}

	The experimental sequence began with $\sim10^8$ atoms in the metastable $2\triplet S_1$ state, cooled to  $\sim1~\textrm{mK}$ in a magneto-optical trap, which were then transferred to a magnetic trap by switching off the light.
	Next, during the Doppler cooling stage, we illuminated the atoms with $\sim$30$\mu$W$/m^2$ of $\sigma^+$ polarized cooling light, further cooling the atoms to $\sim200~\mu \textrm{K}$.
	Finally, forced evaporative cooling lowers the sample below the critical temperature to form a BEC.
	Each iteration of this procedure produced a BEC of $\sim 5\times10^5$ atoms in a cigar-shaped harmonic trap with trapping frequencies $\omega = 2\pi (425,425,45)$ Hz.
	To perform the absorption measurements, we disturbed the Doppler cooling stage by applying a probe beam tuned near the target transition, resulting in a reduction in the number of atoms cooled to degeneracy.

	At the end of the experimental sequence a pulsed atom laser, described in section \ref{sec:atomlaser}, transfers $\approx$2\% of the trap population at a time to the untrapped $m_J=0$ state \cite{Manning10,Henson18_BCR}.
	The resulting coherent matter-wave pulses fall onto the detector, {depleting the entire trapped population after $\sim$200 pulses over 2 seconds.
	This allows}  the atom number and temperature to be accurately determined without saturating the detector.
	Our data collection protocol consisted of a cycle of one calibration shot (with the probe beam switched off), followed by one measurement shot (with the probe beam on simultaneously with the Doppler cooling light) at each of two magnetic field strengths used in the in-trap cooling stage.
	We denote the atom number measured at the end of the calibration and measurement shots as $N_c$ and $N$, respectively.
	The signal is defined to be the normalized loss of atoms, in the form $(N_c-N)/N_c$.
	The reference population $N_c$ was estimated by linear interpolation between the successive calibration shots either side of a given measurement shot in order to compensate for drift in the trap population over time.
	
	The physical basis of the measurement is the sensitivity of forced evaporative cooling to the initial conditions of the helium atoms.
	The precise effect of photon scattering on the final cloud properties depends on the exact details of the evaporation sequence, and is hard to model exactly.
	Instead, what follows is a qualitative picture of the role evaporative cooling plays in transforming photon scattering to a measurable change in trap population.

	During the Doppler cooling stage of BEC creation, the 1083 nm cooling beam acts as an optical pump and excites atoms to the $2\triplet P_2(m_J=2)$ state.
	From the $2\triplet P_2$ state they may decay, with a lifetime of $\sim$97ns, back to the trapped metastable state or absorb photons from the probe beam and become excited again to the target state.
	Doubly-excited atoms may decay back to the trapped $m_J=+1$ state of the $2\triplet S_1$ level, in which case the photon absorption and emission events add heat to the cloud by imparting a nonzero average impulse to the atoms.
	This leads to an initially hotter cloud, which in turn reduces the efficiency of evaporative cooling, resulting in a higher atom loss during evaporation and a lower final number in the trap. Alternatively, atoms may decay to other untrapped magnetic states of the metastable state, or to the true ground state via a spin-flip transition to the singlet manifold.

	Decay to untrapped states reduces the initial atom number and can even impart heat to the cloud as these atoms leave the trap - via scattering with trapped atoms.
	This heating will be much smaller than in the previous case because the scattering rate will be small in such a dilute gas.
	However, reducing the initial trap population also manifests as a reduced final atom number.
	In both cases the effect of photon scattering manifests as a reduction of the total trapped final number $N$ relative to the final number $N_c$ in the calibration shots.

\begin{figure}
      \includegraphics[width=\textwidth]{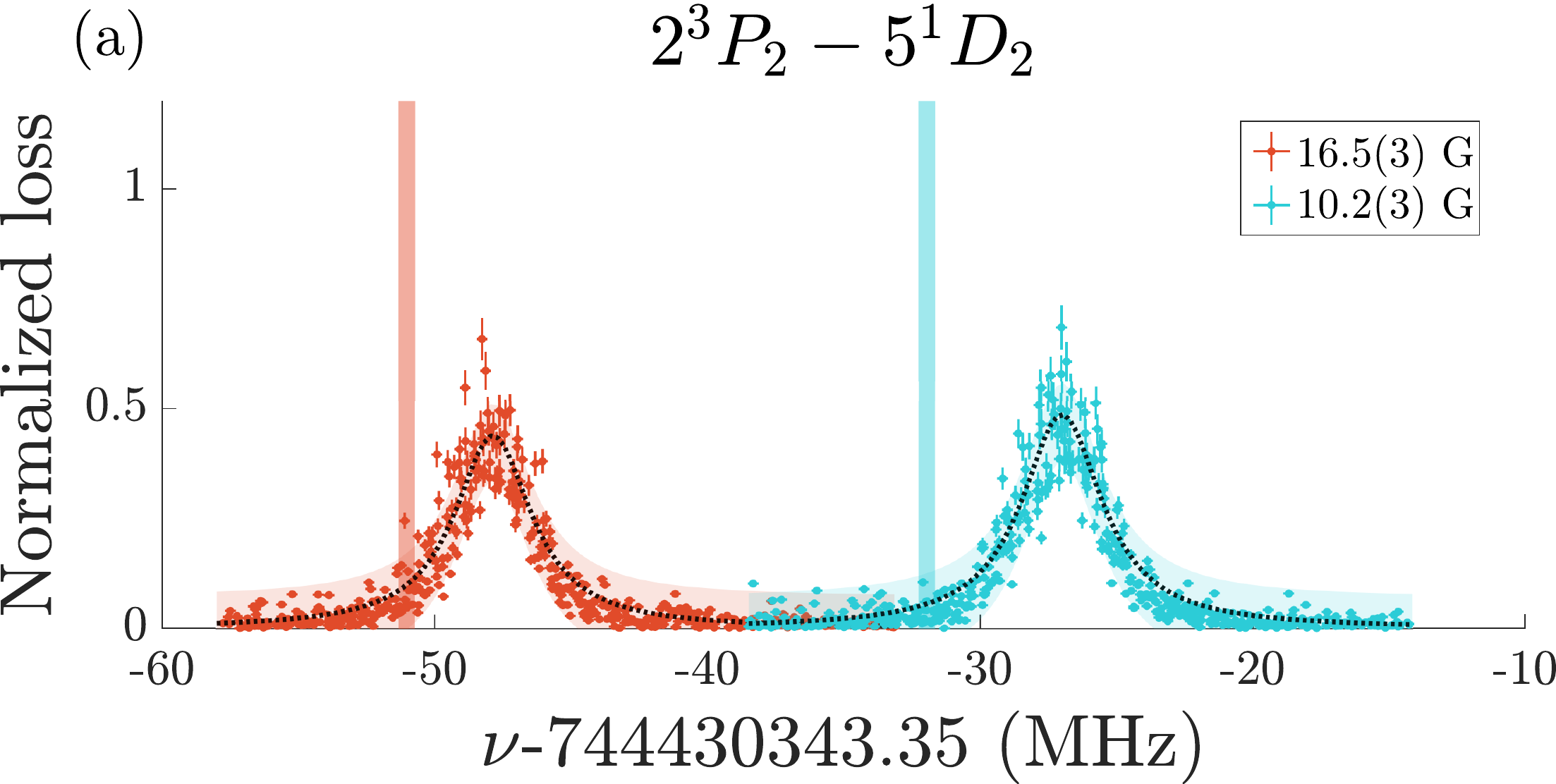}
    \includegraphics[width=\textwidth]{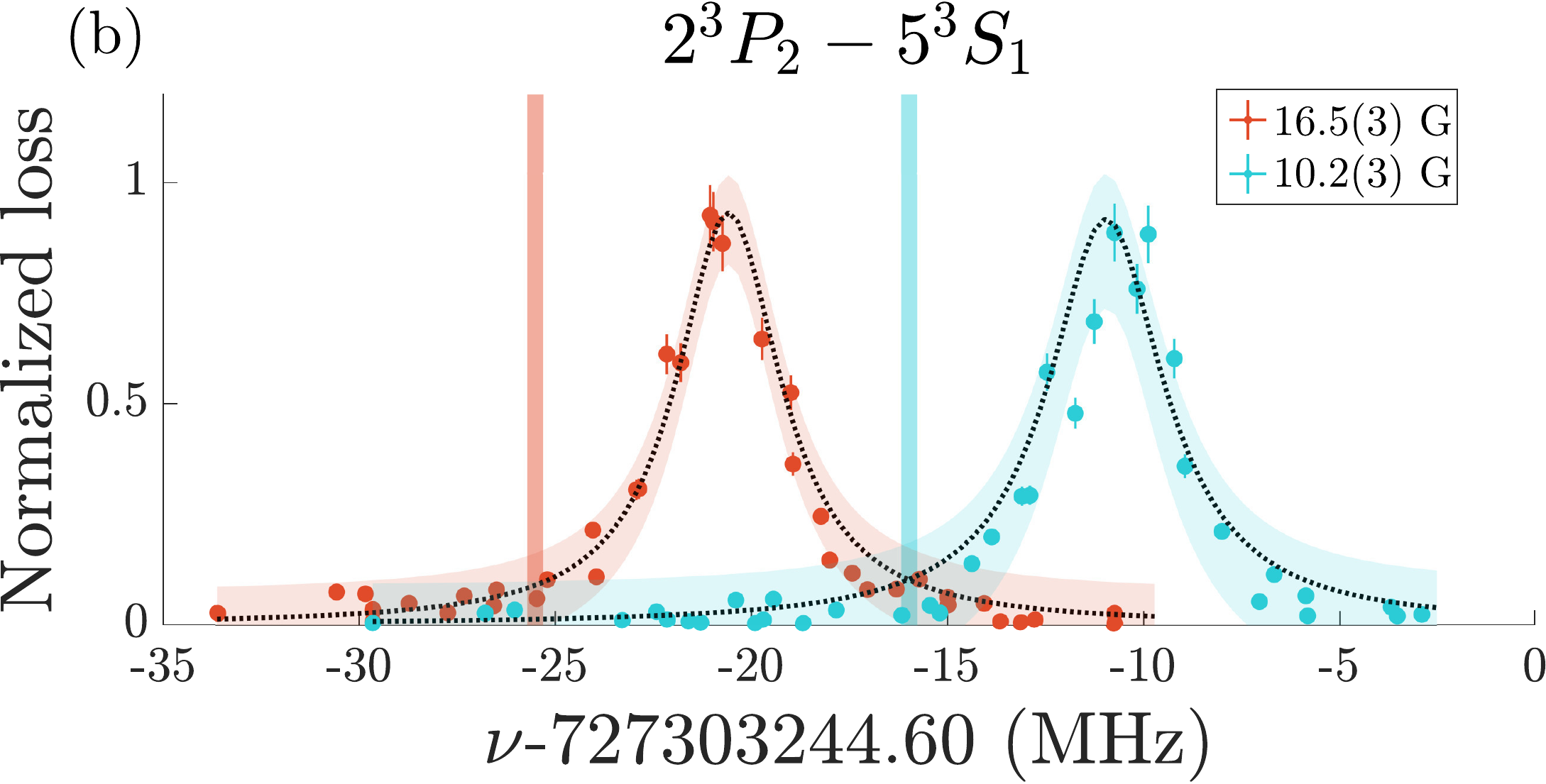}
   \caption{Line profile for (a) the spin-forbidden $\PStateManifold_2 -  5^{1\!}D_2$  and (b) the $2\triplet P_2 - 5\triplet S_1$  resonances, showing normalized atom number loss versus probe laser frequency $\nu$, as measured in a {16.5(3)} G (red) and {10.2(3)} G (blue) background field, with Lorentzian fits (black dotted line, with the observation interval shaded - see also section \ref{sec:fitting}).
	Error bars account for detector efficiency and calibration model uncertainty. 
	Vertical bars indicate theoretical predictions for the line centres, with the uncertainty (mostly due to uncertainty in the magnetic field) indicated by the width of the bars.
	For comparison, theoretical predictions (vertical bars) Zeeman shifted from the predicted zero-field value \cite{Drake07} according to the field calibration, whose uncertainty (shaded width) is dominated by background field measurements.}
    \label{fig:simple_lines}

\end{figure}

	We illuminated the atoms with the probe light (generated by the system described in section \ref{sec:spec_laser}) during the Doppler cooling stage, then measured the relative number loss as a function of frequency.
	For the $5\singlet D_2$ and $5\triplet S_1$ states, the exposure time was of order 100 ms. 
	The exact duration differed for each line to obtain a good SNR without saturating the atom loss.
	The light was $\sigma^-$ polarized to drive transitions to the $m_J=1$ states of the upper levels.
	For the forbidden $5\singlet D_2$ transition the beam was focused on the atom cloud with a waist of approximately $100~\mu$m and a peak intensity of order $5\times 10^3$ W/m$^2$.
	For all other measurements the beam was collimated with a peak intensity of order $ 5$ W/m$^2$.
	The Doppler cooling stage uses two magnetic field strengths and so measurements could be made with bias field strengths of {16.5(3)} and {10.2(3)} Gauss, which were calibrated independently by RF spectroscopy.
	For each field strength, the atom loss (with respect to calibration shots) versus probe laser frequency can be fitted by a Lorentzian lineshape to obtain the centre frequency and full-width at half-maximum (FWHM).
	Results of these measurements are shown in Tab.	\ref{fig:simple_lines}, after corrections for the AOM and vapor cell shifts (see the supplementary material of \cite{Thomas20} for details on the latter).
	The functional relationship between photon scattering and atom loss via evaporative cooling is complicated and not linear. 
	However, fits using a nonlinear function of the Lorentzian lineshape differed from the simple Lorentzian fit by substantially less than the statistical uncertainty.
	It would be interesting to study the relationship between the signal strength and photon scattering rate as a means to measure the oscillator strength of these transitions.
	This study would also need to characterize the population of the $2\triplet P_2$ in response to the probe beam. One approach would be to start by constructing a model of this intermediate process. It would be necessary to anchor this model to empirical reality, which may not be straightforward.

	After correcting for the linear Zeeman shift, the field-free transition frequencies are fixed with sub-MHz statistical uncertainty.
	This determines the $2\triplet P_2-5\singlet D_2$ and $2\triplet P_2-5\triplet S_1$ transition energies to be 3 MHz and 5 MHz larger, respectively, than the predictions presented in \cite{Drake07}.
	However, the absolute accuracy of these measurements is limited by the instrumentation.
	The results (Tab.	\ref{tab:spec_results}) are consistent with current predictions \cite{Drake07} within $2\sigma$ after accounting for all systematic uncertainties (Tab.	\ref{tab:errors}).

\section{5$\triplet$D fine structure}

\begin{figure}
	\centering
	\includegraphics[width=\textwidth]{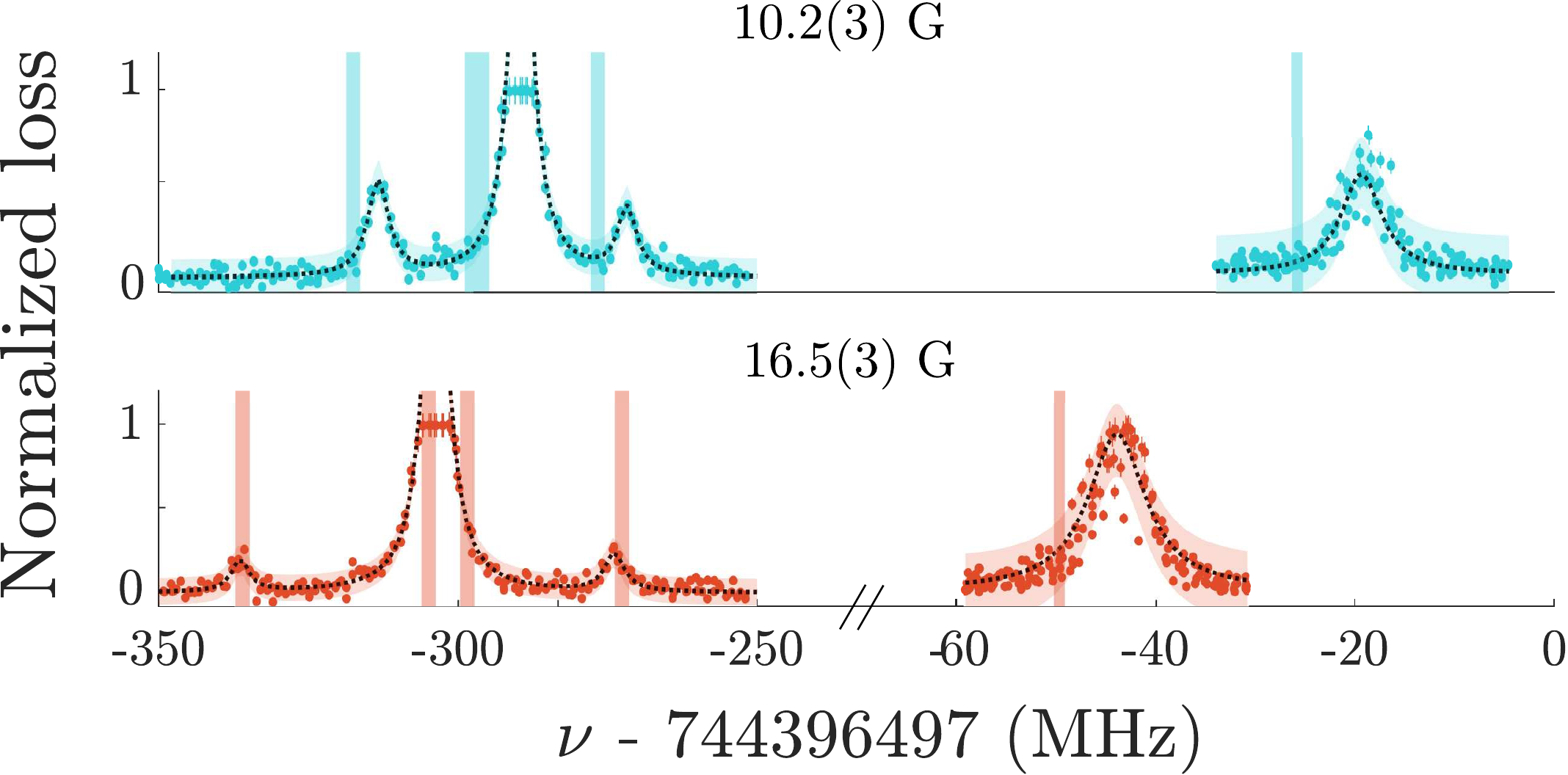}
   \caption{Line profiles for the $\PStateManifold_2 -  5^{3\!}D$ transitions, shown as for Fig.
	\ref{fig:simple_lines}.
	The normalized loss is shown versus probe laser frequency for the two different field strengths with a common horizontal scale.
	Theory lines indicate predictions from \cite{Drake07} after applying the relevant Zeeman shift.
	N.B.	the scale break here and in Fig.	\ref{fig:fitting_3D} coincide.}
    \label{fig:combined_5D_lines}
\end{figure}

	Unlike the $5\singlet D_2$ and $5\triplet S_1$ levels, the $5\triplet D$ manifold splits into fine structure sublevels, leading to multiple absorption peaks and requiring a more involved analysis.
	For this measurement, the final quarter-wave plate in the insertion optics was rotated to give a combination of $\pi$ and $\sigma^-$ polarized light (in the atomic frame) as opposed to pure $\sigma^-$ light.
	This light drove transitions to the $5\triplet D_J, J\in\{1,2,3\}$ levels and obtained four peaks, as shown in Fig.	\ref{fig:combined_5D_lines}.
	The saturated peak near -300 MHz (relative to the predicted $2\triplet P_2 - 5\triplet D_1$ interval) is in fact two peaks corresponding to the $5\triplet D_2(m_J=1)$ and $5\triplet D_3(m_J=2)$ states, which are separated by less than their linewidth. These peaks are excluded from the analysis below.
	This illustrates a shortcoming of the technique, namely the limited dynamic range.
	Specifically, that the sensitivity (roughly, the smallest detectable signal from a given number of measurements) is limited by the shot noise (the shot-to-shot variation) in atom number, whereas the maximum detectable signal is one that completely depletes the condensate. Therefore, when two peaks are close by, they both contribute to absorption and if they would otherwise individually deplete most of the condensate, their combined absorption peak will be saturated.

	For measurements of single peaks this is not an issue as the total irradiated energy can be adjusted to obtain a good signal-to-noise ratio without completely depleting the BEC.
	In this case, however, there is a trade-off between keeping the small peaks above the noise floor and preventing the superposed peaks from saturating.
	This limitation could be eased with a larger initial condensate because the dynamic range is essentially limited by the atom loss, with much larger magnetic field strengths which would ensure the lines are separated by much more than their linewidth, or by measuring the neighbouring peaks separately with a lower laser intensity (the cost of reduced accuracy in fitting the better-distinguished peaks could be recouped by the present measurement). 
	The other peaks correspond to transitions to the $5\triplet D_2(m_J=2)$, $5\triplet D_2(m_J=1)$, and $5\triplet D_1(m_J=1)$ states, which are used in the analysis described below.

	The Zeeman shift of the $J=2$ and $J=3$ levels is comparable to the interval between them, and so the mixing of levels means the correction is no longer proportional to $B$.
	Instead, the Zeeman shift can be corrected for by solving the eigenvalue optimization problem 
\begin{equation}
\textrm{min}_{E_{\textrm{fs}}} \sum_{J,m_J} \left(\nu_{{J,m_J}}^{\textrm{{pred}}}(E_{\textrm{fs}},B) - \nu_{{J,m_J,B}}^{\textrm{{obs}}}\right)^2,
\label{eqn:opt-problem}
\end{equation}
which minimizes the squared error between observed and predicted transition frequencies ($\nu^{\textrm{{obs}}}$ and $\nu^{{\textrm{pred}}}$ respectively), summed over all relevant $|J,m_J\rangle$ states and magnetic field strengths $B$.
	The optimized variable $E_{\textrm{fs}}=(E_1,E_2,E_3)$ is the bare fine-structure splitting of the $5\triplet D$ levels.
	In the argument below, the only assumption is the formalism of atomic structure theory and the data from the experiment.
	To determine the bare $5\triplet D$ transition energies from the data, consider the Hamiltonian
\begin{equation}
    \hat{H}(B) = \hat{H}_{\textrm{fs}} - B\hat{\mu}_z,
    \label{eqn:hamiltonian}
\end{equation}

	\noindent where $\hat{\mu}_z = \mu_B(\hat{L}_z + g_s \hat{S}_z)/\hbar$ is the coupling of the orbital and spin angular momenta of the electron with a magnetic field of strength B pointing in the $z$-direction, $\mu_B$ is the Bohr magneton and $g_s$ is the electron spin $g$-factor.
		The fine structure Hamiltonian $\hat{H}_{\textrm{fs}}$ is diagonal in the $|LSJ m_J\rangle$ basis with eigenvalues $E_{\textrm{fs}}$,
	\begin{equation}
	  \hat{H}_{\textrm{fs}}|LSJm_J\rangle = E_{\textrm{fs},LSJ}|LSJm_J\rangle,
	\end{equation}
	which are degenerate for all $m_J$ with fixed $J$.
		The magnetic moment $\hat{\mu}_z$ couples states of different $J$, and is instead diagonal in the $|L m_L S m_S\rangle$ basis.
		In the $|LSJ m_J\rangle$ basis the matrix elements of $\hat{H}(B)$ are, with abbreviated notation,
	\begin{equation}
	\begin{aligned}
	H_{J',J} =& \langle J'|\hat{H}|J \rangle\\
	  =& E_{\textrm{fs},J} - B \frac{\mu_B}{\hbar} \sum_{m_L} (2m_J - m_L)C_{J,'m_L}C_{J,m_L},\\
	\end{aligned}
	\end{equation}
	where $C_{J,m_L} = \langle LSJ m_J|L m_L S m_S\rangle$ is shorthand for the Clebsch-Gordan coefficients.
		For $B>0$, the contribution of $\hat{\mu}_z$ breaks the degeneracy of $\hat{H}_{\textrm{fs}}$, giving rise to the Zeeman shift.

	\begin{figure}
	\centering
		\includegraphics[width=0.9\textwidth]{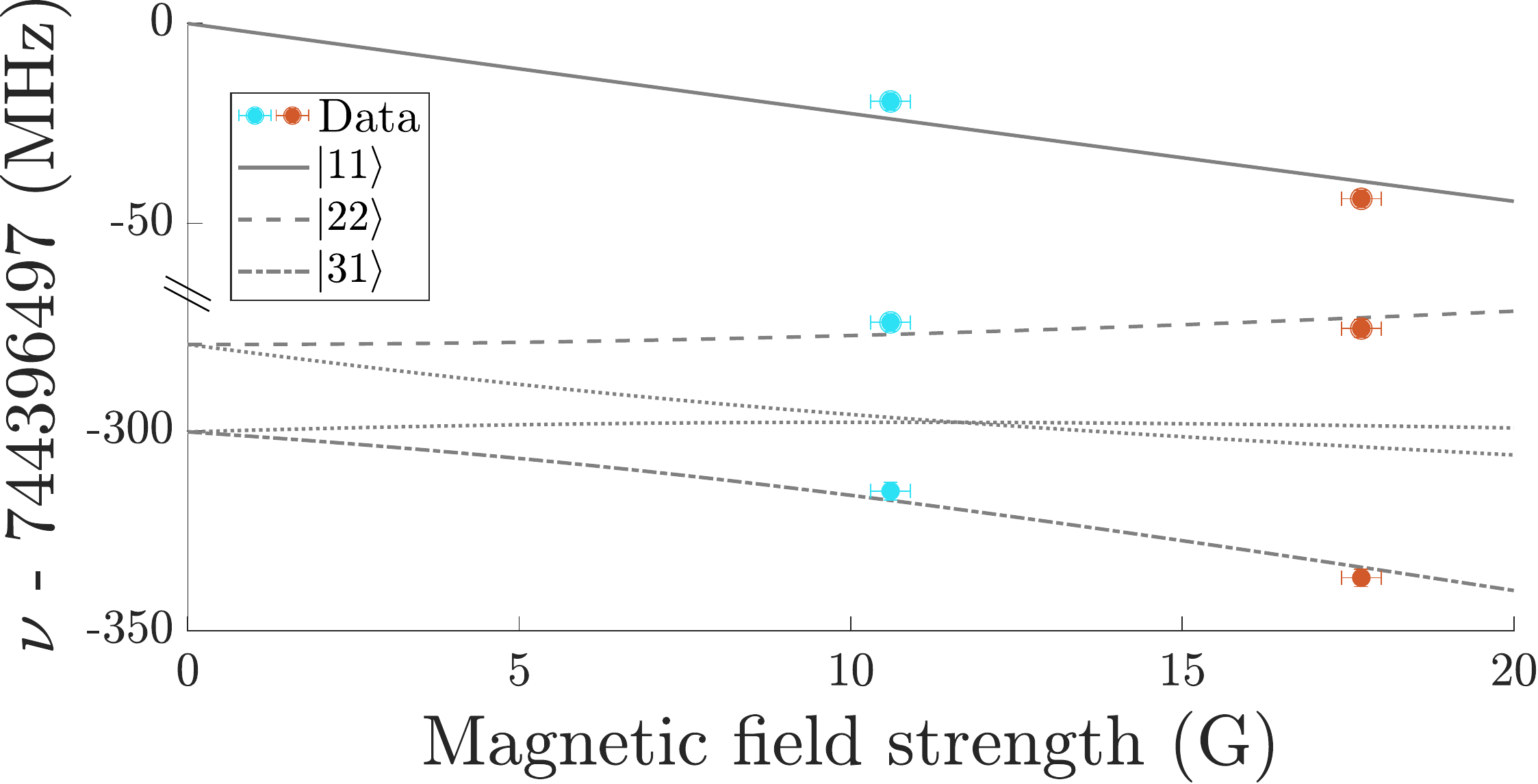}
		\caption{Determining the $5\triplet D$ fine-structure splitting.
		The values for the $|J,m_J\rangle=5\triplet D_J(m_J)$ levels (grey lines) at $B=0$ are fixed by solving the optimization problem (Eqn.
		\ref{eqn:opt-problem}), constrained by the fitted peak centres (filled circles).
		The saturated peaks are not used to constrain the levels and are not shown here. \com{}
		The corresponding frequencies predicted by the optimization method are shown in dotted lines.}
	    \label{fig:fitting_3D}
	\end{figure}

	The solution of Eqn.
	\ref{eqn:opt-problem} is illustrated in Fig.
	\ref{fig:fitting_3D}.
	By defining the energies $E_{\textrm{fs}}$ relative to the $2\triplet P_2(m_J=2)$ state,  the predicted transition frequencies can be read directly from the eigenvalues of $\hat{H}$ via $\nu_{J,m_J}^{\textrm{pred}}=E_{J,mJ}(B)/h$, where $h$ is Planck's constant.
	The observed frequencies $\nu_{J,m_J,B}^{\textrm{obs}}$ used in this procedure exclude the overlapping saturated peaks because their centre frequencies cannot be determined with satisfactory accuracy.
	The triple $E_{\textrm{fs}}$ which minimizes the cost function (Eqn.
	\ref{eqn:opt-problem}) corresponds to the $2\triplet P_2 - 5\triplet D_J$ intervals at $B=0$, as listed in Tab.
	\ref{tab:spec_results}.
	Again, the difference in this determination of the field-free splitting is consistent within 2$\sigma$ predictions in \cite{Drake07} after accounting for systematic effects (Tab.
	\ref{tab:errors}).

\begin{table*}
\centering

    \begin{tabular}{c c c c c c c c c c c}
      \hline\hline
      Transition                        &  $f_\textrm{exp}$ &  $f_\textrm{theory}$ & Diff.
		  &  $\textrm{FWHM}_{\textrm{exp}}$  &  $\textrm{FWHM}_{{\textrm{pred}}}$ \\
      \hline
        $2\triplet P_2 - 5^3\mathrm{S}_1$ &  {727,303,248(3)} &   727,303,244.6(4)   &  {3(3)}      &  3.4(5)  &  1.5\\ 
        $2\triplet P_2 - 5^3\mathrm{D}_1$ &  {744,396,496(7)} &   744,396,511.1(7)   &  {-16(7)}     &  5.8(6)  &  2.6\\
        $2\triplet P_2 - 5^3\mathrm{D}_2$ &  {744,396,220(7)} &   744,396,227.6(7)   &  {-8(7)}      &  4.2(5)  &  2.6\\
        $2\triplet P_2 - 5^3\mathrm{D}_3$ &  {744,396,194(7)} &   744,396,208.3(7)   &  {-14(7)}     &  4.0(1)  &  2.6\\
        $2\triplet P_2 - 5^1\mathrm{D}_2$ &  {744,430,343(7)} &   744,430,343.1(7)   &  {0(7)}      &  3.2(1)  &  2.2\\  
      \hline\hline
    \end{tabular}
\caption{Summary of results for each transition.
	After correcting for the AOM and vapor cell shifts, the centre frequencies obtained from Lorentzian fits have statistical error at the 10 kHz level.
	The field-free energies follow after correcting for Zeeman shifts, shown with theoretical predictions in the row below.
	The difference between our measurements and theoretical predictions is comparable with the experimental error.
	Observed full width at half maximum line widths (FWHM) of the Lorentzian fit to each line are shown in comparison to predicted linewidths as given in \cite{Drake07}.
	$f_\textrm{theory}$ for the $2\triplet P_2 - 5^3\mathrm{S}_1$ line comes from \cite{Drake07}, all others from \cite{Yerokhin20}
	All values are in MHz with uncertainty in the final digit in parentheses.}
  \label{tab:spec_results}
\end{table*}

\section{Shifts, broadening, and errors}

	The accuracy of these determinations of the field-free transition energies is limited by the absolute accuracy of the wavemeter.
	High Finesse specifies \cite{HighFinesseDoc} a 3$\sigma$ accuracy of 2 MHz within 2 nm of a calibration line (as in the transition to the $5\triplet S_1$ state), and 10 MHz for all other lines measured.
	Because the wavemeter is used to lock the seed light for the doubler, the uncertainty is doubled in determinations of the absolute probe frequency.
	In Table \ref{tab:spec_results} the corresponding 1$\sigma$ accuracy is shown in order to be consistent with other terms in our error budget, as displayed in Table \ref{tab:errors}.
	Note that this specification does not depend on the specific difference between the calibration and measured wavelengths, and may vary due to nonlinear dispersion of the wavemeter optics.
	Without an independent calibration this error cannot be rigorously constrained, which would be overcome with the instrumental improvements discussed below.
	As such, the 1-sigma errors determined in this way are presented with the caveat that they may be slightly underestimated.
	Still, all measured frequencies are consistent with predictions to within 2.1$\sigma$.
	Finally, the $5\triplet S_1$ transition line is 1.9 nm away from the calibration line, and as such the 2 MHz uncertainty may again be a slight underestimate.

\begin{table}
\centering
  \begin{tabular}{c c c}
      \hline\hline
          Source & Shift & Broadening  \\
      \hline
          Wavemeter ($5\triplet S_1$)& 0(1.3) & - \\
          Wavemeter (all other lines)& 0(6.7) & - \\
          Pump lock & - & 4$\times10^{-2}$ \\
          Pump AOM & - & 0.3 \\
          Probe lock & - & 0.3\\
          Probe AOM & -189 & 1$\times10^{-6}$\\
          Zeeman & Variable & Variable \\
          Recoil & - & 1.4$\times 10^{-3}$ \\ 
          Doppler & - & 2.7(4) \\
          Interference effects & 0.5 & - \\ 
          Cs cell & -1.9 & 0.4 \\
          \textbf{Total} ($5\triplet S_1$ level) & -190.9(1.7)+ZS& 2.2\\
          \textbf{Total} (all other levels) & -190.9(6.7)+ZS& 2.2\\
      \hline\hline
  \end{tabular}
\caption{Error budget for the determination of the peak centre frequencies.
	 The master laser for the pump beam is described in \cite{Shin16}.
	AOM stabilities were checked with an RF spectrum analyser.
	See \cite{Thomas20} for measurement of the Cesium cell shift and probe beam lock drift.
	The shift and uncertainty from the Zeeman shift (ZS) varies between the lines, so these contributions are omitted from the total.
	All values are in MHz.}
  \label{tab:errors}
  
\end{table}

	The linewidth of the pump and probe laser sources are 40 kHz \cite{Shin16} and 200 kHz \cite{Thomas20}, respectively.
	The laser lock error has a standard deviation of $100$ kHz.
	The additional contribution from the pump and probe AOMs are 300 kHz and 1 Hz, respectively, as determined with an RF spectrum analyser.
	In the first case, the linewidth is a result of frequency instability in the RF drive generation system. In the second, a newer drive system was in use which afforded better performance.

\begin{figure}
	\centering
  \includegraphics[width=\textwidth]{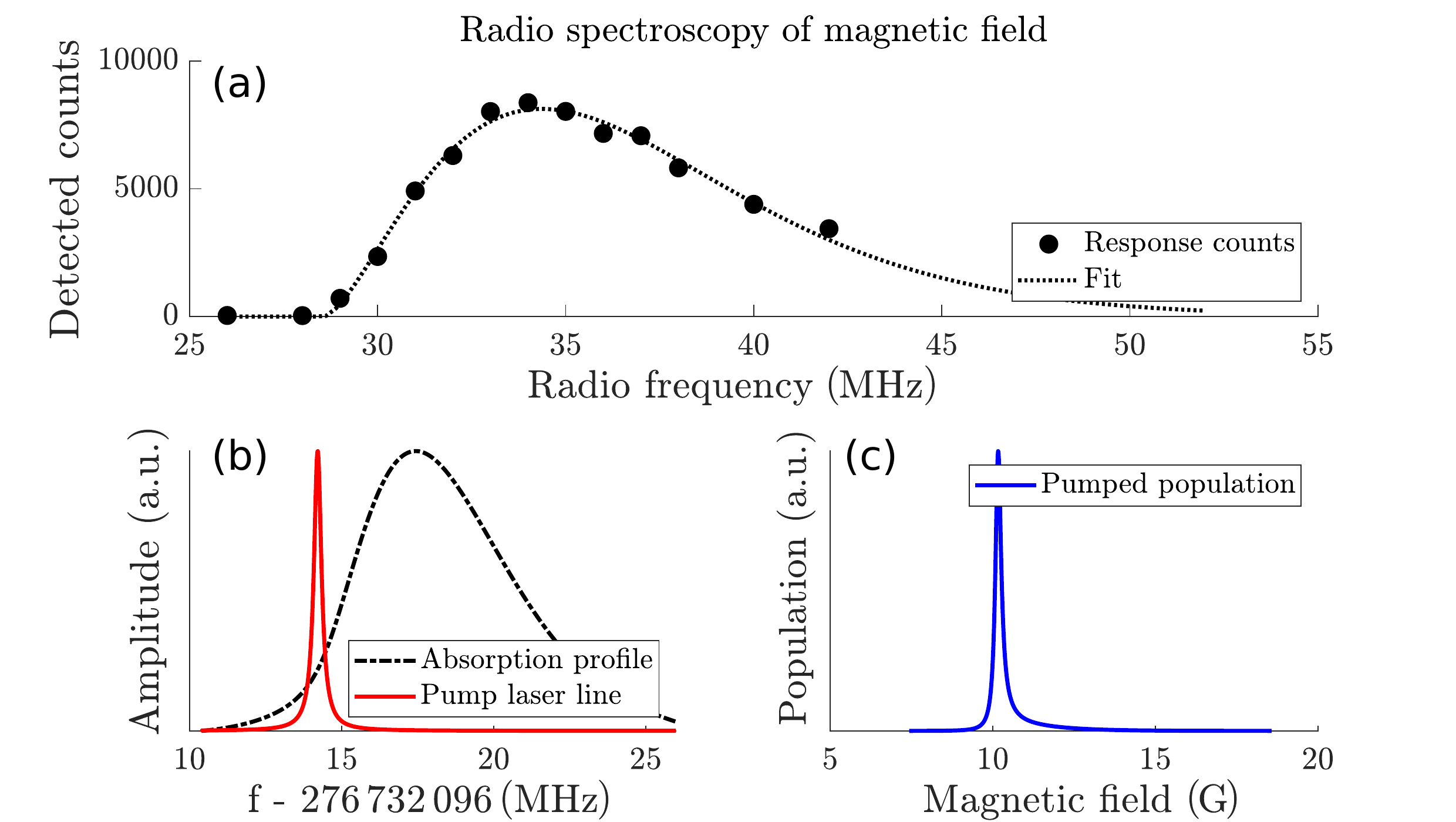}
  \caption{Determination of magnetic field for Zeeman shift correction.
	The number $n_\text{D}$ of atoms detected after probing the trap with 300 ms of RF radiation is shown in (a) versus the frequency of the applied radiation.
	Each point is an average of three shots.
	A Bose-Einstein fit models the population density $n_\text{T}$ of the $2\triplet S_1$ state at a given field strength.
	In (b) the calculated absorption profile of the gas in the vicinity of the 1083 nm pump transition is shown, along with the spectral profile of the pump laser. N.B. the frequency axis is in units of MHz, relative to the resonance as specified in THz.
	The pump laser selectively excites atoms with a certain Zeeman splitting, leading to a population of atoms in the $2\triplet P_2$ state (c) that is concentrated around a specific magnetic field strength.
	The resulting Zeeman broadening is dominated by the probe beam linewidth.}
  \label{fig:RF_spec}
\end{figure}

{I obtained the magnetic field calibrations by measuring the number of atoms detected by the MCP-DLD after applying 300 ms of RF radiation at variable frequencies (see Fig. \ref{fig:RF_spec}).
The number $n_\delta(\nu)$ of detections probes the population of trapped atoms $n_\text{T}(B)$ subject to a magnetic field strength $B$ through the relation $2 \mu_B B = h \nu$.
	
As shown in Fig.
	\ref{fig:RF_spec}, $n_\text{D}(B)$ follows a Bose-Einstein distribution with a chemical potential $g \mu_B B_0$, where $B_0$ is the bias in the magnetic field strength and $g\approx2$ is the g-factor of the $2\triplet S_1$ state.
	The Bose-Einstein fit provides a model for $n_\text{T}(B)$ and an estimate of the temperature of the cloud at each stage of the cooling process (and thus at each field strength).
	We can use the average temperature of $\sim130(20)$ $\mu$K to estimate  Doppler broadenings of 100(10) kHz and 2.6(3) MHz for the pump and probe transitions, respectively.

	Kinetic effects do not contribute any significant uncertainty in these frequency measurements, rather they just broaden the observed peaks.
	The pump light was applied by two counterpropagating beams, subtending angles of 15$^\circ$ and 195$^\circ$ relative to the direction of propagation of the probe beam.
	Photon absorption events contribute a recoil velocity of magnitude $\cos(15^\circ)\cdot\hbar k/m\approx6$mm/s, imparting a Doppler shift of order 1.4 kHz.
	Atoms may absorb probe light after absorbing a photon from the pump beams, but not after decaying again, so the decay events do not contribute.
	Because there were two counterpropagating pump beams, the resulting contribution is a negligible broadening, especially in comparison with the thermal broadening.
	The 1-3 MHz difference between predicted and observed line widths is well accounted for by these broadening effects, mainly by Doppler broadening.

	The optical absorption profile near the 1083 nm pump transition can be calculated by convolving $n_\text{T}(B)$ with Zeeman-shifted absorption profile of the 1083nm transition, which is a Lorentzian $\mathcal{L}_\text{abs}(f,B)$ with a 1.6 MHz FWHM \cite{Drake07}.
	The pumping rate at a given field strength is given by the convolution of $\mathcal{L}_\text{abs}(f,B)$ with the pump laser line $\mathcal{L}_\text{L}(f)$.
	Hence, the range of magnetic field strengths $B$ at which atoms were pumped to the $2\triplet P_2$ state were concentrated at field strengths of 10.2(3) G and 16.5(3) G for the two measurement stages.

	The Zeeman shift of the $2\triplet P_2 - 5L, L\neq D$ transitions is given by $\Delta E = \mu B (g_e m_e-g_g m_g)$, whose error is obtained by standard propagation of uncertainties.
	
	For the $5\triplet D$ states, the uncertainty in the iterative method described above follows from varying the magnetic field constraints within the range of experimental uncertainty ($0.3$ G). }

\begin{figure}
  \begin{minipage}[t]{0.6\textwidth}
  \vspace{0pt}
  \includegraphics[width=\textwidth]{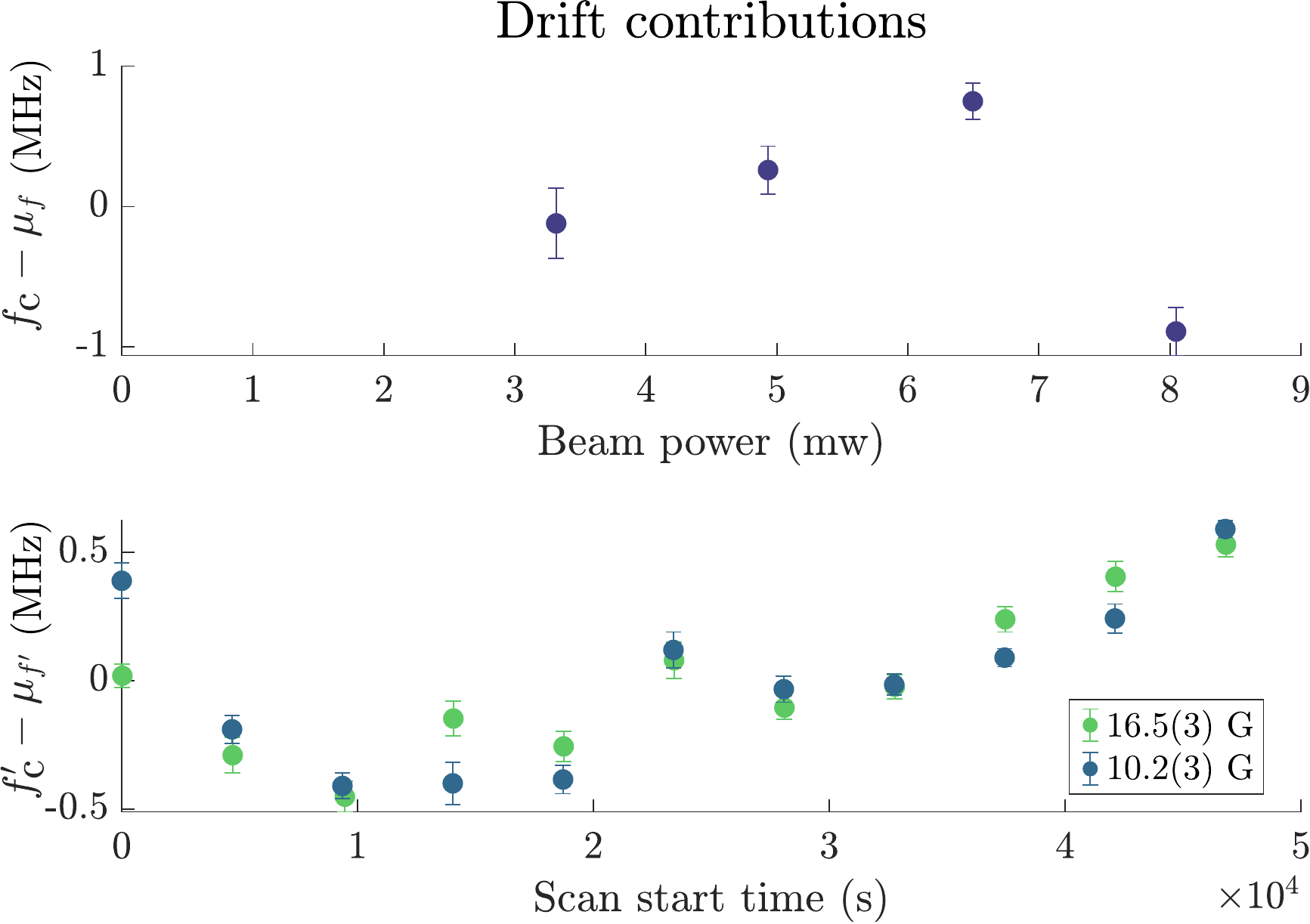}
  \end{minipage}\hfill
  \begin{minipage}[t]{0.38\textwidth}
  \vspace{0pt}
\caption{Top: Variation in fitted centre frequency for single scans across the $5\triplet D_1$ line versus applied laser power.
	The measurements at increasing beam power were not taken in chronological order Bottom: Variation in fit center frequency for the $2\triplet P_2 - 5\singlet D_2$ between scans.
	The value of the fitted peak centre $f_\textrm{c}$ is shown for each field strength, relative to the mean $\mu$ of all values for that field strength.}
\label{fig:power_drift_combined}
  \end{minipage}
\end{figure}

	Other precision measurements of transition frequencies have been shown to be subject to line pulling effects \cite{Marsman15,Marsman15PRA}.
	These effects arise in multilevel transitions because of interference between the laser-driven transition path and off-resonant driving through transitions with neighbouring intermediate states.
	The worst-case shift can be approximated by $w_{\text{pump}}^2/\Delta_{2P} + w_{\text{probe}}^2/\Delta_{5L}$ \cite{Marsman15,Marsman15PRA}, where the $w$ terms are the linewidths of the pump and probe transitions, and the $\Delta$ terms are the Zeeman splittings between the sublevels of the pump and target states.
	The largest estimate among all the reported transitions is 500 kHz.
	While this uncertainty is dominated by other effects in our experiment, it may be important to understand them for improved measurements in the future.

	There was no significant detectable contribution from the AC Stark effect.
	This can be seen in Fig.	\ref{fig:power_drift_combined}: The rightmost data point in the top panel was taken first, followed by the others in increasing order. Two conclusions are apparent: First, the center frequency drifted up during the course of the four scans, and second, that such drift dominated any effect the beam power has on the center frequency. We can characterize the time-dependent drift by examining the drift in center frequency over time with fixed beam power, as shown in the bottom panel, which depicts the individual results of each scan of the $5\triplet D_1$ transition. A drift on the order of 1 MHz per hour is seen, which is comparable to that found in the data shown in the top panel. Separately, after locking the spectroscopic laser system to the Cs reference transition, the inherent drift of the wavemeter can thus be determined, and explains the variation in centre frequency shown in Fig.	\ref{fig:power_drift_combined}.

	The noise floor is ultimately determined by two main factors: The variation in population of subsequent realizations of BEC (shot noise) and the Bernoulli process resulting from imperfect detection efficiency.
	The shot noise is by far the dominant noise process, with a standard deviation at the level of $\approx 10\%$, on the order of $5\times10^4$ atoms in the worst case. 
	For comparison, the relative error induced by imperfect detection can be modeled as a Bernoulli process. 
	Assuming a detection efficiency of $\eta=8\%$, the respective relative error is given by $\sqrt{\textrm{Var}_N}/N=\sqrt{N\eta(1-\eta)}/N\approx0.86/\sqrt{N}$, where $N$ is the atom number, which is below 1\% for the largest clouds. 
	The large-$N$ limit determines the noise floor as the signal is calculated by the relative difference between measurement (probe-on) and calibration (probe-off) shots.
	It should be noted, then, that an improvement in the precision of the number detection would not lead to a consequential reduction in the noise floor, as it is the underlying shot noise itself that is the limiting factor.
	Therefore there is no distinct advantage of using either the MCP-DLD detector or optical absorption imaging, as the latter can also obtain sub-percent precision in single shots \cite{Ockeloen10}.

\section{Discussion}

	This chapter described multilevel laser absorption spectroscopy of excited state transitions in ultracold helium.
	The observations include the first observation (to my knowledge) of the forbidden $2\triplet P_2 - 5\singlet D_2$ transition.
	These measurements agree with current predictions within the error budget and suggest that the $93\sigma$ difference between previous measurements \cite{Martin60} and predictions \cite{Morton06} of the $\PStateManifold_2  -  \UpperS$ and $\PStateManifold_2  -  \SpecUpperStateManifold$ intervals are due to an unknown systematic error\footnote{As a historical note on the advancement of experimental methods, Martin's measurements were made using a nitrogen-cooled helium discharge lamp fed through an in-vacuum prism onto photographic plates, on which the line separations were measured by hand with a ruler.}.
	This work offers five contributions to the NIST database of atomic spectral lines.

	The techniques described here are readily extensible to other opportunities in $^4$He structure measurements.
	For example, while there is an outstanding 7.4$\sigma$ disagreement between the predicted and observed singlet-triplet interval for the n=3 level in $^3$He \cite{Morton06,Derouard80}, the corresponding transition in $^4$He has never been directly measured.
	An indirect measurement in $^4$He could be made with the techniques described here by taking the difference betweeen the $2\triplet P_2 - 3\triplet D_2$ and $2\triplet P_2 - 3\singlet D_2$ transitions near 587.6 nm and 587.4 nm.
	While the latter transition is also spin-forbidden, it is predicted to be an order of magnitude stronger than the $2\triplet P_2 - 5\singlet D_2$ transition reported here \cite{Morton06}.

	{Further, energies of other $2L-nD$ transitions in $^4$He are a few MHz larger than predicted, apparently independent of $L$ \cite{Wienczek19,Yerokhin20}.
	The results here deviate from this trend, and invite independent verification.
	Further study of transitions between states from different shells to MHz precision or better, in particular the prospective study of the $2\triplet P - 3 D$ intervals, would also provide further clarification.}

	Simply exchanging the light source would suffice to make these measurements, but a definitive comparison with theory would require an improved frequency reference.
	The associated theoretical uncertainties are dominated by the 700 kHz uncertainty in the lower state \cite{Pachucki17,Wienczek19}.
	As the $\alpha^7$ terms could improve the theoretical accuracy to as little as 10 kHz \cite{Pachucki17}, this more challenging precision appears to be a more appropriate budget, and readily achievable with current methods.
	
	Reference-locked optical frequency combs can readily achieve kHz accuracy or better \cite{Luo15,Rengelink18}.
	The comb would need to be integrated in a similar technique to that in Ref. \cite{Rengelink18}, which achieved a 200 Hz uncertainty in their measurement of the $2\triplet s_1 \rightarrow 2\singlet S_0$ transition.
	In a suitable setup, a reference-locked comb would be used in a phase-locked loop to transfer-lock the probe light, or at least light that could be fed into a doubling cavity.
	Such an upgrade would eliminate the systematic uncertainties due to the wavemeter, Cesium cell, and probe lock loop.
	It would also be prudent to replace the RF driver on the pump beam AOM, whose RF linewidth was on the order of 300 kHz, with an oscillator of comparable quality to that in the probe AOM (with a 1 Hz linewidth, as is also used in the main cooling laser lock).
	The next leading uncertainty is associated with and Zeeman effect.
	Magnetic field strengths can be determined by RF spectroscopy with sub-kHz accuracy and so would not present a serious limitation.
	One could also consider illuminating a trapped cloud with an exceptionally weak pump beam on resonance, and scan the probe beam across the target transition while measuring the difference in heating rate (similar to our measurement of the  427 nm forbidden transition \cite{Thomas20}), without using the evaporative cooling ramp. 

	Extending these methods to direct measurements on $^3$He would also permit isotope shift measurements from forbidden excited-state transitions in $\triplet$He.
	Theoretical calculations of isotope shifts are already accurate to the sub-kHz level, so such measurements would be even more demanding than the prospects above.
	Existing demonstrations of comparable accuracy \cite{Rengelink18} show such measurements are worthy challenges whose completion can access nuclear structure information at the femtometre scale via optical atomic spectroscopy.

\vfill

\begin{flushright}
\singlespacing
\emph{
``Immediately you would like to know where this \\
number for a coupling comes from: is it related to pi\\
or perhaps to the base of natural logarithms? \\
Nobody knows.\\
It's one of the greatest damn mysteries of physics: a magic number \\
that comes to us with no understanding by man. You might say \\
the ``hand of God" wrote that number, and ``we don't know how\\
He pushed his pencil." We know what kind of a dance to\\
do experimentally to measure this number very accurately,\\
but we don't know what kind of dance to do on the computer\\
to make this number come out, without putting it in secretly!"}\\
- Richard P.
Feynman \footnote{\emph{QED: The Strange Theory of Light and Matter}.
	Princeton University Press.	p.	129.	(1985)}
\end{flushright}
\onehalfspacing

%% file: latex/22_tuneout.tex
\chapter{Precision measurement of the 413~nm tune-out point}
\markboth{\thechapter. MEASUREMENT OF THE 413 nm TUNE-OUT}{}
\label{chap:tuneout}

\blankfootnote{\noindent The contents of this chapter relate to the work published in \textbf{Measurement of a helium tune-out frequency: an independent test of quantum electrodynamics} by B. M. Henson$^\dagger$, {J. A. Ross}$^\dagger$, K. F. Thomas, C. N. Kuhn, D. K. Shin, S. S. Hodgman, Y. H. Zhang, L. Y. Tang, G. W. F. Drake, A. T. Bondy, A. G. Truscott, K. G. H. Baldwin, \href{https://www.science.org/doi/10.1126/science.abk2502}{\emph{Science} \textbf{376}} (2022) ({$^\dagger$\emph{Equally-contributing authors})}}

	\begin{flushright}
	\singlespacing
	{\emph{``Turn on\\
			Tune in\\
			Drop out"\\} 
	Tim Leary \cite{LearyNote}.}
	\end{flushright}
	\onehalfspacing
	\vspace{1cm}

	\noindent{Nothing} in the universe is truly motionless, as imposed by the ineradicable zero-point energy and uncertainty principle.
	And yet, atoms may be unmoved by an electric field if the latter oscillates at a \emph{tune-out} point.
	In vacuum conditions, the specification of the wavelength, frequency, or photon energy of light are all equivalent. 
	Thus, I use the encompassing term `tune-out point' where the context does not favour any particular unit of measurement.
	However, in keeping with the tendency in metrology, the quantity of interest in this chapter is the tune-out \emph{frequency}, or simply \emph{tune-out} for brevity. 
	In the publication pertaining to this chapter (Ref. \cite{Henson22}) we introduced the notation $L-U_1/U_2$ to specify a tune-out point by the occupied state $L$ followed by the two transitions $U_1$,$U_2$ which dominate the polarizability at the specified tune-out. 
	This chapter concerns the measurement of the frequency of the $\MetastableState-2\triplet P/3\triplet P$ tune-out point near 413 nm (726 THz) \cite{Henson15,Mitroy13} and a comparison with state-of-the-art atomic structure calculations as a test of quantum electrodynamics.
	Interestingly, the tune-out point is usually determined by the position and strength of the nearest transition frequencies. 
	However, in this case the tune-out is principally controlled by the $2\triplet P$ and $3\triplet P$ levels (with transitions at 277 and 771 THz, respectively) despite the $3 \triplet S_1$ being only $701$ THz away. This is because the  $\MetastableState \rightarrow 3^{3\!}S_1 $ transition is extremely weak\footnote{We detected this transition during the laser spectroscopic campaign which included the present and prior chapters. As reported in Ref. \cite{Thomas20}, this is at present the weakest transition detected in a neutral atom.}.

	As this work constitutes the most detailed work undertaken in the course of study, and the longest chapter, a brief mention is warranted of the way ahead.
	Section \ref{sec:TO_bg} includes a detailed definition of a tune-out point with aid of semiclassical models along the way. This section also discusses the motivation for the work and a short summary of the theoretical progress made in parallel to the experimental measurement, which is reported in section \ref{sec:trap_freq_measure}. The foci of section \ref{sec:trap_freq_measure} are the description of the physical process underpinning the probe and measurement methods, including a fine-grained discussion of the effects of the frequency and polarization of the probe laser at the interaction zone. Having established the working principles of the measurement, section \ref{sec:TO_analysis} gives an end-to-end account of the data analysis. This section covers the criteria for diagnosing failure modes and discarding aberrant data files, the extraction of a final measured value and quantifying its statistical confidence, and some sanity checks to assure ourselves that our model maps well onto the ground truth. Section \ref{sec:systematic_effects} enumerates the systematic effects that we quantified in order to determine the ultimate accuracy of the measurement. Finally, section \ref{sec:TO_discussion} summarizes the findings of the experiment, including the establishment of a handful of figures-of-merit and a claim to a new sensitivity record.

\section{Background}
\label{sec:TO_bg}

\subsection*{Feynman's jewel}
	
	Quantum electrodynamics describes the interaction between matter and light, an interplay so ubiquitous that the theory is regarded a cornerstone of modern physics.
	QED has yielded extraordinarily accurate predictions about fundamental processes such as spontaneous emission rates of photons from atoms and the anomalous electron magnetic moment \cite{Aoyama15}.
	However, despite withstanding decades of stringent testing against high-precision measurements, recent results in atomic spectroscopy have revealed discrepancies between experiment and theory.  
	With the precision of atomic spectroscopy approaching the part-per-trillion level, the `proton radius puzzle' has emerged, wherein determinations of the proton radius from spectroscopic measurements (of muonic \cite{Pohl10} and eletronic hydrogen \cite{Bezginov19,Beyer17}, and of muonic deuterium \cite{Pohl16}) disagree by up to five standard deviations with other approaches (electron-proton scattering \cite{Zhan11} and hydrogen spectroscopy\cite{Fleurbaey18}).

	Helium is an exemplary system for complementary tests of QED thanks to its simple two-electron structure, which makes high-precision predictions tractable and testable. 
	Notably, helium presents a nuclear `puzzle' of its own: precision measurements of the isotope shifts of the \(\MetastableState \rightarrow \LowerStates \) \cite{Zheng17} and \(\MetastableState \rightarrow \SingletState \) \cite{Rengelink18} transitions disagree by two standard deviations in the derived nuclear radius shift $r(^4\textrm{He})^2-r(^3\textrm{He})^2$. 
	Further, recent measurements of the ionisation energy for the helium $2^{1\!}S_0$ state \cite{Clausen21} confirm similar discrepancies in the Lamb shift to those recently revealed theoretically \cite{Patkos21}.
	These puzzles raise the possibility that QED itself may be a flawed jewel \cite{Hill17} - and that some new physics may be within experimental reach in the atomic physics laboratory. 

	The tune-out frequency, where an atom feels no force resulting from applied laser light, is an observable that tests QED independently of the conventional measurement of energy level differences.
	We selected the $\MetastableState-2\triplet P/3\triplet P$ tune-out as our target quantity because the nearest two transitions are separated by more than an octave, and thus the atomic polarizability must vary slowly with optical frequency near the tune-out. 
	Therefore, this tune-out is especially sensitive to higher order QED effects. 
	We achieve a 15-fold improvement in the precision over the sole previous measurement (which was also undertaken in the ANU \mhe lab \cite{Henson15}) and make a definitive comparison with theory thanks to a new theoretical prediction of the \(\TO\) tune-out (summarized in section \ref{sec:to_theory}).
	The experimental value of 725\,736\,700\,$(40_{\mathrm{stat}},260_{\mathrm{syst}})$~MHz is within \({\sim} 2.5\sigma\) of the predicted frequency (725\,736\,053(9)~MHz),   resolving contributions from QED effects (\({\sim} 30 \sigma\)) and probes finite-wavelength retardation corrections (\({\sim} 2 \sigma\)) for the first time, albeit with lower confidence.

\subsection*{Tune-out points}
\label{sec:TO_points}
\begin{figure}
		\centering
		\includegraphics{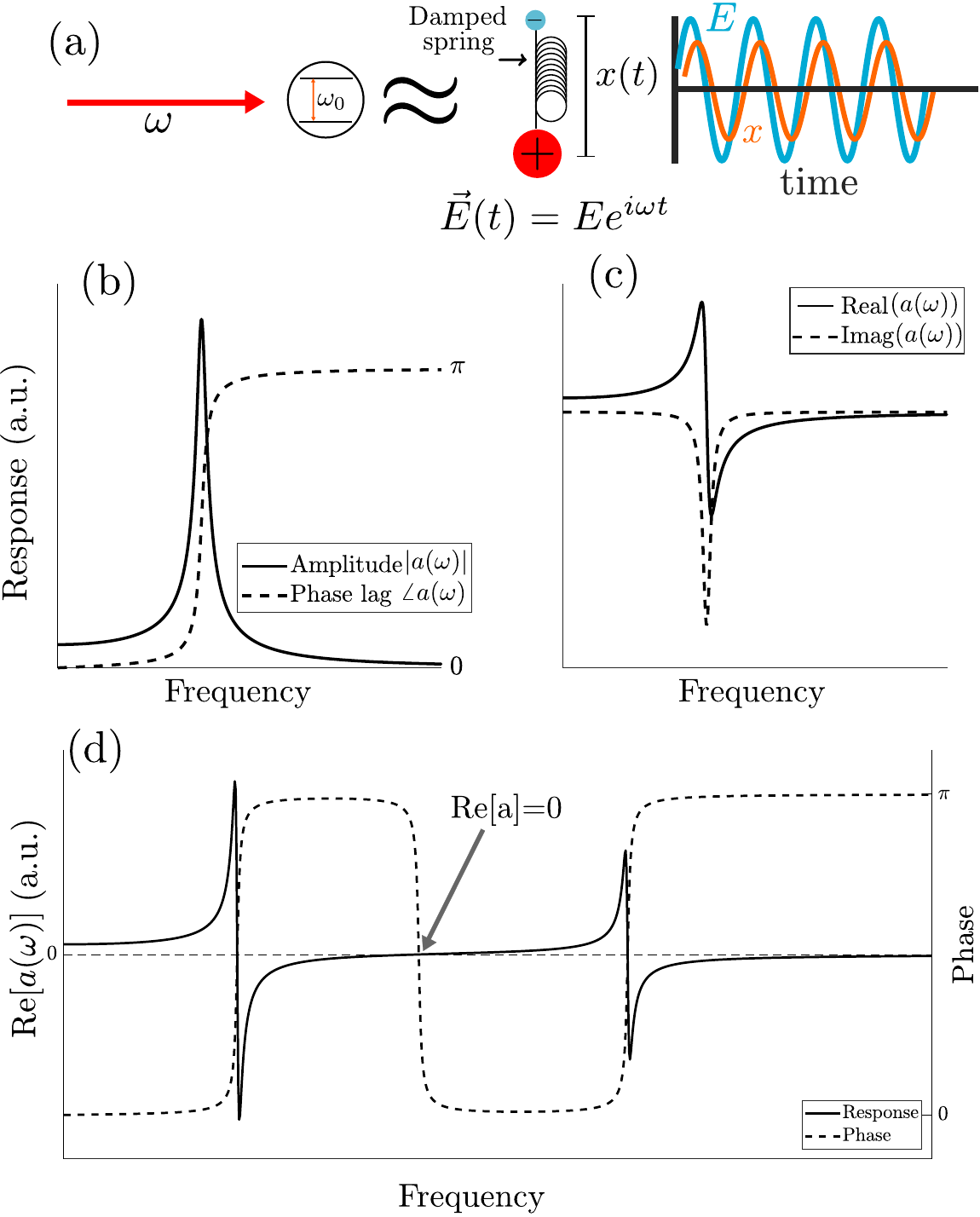}
		\caption{(a) Illustration of the Lorentz oscillator model of polarizability, wherein the dipole moment of atom in an off-resonant light field is approximated by a periodically driven mass-spring system. The response function (b) has a peak at the resonant frequency $\omega_0$, about which the phase of the response flips from 0 to $\pi$. The spectrum (c) shows that at the zero-crossing of the phase, the steady-state amplitude (real part) of the response is zero but the imaginary part is negative, indicating absorption of energy from the driving force (resonance). A multi-level atom can be approximated by the superposition of such oscillators which give rise to multiple resonances (d). Between the resonances there is also a sign-flip of the phase, coincident with a zero-crossing of the real part. The position of these 'tune-out' points depends on the positions and relative strengths of the resonances.}
		\label{fig:lorentz}
	\end{figure}
	
	Recall from chapter \ref{chap:theory} that an atom in an oscillating electric field experiences an energy shift.
	The magnitude of this shift is proportional to the real part of the dynamic polarizability $\alpha(f)$, a fundamental atomic property determined by the frequency $f$ of the oscillating electric field, the spacing of energy levels, and the strengths of the respective dipole matrix elements.
	This can be illustrated by the classical Lorentz model which approximates the atomic dipole moment by a fixed positive charge plus a negatively-charged particle attached to the `nucleus' by a spring (illustrated in Fig. \ref{fig:lorentz})\footnote{An equivalent model can be constructed using an LC circuit.}.
	The object of ultimate interest is the equation of motion, which in the Lorentz model is 
	\begin{equation}
		\ddot{x} + \Gamma \dot{x} + \omega_0^2x = -e\vec{E}/m.
	\end{equation}
	Here, $m$ is the mass of the `electron' and the spring has resonant frequency $\omega_0$ and damping rate $\Gamma$, as illustrated in Fig. \ref{fig:lorentz}.
	The general solution for the equation of motion has the form $x(t) = a(\omega)|\vec{E}|e^{i\omega t}$, whose frequency spectrum has a Lorentzian profile,
	\begin{equation}
		a(\omega) = -\frac{e}{m}\frac{1}{(\omega_0^2-\omega^2)+i\Gamma\omega},
	\end{equation}
	where the real part gives the steady-state amplitude of the induced oscillation, and the imaginary part relates to the exchange of energy from the electric field.
	In fact, a real atom possesses many resonances and to first order (in the linear-response regime) the total response is the sum over all resonances, as in $a(\omega) = \sum_i A_i a_i(\omega)$. 
	Thus there generally exist points in the spectrum where the real part of $a(\omega)$ vanishes, which are determined by the resonant frequency and oscillator strength of each possible transition from the current state of the atom.
	
\begin{figure} 
		\centering
		\includegraphics[width=\textwidth]{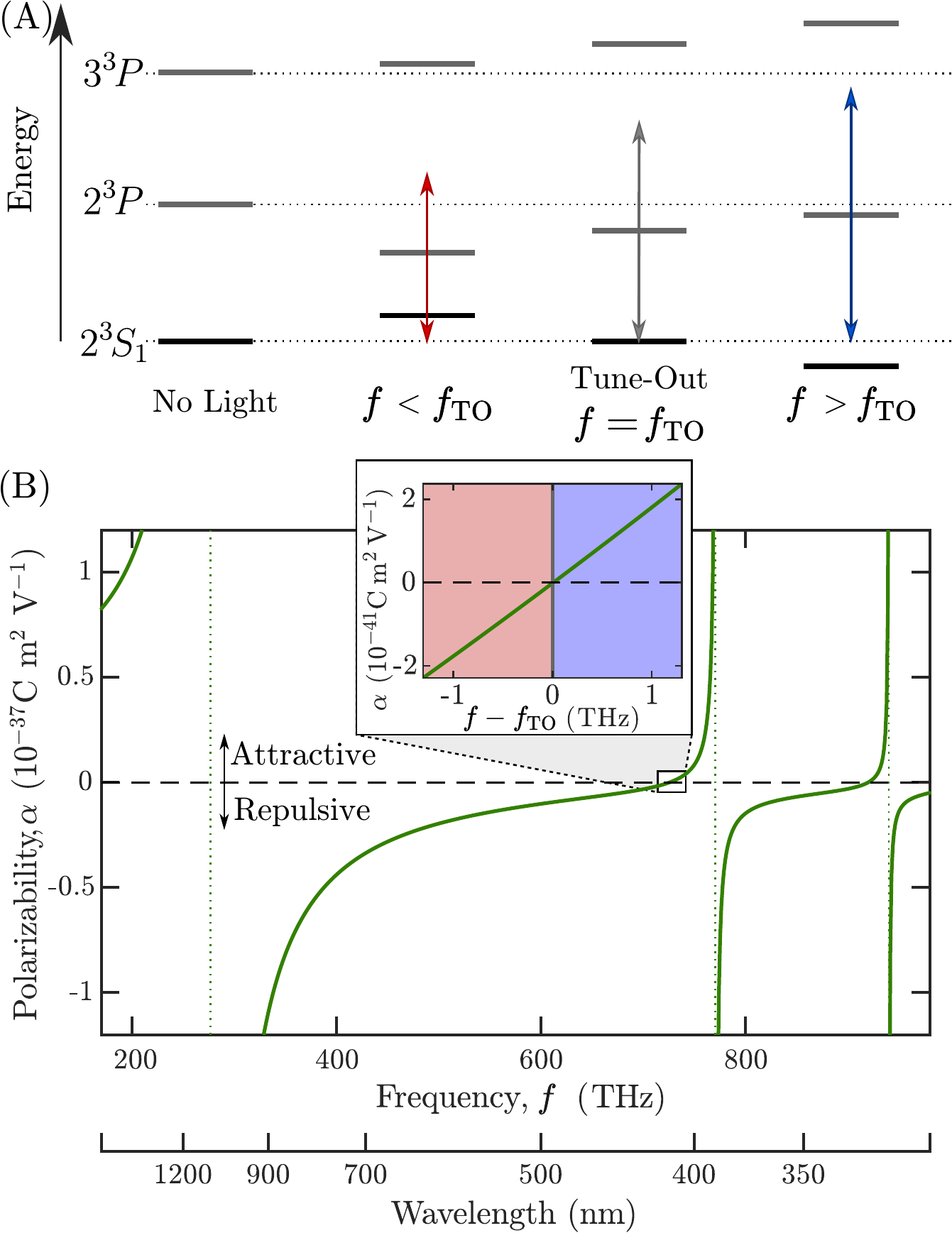}
		\caption{Tune-out in atomic helium:
		(A) Atomic energy level shift of the dominant state (manifolds) about the tune-out.  When an optical field of frequency $f$ (arrows) is applied to the atom the individual
		levels shift dependent on the difference between $f$ and the transition frequency. At the tune-out frequency $f_{\mathrm{TO}}$ (middle right), the shifts to the $\MetastableState$ state energy cancel.
		Energy spacing and shifts not to scale.
		(B) Theoretical frequency dependent polarizability of $\MetastableState$ helium, for a constant light polarisation, indicating that the polarizability vanishes near 726~THz, - the tune-out frequency measured in this paper. 
		Vertical dotted lines show, from left to right, the transitions to the  $\TOLowerStateManifold$, $\TOUpperStateManifold$,$4^{3\!}P$ manifolds. Inset shows the approximately linear polarizability with frequency about the tune-out.
		}
		\label{fig:he_polz} 
	\end{figure}

	In the quantum picture a similar correspondence holds: The real part of the polarizability $\alpha(\omega)$ gives the energy shift as a result of a perturbing light beam, and the imaginary part fixes the photon scattering rate  via $\Gamma_{sc} = \textrm{Im}(\alpha)I/(\hbar\varepsilon_0 c)$ in a monochromatic light field with intensity $I$. 
	For a multilevel atom in the state $i$ under the dipole- and rotating-wave approximations \cite{Grimm00} the Stark shift is
	
	\begin{align}
		\Delta E &= -\frac{1}{2\epsilon_0 c}\mathrm{Re}(\alpha(f))\\
		 	&= -\sum_{j\neq i}\frac{|\bra{j}\hat{\mu}\cdot\mathrm{\textbf{E}}\ket{i}|^2}{E_i-E_j}
		\label{eqn:dipole_shift}
	\end{align}
	
	\noindent where the $E_i$ are the unperturbed energies, and $\hat{\mu}\cdot\mathrm{\textbf{E}}$ is the dipole interaction energy.
	In the quantum-mechanical framework the strength of these resonances is characterized by the oscillator strengths $f_{ij}$, which are proportional to the (squared) matrix elements $|\bra{j}\hat{\mu}\cdot\mathrm{\textbf{E}}\ket{i}|^2$.
	A ‘tune-out’ frequency (labeled $f_\mathrm{TO}$) occurs between transition frequencies at the point where the contributions to the dynamic polarizability\footnote{For brevity, and as we are concerned solely with the real part of the polarizability, we will simply use the notation $\alpha(f)$ to refer to the \emph{real} part of the polarizability.} $\alpha(f)$ from transitions below that frequency are balanced by those above it (i.e. $\alpha(f)=0$) \cite{LeBlanc07}. 
	This balance point, illustrated in Fig. \ref{fig:he_polz} is therefore fixed by the strength (via the numerator in the RHS of Eqn \ref{eqn:dipole_shift}) and frequency (via the denominator) of every transition in the atomic spectrum and thus provides a precise constraint on the ratio of transition dipole matrix elements. 
	The frequency-dependend polarizability for \mhe~is also shown in Fig. \ref{fig:he_polz} spanning just under 2.5 octaves.

\subsection{Polarizability near $\fto$}
\label{sec:polz_linearization}

	The atomic polarizability has a complex profile, as shown in Fig. \ref{fig:he_polz}, but as we are concerned with measurements in a small region around a tune-out point, we can simplify the analysis below by linearizing the polarizability with respect to frequency. This assumption breaks down if one were to make measurements over much wider frequency intervals, and the magnitude of the loss of accuracy is quantified in section \ref{sec:systematic_effects}.
	In the rest of this section, I describe how the tune-out depends on the frequency and polarization of the laser light, and how this allows us to select a single figure of merit for comparison with theoretical clculations.

	Consider the interaction of an atom with a (real-valued) electric field with the form
	\begin{equation}
		\mathbf{E} = \frac{1}{2}\left(\vec{\mathcal{E}}e^{-2\pi i f t}+\vec{\mathcal{E}}^*e^{2\pi i f t}\right),
	\end{equation}
	where $\vec{\mathcal{E}}=\mathcal{E}\hat{\mathbf{u}}$ is the electric field envelope with magnitude $\mathcal{E}$, polarization vector $\mathbf{u}$ (which may be complex), and oscillation frequency $f$.
	Hereafter in this section, we will consider the atomic reference frame defined by a quantization axis $z$. For a far-off-resonant light field, the Stark shift will be small compared with the fine-structure level splitting. 
	In this case, and working in the dipole appoximation \cite{LeKien13}, the interaction energy between the atomic dipole $\mathbf{d}$ and the light field is captured by the operator
	\begin{equation}
		\Delta E_\mathrm{Stark} = -\frac{1}{2}\left(\mathcal{E}\mathbf{u}\cdot\mathbf{d}e^{-2\pi i f t} - \mathcal{E}^*\mathbf{u}^*\cdot\mathbf{d}e^{2\pi i f t}\right),
	\end{equation}
	which, after a lengthy derivation \cite{LeKien13} can be written as
	\begin{align}
		\Delta E_\mathrm{Stark} &= -\frac{1}{2\epsilon_0 c} \mathrm{Re}(\alpha(f)) I
	\end{align}
	where $I= \frac{\epsilon_0 c}{2} |E|^2$ is the electric field intensity and
	\begin{equation}
		\alpha(f) = \alpha^S + C\alpha^V \frac{m_J}{2J} + D\alpha^T\frac{3m{_J}^{2}-J(J+1)}{2J(2J-1)}
		\label{eqn:full_polz}
	\end{equation}
	is the frequency-dependent polarizability for an atom in a state with total angular momentum $J$ (and a $z$-component of $m_J$).  
	The frequency-dependent scalar, vector, and tensor polarizabilities are $\alpha^S$, $\alpha^V$, and $\alpha^T$, respectively, and the coefficients $C=2\mathrm{Im}(\mb{u}_{x}^{*}\mb{u}_{y})$ and $D= 3|\mb{u}_z|^2-1$ depend on the polarization vector $\mathbf{u}$. More will be said about the polarization effects in section \ref{sec:polz_dep}, but for now let us return to the frequency-dependent effects.
	From the preceding equation \ref{eqn:full_polz} one then obtains the polarizability for the $2\triplet S_1(m_J=1)$ state:
	\begin{equation}
		 \alpha(f) = \alpha^S(f) + \frac{1}{2}\left(C\alpha^V(f)  + D\alpha^T(f)\right).
		 \label{eqn:master_polz}
	\end{equation}
	Expanding the preceding equation about some frequency $f_0$ near $\fto$ we and truncating the Taylor expansion we have
	\begin{align}
		 \alpha(f)  \approx & \alpha^S(f_0) + (f-f_0)\frac{d\alpha^S}{df}\Bigr|_{f=f_0} \\
		 &+ \frac{C}{2}\left(\alpha^V(f_0) + (f-f_0)\frac{d\alpha^V}{df}\Bigr|_{f=f_0} \right)\\
		 & +\frac{D}{2}\left(\alpha^T(f_0) + (f-f_0)\frac{d\alpha^T}{df}\Bigr|_{f=f_0} \right) + \cdots
	\end{align}
	It is expected on theoretical grounds that the vector and tensor polarizabilities are nearly constant over the scan ranges we use in this experiment (and we can check this experimentally - indeed we do find that the gradient of the fits $\propto d\alpha /d f$ is independent on the polarization, thus only the $\propto d\alpha^S /d f$ is significant).	Hence, we can drop the respective derivative terms and arrive at

	\begin{align}
		 \alpha(f)  \approx & \alpha^S(f_0) + \frac{d\alpha^S}{df}\Bigr|_{f=f_0} (f-f_0) + \frac{C}{2}\alpha^V(f_0) + \frac{D}{2}\alpha^T(f_0).
		 \label{eqn:polz_taylor}
	\end{align}
	The condition that $\alpha(\fto)=0$ implies
	\begin{align}
		 \frac{d\alpha^S}{df}\Bigr|_{f=f_0} (\fto-f_0) \approx & -\alpha^S(f_0) - \frac{C}{2}\alpha^V(f_0) - \frac{D}{2}\alpha^T(f_0).
	\end{align}
	We can simplify this result by setting $f_0$ to be the tune-out for the scalar polarizability, i.e. $f_0=\fto^S$ s.t. $\alpha^S(\fto^S)=0$, thus arriving at
	\begin{align}
		 \fto \approx & \fto^S - \frac{C}{2}\beta^V(\fto^S) - \frac{D}{2}\beta^T(\fto^S)
		 \label{eqn:redpolz}
	\end{align}
	\begin{equation}
		\beta^X(f) = \alpha^X(f)\left(\frac{d\alpha^S}{df}\Bigr|_{f=f_0}\right)^{-1}
	\end{equation}
	for $X\in\{V,T\}$.

	The main conclusions of the preceding calculations is that the polarizability is well-approximated by the linearized equation (\ref{eqn:polz_taylor}) for experimentally relevant detunings from the tune-out. The change in polarizability with respect to frequency is dominated by the scalar term $\alpha^S(f)$, but the zero-crossing of the total polarizability $\alpha(f)$ is modified by the polarization-dependent terms ($\frac{C}{2}\beta^V(\fto^S)$ and $\frac{D}{2}\beta^T(\fto^S)$). Therefore, the tune-out frequency depends on the polarization of the laser beam (Eqn. (\ref{eqn:redpolz})), and so we need to establish some terminology to relate the beam parameters to the atomic dipole response. As we will see below, this also permits us to define a single measurable quantity that can be compared directly with theory.

\subsection{Polarization dependence of the tune-out}
	\label{sec:polz_dep}
\begin{figure}
	\centering
	\includegraphics[width=\textwidth]{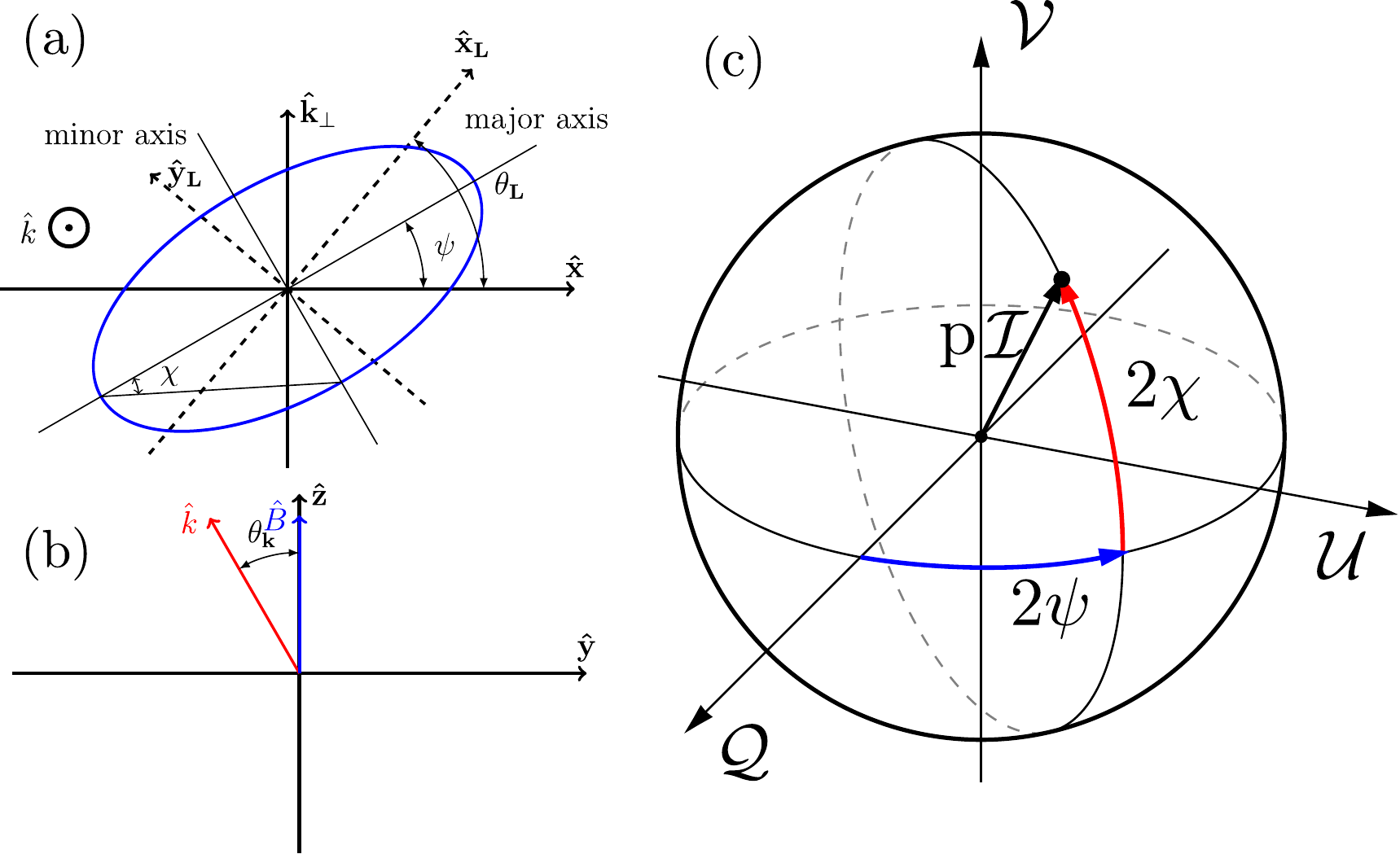}
	\caption{Aspects of light polarization in relation to the experiment. The polarization vector traces out an ellipse in the plane transverse to the light wavevector $\mathbf{k}$, depicted in (a). The ellipse traced in the plane normal to the propagation vector $\hat{k}$ is characterized by the angle $\psi$ between the major axis and the horizontal, and the angle $\chi$ determines the ellipticity (i.e. degree of circular polarization). The lab frame ($\hat{x}_L$) is rotated by an angle $\theta_L$ with respect to the atomic frame.  The light wave-vector is also potentially misaligned with the magnetic field vector, which defines the quantization axis of atomic angular momentum, as shown in (b). Such a possible misalignment factors into the transformation through the angle $\theta_k$. The polarization state can be treated as a real vector on the Poincar\'{e} sphere (c), on which the relationships between the elliptical angles are marked ($\chi$, the degree of circular polarization, gives the elevation angle in red. $\psi$ gives the polar angle, shown in blue). The position of the Stokes vector specifies a polarization state in terms of the Stokes parameters ($\st{Q,U,V}$), as described in the text. For example, circularly polarized light has an angle $\chi=\pi/4$ and thus an elevation angle $\pi/2$, thus living on the `north' or `south' poles of the Poincar\'{e} sphere ($|\st{V}|=1$). 
	}
	\label{fig:polz_figure}
	\end{figure}

	We can now turn attention back to the $C$ and $D$ coefficients. These are most easily expressed in terms of the polarization vector $\mathbf{u}$ and have the form
	\begin{align}
		C &= 2\mathrm{Im}(\mathbf{u}_{x'}^*\mathbf{u}_{y'}),\\
		D &= 3|\mathbf{u}_{z'}|^2-1,
	\end{align}
	where each element can be decomposed as $\mathbf{u}_{x'} = u_{x'} e^{i\phi_{x'}}$, etc.
	The dashed coordinate labels correspond to the atomic reference frame defined by $\hat{z'} = \mathbf{B}/|\mathbf{B}|$, see Fig. \ref{fig:polz_figure}.
	However, the magnetic field vector at the interrogation zone is not necessarily perfectly aligned with the laser beam axis and thus we need a means to translate between different coordinate systems.
	Therefore, before we return to the tune-out equation \ref{eqn:redpolz} we take a short detour here to build up the language necessary to discuss the role of polarization in determining the tune-out frequency.
	The bulk of this section is devoted to introducing the detailed theory of light polarization. For an operational reading of this section, one can consult Fig. \ref{fig:polz_figure} and skip ahead to section \ref{ssec:connection_to_experiment}

	\subsubsection{Theory of polarized light}

	A monochromatic electric field takes on a classical state as
	\begin{equation}
		\vec{E}(\mathbf{r},t) = \mathcal{E}(\mathbf{r},t)\hat{\mathbf{u}}(\mathbf{r}),
		\label{eqn:lightfield}
	\end{equation}
	where $\hat{\mathbf{u}}$ is the unit polarization vector. The electric field propagates in the direction of the Poynting vector $\mathbf{S} = \mathbf{E}\times\mathbf{B}$, and hence we have $\hat{\mathbf{u}}\perp\mathbf{S}$ and $\hat{\mathbf{u}}\perp\vec{k}$, where $\vec{k}$ is the wave-vector of the light. Therefore, if we fix $z'$ as the propagation direction (i.e. along the beam axis, which we will relate the dashed coordinates to the lab frame shortly) then we can write the electric field in the coordinates $(x',y',z')$ as 
	\begin{equation}
		\vec{E}(\mathbf{r},t) = \begin{bmatrix}u_{x'}e^{i\phi_{x'}}\\
												u_{y'}e^{i\phi_{y'}}\\
												0
												\end{bmatrix}e^{i(kz'-2\pi f t)}.
	\end{equation}
	One can then consider the first two components, also known as the \emph{Jones vector} which captures the amplitude and relative phases in the $x'$ and $y'$ direction.
	We can write the Jones vector as a scalar multiplied by a unit-length vector $\hat{\mathbf{j}}$, and following a rotation by an angle $-\phi_x'$ we obtain
	\begin{align}
		\mathbf{j} &= |E|~\hat{\mathbf{j}}e^{-2\pi i f t},~\mathrm{where}\\
				\hat{\mathbf{j}}&= \begin{bmatrix}u_{x'}\\
								u_{y'}e^{i(\phi_{y'}-\phi_{x'})}\end{bmatrix}
	\end{align}
	and $u_{x'}={E_{x'}}/{|E|}$ (and similarly for $y'$).
	As the polarization vector precesses in time (as per the $t$-dependence in Eqn. \ref{eqn:lightfield}) it traces out a locus in the $x',y'$ plane normal to $\vec{k}$.
	For purely-polarized monochromatic light this locus is an ellipse, or rather the Lissajous figure whose $x'$ ($y'$) amplitude is given by $u_{x'}$ ($u_{y'}$) and the relative phase is $\phi_{y'}-\phi_{x'}$.
	Thus we have, for example, circularly polarized light when $u_{x'}=u_{y'}$ and $\phi_{y'}-\phi_{x'}=\pi/2$, or $\hat{\mathbf{j}} = (1,i)/\sqrt{2}$. On the other hand, if $\phi_{y'}-\phi_{x'}=0$ the light is linearly polarized along the axis given by $(u_{x'},u_{v'})$. 
	Up to a factor corresponding to the global phase, the Jones vector notation can be used to express the common polarization states of light in the following forms:

	\begin{minipage}{0.5\textwidth}
	\begin{align*}
	\mathbf{h} &= \begin{bmatrix}1\\0\end{bmatrix} \\
	\mathbf{d} =& \frac{1}{\sqrt{2}}\begin{bmatrix}1\\1\end{bmatrix} \\
	\mathbf{l} =& \frac{1}{\sqrt{2}}\begin{bmatrix}1\\i\end{bmatrix} 
	\end{align*}
	\end{minipage}
	\begin{minipage}{0.5\textwidth}
	\begin{align*}
		\mathbf{v} &= \begin{bmatrix}0\\1\end{bmatrix}\\
		\mathbf{a} =& \frac{1}{\sqrt{2}}\begin{bmatrix}1\\-1\end{bmatrix}\\
		\mathbf{r} =& \frac{1}{\sqrt{2}}\begin{bmatrix}1\\-i\end{bmatrix},
	\end{align*}
	\end{minipage}
	\hfill

	\noindent where the row-wise pairs correspond to the orthogonal pairs of horizontal/vertical, diagonal/antidiagonal, and left/right-circular polarization, respectively.
	Indeed any $\hat{\mathbf{j}}$ can be written as a linear combination of any row-wise pair of these vectors with complex coefficients.
	We can take a moment to notice that $\hat{\mathbf{j}}$ is a two-dimensional complex vector of length 1, and thus inhabits the same space as any two-level quantum system - i.e. a qubit.
	The coherence matrix\footnote{Elsewhere called the coherency matrix.} can thus be defined through the outer product
	\begin{align}
		\mathbf{\chi} &= \mathbf{j}^\dagger\otimes\mathbf{j} \\
		 &= \st{I} \begin{bmatrix} u_{x'}^2 & u_{x'} u_{y'} e^{-i(\phi_{x'}-\phi_{y'})} \\ u_{x'} u_{y'} e^{i(\phi_{x'}-\phi_{y'})} & u_{y'}^2\end{bmatrix},
	\end{align}
	where $I=|E|^2$ (which, up to some physical constants, is the intensity of the light field), and the off-diagonal elements are called the \emph{coherences}. 
	The outer product also maps the conventional polarization basis vectors to their respective matrix forms:

	\begin{minipage}{0.5\textwidth}
	\vspace{0pt}
	\begin{align*}
		{H} &= \begin{bmatrix}1&0\\0&0\end{bmatrix} \\
		{D} =& \frac{1}{{2}}\begin{bmatrix}1&1\\1&1\end{bmatrix}\\
		{L} =& \frac{1}{{2}}\begin{bmatrix}1&-i\\i&1\end{bmatrix}
	\end{align*}
	\end{minipage}
	\begin{minipage}{0.5\textwidth}
	\vspace{0pt}
	\begin{align*}
		{V} &= \begin{bmatrix}0&0\\0&1\end{bmatrix}\\
		{A} =& \frac{1}{{2}}\begin{bmatrix}1&-1\\-1&1\end{bmatrix}\\
		{R} =& \frac{1}{{2}}\begin{bmatrix}1&i\\-i&1\end{bmatrix}.
	\end{align*}
	\end{minipage}

	The coherence matrix contains the full polarization information for a classical light-field and can be written in the Pauli basis as $\mathbf{\chi} = \sum_{i=0}^3 a_i\sigma^{i}$, where

	\begin{minipage}{0.5\textwidth}
	\vspace{0pt}
	\begin{align*}
	\sigma_0 &= \begin{bmatrix}1&0\\0&1\end{bmatrix}\\
	\sigma_2 &= \begin{bmatrix}0&1\\1&0\end{bmatrix}
	\end{align*}
	\end{minipage}
	\hfill
	\begin{minipage}{0.5\textwidth}
	\vspace{0pt}
	\begin{align*}
		\sigma_1 &= \begin{bmatrix}1&0\\0&-1\end{bmatrix}\\
		\sigma_3 &= \begin{bmatrix}0&-i\\i&0\end{bmatrix}.
	\end{align*}
	\end{minipage}

	thus the coherence matrix can generally be written in the form $\chi = \frac{p\st{I}}{2}\left(\mb{1} + p\mathbf{s}\cdot\mathbf{\sigma}\right)$ where $\mathbf{s}=(\st{Q},\st{U},\st{V})\in\mathbb{R}^3$ and $|\mathbf{s}|=1$, analogous to the expression of a qubit state as a Bloch vector, where $\mb{1}=\sigma_0$ is the identity matrix\footnote{Note that the subscript indices differ from the Bloch sphere convention, opting instead for the convention consistent with the Stokes parameters.}.
	Note also that, as for the density matrix in the context of quantum states, the coherence matrix can be used to deal with statistical mixtures of polarization states. In this case, the purity $p\leq1$ can be introduced as a multiplying factor of the length of the (unit) Stokes vector $\mb{s}$. Fully incoherent light has $p=0$ and corresponds to the zero vector in the Stokes picture. Partially coherent light, for example an equal statistical mixture of $\mb{a}$ and $\mb{r}$, has a nonzero Stokes vector ($\frac{1}{2}(\mb{a}+\mb{r})=\frac{1}{2\sqrt{2}}(2,-(1+i))$) which lies inside the Poincar\'{e} sphere, while purely-polarized light lives on the surface of the sphere.
	
	By convention, the full form of the Stokes vector is $\mathbf{S} = (p\st{I},\st{Q},\st{U},\st{V})$ where $I$ corresponds to the beam intensity and the latter three components are the degrees of linear polarization ($\st{Q}$=$\pm$1 for horizontally and vertically polarized light), (anti-) diagonal polarization (accordingly, $\st{U}=\pm1$), and circular polarization (where we pick $\st{V}=\pm1$ for left- and right-circular polarization in the frame defined by the beam axis). Hereafter, we consider the case of purely-polarized light with $p=1$.
	Further, in the coherence matrix picture, the degree of polarization along vector $P$ (expressible as a linear combination of $\{H,V,D,A,R,L\}$) is given by the inner product Tr($\chi P$), as the algebra of the inner product is preserved under the linear action of the tensor product. 
	Completing the analogy, the pairs of orthogonal polarization are related by the transformation $\mathbf{s}\rightarrow-\mathbf{s}$ (just as orthogonal states of a qubit are at antipodal points on the Bloch sphere).
	
	The Stokes vector is illustrated on the Poincar\'{e} sphere, in relation to the polarization ellipse, in Fig. \ref{fig:polz_figure}.
	Notice that in taking the outer product, the factor $e^{\pm2\pi i f t}$ has vanished - the coherence matrix, just as for the density matrix, discards the global phase information by construction. 
	This can be a problem in applications concerning the coherent nature of laser light, wherein the Jones calculus is preferred for its preservation of phase information.
	However, in our case we are concerned with the time-averaged behaviour of the atom-light interaction (c.f. section \ref{sec:atoms_and_light}) and so this loss is not critical.
	For our treatment, we are considering purely polarized light in the experimental context of applying wave-plates after the output from a polarizing beamsplitter, and can do without this parameter.

\subsubsection*{Connection to experiment}
	\label{ssec:connection_to_experiment}

	A full definition of the Stokes parameters in terms of the unit electric field polarization vector $\mb{u}$ the $x',y'$ basis is:

	\begin{align}
	    \mathcal{I} &= |\mathbf{u}_{x'}|^2 + |\mathbf{u}_{y'}|^2 \\ 
	    \mathcal{Q} &= |\mathbf{u}_{x'}|^2 - |\mathbf{u}_{y'}|^2 \\
	    \mathcal{U} &= 2 \text{Re} \left(\mathbf{u}_{x'} \mathbf{u}_{y'}^*\right) \\
	    \mathcal{V} &= -2 \text{Im}\left(\mathbf{u}_{x'} \mathbf{u}_{y'}^*\right)
	\end{align}
	Which allows an expression in the form
	\begin{equation}
		\mathbf{S} = A \mathcal{J},
	\end{equation}
	where
	
	\begin{equation}
		A=\begin{bmatrix}
		 1 & 0 & 0 & 1 \\
		 1 & 0 & 0 & -1 \\
		 0 & 1 & 1 & 0 \\
		 0 & -i & i & 0 \\
		\end{bmatrix}
	\end{equation}
	and 
	
	\begin{align}
		\mathcal{J} &= \mathbf{j}^*\otimes\mathbf{j}\\
					&= (u_{x'}^2,u_{x'}u_{y'}e^{-i(\phi_{y'}-\phi_{x'})},u_{x'}u_{y'}e^{i(\phi_{y'}-\phi_{x'})},u_{y'}^2)^\mathrm{T},
	\end{align}
	Therefore the electric field components can be recovered, up to a global phase, via
	\begin{align}
		 \mathcal{J} &= A^{-1} \mathbf{S}\\
					 &= \frac{1}{2}\begin{bmatrix} 
					 				\st{I}+\st{Q}\\\st{U}+i\st{V}\\\st{U}-i\st{V}\\\st{I}-\st{Q}
								 	\end{bmatrix}.
	\end{align}
	thus the electric field components and their relative phase can be recovered through
	\begin{equation}
		 u_{x'} = \sqrt{\frac{\st{I}+\st{Q}}{2}},~u_{y'} = \sqrt{\frac{\st{I}-\st{Q}}{2}},~\phi_{y'}-\phi_{x'} = \mathrm{arctan}(\st{V}/\sqrt{1-\st{V}^2}),
	\end{equation}
	noting that, as expected, we are unable to retrieve the global phase information. For purely-polarized light, whose amplitude is obtained by $\sqrt{\st{I}}$, the real three-dimensional Stokes vector only contains enough parameters to specify the Jones vector. 
	However, we introduce the Stokes parameters for their utility in capturing the polarization-dependence of the atomic polarizability.

	We can now examine the $C$ and $D$ coefficients directly.
	Writing out the coefficient $C=2\mathrm{Im}(u_{x}^*u_y)$ we see it can be obtained by 
	\begin{equation}
	C=2\mathrm{Im}(u_{x}^*u_y) = -\mathrm{Tr}(\sigma_3\chi) = -\st{V}.
	\end{equation}
	That is, it gives the fourth stokes parameter, the degree of (right-) circular polarization in the atomic frame (indeed this follows directly from the definition of $\st{V}$).
	The $D$ coefficient can be seen to depend on the length of the polarization vector along the $z$ axis in the atomic frame, and can be shown \cite{Henson22} to depend on the second Stokes parameter $Q$ through
	\begin{equation}
		D = \frac{3}{2}\sin^2(\theta_k)(1+\st{Q}) - 1,
	\end{equation}
	where $\theta_k$ is the angle between the magnetic field pointing $\hat{b}$ and $\mathbf{k}$.
	The linearized tune-out equation thus reads
	\begin{align}
		 \fto = & \fto^S + \frac{\st{V}}{2}\beta^V - \frac{1}{2}\beta^T \left(\frac{3}{2}\sin^2(\theta_k)(1+\st{Q}) - 1\right)
		 \label{eqn:fto_stokes_eqn}
	\end{align}
	in terms of the Stokes parameters in the atomic reference frame.

\subsubsection{Magic wavelengths}
	
	It bears noting that tune-out points are associated with \emph{states} of the atom, where the stark shift $\Delta E_\mathrm{S} =0$.
	A distinct but related concept is `magic wavelengths' which are associated with \emph{transitions} between states and are defined by the point where the differential dynamic polarizability (or differential stark shift) vanishes: $\Delta E_\mathrm{S,U}-\Delta E_\mathrm{S,L} = 0$. 
	Magic wavelengths have attracted a great deal of interest for their utility in constructing optical lattice clocks \cite{Takamoto05,Derevianko11}, decoupling the clock transition frequency from the intensity of the trapping beams.

	Magic wavelengths are, of course, also of interest for their potential use in tests of basic atomic structure theory.
	However, most magic wavelength measurements to date have been made using heavy elements like strontium \cite{BilickiThesis}, barium \cite{Chanu20}, mercury \cite{Yi11}, rubidium \cite{Herold12}, and cesium \cite{Yoon19}, whose complex internal structure restricts the accuracy with which these wavelengths can be predicted.
	For instance, while current predictions of magic wavelengths in heavy elements are yet to reach parts-per-million precision, calculations for helium are already at this level \cite{Wu18,Zhang21_magic}.
	
	Highly precise measurements of magic wavelengths have been used to determine transition matrix elements in rubidium to a relative accuracy better than $10^{-3}$ \cite{Herold12}.
	For comparison, tune-out measurements in rubidium have yielded the ratio of oscillator strengths (rather than their absolute values) with accuracy at the $10^{-4}$ level \cite{Leonard15}.
	
	While the contribution of QED effects are included for helium magic wavelengths, their contribution is generally smaller than the theoretical uncertainty associated with the individual energy levels and thus are not yet ready for comparison with experiments.
	However, a very recent work identified the 1335 nm magic wavelength as a candidate for tests of atomic structure theory due to its sensitivity to the finite nuclear mass, and relativistic and QED effects \cite{Zhang21_magic} (contributing to the wavelength at the level of 232, 247, and 21 ppm respectively).
	A measurement of this wavelength to 0.01 nm precision (fractional error $\sim10^{-5}$) would suffice to check the predicted QED contributions.
	
	Indeed, magic wavelengths have already found use in tests of QED with helium, albeit through their role as dipole traps for precision measurements of forbidden transition energies \cite{Rengelink18}.
	A further application to helium was recently proposed wherein one would use a 1265 nm magic wavelength optical trap while another 934 nm magic wavelength beam provides the first of two photons for dichroic excitation of the forbidden $2\triplet S_1\rightarrow 3\triplet S_1$ transition near 427.7 nm \cite{Thomas20, Zhang21_forbidden}.
	This would reduce the effect of the main systematic uncertainty (the AC Stark shift) from 5 MHz to below 100 kHz.

\subsection*{QED prediction of the tune-out point}
\label{sec:to_theory}
	Following the first prediction \cite{Mitroy13} and measurement  of the metastable helium  tune-out near 413 nm \cite{Henson15}, a vigorous campaign of theoretical studies \cite{Zhang16,ManaloThesis,Drake19,Zhang19, Pachucki19} has reduced the uncertainty in the predicted frequency, which previously limited comparison with experiment. In parallel with work at ANU, our collaborators\footnote{Namely Gordon Drake and Li-Yan Tang, with assistance from Aaron Bondy and Yong-Hui Zhang.} improved on the state-of-the-art calculation \cite{Zhang19} of the tune-out frequency.

	These calculations were performed using the non-relativistic QED method (nr-QED) and improve the precision by an order of magnitude (compared to the previous prediction \cite{Mitroy13}) with a total uncertainty at the level of 9 MHz (relative error $1.2\times10^{-8}$) . 
	The results are presented in Table~\ref{tab:theory} with a summary explanation here. A more detailed description is in the full manuscript \cite{Henson22}.
	
	\begin{table}[t]
	\centering
	\begin{tabular}{l r r}
	\hline\hline
	Quantity    &Value (MHz) & Uncertainty (MHz) \\
	\hline
	\multicolumn{3}{c}{\textbf{Nonrelativistic and Relativistic terms}} \\
	Nonrelativistic (NR) & 725\,645\,115& 2       \\
	NR + relativistic scalar $(\alpha^{\rm S})$    & 725\,742\,216&6   \\
	Relativistic tensor $(-\frac12\alpha^{\rm T})$ &   1\,755& \\
	\hline
	Total non-QED        & 725\,743\,950&6     \\
	\hline
	\multicolumn{3}{c}{\textbf{QED terms}} \\
	QED $\alpha^3$       &       -7\,298& 1       \\
	QED $\alpha^4$       &          -127& 6       \\
	\hline
	Total QED            &         -7\,425& 8     \\
	\hline
	Retardation          &          -477 &        \\
	Nuclear size         &             5&          \\
	\hline
	Grand total          & 725\,736\,053& 9       \\
	Experiment           & 725\,736\,700& 260 \\
	\hline
	Difference           &          -647& 260\\
	\hline\hline
	\end{tabular}\\
	\caption{\label{tab:theory}Summary of theoretical contributions to the helium $\TO$ manifold tune-out frequency near 725.7 THz.}
	\label{tab:theory}
	\end{table}

	The entries from nonrelativistic, relativistic, and QED contributions are not strictly additive because changing one effect, such as the relativistic correction, changes the tune-out frequency at which the other effects are evaluated.  Thus the first entry is the nonrelativistic tune-out frequency with finite nuclear mass effects included.  The next entry is the scalar part of the relativistic correction arising from the higher terms in the Breit interaction.  The entry from the spin-spin interaction (1755~MHz) is entirely responsible for the tensor part of the tune-out frequency (excluding the Schwinger radiative correction term $\alpha/\pi$).  These terms determine the nonrelativistic and relativistic part of the tune-out frequency.  The remaining terms are small enough that they can be added linearly.  The leading QED correction of order $\alpha^3$ ($-7298(6)$ MHz) includes the anomalous magnetic moment correction (8 MHz).  The terms of order $\alpha^4$ include only the radiative corrections.  The dominant source of uncertainty is thus the remaining nonradiative terms not included in the calculation, but which were previously computed for the $2^3S_1$ state energy \cite{Pachucki06}.  The remaining terms are the retardation correction of $-477$ MHz and a finite nuclear size correction.

\section{Measurement technique}
\label{sec:trap_freq_measure}

	\begin{figure}
	\begin{minipage}{0.5\textwidth}
	\vspace{0pt}
	\includegraphics[width=\textwidth]{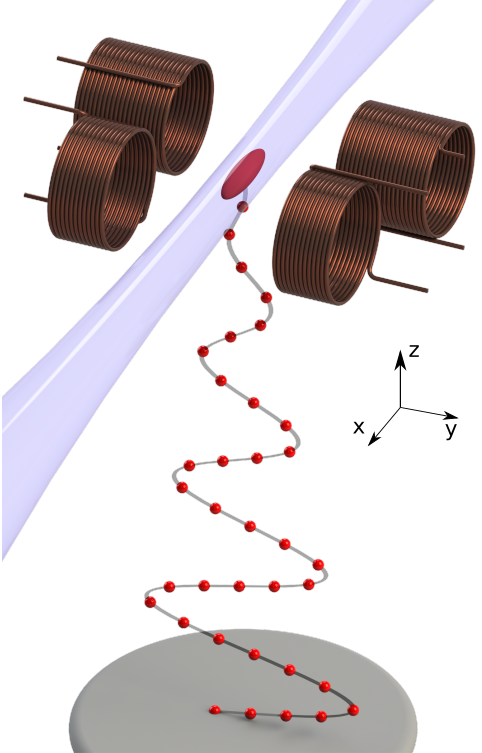}
	\end{minipage}
	\hfill
	\begin{minipage}{0.5\textwidth}
	\vspace{0pt}
	\caption{
	Schematic of the physical system used to determine the tune-out frequency as a function of laser frequency. A magnetically trapped BEC of metastable helium atoms is illuminated with a probe laser beam with an adjustable (optical) frequency. The probe beam is aligned with the weak axis of the trap and oscillations are induced in the BEC centre of mass. The oscillation direction is horizontal and transverse to the laser beam, aligned with the axes of the BiQUIC coils.  A sequence of atom laser pulses (red spheres) is outcoupled from the BEC to sample the oscillation. The oscillation frequency is reconstructed by the anti-aliasing procedure described in section \ref{sec:trap_freq_meas}
	The lab-frame coordinate axes show the oscillation direction $y$, transverse to the beam axis $x$.
	}
	\label{fig:exp_schematic} 
	\end{minipage}
	\end{figure}

	Having defined a tune-out, we now turn our attention to the means we employed to determine the 413 nm tune-out using our apparatus. 
	This section provides a physical description of the measurement technique and establishes the important roles played by the frequency and polarization of the probe beam.
	To determine a tune-out point is essentially to measure the frequency where the optical dipole potential  vanishes.
	In our measurement, we quantified the shift in the oscillation frequency of a harmonically trapped BEC when the magnetic trap was overlapped with a power-stabilized probe laser beam (see Fig. \ref{fig:exp_schematic} and Section \ref{sec:spec_laser}). 
	This is in contrast to previous works (eg. \cite{Rengelink18}) wherein magic-wavelength optical traps were used in conjunction with a resonant probe beam. 
	The main motivations for the use of a magnetic trap are that it permits the pulsed outcoupling method, enabling the highly accurate trap frequency determinations which underpin this measurement. Furthermore, the magnetic field is easily configured to be extremely harmonic at length scales much larger than the BEC or its in-trap motion.

	The position-dependent potential energy in this hybrid trap is the sum of both the harmonic magnetic potential and a Gaussian optical potential,
	\begin{equation}
		U_\mathrm{net} = U_{\mathrm{trap}} + U_\mathrm{dip},
	\end{equation}
	where the magnetic trapping potential takes the form
	\begin{equation}
		U_\mathrm{trap} = \frac{m}{2}\left(\omega_{x}^{2}x^2+\omega_{y}^{2}y^2+\omega_{z}^{2}z^2 \right)
	\end{equation}
	where the $\omega_i$ terms are the oscillator frequencies of the harmonic magnetic trap, and the dipole potential is given in general by  \cite{Grimm00}
	\begin{align}
	    U_{\mathrm{dip}}=-\frac{1}{2 \epsilon_{0} c} \operatorname{Re}(\alpha) I(\rvec),
	\end{align}
	where $I(\rvec)$ is the local intensity of the electric field, $c$ is the speed of light and $\epsilon_0$ is the permittivity of free space.
	For a Gaussian beam which is aligned with the weak ($x$) axis of the trap and focuses $P$ watts of power to a spot size $w_0$, the intensity profile is 
	\begin{equation}
	    I_{\mathrm{dip}} =
	        \frac{2 P}{\pi w_0^2} \left(\frac{w_0}{w(x)}\right)^2 \exp\left( -\frac{1}{2} \frac{(y^2+z^2)}{w(x)^2} \right)
	 \end{equation}
	 where $w(x) = w_0\sqrt{1+(x/x_R)}$ is the waist size at a position $x$ along the beam axis and $x_R$ is the Rayleigh length. 
	 In the small-amplitude oscillation limit we can consider the total potential to be harmonic and retain only the terms up to second order in a Taylor expansion. If we further assume that the trapped BEC oscillates in the $y$ direction only then we can write the potential along the direction of motion as
	 \begin{equation}
	 U_\mathrm{net} \approx -\frac{\mathrm{Re}(\alpha)  P}{\pi 
   c {w_0}^2 \epsilon} + \left(\frac{2 \mathrm{Re}(\alpha)  P}{\pi  c {w_0}^4 \epsilon }+\frac{m {\omega_y}^2}{2}\right) y^2
	 \end{equation}
	 when evaluated about $x=z=0$.
	 Noting that the restoring force for a harmonic oscillator is given by $F_y = -\partial_y U_\mathrm{net} = -k_\mathrm{net} y$, we can identify the perturbed trapping frequency via the spring constant $k_y$:
	 \begin{align}
	 	\Omega_\mathrm{net} &= \frac{1}{2\pi}\sqrt{\frac{k_\mathrm{net}}{m}}\\
	 	&= \frac{1}{2\pi}\sqrt{{\omega_y}^2+\frac{4 P \mathrm{Re}(\alpha)}{\pi m c \epsilon w_{0}^4}}
	 \end{align}
	wherein we can identify the trap and probe frequency contributions by setting $P=0$ and $\omega_y=0$, respectively. 
	We thus obtain 
	\begin{equation}
		\Omega_\mathrm{probe} = \frac{1}{2\pi}\sqrt{\frac{4 \mathrm{Re}(\alpha)  P}{\pi m c \epsilon w_{0}^4}}
		\label{eqn:omega_probe}
	\end{equation}
	and have the relation
	\begin{align}
		\Omega_\mathrm{net}^2 &= \Omega_\mathrm{trap}^2 + \Omega_\mathrm{probe}^2\\
		&= \Omega_\mathrm{trap}^2 + \frac{4P}{\pi m c \epsilon w_{0}^4}\mathrm{Re}(\alpha)
		\label{eq:delta_omega_squared}
	\end{align}
	which shows that the squared shift in trapping frequency $\Omega_\mathrm{probe}^2 = \Omega_\mathrm{net}^2 - \Omega_\mathrm{trap}^2$ is linear in the probe power and the polarizability.
	Note the $\Omega$ terms have units 1/time - the trapping frequencies $\omega_i$ are angular frequencies with units Rad Hz. Thus, $\Omega_\textrm{trap} = \omega_y/2\pi$.
	
	\begin{figure}
	\includegraphics[width=\textwidth]{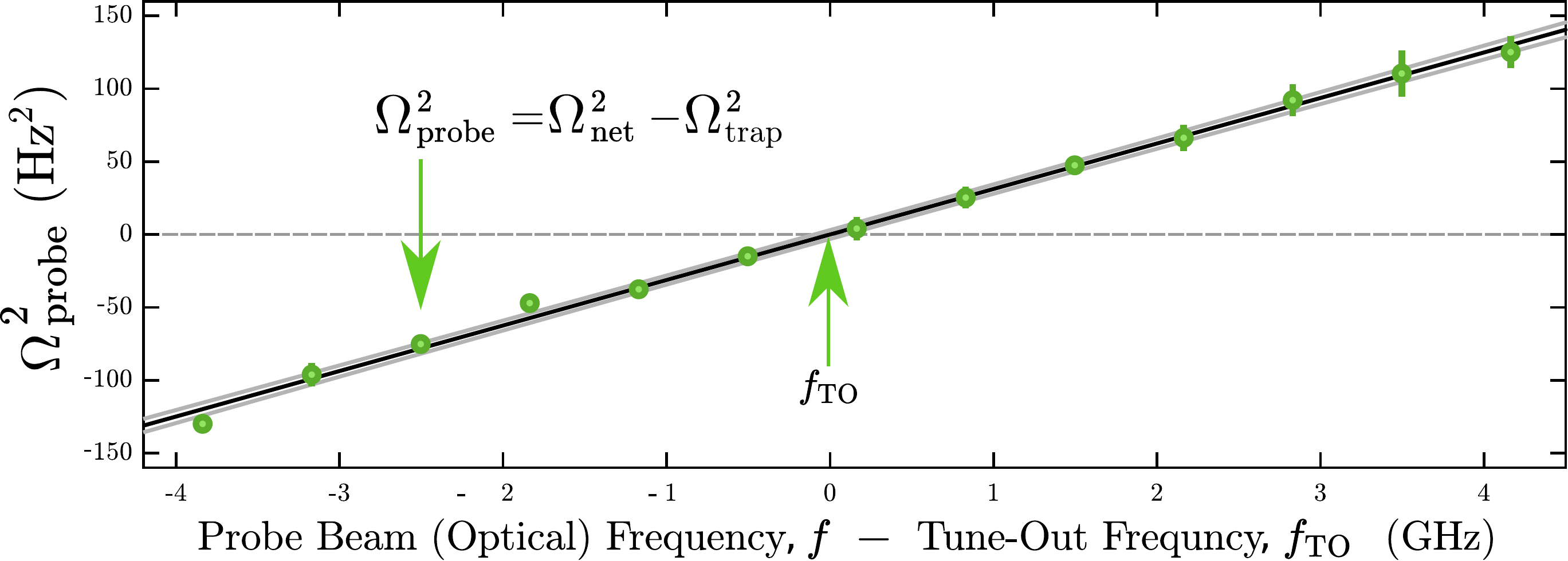}
	\caption{
	Determining the tune-out frequency for fixed polarization. The squared probe beam trap frequency (response) is found by comparison to a measurement of the unperturbed trap frequency. We repeated these measurements over a small frequency interval and extract the tune-out by finding the \textit{x}-intercept of the response. Also shown is the linear fit (black solid line) with light grey lines showing the \(1\sigma\) confidence intervals about the expected value of the mean. All error bars on points represent the standard error in the mean.
	}
	\label{fig:freq_scan} 
	\end{figure}

	An immediate corollary of Eqn. \ref{eq:delta_omega_squared} is that if the probe beam power is stabilized then $\Omega_\mathrm{probe}^2$	is linear solely in $\mathrm{Re}(\alpha)$. 
	It then follows that ascertaining the frequency at which $\Omega_\mathrm{probe}=0$ constitutes a determination of the tune-out point.
	Figure \ref{fig:freq_scan} shows the linear variation of experimentally-determined values of $\Omega_\mathrm{probe}^2$ versus the probe beam frequency $f$, demonstrating that this approximation holds in practise. 
	Indeed, we found that including a quadratic term in a fit to $\Omega_\mathrm{probe}^2(f)$ yielded an estimate of the zero-crossing frequency that was not statistically distinguishable from the zero-crossing estimate obtained from a linear fit.
	Section \ref{sec:implementation} provides more information about the experimental methods we used to obtain these measurements.

\subsection{Experimental implementation}
	\label{sec:implementation}
	We implemented such a hybrid trap using the BiQUIC machine and the tunable laser system described in chapter \ref{chap:apparatus}, especially section \ref{sec:spec_laser}. 
	The beam was tightly focused down to a waist of order 10 $\mu$m at the magnetic trap centre, then aligned with the weak axis of the harmonic magnetic trap as illustrated in Fig. \ref{fig:exp_schematic}.
	The laser system delivers up to 150 mW of power to the vacuum chamber when operating at the wavelengths relevant to this measurement. 
	We then scanned the laser wavelength $\pm4$ GHz about the tune-out in 13 steps ($\sim666.7$ MHz intervals), depicted in Fig. \ref{fig:freq_scan} .	
	This range is chosen in order to minimize the nonlinearities that are present when the optical potential is strong (see the discussion of systematic effects in section \ref{sec:systematic_effects}) while still presenting a sufficient signal-to-noise for interpolation of the linear response.
	
	For a given magnetic or hybrid (magnetic and optical) trap, the trap frequency can be determined by repeatedly sampling the momentum during a single shot using a pulsed atom laser. 
	The trap frequency measurement technique is discussed briefly in the next section, and in more detail in the next chapter (specifically section \ref{sec:n0_cal}). 
	First, I describe the experimental sequence used to realize the test conditions.
	Each measurement began with the production of a BEC of $(6\pm1)\times10^5$ \mhe~atoms\footnote{The variation in atom number reflects the range of atom numbers used in all runs. Generally the atom number in a given data collection run has a standard deviation of no more than 10\% of the mean.} cooled to $\sim80$ nK ($T/T_c\approx 0.1$) to reduce damping via coupling to the thermal part.
	Just prior to the frequency measurement we applied a 50~$\mu$s current pulse to a small coil near the BEC, inducing oscillation in the $y$-axis with initial amplitude (velocity) of $\approx30~\mu$m ($15$ mm/s).
	About 1\% of the atoms in the BEC were outcoupled every 8ms by a 5$\mu$s pulse of RF radiation (resonant with the trap centre at 1.7 MHz), creating a pulsed atom laser.
	The outcoupled atoms are in the $m_J=0$ state which is unaffected by magnetic fields, including the trap, and thereafter freely fall the 852~mm to the MCP-DLD (section \ref{sec:DLD}).
	From the time-of-flight data we reconstruct the initial velocity of each atom.
	The average velocity of atoms in a given pulse is an estimate of the centre-of-mass velocity of the BEC at the time the pulse was applied.
	The in-trap oscillation thus resolves as a variation in the centre-of-mass position on the detector, as shown in Fig. \ref{fig:trap_freq_eg}.
	The underlying trap frequency is then determined by means of the anti-aliasing technique described in the next section and in section \ref{sec:n0_cal}.
	Ultracold clouds below the condensation temperature were used because thermal clouds are more strongly damped in their oscillation, which is the dominant limitation for the precision of the trap frequency measurement described below (see e.g., \cite{Henson22_PAL}). 
	This is a major factor in the improved precision of this measurement over the previous one \cite{Henson15}, thus while thermal clouds can certainly be used, they would compromise the accuracy of the measurement.

\subsubsection{Measuring the trapping frequency}
	\label{sec:trap_freq_meas}

\begin{figure}
	\centering
		\includegraphics[width=0.8\textwidth]{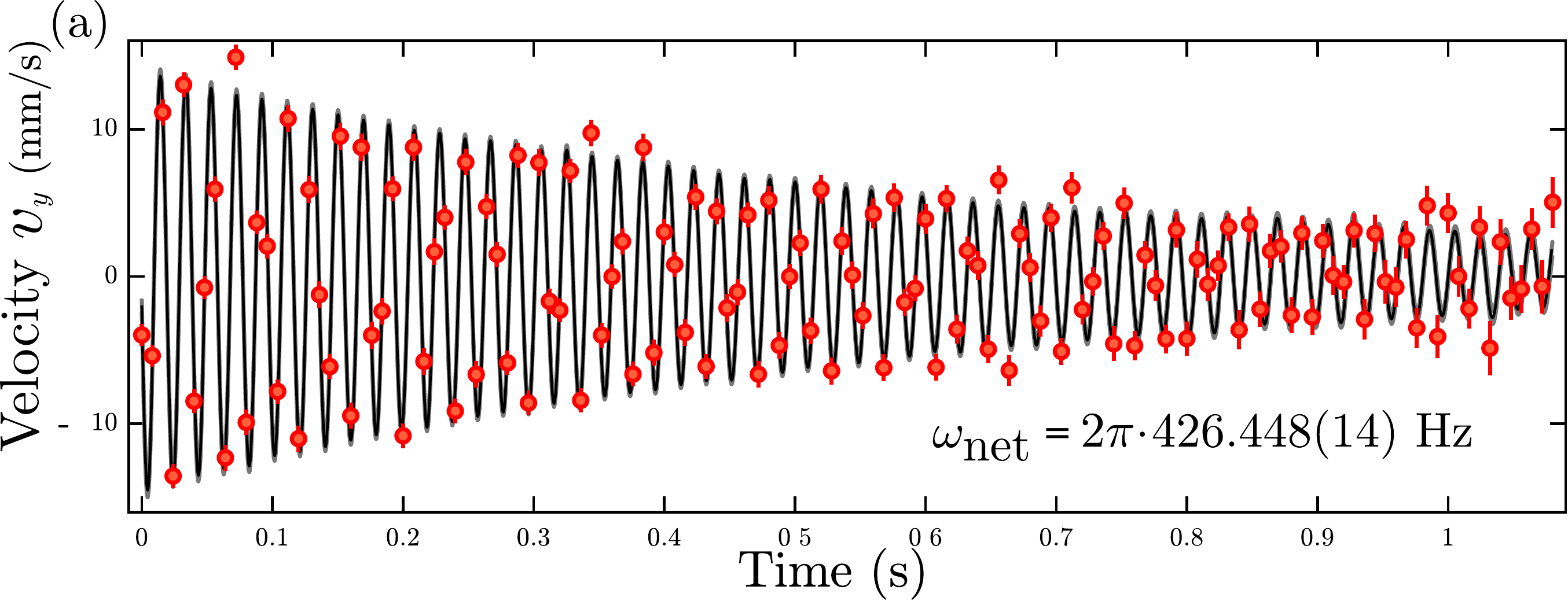}
			\caption{ Determining the trap frequency. The mean velocity of each pulse in the \textit{y}-direction ($v_y$) is used to trace out the oscillation over time (red points) and extract the oscillation frequency with a damped sine wave fit (solid line) for each realization of a BEC. The fitted frequency parameter is used to obtain the underlying trap frequency, as described in the surrounding text. A single experimental realization is shown.}
			\label{fig:trap_freq_eg}
	\end{figure}

	The oscillation frequency can be determined by fitting a damped sine wave to the centre-of-mass velocity as a function of time, as shown in Fig. \ref{fig:trap_freq_eg}.
	The sampling rate of the pulsed atom laser is limited by the velocity width of the BEC in the vertical axis (\(\sim40\)~mm/s, which corresponds to a temporal width at the detector of \(\sim\)4~ms) along with the vertical oscillation amplitude. This presents a challenge for measuring the trap frequency as the oscillation is much faster than the sampling rate. 
	Thus the true oscillation frequency is \emph{aliased} to a frequency below the Nyquist frequency of the sampling (with an 8 ms period, the sampling rate is 125~Hz, and the Nyquist frequency is half this again, about 62 Hz).
	It is possible to reconstruct the true trapping frequency using an anti-aliasing technique which I describe in more detail in section \ref{sec:n0_cal}.
	We find that our magnetic trap is extremely stable, with characterisic frequencies of $(\omega_x,\omega_y,\omega_z)= 2\pi\times(55.4(3),426.6(1),428.41(4))$ Hz (where the values in brackets represent the standard deviation all measurements, which is larger than the typical variation within a single run).
	
	The trap frequency measurement alternates between the combined trap and a purely magnetic trap to account for any drift in the magnetic trapping frequencies. 
	This calibration permits a determination of $\omega_\mathrm{probe}$ in two shots (about 50 seconds). 
	Given a measurement of the frequency $\omega_\mathrm{net}$ of the combined potential, the trap frequency $\omega_\mathrm{trap}$ (at the time of the interrogation) is predicted  by evaluating an interpolated model of the calibration frequency (smoothed by a Gaussian kernel with $\sigma=60$~s).
	This was done to correct for drifts of the magnetic trap frequency and to reduce noise. 
	The calibration data can also be used to obtain an estimate of the error in the trap frequency, which combines (true) trap frequency variation along with measurement error.

\subsubsection{Measuring polarization}

	We can measure the polarization of the light beam by use of a polarizing beamsplitter (PBS) and quater-wave plate.
	For our measurements we used a rotating Glan-Thompson polarizer with a high extinction ratio $>10^{5}$ in the transmitted component. 
	The calcite crystal in this type of polarizer internally reflects the $p$-polarized component (parallel to the \emph{plane of incidence} spanned by $\hat{k}$ and the normal to the optical surface) at the crystal interface, which leaves the $s$-polarized (parallel to the reflective surface) to exit through the transmitting port.
	That is, when the reflecting surface is vertical in the laboratory, the vertically-polarized component is transmitted, and the horizontally-polarized component is reflected.
	The fraction $p_T$ of optical power transmitted and reflected by the PBS are given by the projections $p_T=\Tr(\mathrm{V}\chi)=\frac{1-\st{Q}}{2}$ and $p_R=\Tr(\mathrm{H}\chi)=\frac{1+\st{Q}}{2}$, respectively. 
	Therefore one has $\st{Q} = \Tr(\sigma_1\chi) = p_R - p_T$. In terms of the absolute powers $P_T$ and $P_R$ measured at the beamsplitter ports, one must normalize by the total power and find 
	\begin{equation}
		p_R - p_T = \frac{P_R-P_T}{P_R+P_T}.
	\end{equation}
	which is commonly known as the \emph{contrast}.
	If one rotates the polarizer relative to the beam\footnote{This is usually much easier than rotating the lab relative to the polarizer.} by an angle $\theta$, this is equivalent to transforming the Stokes vector by matrix multiplication with
	\begin{equation}
		R(\theta) = \begin{bmatrix}
					1 & 0 & 0 & 0 \\
					0 & \cos(2\theta) & \sin(2\theta) & 0 \\
					0 & -\sin(2\theta) & \cos(2\theta) & 0 \\
					0 & 0 & 0 & 1
					\end{bmatrix}.
	\end{equation}
	The framework for considering the action of linear-optical elements by expressing them in such 4$\times$4 matrices which act on the Stokes vector is known as the Mueller calculus. Accordingly, the preceding matrix is (an example of) a Mueller matrix. 
	This allows a quantification of the observation that the maximum contrast occurs when the reflecting port of the polarizer is aligned (or orthogonal) with the major axis of the polarization ellipse.
	Denoting, then, the maximum and minimum transmitted powers, the maximum contrast (also called the \emph{visibility}) is $(P_\textrm{max}-P_\textrm{min})/(P_\textrm{max}+P_\textrm{min})$, where $P_\textrm{max}$ (resp. $P_\textrm{min}$) is the maximum (minimum) power transmitted over all polarizer angles $\theta$, which occurs when $\theta=\theta_\textrm{max}$ (resp. $\theta=\theta_\textrm{min}$). 
	Due to energy conservation considerations we should have $p_\textrm{max} = 1 - p_\textrm{min}$. 
	Further, if the major axis of the ellipse is at some angle $\theta_L$ from the laboratory $x$-axis, then the transmitted power will be minimized when the polarizer angle is at $\theta_L/2$ relative to the horizontal, c.f. the trigonometric terms in $R(\theta)$.
	Therefore we can obtain the laboratory $\st{Q}$ component via 
	\begin{equation}
		\mathcal{Q_{L}} =\frac{P_{\mathrm{max}}-P_{\mathrm{min}}}{P_{\mathrm{max}}+P_{\mathrm{min}}} \cos(2\theta_{\mathrm{min}}).
	\end{equation}
	We see also that the rotation matrix $R(\theta)$ does not mix the linear polarization components with $\st{V}$; the relative phases of the orthogonal oscillations are invariant.
	However, a similar argument to that just given can show that 
	\begin{equation}
		\mathcal{U_{L}} =\frac{P_{\mathrm{max}}-P_{\mathrm{min}}}{P_{\mathrm{max}}+P_{\mathrm{min}}} \sin(2\theta_{\mathrm{min}}),
	\end{equation}
	and so it follows from $|\mathbf{s}|=1$ that
	\begin{equation}
		\mathcal{V_{L}}^2 =\frac{4 P_\textrm{max}P_\textrm{min}}{(P_\textrm{max}P_\textrm{min})^2},
	\end{equation}
	which determines the degree, but not handedness, of the circular polarization in the lab frame.
	However, as we will see in section \ref{sec:fitting}, the sign of $\st{V}$ need not be determined directly as either sign can be made consistent with the tune-out equation within our analytical procedure.

\subsubsection*{Polarization at the atoms}

	The transformation of $\mathcal{Q}$ from the laboratory to the atomic reference frame is given by a rotation around the probe beam axis by the angle \(\mathcal{\theta_{L}}\),
	and hence can be written in terms of the lab-measured parameters as
	\begin{equation}
	 \mathcal{Q_{A}} =\frac{P_{\mathrm{max}}-P_{\mathrm{min}}}{P_{\mathrm{max}}+P_{\mathrm{min}}} \cos(2(\mathcal{\theta_{L}}+\theta_{\mathrm{min}})).
	 \label{eqn:q_transform}
	\end{equation}
	This can also be interpreted as the rotation around the beam axis which aligns the laboratory polarizer angle origin with the component of the magnetic field vector which is orthogonal to $\mathbf{k}$. 
	Note that the circular component of the light is unchanged by the coordinate transformation, thus let us denote \(\mathcal{V_{L}}=\mathcal{V_{A}}\) by  \(\mathcal{V}\) hereafter.
	However, the rotation between the coordinate axes in the lab and atom frames mean that the Stokes parameters in the atomic frame are given by their respective projections, quantified in trigonometric coefficients.
	Thus, the final form of the tune-out equation is
	\begin{align}
		 \fto(\st{Q},\st{V}) = & \fto^S + \frac{\st{V}}{2}\cos{\theta_k}\beta^V - \frac{1}{2}\beta^T \left(\frac{3}{2}\sin^2(\theta_k)(1+\st{Q}) - 1\right)
		 \label{eqn:tune_out_eq}
	\end{align}
	where $\st{Q}=\st{Q_A}(\st{Q_L},\theta_L)$ in the atomic frame, as defined in Eqn. \ref{eqn:q_transform}. 
	The tune-out value $\fto(\st{Q}=-1,\st{V}=0)= \fto^S + \frac{1}{2}\beta^T$ is the sole option that eliminates any dependence on $\theta_k$, and we can leave $\theta_L$ as a fit parameter.

\section{Determination of the tune-out frequency}

	We obtain \(f_{\mathrm{TO}}(-1,0)\) through three stages of measurements. 
	First, as described above, we fix the probe beam polarization $(\mathcal{Q_A},\mathcal{V})$ and frequency $f$, and then measure the atomic polarizability via the squared shift of the centre-of-mass oscillation frequency of a BEC in the combined optical dipole and magnetic potentials ($\Omega_\mathrm{probe}^{2}(f)$).
	Second, we take many such measurements while scanning the probe beam frequency and find the optical frequency where the polarizability is zero for this probe beam polarization (i.e. the tune-out frequency \(f_{TO}(\mathcal{Q_A},\mathcal{V})\)) . 
	Finally, we measure $\fto(\mathcal{Q_A},\mathcal{V})$ for many settings of the probe beam polarization and extract \(f_{\mathrm{TO}}(-1,0)\) through the procedure described in this section.

\label{sec:TO_analysis}

\subsection{Exclusion criteria}

	An experiment this complex is susceptible to many potential failure modes, and it is prudent for the experimentalist to mitigate against them.
	To ensure the integrity of the data set used for analysis we complemented the aforementioned control systems with a veto checklist in the data preprocessing pipeline.
	This consisted of a software screening procedure which discarded any measurements where any of the following essential conditions failed to hold. These conditions result in a loss of only one shot per $10^{4}$ across the whole dataset of 43500 shots.

	\textbf{Single-mode laser:}	On rare occasions we observed that the Ti:S laser would spontaneously run in a mode where multiple optical frequencies were resonant in the cavity and present in the output, a so-called multi-mode regime (strictly a \emph{temporal} multi-mode regime, as distinct from the case where multiple spatial modes are present in the beam profile). This behaviour would cause the wavemeter to provide inaccurate frequency measurements, and also fail to produce the desired test conditions at the interaction zone. We eliminated runs where this happened by recording the output of a scanning Fabry-Perot cavity (Thorlabs SA200-5B). We scanned the cavity over twice its full free-spectral range in a sawtooth manner by scanning the voltage across a piezoelectric actuator, which modulates the cavity length, at frequency of 20 Hz. 
	We recorded the piezo and photodiode voltage during the interrogation and also in a $\sim0.2$~s period before and after.
	The multimode condition was triggered when the separation between peaks detected in the photodiode voltage were less than the free-spectral range, indicating the presence of multiple optical frequencies.
	To detect these peaks we processed the voltage recordings from the photodiode at the output of the cavity using a peak-detection algorithm. The threshold for detection was a peak height that was $1.5\times10^{-3}$ times the peak transmission intensity, which could distinguish spurious (multimode) peaks from electronic noise (all multimode conditions we  observed had peak intensities some orders of magnitude larger than this).

	\textbf{Stable probe power:}		We employ a check on the probe beam power as measured by a photodiode that monitors a sample of the beam just before it enters the experimental chamber, as shown in Fig.~\ref{fig:tunable_laser}. This ensures that the power feedback system is operating at set point in all measurements used in the analysis. Restrictive thresholds are set on the mean ($<0.02$ fractional difference to set point), standard deviation ($<0.05$ of set point) and maximum difference between the mean and the set point ($<0.03$ of set point) of the power during interrogation.

	\textbf{Stable probe frequency:}		The probe beam frequency at each measurement is taken to be the average optical frequency returned by the wavemeter during the interrogation. Shots were rejected if the frequency records in the lock loop's log file had a standard deviation greater than $3$~MHz or a range greater than $5$~MHz. These thresholds were based on our observations of typical locked performance, wherein the wavemeter reading displayed excursions from the mean that were smaller than the aforementioned ranges.

	\textbf{High atom number:}		Shots with very low atom number would lead to inaccurate determinations of the trapping frequency because the damped-sine fit would not be able to fit to many points. We mitigated this by screening out shots with less than $2\times10^{5}$ detected atoms. We found that the choice of threshold introduces a negligible systematic shift in the final tune-out determination, hence we could use shots with a wide range of atom numbers.

	In the end, it is likely that most of the failed shots would have been detected by the outlier-removal step in the fitting procedure, and thus it is plausible that these exclusion rules were of marginal benefit. However, we did not compare the results of running the analysis with and without the exclusion.

\subsection{Regressing the tune-out against polarization}
 \label{sec:fitting}

\begin{figure}
	    \centering
	    \includegraphics[width=\textwidth]{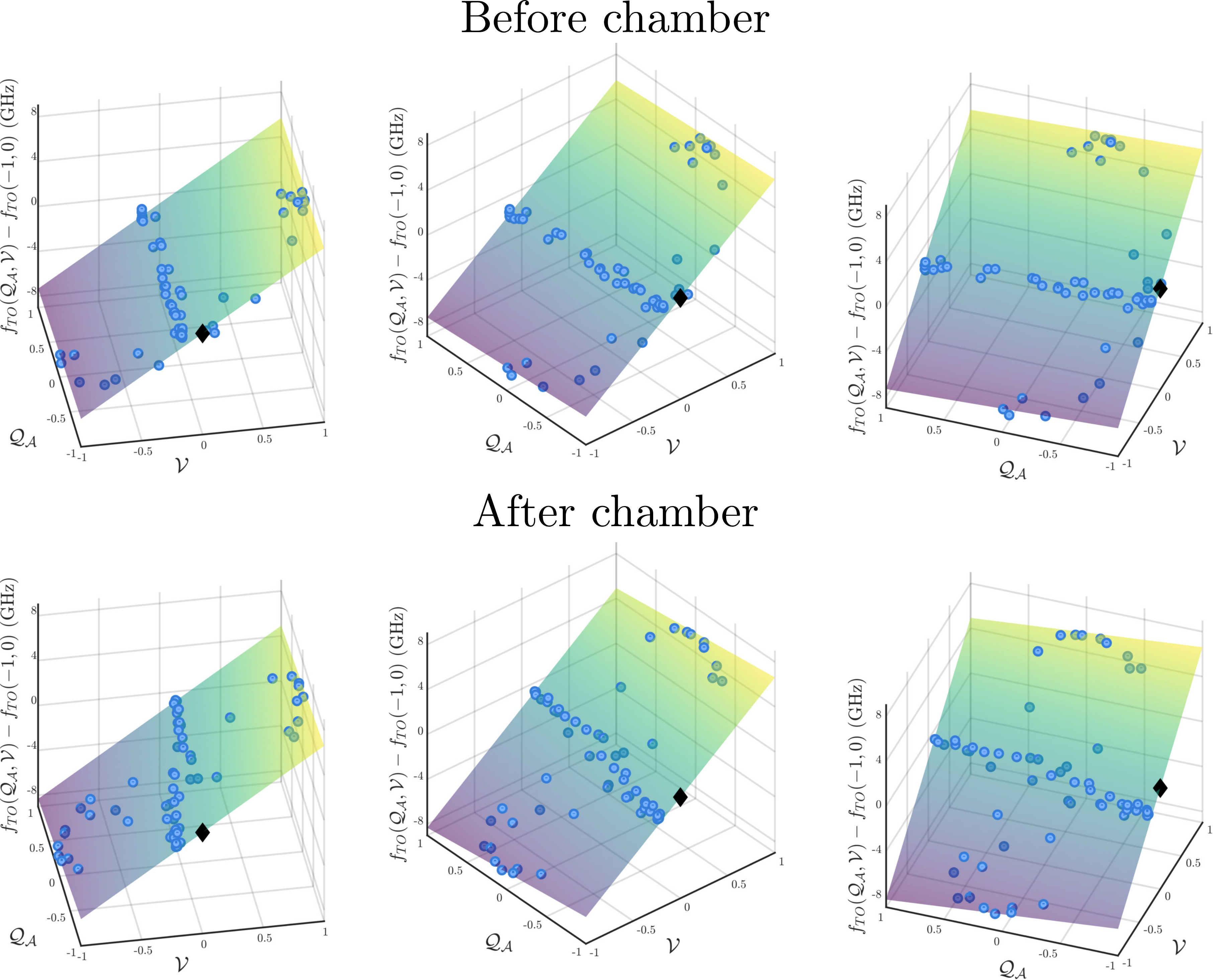}
	\caption{Visualizing the fit to $f_{TO}$ as a function of the polarization parameters $\mathcal{Q_{A}}$,\, $\mathcal{V}$  (see Eq.~(\ref{eqn:tune_out_eq})). The top row shows a fit using the polarization data taken before the vacuum chamber. The bottom row uses data taken after the vacuum chamber.  Each tuneout measurement is shown as a blue point. Note that (a-c) and (d-f) show, respectively, different viewing angles on the same data. The black diamond shows the value for \(f_{\mathrm{TO}}(-1,0)\) obtained in each case: \(f_{\mathrm{TO}}(-1,0)=725\, 736\, 810(40)\)~MHz in the top row, and \(f_{\mathrm{TO}}(-1,0)=725\, 736\, 610(40)\)~MHz in the bottom row. The other fit parameters are \(\beta^V \cos(\theta_k)=13240(70)\)~MHz, and \(\beta^T \sin^2(\theta_k)=1140(20)\)~MHz.}
	
	\label{fig:full_tune_out}
	\end{figure}

\begin{figure}
	\centering
	\includegraphics[width=\textwidth]{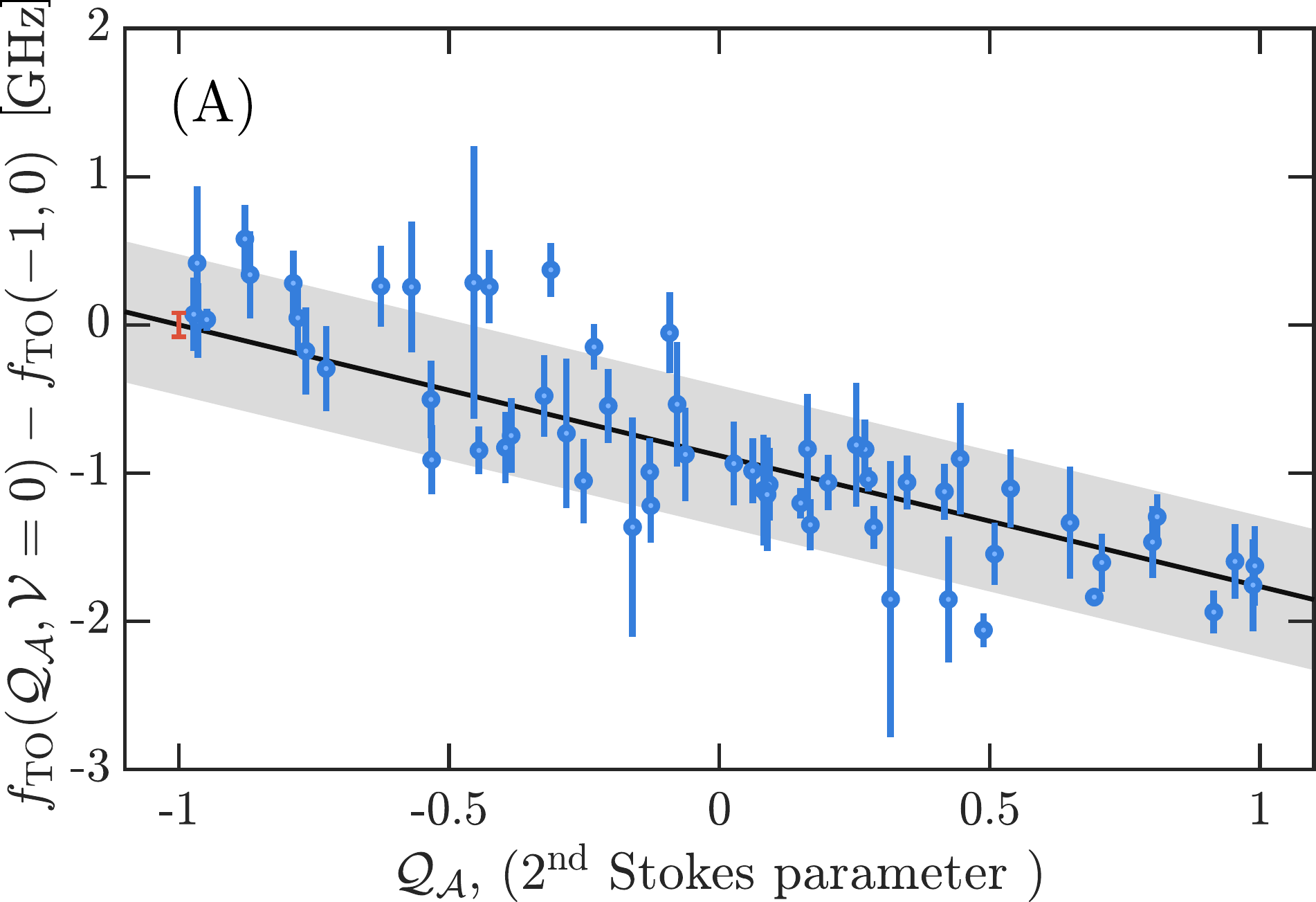}
	\includegraphics[width=\textwidth]{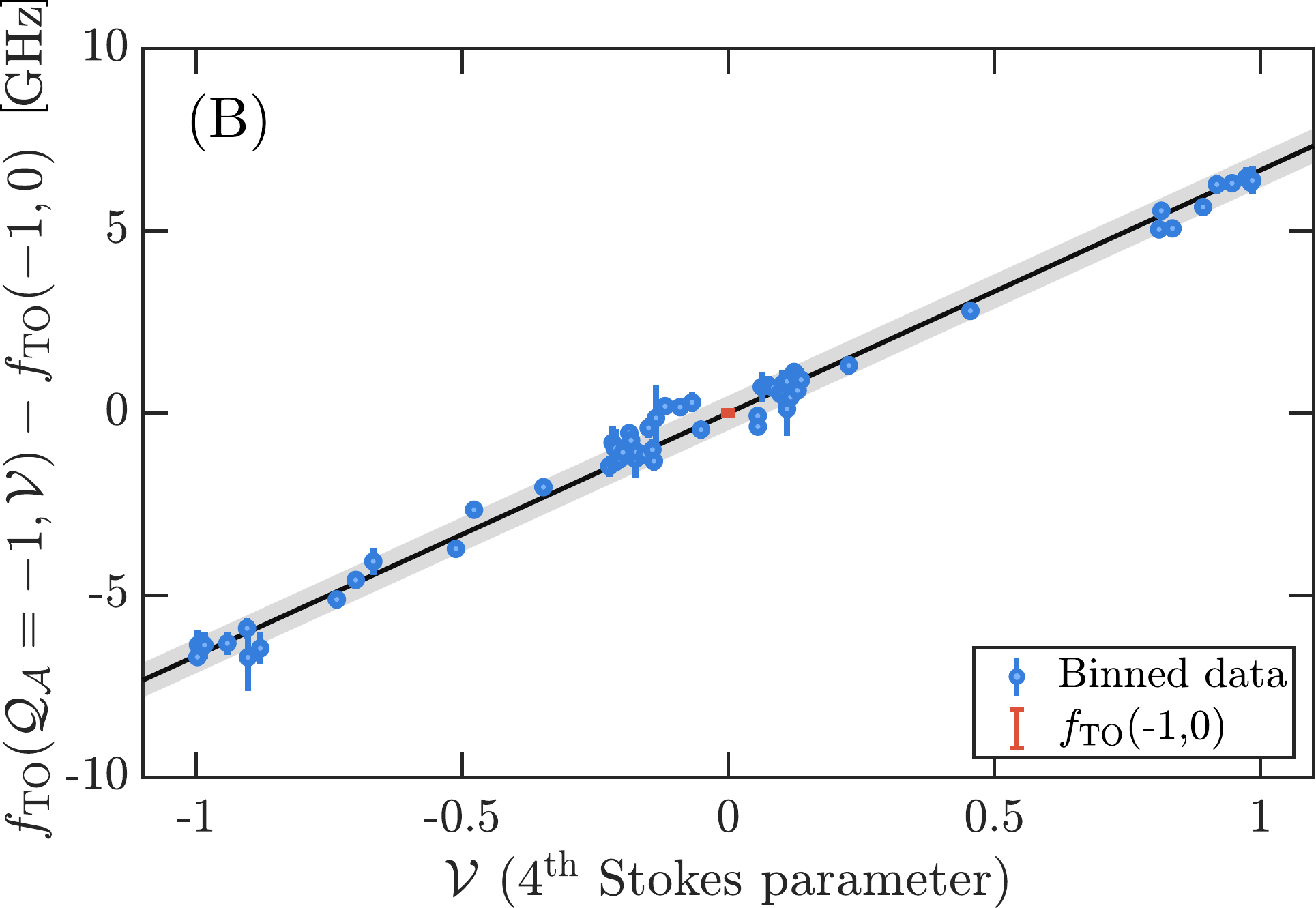}
	\caption{
	Tune-out dependence on probe beam polarisation.
	 In (A), the data points show the $\omega_{TO}$ for each measurement after interpolating to the polarization state  ($\mathcal{Q_{A}},\mathcal{V}=0$). 
	Similarly, (B) shows the interpolation to  ($\mathcal{Q_{A}}=-1,\mathcal{V}$). 
	Error bars on the blue points are the estimated standard error in the mean value for a given polarization. The light-shaded and dark-shaded areas are the $1\sigma$ observation and confidence intervals, respectively. 
	The distinguished point is the predicted $\fto(-1,0)$, with the error bar indicating the (non-simultaneous) confidence interval.
	}
	\label{fig:pol_TO} 
	\end{figure}

	We measured the tune-out frequency for 75 (\(\mathcal{Q_{L}},\mathcal{V}\)) pairs, comprising a total of $736$ tune-out measurements.
	We then fit the linearized model (Eq.~(\ref{eqn:tune_out_eq})) to the data, regressing $\fto$ against the measured $\mathcal{Q_{L}}$ and $\mathcal{V}$ values using five free parameters: $f^{S}_{TO}$ (the frequency at which the scalar polarizability $\alpha^S(f)=0$), the angle $\theta_{\mathcal{L}}$ between the lab and atomic reference frames, the angle $\theta_{k}$ between the beam propagation vector $\hat{k}$ and the magnetic field pointing, and the reduced vector- and tensor-polarizabilities $\beta^V$ and $\beta^T$ (see section \ref{sec:polz_linearization}).
	We can then predict a value for $\fto(-1,0)$, despite the lack of precise knowledge of the magnetic field pointing, by evaluating the resulting model at \(\mathcal{Q_{A}}=-1,\mathcal{V}=0\) to find \(f_{\mathrm{TO}}(-1,0)\). 
	The statistical error in the calculated \(f_{\mathrm{TO}}(-1,0)\) is determined with a bootstrapping technique\footnote{In the bootstrap method, one applies an analysis procedure on subsets of the data. The variance of the resulting findings provides an estimate of the sample variance, quantifying the statistical error in the fit to the entire data set. Source code is available at \cite{bryce_bootstrap_error_code}.}. 
	Figure \ref{fig:full_tune_out} displays the $\mathcal{Q_{A}}$ value calculated from the measured $\mathcal{Q_{L}}$ using the fitted $\theta_{\mathcal{L}}$. The measured \(f_{TO}\) is displayed as a function of $\mathcal{Q_{A}}$, $\mathcal{V}$. Figure ~\ref{fig:full_tune_out} also shows the tune-out calculated using polarization measurements taken before and after the vacuum chamber.
	In Fig. \ref{fig:pol_TO}, the measurements are displayed by interpolating the tune-out measurements to the one-dimensional subspaces defined by $\st{V}=0$ (a) and $\st{Q}=0$ (b).
	
	The preceding procedure amounts to finding the plane of best fit for trivariate data (i.e. triples of the form $(\st{Q},\st{V},f_{\mathrm{TO}}(\st{Q,V}))$), which requires only three parameters for complete specification. The use of five free parameters is an over-parametrization of the plane and thus the fit is overdetermined and cannot uniquely fix the fit parameters without additional measurements of at least one such quantity.	
	Further, the fit gives equal agreement between either polarity of \( \beta^T \), compensating by finding a different optimal value for $\theta_L$, and thus preventing a determination of \(f_{\mathrm{TO}}(-1,0)\). 
	To eliminate this issue, we constrain the sign of \( \beta^T \) using measurements and simulations of the magnetic field pointing and theoretical atomic structure calculations, both of which agree with the sign constraint \( \beta^T >0 \).
	We can also see here that the sign of $\st{V}$ is not critical; the cosine term can take on either sign, whereas the sine-squared term will be unchanged by a shift $\theta_k\mapsto\pi-\theta_k$.
	
	The presentation of the fit results in the remainder of this chapter include two distinct quantifications of statistical uncertainty intervals. The \emph{observation interval} quantifies the variance in data that is not explained by the predictor variables $\st{Q}$ and $\st{V}$ (\emph{i.e.} the noise around the mean value). This can also be interpreted as a measure of the expected variance of future observations at a given polarization value. In contrast, \emph{confidence intervals} quantify the uncertainty in the estimates of the model parameters. The  interval for a given confidence level $C$ denotes the range of values wherein a model parameter is expected to lie with a probability of $C$. 
	The confidence interval is similar to the standard error in that additional data can reduce the width of the confidence interval, whereas the observation interval quantifies the systematic variance and does not generally decrease with the addition of more data.
	The stated values for the fit parameters throughout this chapter are generally the 95\% confidence intervals, and when we predict the value of $f_\mathrm{TO}(-1,0)$, the relevant value is the mean value of tune-out measurements at this polarization value, and as such the relevant uncertainty interval is the confidence interval. The 1$\sigma$ observation interval ($C\approx 0.68$) is used to illustrate the variance of the data.

	The validity of the observation interval can be quantified by the reduced chi-squared ($\chi^2$) statistic, defined as follows.
	Given a model with input predictor- and outcome-data tuples $(x_i,y_i)$, the residuals are the difference $r_i=\hat{y}_i - y_i$ between the predicted value $\hat{y}_i$ and the true value. 
	If the estimated standard deviation of the model error (i.e. the 1$\sigma$ observation interval) is $\sigma_i$ (which may be defined at each $x_i$), then the chi-squared statistic is 
	\begin{equation}
	\chi^2 = \sum_i \left(\frac{r_i}{\sigma_i}\right)^2.
	\end{equation}
	The reduced chi-squared statistic is $\chi^2/d$, where $d$ is the number of error degrees of freedom (that is, the number of observations minus the number of fit parameters).
	If the model error estimates are accurate (and one has normally-distributed residuals) then the reduced chi-squared statistic, also written $\chi^2$/dof, will be close to 1.
	On the other hand, a $\chi^2$/dof much larger than 1 indicates the error term is too small or that the model is overfitted (i.e. underestimates the underlying noise); a $\chi^2$/dof less than 1 indicates the model error is too large. In both extreme cases, the model is unlikely to generalize well to new data, but when $\chi^2$/dof $\approx$1, the predictions are expected to be more reliable. The $\chi^2$/dof is given in figure captions where relevant.

\subsection{Validating linearity against theory}
 
We can use the values of the fit parameters to estimate the size of the polarization-dependent contributions to the measurement. That is, as a sanity check, we can make a low-precision comparison between theoretical predictions for the coefficients $\beta^V$ and $\beta^T$, and also compare the predicted and experimental value of the polarizability gradient $d\alpha/df$. These values are summarized in Tab. \ref{tab:param_compare} and the means by which we estimate them are given below. These values are not the primary interest of this work and accordingly we do not expect precise values to come from the experiment. However, we should expect that the range of values obtained from the experiment should at least include the theoretical prediction. 


	First, we attend to an estimate of the gradient $d\alpha/df$, starting with its relation to the probe-beam induced shift $\Omega_\textrm{probe}^2$ in the (squared) trapping frequency (c.f. Eqn. \ref{eqn:omega_probe}),
	\begin{align}
	\frac{d~ Re(\alpha)}{d~f}&=A^{-1}\frac{d~\Omega_{\text{probe}}^2}{d~f},~\textrm{where}\\
	    A&=\frac{P}{m \epsilon_{0} c \pi^3  w_0^4}=\frac{1}{ \pi^2 m \epsilon_{0} c }\frac{I}{2 w_0^2}.
	    \label{eqn:da_df}
	\end{align}
	Figure \ref{fig:freq_scan} shows a response with a slope of approximately $d\Omega_\textrm{probe}^2/df \approx 30$ Hz$^2$/GHz, which is typical of the data collection runs. 
	We then need to compute the conversion factor, which in turn demands an estimate of the power and spot size of the laser beam at the interaction zone.
	We recorded of the probe beam power and profile obtained from measurements outside the vacuum chamber, but the vacuum window (through which the beam passes) subtly alters the beam profile and focal point.
	However, the arguments leading to Eqn. (\ref{eqn:omega_probe}) show that the tune-out is independent of the probe beam intensity (to first order) and therefore regular measurements of the beam profile were not necessary. 
	Hence, based on the measurements we did take of the profile through the course of the experiment, we adopt a conservative margin of error for the beam waist, using the range 15(5) $\mu$m. 
	Similarly we assume knowledge of the beam power at the trap to within about 20\%, i.e. $P\approx140(30)$ mW. The conversion factor $A=\frac{P}{m \epsilon_{0} c \pi^3  w_0^4}$ consistent with these values is $A\approx5(7)\times10^{45}$ kg A$^{-2}$  s$^{-6}$. 
	The uncertainty in this estimate is obtained by simple propagation of errors, and is dominated by the strong dependence on the beam waist that converts a 20\% error in $w_0$ into a $\sim140$\% uncertainty in $A$. 
	Note that Eqn. (\ref{eqn:da_df}) shows that the value $A$ must be positive if $d\Omega_{\textrm{Probe}}^2/df$ is positive (which it is in all shots included in the analysis), hence we can specify the uncertainty could be stated more strictly as $A=5^{+7}_{-5}\times10^{45}$.
	However, as this quantity is calculated for the purposes of a sanity-check, this detail is omitted from the remainder of the calculation.
	Thus the experimental value of the polarizability gradient is obtained by $d\Omega_{\textrm{Probe}}^2/df A^{-1}$ and has the value $5(7)\times10^{-54}~\textrm{C}~\textrm{m}^2~\textrm{V}^{-1}~\textrm{Hz}^{-1}$,
	 where the bracketed digit includes the statistical variation across all runs used in the analysis as well as the uncertainty due to imperfect knowledge of the beam parameters at the interrogation zone. 
	This value is included, with associated uncertainty interval, in Tab. \ref{tab:param_compare} and is broadly consistent with the theoretical value in that the theoretical value lies within the margins of error.
	The accuracy of this estimate is fundamentally limited by the quartic dependence on the beam waist, which was not precisely quantified as it is not a critical parameter for the purposes of measuring the tune-out frequency (Eqn. \ref{eqn:omega_probe} shows that the waist must only be stable, not precisely measured, to determine the tune-out frequency). However, as we discuss below in the context of the hyperpolarizability, the error induced by imperfect knowledge of the probe beam intensity is not significant.

	As for the polarization-dependent effects, our nonlinear fitting procedure returns $\beta^T\sin^2(\theta_k)=1140(20)$ MHz but, as noted above, unique values of $\beta^T$ and $\theta_k$ are not obtainable because the physical model is over-articulated (i.e. has more parameters than degrees of freedom due to physical constraints). 
	We can estimate the angle $\theta_k\approx 30^\circ$ from simulations of our magnetic trap, which is consistent with the fitted value of $25^\circ$. 
	Working with an estimate of $\theta_k = 27.5\pm(5)^\circ$, we determine the values shown in Tab. \ref{tab:param_compare}. 
	These calculations indicate effect sizes that are of the same order of magnitude as theoretical predictions, and whose uncertainty bounds also include the predicted value.
	Similarly, the theoretical value for $\beta^V$ is of comparable scale to the experimental value.
	The conclusion of the working in this section is that our linearized model for the tune-out is a good approximation for the system under relevant test conditions because our (approximate) inferences of relevant parameters agree with theoretical expectations.

	\begin{table}
	    \centering
	    \begin{tabular}{l c c c c c}
	        \hline\hline
	         Quantity & $d\alpha/df$ (C m$^2$ V$^{-1}$ Hz$^{-1}$) && $\beta^V$ (MHz) &&  $\beta^T$ (MHz)\\
	         \hline
	         Experiment & $5(7)\times10^{-54}$ &&  $1.5(1)\times10^4$ && $2.6^{+1.2}_{-0.6}\times10^3$\\
	         Theory & $1.78\times10^{-53}$ && $1.22\times10^4$ && $3.4\times10^3$\\
	         Ratio exp./thr. & 0.3(4) && 1.2(1) && \(0.8^{+0.4}_{-0.2}\)\\
	         \hline\hline
	    \end{tabular}
	    \caption{Comparison between the theoretical and experimentally-determined values for the gradient of the polarizability, and the reduced-vector and reduced-tensor polarizabilities. Note that the latter two are coupled via the fit model (through $\theta_k$) and hence this determination is not unique. Nonetheless, these comparisons show that the experimental values are generally consistent with the theoretical predictions. The parentheses denote the 1$\sigma$ uncertainty in the final digit. Uneven uncertainty intervals are quantified by the upper (superscript) and lower (subscript) values. All experimental values are within 2$\sigma$ of the predicted values.}
	    \label{tab:param_compare}
	\end{table}
\section{Systematic effects}
\label{sec:systematic_effects}

\subsection{Polarization}

	 The extraction of \(\fto(-1,0)\) needs an accurate determination of the probe beam polarization at the point of interaction with the atoms. 
	 As we did not have polarization optics mounted in vacuo, we inferred it from measurements outside the vacuum system. 
	 We estimated the contribution of two effects: the birefringence of the vacuum windows and the variation of polarization across the beam profile.

\subsubsection{Birefringence}
	
	The polarization of the light may be altered as it passes through the final optical element: The glass window into the vacuum chamber.
	This means that the in-vacuum polarization may be different from the polarization set by the pre-insertion optics and measured prior to vacuum entry.
 	We estimated the size of this effect by measuring the probe beam polarization before it entered the vacuum system, and again after it exited through a second window\footnote{A portion of the probe beam escaped back through the in-vacuum mirror at the LVIS and then the insertion window for the horizontal trapping beam of the first MOT, making this possible.}.
 	We determined the error by running the $\fto(-1,0)$ analysis, as above, using both sets of polarization measurements (see Fig.~\ref{fig:full_tune_out}). 
 	We found that these values agreed within \(200\)~MHz, and hence used the average of the two as the final measured value and, and the difference is taken to be the uncertainty associated with the window birefringence. 

\subsubsection{Polarization across the beam}

	We also observed a small variation in the polarization across the transverse profile of the beam, as measured through a 1mm-diameter aperture.
	We determined that the mirrors in the pre-vacuum optics were the culprit.
	We characterized this by measuring polarization at several points across the beam.
	Using the polarization measured far from the beam axis produces a value of \(f_{\mathrm{TO}}(-1,0)\) that is up to \(400\)~MHz different from using the polarization at the beam center. 
	However, these contributions from differently-polarized light are weighted by their respective intensities, and accordingly the  weighted values produce a \(150\)~MHz uncertainty in $\fto$.

\subsection{Linearity} \label{sec:syst.subsec:lin}
	There are two instances in the preceding discussion where a linear approximation has been made in order to render the analysis tractable.
	Here I quantify the possible systematic shifts that could arise from these approximations.

\subsubsection{Linearity of the frequency shift}
	The first instance is the assumption that the perturbation $\Omega_\mathrm{probe}^2$ is linear with respect to the probe beam frequency, which leads to the use of linear regressions on the individual scans to obtain \(f_{\mathrm{TO}}(\st{Q,V})\).
	This is only approximately true: The polarizability is a nonlinear function of laser frequency at larger distances from $\fto$ (on the THz scale). 
	Nonetheless, quadratic (and higher) terms included in the fits to the laser scans were not significant.
	Using a theoretical model of the polarizability, we found that a linear fit spanning 4~GHz either side of the tune-out would return a zero-crossing only $-88(1)$~kHz from the true tune-out. 
	If we were to measure out 40 GHz either side of the tune-out, this shift would increase to $-9.6(2)$~MHz because the quadratic (and higher-order) terms become more significant, reducing the accuracy of the linear fit.
	However, these shifts are reduced to 0.6~kHz and 40~kHz in the 4~GHz and 40~GHz cases, respectively, by including a quadratic term.
	In comparison to the other systematic effects, this linearization introduces a negligible error.

\subsubsection{Linearity of hybrid potential}
	We also made a linear approximation when deriving the relation $\Omega_{\text{net}}^2=\Omega_{\text{trap}}^2+\Omega_{\text{probe}}^2$, which was derived assuming the probe beam created a harmonic potential, ignoring higher-order derivatives of the Gaussian optical potential.
	The overlapping combination of harmonic traps is also harmonic, thus the equations of motion of an oscillating BEC are given by linear second-order ODEs.
	In contrast, a feature of oscillatory motion in anharmonic potentials is that the period of oscillation depends on the total energy (\emph{i.e.} on the amplitude).
	We checked for this by fitting the damped-sine model to different (contiguous) subsets of the atom laser pulses.
	While we did observe that the largest oscillations were 0.3 Hz slower than the smallest, there was no discernable effect on the zero crossing $\fto$.  
	Hence, again, this was found to be a suitable approximation.

\subsubsection{Linearity in the beam intensity}
	
\begin{figure}
	    \begin{minipage}{0.58\textwidth}
	    \vspace{0pt}
	    \includegraphics[width=\textwidth]{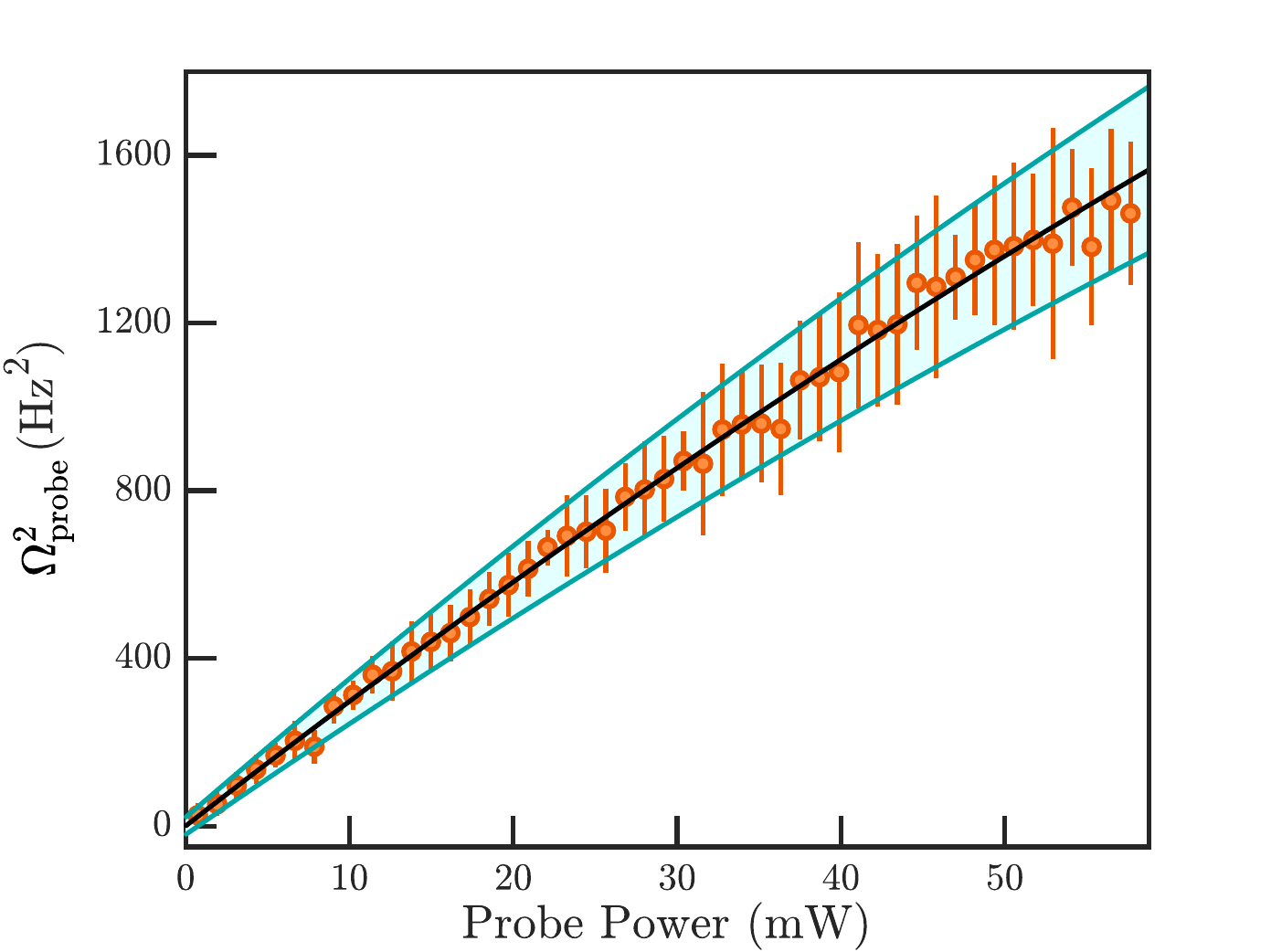}
	    \end{minipage}
	    \hfill
	    \begin{minipage}{0.4\textwidth}
	    \vspace{0pt}
	    \caption{Testing for nonlinear response versus power. We set the  probe frequency 20~GHz blue of $\fto$,  producing a strong potential, and fit the response against the laser power P with the model \( \Omega_{\mathrm{probe}}^2 = a  P - b P^2 \).  The fit parameters are \(a=\text{30.3(1)}\times 10^{-3}\text{~Hz}^2\text{W}^{-1}\), \(b= \text{0.06(2)} \times 10^{-6}\text{~Hz}^2\text{W}^{-2}\), and $\chi^2$/dof=0.992. Higher-order terms  were not statistically significant. The \(1\sigma\) (observation) confidence interval is shaded. 	    }
	    \label{fig:probe_beam_linearity}
	    \end{minipage}
	\end{figure}

	We measured \(\Omega_{\mathrm{probe}}\) as a function of probe beam power, with the polarization and frequency fixed, to constrain any shift with respect to probe beam intensity. A second-degree polynomial was sufficient to describe the response, illustrated in Fig.~\ref{fig:probe_beam_linearity}. 
	We deduced that using a linear fit at experimentally-relevant beam powers, rather than a quadratic fit, would lead to a  \(-24(30)\)~MHz change in the determined tune-out.

\subsection{Hyperpolarizability}

	The working above also assumes that the metastable-state energy shift due to a nonzero polarizability is a linear function of the light field intensity. 
	We combined theoretical predictions and experimental measurements to obtain an estimate for this error.

	\begin{figure}[t]
	    \centering
	    \includegraphics[width=\textwidth]{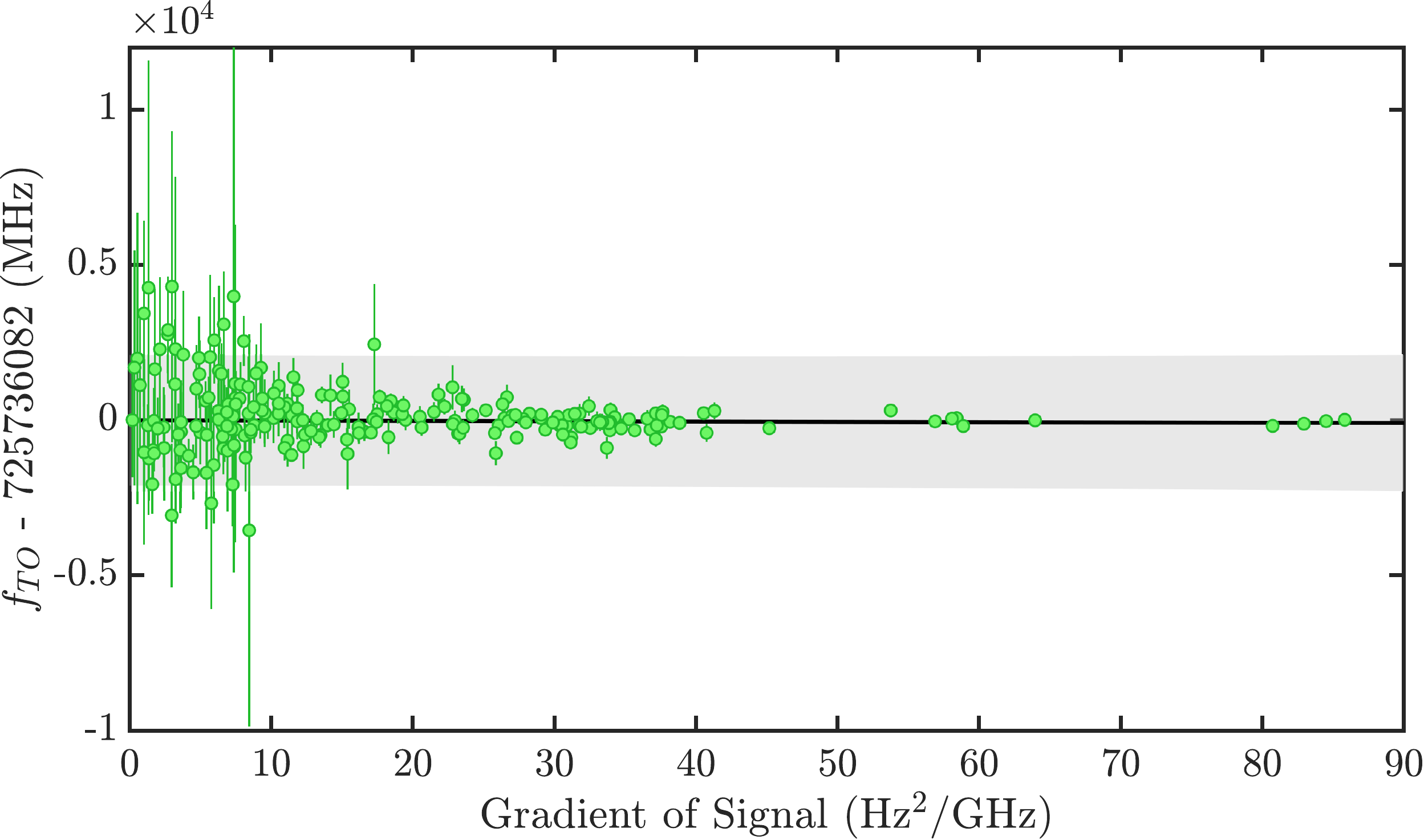}
		\caption{Measured tune-out dependence on probe beam intensity. The highest signal gradient (right) corresponds to a peak light field intensity of \(\sim4\times10^{8}\: \mathrm{W}\cdot \mathrm{m}^{-2}\). Error bars show the variance of binned data. The shaded region indicates the \(1\sigma\) observation interval, quantifying the variance in the data.  An uncertainty-weighted linear fit to the full data set data with parameters offset\(=725\,736\,082(3)\)~MHz and gradient -1(2) MHz/(Hz\(^2\)/GHz) is shown. 95\% confidence intervals in parameters are given in parentheses. This fit has $\chi^2$/dof=0.987 and determines that the gradient dependent tune-out shift is $30(50)$~MHz for the power used in the main measurement, which is statistically consistent with zero.
		    }
	    \label{fig:hyperpolarizability}
	\end{figure}

\subsubsection{Theoretical treatment}

	The energy of an atom in the presence of an electric field \(E\) oscillating at a frequency \(f\) is 
	\begin{equation}
	\mathcal{E}=\mathcal{E}_0 - \frac{1}{2} \alpha(f) E^2 - \frac{1}{24} \gamma(f) E^4 + \ldots \, , \label{eqn:energy_shift}
	\end{equation}
	where \(\mathcal{E}_0\) is the energy of the unperturbed atom, \(\alpha(f)\) is the dynamic polarizability, and \(\gamma(f)\) is the (second) hyperpolarizability.  
	For a monochromatic light field, the time-averaged electric field amplitude is related to the intensity $I$ via
	\begin{equation}
	    E^2=\frac{2 I}{c \epsilon_0},
	\end{equation}
	where \(c\) is the speed of light and \(\epsilon_0\) is the permittivity of free space. 
	A tune-out measurement may be shifted due to the dynamic polarizability cancelling any contribution from the hyperpolarizability. 
	We can estimate this shift by noting that, by definition, \(\mathcal{E}=\mathcal{E}_0\) at the measured tune-out, and expanding the polarizability to first order about the tune-out.
	Thus Eq.~(\ref{eqn:energy_shift}) gives 
	
	\begin{equation}
	 (f-f_\mathrm{TO} ) = - \frac{1}{12} \gamma(f) \left(\frac{2 I}{c \epsilon_0}\right) \left(1\left/ \frac{\partial\alpha}{\partial f}\bigg|_{f=f_\mathrm{TO}} \right. \right).
	\end{equation}
	From the theoretical calculations, the dynamic hyperpolarizability  is $6.964\times10^{-58}$ $\mathrm{C}^4\mathrm{m}^4\mathrm{J}^{-3}$ (about $-1.12\times10^{7}$ a.u.) at the tune-out. The probe beam intensity in the experiment was less than $10^{9}\: \mathrm{W} \mathrm{m}^{-2}$ and thus a shift due to the hyperpolarizability is  constrained to be below 1.5~MHz, which is dominated by other systematic effects.

\subsubsection{Experimental treatment}

	We also constrained the effect of the hyperpolarizability using an independent experimental approach, by studying the effect of the light intensity on the the measured tune-out. 
	The gradient of \(\Omega_{\text{probe}}^2\) with respect to the laser frequency gives an indirect measurement of the intensity in the region of interaction, and the effect of beam power on the tune-out frequency (via this proxy) is shown in Fig.~\ref{fig:hyperpolarizability}.
	We evaluate the linear fit to the data at the experimentally-relevant gradient of $\approx 30$ Hz$^2$/GHz and find that the resulting shift is 30(50) MHz, which is statistically consistent with zero and with the (smaller) estimate made above on theoretical grounds.

\subsubsection{DC Electric Field}

	A worst-case estimate of any DC electric field background is $2~\text{kV}\cdot\text{m}^{-1}$. 
	We can use an approach similar to that of the hyperpolarizability  and find a worst-case shift of \(10^{-2}\)~MHz.

\subsection{Broadband Light}
	\label{sec:BBL}
	\begin{figure}
	    \centering
	    \includegraphics[width=\textwidth]{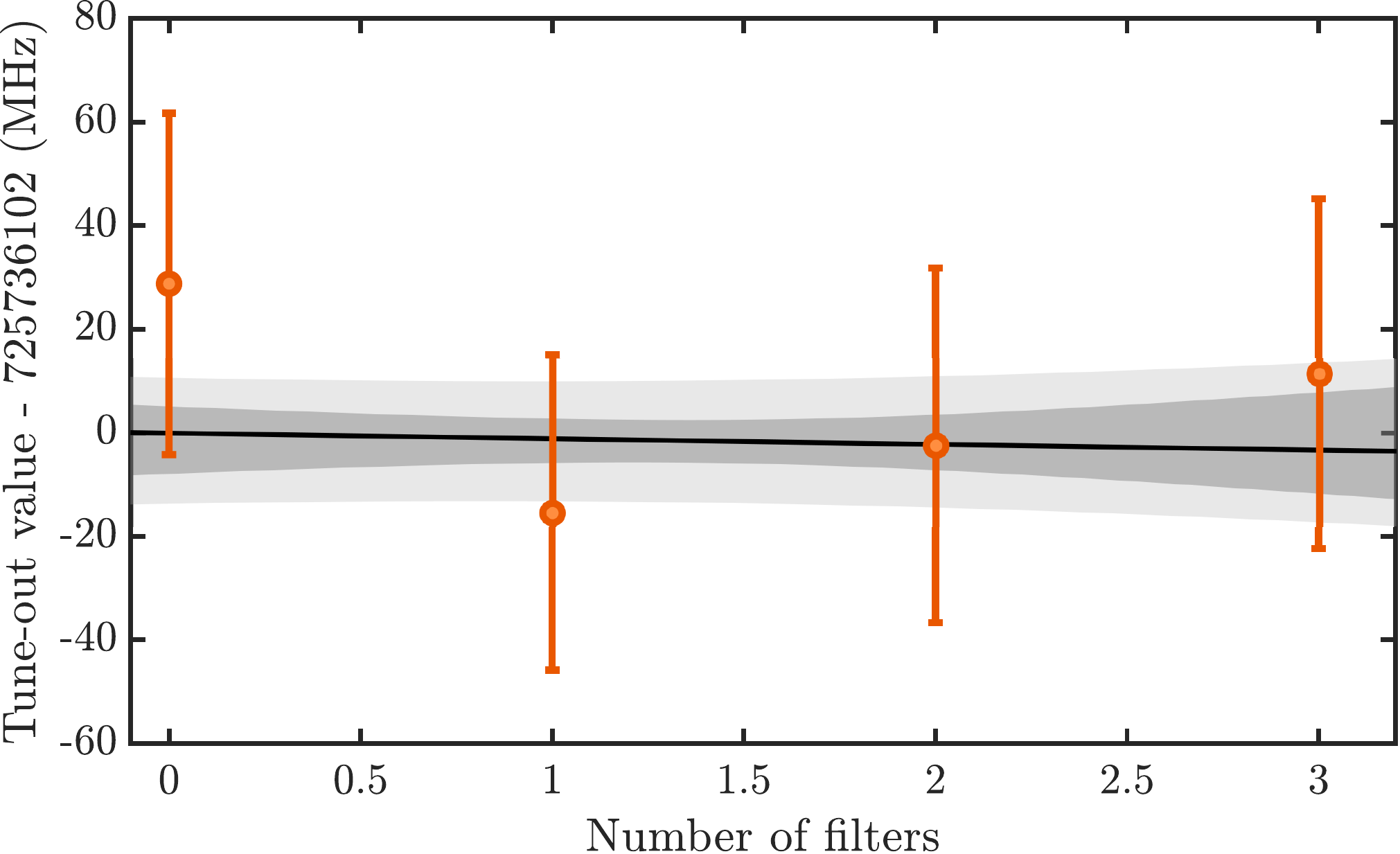}
	    \caption{The tune-out frequency as measured with fixed probe beam polarization and power, as a function of the number of filters in the beam path. The gradient of this dependence is -1(5)~MHz/filter, and hence zero within error. The linear fit has $\chi^2$/dof=0.6406. The light region shows the observation interval, and the dark region shows the confidence interval.
	    }
	    \label{fig:broadband light dependence}
	\end{figure}

	Broadband background light could be produced, for example, by  spontaneous emission in the Ti:S laser which is then amplified in the laser cavity. 
	In the presence of weak broadband light, the energy of the atom shifts by the (integrated) pointwise product of the spectral power density and the polarizability. 
	Because the magnitude of the polarizability increases superlinearly with the detuning from the tune-out ($|\alpha(f)| > \alpha'(f) (f-\fto)$ for large detunings), this effect puts a large weight on the tails of the laser line profile $g(f,\Gamma)$ and may shift the apparent tune-out from its actual value by contributing an energy proportional to $\int g(f,\Gamma)\alpha(f)~df$. 
	This places demanding constraints on the tolerable spectral background of the probe laser. 
	This has been a challenge for measurements in other species \cite{HolmgrenThesis}. 
	
	Our use of a frequency doubled laser system provides some protection from this because the doubling cavity suppresses broadband light that might originate in the Ti:S, unless it happens to have a wavelength equal to a multiple of the doubler's free spectral range. 
	Futher suppression is obtained by passing the probe beam through a series of optical filters: a 450~nm shortpass filter (Thorlabs FESH0450, optical density $>5$ between 450~nm and 1200~nm), a 415~nm band-pass filter (Semrock FF01-415/10-25, with a 27~THZ (15.3~nm) FWHM and optical density $>4$ between 250-399~nm and 431-1100~nm), and finally an angle-tunable filter with a 0.9~THz (0.5~nm) FWHM, which we centred on \(\sim 413\)~nm.

	In principle, one could use a spectrometer to measure the spectral power density of the background, but the requisite dynamic range to observe such small effects near a laser peak makes this unfeasible.
	Instead we used a scheme similar the one described in \cite{Leonard15}, and measure the tune-out as a function of the number of (progressively narrowing) filtering stages.
	We can then estimate the shift in the  final measurement. 
	As shown in Fig.~\ref{fig:broadband light dependence},  the measured tune-out frequency is independent of number of filters (within statistical error). 
	Therefore we take the standard deviation between the various filters (\(30\)~MHz)  as the uncertainty associated with the spectral background.

	\begin{table}
	\centering
	\begin{tabular}{l|r|r}
	\hline\hline
	Term              & Value &  Uncertainty \\
	\hline
	Measured Value      & 725\,736\,810             & 40      \\
	Polarization        & & \\
	\, \, - Birefringence & -100                   & 200   \\
	\, \, - Beam Anisotropy & 0                   & 150   \\
	Method Linearity    & 24                   & 30       \\
	Hyperpolarizability           & -30                   & 50   \\
	Broadband Light     & 0                     & 30      \\
	DC Electric field   & 0                     & \(\ll 1\) \\
	Wave-meter          & 0                     & 4    \\
	\hline
	Total               &  725\,736\,700            &  260\\
	\hline\hline
	\end{tabular}
	\caption{Contributions to measured tune-out frequency with associated systematic uncertainties (MHz). The measured value is found using only polarization data measured after the vacuum chamber. The polarization row gives the average of the tune-out frequencies calculated using polarization data pre and post vacuum chamber relative to the measured value (equivalent to assuming the shift from one window is half of the shift calculated using the measurements after both windows), and the uncertainty is equal to the difference between these values. Note that uncertainties are added in quadrature.}
	\label{tab:results}
	\end{table}

	A summary of the systematic shifts is shown in Tab. \ref{tab:results}. 
	The main limiting factor in the accuracy of our value for $\fto$ is  the polarization of the laser light.

\section{Discussion}
\label{sec:TO_discussion}
\begin{figure}[t]
	    \centering
	    \includegraphics[width=\textwidth]{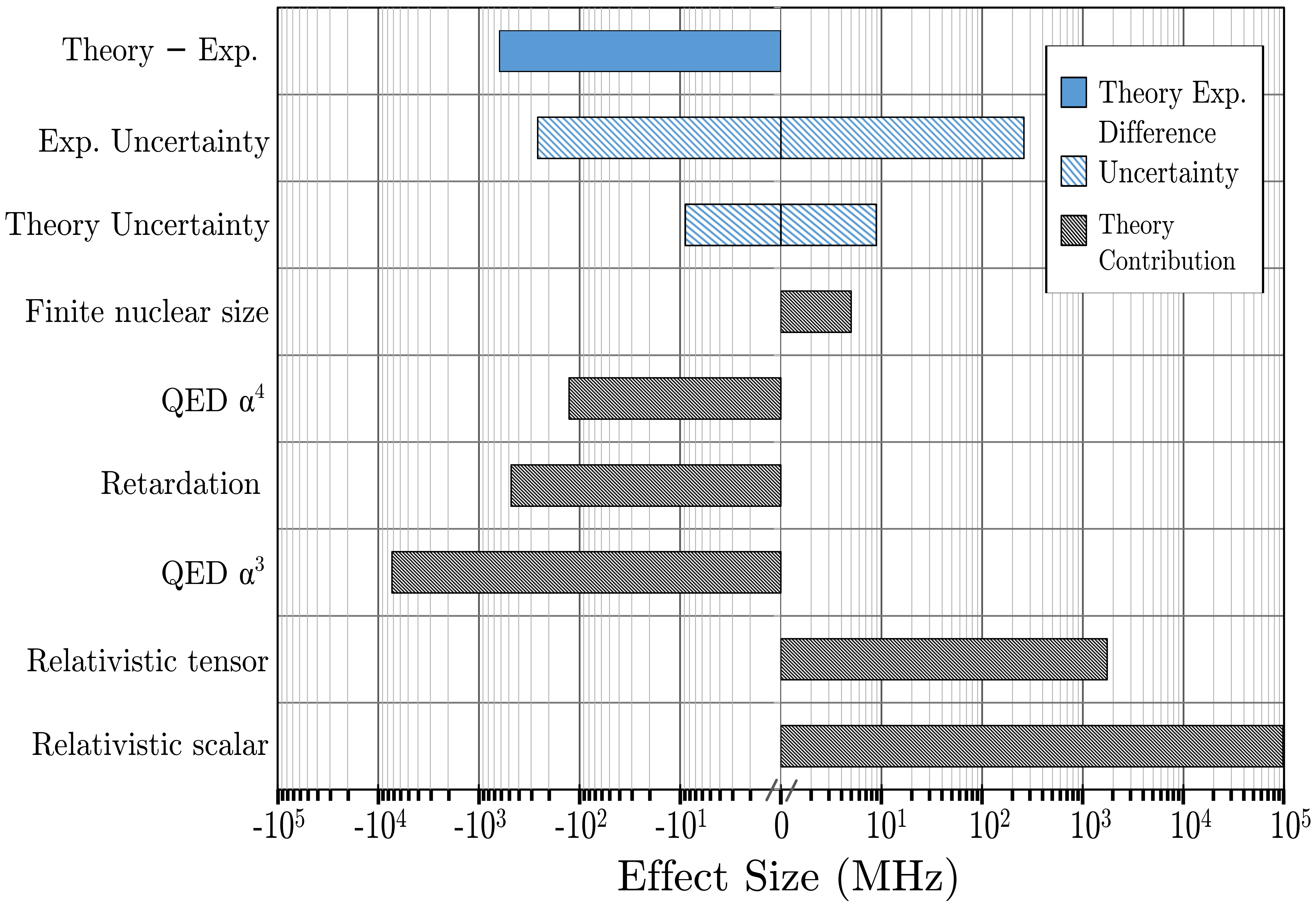}
	    \caption{Theoretical contributions to the tune-out value in comparison to uncertainties in the theoretical and experimental determinations of the \(\TO\) tune-out frequency.}
	    \label{fig:contributions}
	\end{figure}

	Figure~\ref{fig:contributions} shows a comparison of contributions to the theoretical value of the tune-out and the uncertainties in both the experimental and theoretical determinations.
	The combined theoretical and experimental uncertainties yield a  \(\sigma{\sim} 260\)~MHz precision which is much smaller than the contribution of QED effects (\({\sim} 30 \sigma\)), and thus our measurement validates these calculations.
	The retardation correction to the dipole interaction is slightly larger than the combined uncertainty (\({\sim} 2 \sigma\)), but the contribution of finite nuclear size effects (\(5\)~MHz) are not resolvable at this precision. 
	 
	This is the first measurement with sufficient precision to probe the retardation correction, which is significant as they are typically omitted from calculations of the frequency-dependent polarizability \cite{Drake19, Pachucki19}. 
	Overall, we find a \({\sim} 2.5 \sigma\) difference between experiment and theory, including an estimate of the uncertainty due to terms excluded from the theoretical calculation.  
	Notably, if we ignore the retardation correction (as proposed in Ref.~\cite{Pachucki19} and implemented in tune-out frequency calculations for the first time in this work), then the difference is only \({\sim} 0.5 \sigma\).  
	The retardation contribution could be subject to a more stringent test if the measurement precision could be improved by an order of magnitude (see section \ref{sec:TO_conc} for comments towards this).
	
	Using this method we can infer the peak value of the energy shift induced in the atoms by the probe beam. 
	Through Eq.~(\ref{eqn:omega_probe}), a measurement of the shift in trap frequency determines the curvature of peak of the Gaussian optical potential energy surface. 
	Along with knowledge of the beam spot size at the focus, the inferred curvature completely specifies the geometry of the optical potential. 
	We typically achieve an uncertainty in the tune-out of approximately 1.2~$\text{GHz}$ per $\sqrt{N_\text{shots}}$, where \(N_\text{shots}\) refer to the number of BEC's used (including calibration and probe beam measurements). 
	A single scan takes approximately 700~s (12~minutes) and consists of 26 trap frequency measurements (BEC productions). 
	As such, one full scan (26 trap frequency measurements) gives an uncertainty of $\sim 1.2\;\mathrm{GHz}/\sqrt{26}=235$~MHz\footnote{The number of measurements taken to find the tune-out for a given polarization varies from 50 to a few thousand for the data presented here.}.
	Thus, we can indirectly measure the absolute energy shift in the $\MetastableState$ state with a sensitivity of $1.7\cdot10^{-33}\mathrm{J}/\sqrt{\mathrm{sec}}$, where the time is the probe beam interrogation time. 
	In the case of the measurement with the lowest frequency uncertainty in $f_\textrm{TO}$, (30~MHz), the minimum potential energy peak we can thus discern is approximately $10^{-35}\mathrm{J}$ ($U/k_B=3$~pK).

\subsection{Improvements upon past measurements}

	The past measurement of this tune-out frequency \cite{Henson15} obtained a 4 GHz uncertainty in the determined value of $\fto = 725.7249$ THz (6ppm). This was comprised of a 1.5 GHz statistical uncertainty and a 3.5 GHz systematic uncertainty. 
	The 263~MHz total uncertainty presented here amounts to relative precision of 0.4ppm, a 15-fold improvement. 
	The 40~MHz statistical error in the new measurement is 40-fold smaller than that of previous measurement, due mainly to three factors mentioned below. This is significant progress, but the ultimate precision is limited by the systematic uncertainty of 260 MHz (which is improved roughly 14-fold over the prior measurement).	There are two main factors underpinning the improved systematic uncertainty.

	\begin{itemize}	
		\item \textbf{Purity of probe laser light:} In conventional spectroscopy, one is typically interested in the resonances (i.e. peaks) in some spectrum. In this setting, the finite linewidth of the system's response provides some insurance against the possibility that the probe beam could be polluted by spectral power at wavelengths other than the desired driving frequency. The nature of the tune-out measurement is the opposite: Because the polarizability increases with detuning from the tune-out, spectral impurities become more problematic the further detuned they are. This was the leading systematic effect in the previous work \cite{Henson15}. Recalling section \ref{sec:BBL}, this effect is constrained to the level of 30~MHz in the present work, nearly 120-fold better than the first measurement. The use of a frequency-doubling cavity was a significant factor here. The cavity modes act as narrow filters (limited by the finesse of the cavity), and the nonlinear gain medium results in a quadratic increase in the suppression of off-resonant light.
		\item \textbf{Characterisation of polarization effects:}  The atomic tune-out point depends strongly on the polarization of the electric field with respect to the atom's quantization axis (see e.g. section \ref{sec:polz_dep}, especially Eqn. \ref{eqn:fto_stokes_eqn}). Accordingly, this work paid careful attention to measuring this dependence in order to obtain a single measured  value ($\fto(-1,0)$). In contrast, the previous measurement used only linearly-polarized light, and the direction of polarization relative to the atomic quantization is not known. As can be seen from Fig. \ref{fig:pol_TO}, this amounts to some 5 GHz in systematic uncertainty. However, the previous analaysis did not account for the dependence on the polarization of the laser.  The 4 GHz total uncertainty that was included in the prior measurement (mainly due to the prior point) can be combined with this estimate to re-evaluate the the total systematic error of the prior work at the level of 9 GHz, which is comparable to the 11 GHz difference between the measurements. 
	\end{itemize}	

	Despite a more thorough examination of the latter point in this chapter, the polarization of light remains the leading source of uncertainty. The main contribution, as discussed in section \ref{sec:systematic_effects}, is that the polarization of the light \emph{at the atoms} is not directly measurable. The improvement in statistical precision is a result of three technical differences:

	\begin{itemize}
		\item \textbf{Laser power:} The present measurement applied a 150 mW probe beam at the point of vacuum entry, whereas the original measurement used a 3 mW laser diode (subject also to losses when passing through the focusing optics). 
		\item \textbf{Probe wavelength precision}: The High Finesse WS-8 wavemeter, obtained especially for this measurement, affords a locking of the probe beam with $\approx$175-fold greater accuracy. Moreover, the day-to-day drift of the wavemeter in the previous measurement was up to 350 MHz \cite{HensonHonsThesis}, larger than the drifts relevant to the present measurement by a similar factor. 
		\item \textbf{Optical dipole potential determination:} Previously, the optical potential was detected by modulating the net trapping potential with a probe beam and using a `lock-in' technique \cite{HensonHonsThesis} based on Fourier analysis of the outcoupling signal of a continuous atom laser. The idea is that the outcoupling efficiency would be modulated at the probe beam modulation frequency, and the phasor of the response would cross zero amplitude (with a phase flip) at the tune-out point. However, this signal exhibited a very high variance and thus the final linear fit to determine the tuneout was subject to significant statistical uncertainty. In contrast, the method employed in this work provides trap frequency measurement accurate to a few dozen ppm \cite{Henson22_PAL}, drastically improving the accuracy of $\fto$ as determined by the fitting procedures.
	\end{itemize}

	The effect of these factors can be seen in the following description of the sensitivity of the new method.
	The gradient of the linearized tune-out ecd dcdquation is
	\begin{equation}
		\frac{\partial \Omega_\textrm{net}^{2}}{\partial f} = \frac{4P}{\pi m c \epsilon w_0^4}\frac{\partial \textrm{Re}(\alpha(f))}{\partial f}
	\end{equation}
	Without loss of generality (within the assumption of linearity), we consider the problem of determining the detuning $\delta = f-\fto$ given a measurement of $\Omega_{net}$ at some laser frequency $f$.  Noting that the underlying trap frequency $\Omega_\textrm{trap}$ is known to within some parts in $10^5$, let us assume the uncertainty $\Delta$ in $\Omega_\textrm{net}$ is $\Delta_{\Omega_{net}^2}\approx\Delta_{\Omega_{probe}^2}$. We can then write
	\begin{equation}
		\Delta_{\delta} = \frac{\pi m c \epsilon w_0^4}{4P\alpha'}\Delta_{\Omega_{\textrm{probe}}^{2}},
	\end{equation}
	where $\alpha'$ is shorthand for the gradient with respect to frequency. It is evident that increasing laser power has a favourable effect on the precision. The improved wavelength precision reduces the systematic error in the means of generating the perturbing potential, and the improved trap frequency measurement reduces the systematic uncertainty in the actual response of the system, which both lead to improvements in $\Delta_{\Omega_{\textrm{probe}}^{2}}$

\subsection{Comparison With Previous Oscillator Strength Ratio Measurements}
	
	We can extend the approach taken in Ref. \cite{Mitroy13} to a three-level atom (c.f. section \ref{sec:TO_points}) and write the polarizability of an atom in the state $\ket{1}$ as 
	\begin{equation}
	\alpha_1(f) = \frac{\mathcal{F}_{12}}{E_{21}^2-h^2 f^2}+\frac{\mathcal{F}_{13}}{E_{31}^2-h^2 f^2}
	\end{equation}
	where $\mathcal{F}_{12}$ and $\mathcal{F}_{13}$ are the oscillator strengths, $E_{12}$ and $E_{13}$ are the excitation energies of the dominant transitions, $f$ is the photon frequency, and \(h\) is Planck's constant. The condition $\alpha(\fto)=0$ implies that
	\begin{equation}
		 \frac{\mathcal{F}_{12}}{E_{21}^2-h^2 \fto^2} = -\frac{\mathcal{F}_{13}}{E_{31}^2-h^2 \fto^2},
	\end{equation}
	and hence
	\begin{equation}
		\frac{\mathcal{F}_{13}^2}{f_{12}^2} = \left(\frac{E_{31}^2-h^2 \fto^2}{E_{21}^2-h^2 \fto^2}\right)^2.
	\end{equation}
	The fractional sensitivity of the ratio $X=\frac{\mathcal{F}_{13}^2}{\mathcal{F}_{12}^2}$ to the tune-out is
	\begin{align}
	    \frac{\delta X}{X} &= \frac{1}{X} \cdot  \frac{\partial X} {\partial f_{TO}} \delta f_{TO} \\
				    &=\frac{2  h^2 f_{TO} (E_{12}^2-E_{13}^2)}{(E_{12}^2-h^2 f_{TO}^2) (-E_{13}^2 + h^2 f_{TO}^2 )} \cdot \delta  f_{TO}\\
				    & = \frac{-2 f_{TO}^2 (f_1^2-f_2^2)}{(f_1^2-f_{TO}^2)(f_2^2-f_{TO}^2)} \frac{\delta  f_{TO}}{f_{TO}}
	\end{align}
	where $f_i=E_{1i}/h$.
	Given the dominant transition manifolds at 276.7465 THz (\(\MetastableState \to \TOUpperStateManifold \)) and 770.7298 THz (\(\MetastableState \to \TOLowerStateManifold\)), as well as our value for the tune-out frequency of \(f_{TO}=725.73670\) THz, we reach a fractional uncertainty in the oscillator strength ratio of 6~ppm.
	We can compare our measurement to others in the literature by using frequency of the dominant transitions and the measured tune-out value to estimate the sensitivity to the ratio of transition strengths. 
	We find that our measurement improves against the previous record of 15~ppm, reported in Ref.~\cite{Leonard15}. 
	We note that the relative uncertainty in $X$ equals the relative uncertainty in the ratio of transition matrix elements, which were the quantity calculated in Ref.~\cite{Leonard15}.
	We further note that this method is approximate and neglects the contribution from the DC polarizability, however this is a small effect and not needed for such coarse comparison of sensitivity. 
	Thus, we claim that our measurement of the $\TO$ tune-out wavelength produces the most precise constraint of relative transition rates in any atomic system to date.

\subsection{Conclusion}
	\label{sec:TO_conc}
	Our experimental determination of 725\,736\,700\,$(40_{\mathrm{stat}},260_{\mathrm{syst}})$ MHz has a relative precision of $4\times 10^{-7}$, constituting the most precise measurement of atomic transition rate ratios to date \cite{Mitroy13}, and is 2.5$\sigma$ larger than the theoretical prediction. 
	This measurement determines the ratio of oscillator strengths to 6 ppm, which is a factor of two improvement on the previous record. 
	Furthermore, our novel method for measuring the dipole potential is able to discern a peak potential energy of as little as 10$^{-35}$ J. This is, to our knowledge, the highest precision in a potential energy measurement reported to date \cite{Henson22}.
	 
	In future precision measurements of tune-out points, significant improvements would follow from more precise calibration of the light polarization.
	One solution would be to use in-vacuum optics and systematically vary the angle of the beam axis relative to the magnetic field.
	These improvements would allow for independent tests of the scalar, vector, and tensor polarizabilities, thus yielding more information about the structure of the helium atom and QED itself.

	The method above could be used to measure other tune-out frequencies in helium or other species.
	It could also serve as a means to investigate other issues relating to QED. 
	If a future measurement could be made with MHz-level precision,  this tune-out could be used to calculate the nuclear charge radius of helium. 
	Thus this method, and its future improvements, may continue to  clarify the `jewel of physics.'

\vfill

\begin{flushright}
\singlespacing
{\emph{
``Imagination reaches out repeatedly trying to achieve some \\
higher level of understanding, until suddenly I find myself \\
momentarily alone before one new corner of nature’s \\
pattern of beauty and true majesty revealed. \\
That was my reward."}\\ 
- Richard P Feynman\footnote{Nobel banquet speech, 1965}}
\end{flushright}
\onehalfspacing

%% file: latex/23_depletion.tex
\chapter{Quantum depletion of a harmonically trapped Bose gas}
\markboth{\thechapter. QUANTUM DEPLETION OF AN EXPANDING CONDENSATE}{}
\label{chap:QD}

\blankfootnote{\noindent The contents of this chapter relate to the work published in \textbf{On the survival of the quantum depletion of a
condensate after release from a magnetic trap} by J. A. Ross, P. Deuar, D. K. Shin, K. F. Thomas, B. M. Henson, S. S. Hodgman, A. Truscott, \emph{Accepted for publication in Nature Scientific Reports (2022)}, \href{https://arxiv.org/abs/2103.15283}{\emph{ArXiv}} \textbf{2103.15238}}

\begin{adjustwidth}{2cm}{0cm}
\begin{flushright}
\singlespacing
\emph{``Do not be content with the answer that is almost right; \\
	seek one that is exactly right... The Way is a precise Art.\\
	Do not walk to the truth, but dance."} 
	\\- Eliezer Yudkowsky\footnote{\url{http://yudkowsky.net/rational/virtues/}}
\end{flushright}
\end{adjustwidth}
\onehalfspacing
\vspace{1cm}

	{One} of Nikolay Bogoliubov's seminal contributions was to recognize the Bose-Einstein condensation of collective excitations as the mechanism underlying the physics of superfluidity \cite{Bogoliubov47}.
	The population of excited quasiparticle modes makes up the normal component in the Landau two-fluid model while the 
	quasiparticle ground state comprises the superfluid part. {The latter is composed of a macroscopically-occupied condensate and (in the presence of s-wave interactions between constituent particles) correlated particle pairs \cite{Bogoliubov47,Vogels02}}.
	{A consequence of these pairs is  that excited} single-particle modes are populated even at zero temperature.
	This is the \emph{quantum depletion} of the condensate and presents as an occupation of single particle modes with large momentum $p$ that decays like $p^{-4}$ \cite{PitaevskiiStringari,Decamp18}.

	The amount of quantum depletion, usually quantified either as the total population of depleted modes or the depleted fraction, increases with the s-wave scattering length $a$ and, crucially, with the density of the gas. 
	However, it has been argued on various grounds (discussed below, see \cite{Xu06,Qu16}) that the quantum depletion should \emph{decrease} during the time-evolution of a condensate released from a trap and therefore not be detectable in dilute gases.
	In this chapter I present observations of quantum depletion in expanding condensates released from a harmonic trap. 
	In the experiments discussed herein, the single-particle momenta of the constituent atoms are destructively measured after the condensate has expanded for $\approx$0.4 seconds during freefall.
	On this timescale, the condensate wavefunction expands from its initial trapped size, characterized by the initial Thomas-Fermi radii of  $\mathcal{O}(100)$ $\mu$m and $\mathcal{O}(10)$ $\mu$m in the weak and strong trapping axes, respectively, into an ellipsoidal volume with half-widths of $\approx$2 and $\approx$0.5 cm. 
	The density of the condensate in this setting is determined predominantly by the initial momentum distribution and the dispersal of the mean-field interaction energy into kinetic energy.
	In this regime, called the \emph{far-field regime}, the effect of the initial finite size of the condensate is dominated by the just-mentioned factors.

	The measurements in this chapter corroborate prior observations \cite{Chang16} of slowly-decaying tails in the far-field beyond the thermal component that exhibit features consistent with the survival of the quantum depletion. 
	The results of this experiment undermine the hypothesis that the depletion is reabsorbed during the expansion. 
	Indeed, the depletion appears even stronger in the far-field than the Bogoliubov theory predicts for the trapped condensate. 
	This result is in conflict with the hydrodynamic theory which predicts that the in-situ depletion does not survive when atoms are released from a trap.  
	I also describe simulations of the experiment that show how the depletion could survive into the far field and become stronger\footnote{Much gratitude to Piotr Deuar for undertaking these simulations.}. 
	However, while in qualitative agreement, the final depletion apparent in the experiment is larger than in the simulation.

	Before describing the experiment, I present some additional background including the relevant theory of interacting condensates, the connection to Tan's theory of contact interactions, and milestone experiments which validate the respective theories. Then I present the details of the methods used to probe the quantum depletion in the far-field. In this context I describe some of the pitfalls of least-squares nonlinear fitting in the context of power law analysis, and then apply a more parsimonious method which clarifies what can and cannot be said about this regime. After presenting some additional details about the technical aspects of the experiment, I summarize the findings of the simulations. Finally, I review the findings, discuss our leading interpretation, and present a couple of possible ways forward.

\section{Background}

	The Hamiltonian of a homogeneous system of interacting bosons can be written in terms of plane-wave field operators $\hat{a_\kvec}$, labelled by the wavevector $\kvec=\textbf{p}/\hbar$, as

	\begin{equation}
		\hat{H} = \sum_{\kvec} \frac{\hbar^2k^2}{2m}\hat{a}_{\kvec}^\dagger \hat{a}_\kvec + \frac{g n}{2}\sum_{\kvec,\kvec',{\bf q}}\hat{a}_{\kvec+{\bf q}}^\dagger\hat{a}_{\kvec'-{\bf q}}^\dagger \hat{a}_{\kvec'}\hat{a}_{\kvec},
	\end{equation}
	in terms of the particle density $n$ and the effective interaction strength $g=4\pi\hbar^2a/m$, where $a$ is the s-wave scattering length and $m$ is the atomic mass \cite{PitaevskiiStringari,PethickSmith}.
   
    This Hamiltonian can be diagonalized by the Bogoliubov transformation to a free Bose gas of collective excitations through the operator transformation $\hat{b}_{\kvec}^\dagger = u_k \hat{a}_\kvec^\dagger + v_k \hat{a}_{-\kvec}$ \cite{Bogoliubov47,PethickSmith}, wherein the collective excitations are superpositions of particles with opposite momenta \cite{Vogels02}. 
    In this representation, the Hamiltonian takes on the diagonal form 
    \begin{equation}
    	\hat{H} = E_0 + \sum_\kvec \epsilon(\kvec) \hat{b}^{\dagger}_{\kvec} \hat{b}_\kvec
    \end{equation}
	where the quasiparticle dispersion relation is
	\begin{equation}
		\epsilon(k) = \sqrt{\left(\frac{\hbar^2k^2}{2m}\right)^2 + gn\frac{ \hbar^2k^2}{m}}.
	\end{equation}
    The $u_k$ and $v_k$ coefficients are given by
	\begin{align}
		u_{k}^2 &= \frac{1}{2}\left(\frac{\hbar^2k^2/2m + gn}{\epsilon(k)} + 1\right)~\textrm{and}\\
		v_{k}^2 &= \frac{1}{2}\left(\frac{\hbar^2k^2/2m + gn}{\epsilon(k)} - 1\right),
	\end{align}
	In the non-interacting ($a\rightarrow0$) limit, $u_k=1$ and $v_k=0$, so the transformation reduces to the identity and the dispersion is that of free particles. 

	Since the realization of atomic Bose-Einstein condensates there has been considerable experimental \cite{Stewart10,Wild12,Chang16,Makotyn14,Eigen18,Xu06,Vogels02,Pieczarka20,Lopes17_depletion,Cayla20,Kuhnle11,Sagi12,Fletcher17,Lopes17_quasiparticle,Mukherjee19,Carcy19} and 
	theoretical \cite{Colussi20,Kira15_coherent,Decamp18,Smith14,Qu16,Braaten10,Braaten11,Rakhimov20,Braaten08,Zhang09,Combescot09,Werner12_boson,Werner12_fermion,Sinatra00,Deuar11} interest in the 	Bogoliubov theory \cite{Vogels02,Steinhauer03,Lopes17_quasiparticle,Sinatra00,Deuar11} (and quantum depletion in particular \cite{Lopes17_depletion,Chang16,Xu06,Pieczarka20,Cayla20}).
	Notably, the foundational predictions of  the form of the dispersion relation \cite{Steinhauer03} and single-particle decomposition \cite{Vogels02} have been shown to hold.
	The total fraction of condensed atoms in the depletion is $\approx 1.5\sqrt{n a^3}$ \cite{Bogoliubov47} in the Bogoliubov theory, which has borne out in experiments using ultracold atomic Bose-Einstein condensates (BECs) \cite{Xu06,Lopes17_depletion} and exciton-polariton condensates in solid substrates \cite{Pieczarka20}.

	The occupation of single-particle momentum modes can be found using the inverse transformation and is given by
	 \begin{align}
	 \rho(\kvec) &= \langle\hat{a}_\kvec^\dagger\hat{a}_\kvec\rangle\\
		 &=\left(u_{k}^{2}+v_{k}^{2}\right)\langle b_{\kvec}^{\dagger}b_{\kvec}\rangle + v_{k}^{2},
		 \label{eqn:popstats}
	 \end{align}
	wherein the bosonic quasiparticle population statistics follow the canonical ensemble as $\langle \hat{b}^\dagger_\kvec\hat{b}_\kvec\rangle = (\exp[\epsilon(k)/k_B T]-1)^{-1}$ \cite{PitaevskiiStringari,Chang16}. At finite temperatures, quasiparticle modes are thermally populated and deplete the condensate.  Even at zero temperature, when the thermal fraction vanishes, the $v_k^2$ term in Eqn (\ref{eqn:popstats}) persists giving a zero-temperature population of excited 
	{particles \cite{Olshanii03,Decamp18,Chang16}} which decays as $\lim_{k\rightarrow\infty}\rho(\kvec)\propto k^{-4}$ \cite{PethickSmith,PitaevskiiStringari,Chang16}. 
	In the case of a harmonically trapped gas, one can employ the local-density approximation (LDA) to compute the amplitude of the $k^{-4}$ tail by  integrating $v_k^2$ across a Thomas-Fermi distribution \cite{Chang16}. 
	
	A more straightforward derivation can be made through the theory of contact intractions initiated in 2008 by Shina Tan \cite{Tan08_momentum,Tan08_virial,Tan08_energetics}.
	Tan proved, among other important theorems, that the amplitude of the $p^{-4}$ tail is exactly the quantity called the \emph{contact} \cite{Tan08_momentum, Braaten11}.
	The contact determines the amplitude of the power-law decay of the momentum density in terms of the gas density and s-wave scattering length, which fully determine collisional dynamics in ultracold dilute gases. 
	The two-body \emph{contact intensity} is defined by \cite{Tan08_momentum,Braaten11}
	\begin{equation}
		C = \lim_{k\rightarrow\infty}k^4\rho(k),
		\label{eqn:MomentumDef}
	\end{equation}
	which is related to the total contact (or just \emph{contact})  $\mathcal{C} = \int C(r) d^3 r$.
	The contact can be derived from the total energy $E$ through the \emph{adiabatic sweep theorem} \cite{Tan08_energetics},
	\begin{equation}
		\mathcal{C} = \frac{8\pi m a^2}{\hbar^2}\frac{\partial E}{\partial a}.
		\label{eqn:sweep_theorem}
	\end{equation}
	In the Thomas-Fermi approximation, the energy of $N_0$ condensed bosonic atoms is related to the chemical potential via
	\begin{equation}
		\frac{E}{N_0} = \frac{5}{7}\mu = \frac{5}{7} \frac{\hbar \bar{\omega}}{2} \left(\frac{15 N_0 a}{a_\textrm{HO}}\right)^{2/5},
		\label{mu}
	\end{equation}
	where $a_\textrm{HO} = \sqrt{\hbar/(m \bar{\omega})}$ is the harmonic oscillator length and $\bar{\omega}=\sqrt[\uproot{2}\scriptstyle 3]{\omega_x \omega_y \omega_z}$ is the geometric trapping frequency \cite{PitaevskiiStringari,PethickSmith}. The sweep theorem yields
	\begin{equation}
		\mathcal{C} = \frac{8\pi}{7} \left(15^{2}(a N_0)^{7} \left(\frac{m \bar{\omega}}{\hbar}\right)^{6}\right)^{1/5},
		\label{eqn:TotalHarmonicContact}
	\end{equation}
	which can be simplified as $\mathcal{C} = 64\pi^2a^2 N_0 n_0/7$ by dividing out the peak density of a harmonically trapped condensate,
	\begin{equation}
		n_0 = \frac{1}{8 \pi}\left( (15N_0)^2 \left(\frac{m \bar{\omega}}{\hbar\sqrt{a}}\right)	 ^{6}\right)^{1/5}.
		\label{eqn:n0}
	\end{equation}
	Finally, one can compute the asymptotic momentum (density) distribution $n(k)$ of a harmonically trapped gas of spin-polarized bosonic atoms through
	\begin{equation}
		\lim_{k\rightarrow\infty} n(k) = {\frac{\mathcal{C}}{k^4}} = \frac{64\pi^2a^2}{7} \frac{N_0n_0}{k^4}.
		\label{eqn:pred_scaling}
	\end{equation}
	Note that hereon I refer to the momentum distribution $n(k)$ rather than the occupation numbers $\rho(k) = n(k) d^3k/(2\pi)^3$, and that
    the total number of atoms in this normalisation is $N=\frac{1}{(2\pi)^3}\int d^3 k\, n(k)$.

	The contact has also attracted considerable attention in the intervening decade (for a sample of works regarding contact measurements in ultracold gases consult Refs. \cite{Stewart10,Tan08_momentum,Tan08_energetics,Tan08_virial, Braaten10,Braaten11,Colussi20,Makotyn14,Eigen18,Decamp18,Smith14,Chang16,Qu16,Wild12,Hoinka15,Rakhimov20,Braaten08,Smith14,Kuhnle11,Sagi12,Fletcher17,Mukherjee19,Carcy19,Zhang09,Combescot09,Werner12_boson,Werner12_fermion}. This list is non-exhaustive and excludes studies of the contact in other systems such as nuclear matter.)
	Two central properties of the contact, known as the adiabatic sweep theorem and generalized virial theorem \cite{Tan08_momentum,Tan08_virial}, have been verified via radio spectroscopy \cite{Baym07,Punk07,Braaten10} of degenerate Bose \cite{Wild12} and Fermi gases \cite{Stewart10,Sagi12}. 
	While the Bogoliubov prescription breaks down in strongly-correlated systems \cite{Lopes17_quasiparticle}, Tan’s theory applies for any density, temperature, and geometry, and so both theories are expected to agree in the weakly interacting regime.

	In contrast to the case of liquid helium, where the depleted fraction is large (of order 93\% of the fluid \cite{Dmowski17,Glyde00,Moroni04}) due to the strong interparticle interactions, the depletion is generally very small (less than 1\% \cite{Lopes17_depletion,Chang16}) in weakly-interacting dilute gases.
	Observations of the large-momentum tails thus typically employ Feshbach resonances to enhance interactions in ultracold gases and produce a depleted fraction visible with optical imaging techniques, but the power-law tails have proven elusive \cite{Makotyn14,Eigen18} in this regime. 
	A handful of theories have emerged \cite{Kira15_coherent,Colussi20,Smith14} which elucidate the role played by many-body interactions in the evolution following a quench to a large scattering length, and thus modify the momentum distribution.

	However, even measurements in the weakly-interacting regime have returned unexpected results.
	A previous experiment reported the presence of power-law-like tails in the far-field momentum distribution after releasing a BEC of metastable helium from a harmonic optical trap \cite{Chang16}.
	This was surprising because conventional wisdom argues that the density decreases adiabatically during expansion, motivating a hydrodynamic approximation wherein the tails are predicted to vanish \cite{Xu06}. 
	Moreover, the tails were reported to be approximately sixfold heavier than predicted by Bogoliubov theory.

	It is important to verify the anomaly and understand its origin because far-field measurements play a central role in the study of ultracold gases.
	To this end, I measured the momentum distribution of a BEC of metastable helium (\mhe) expanding from a harmonic trap. 
	The experiments cover a range of densities twice as large as the prior work and use a magnetic trap in place of an optical dipole trap, ensuring perfect spin-polarization of the trapped atoms. 
	Tails are visible in the large-momentum part of the condensate wavefunction, whose population agrees qualitatively with the predictions of the Tan and Bogoliubov theory, although a quantitative difference in amplitude remains.
	However, not much can be said with certainty about the exponent of a $p^{-\alpha}$ decay due to inherent difficulties one faces when analysing power-law distributions in general.

\section{Experiment} 
	Information about the momentum distribution of trapped gases is generally obtained by absorption-imaging measurements of the spatial distribution after some finite time of flight. In contrast, metastable helium experiments usually use single-particle detection after a long time of flight (hence in the far-field regime) and thus give direct access to  single-atom momentum information in three dimensions. The metastable $\metastable$ state of helium, denoted He$^*$, is 19.8 eV above the true ground state \cite{Hodgman09_mhe} which enables the use of a multichannel electron multiplier in combination with a delay-line detector (MCP-DLD) \cite{Manning10} for single-atom detection. Such setups have permitted the observation of many-body momentum correlations \cite{Hodgman11,Dall13} and the Hanbury Brown-Twiss effect in both condensed \cite{Schellekens05,Jeltes07,Manning10,Dall11,Perrin07,Perrin12} and quantum depleted atoms \cite{Cayla20}. 
	
	However, investigations of the quantum depletion in \mhe~are challenging because the absence of a known Feshbach resonance precludes control over the contact $\mathcal{C}\propto((a N_0)^7\bar{\omega}^6)^{1/5}$ via the scattering length $a$. 
	Given the small fixed $a=7.512$ nm \cite{Moal06}, Eqn. (\ref{eqn:pred_scaling}) can be tested in the far-field by varying the density of the gas, $n\propto\left(N_{0}\bar{\omega}^3\right)^{2/5}$ (c.f. Eqn (\ref{eqn:pred_scaling})). 
	To facilitate this, I made measurements using two trap configurations with $(\omega_x,\omega_y,\omega_z)\approx 2\pi\cdot(45,425,425)$ Hz (geometric mean $\bar{\omega} = 2\pi \cdot201$ Hz) and 
	$\approx2\pi (71,902,895)$ Hz (geometric mean $\bar{\omega} = 2\pi \cdot393$ Hz), where the (weak) axis of symmetry is horizontal and the frequency is known within 1\% (see section \ref{sec:n0_cal}).
	I varied the endpoint of the evaporative cooling ramp to adjust the number of atoms in the condensate. 
	
\begin{figure}
	    \begin{minipage}{0.4\textwidth}
	    \vspace{0cm}
	    \caption{Sketch of the experimental sequence. A BEC is released from a harmonic trap 
	    (a) and expands during freefall before being split into a superposition of the $m_J\in\{-1,0,1\}$ states (b) by an RF chirp. A magnetic field gradient separates the clouds (c) ensuring that only the magnetically insentitive $m_J=0$ cloud lands on the detector (d), from which the momentum information is reconstructed. The quantum depletion lies in the dilute tails at large momentum (see Fig. \ref{fig:empirical_density}).}
	    \label{fig:sequence}
	    \end{minipage}
	    \hfill
		\begin{minipage}{0.55\textwidth}
		\vspace{0cm}
	    \includegraphics[width=\textwidth]{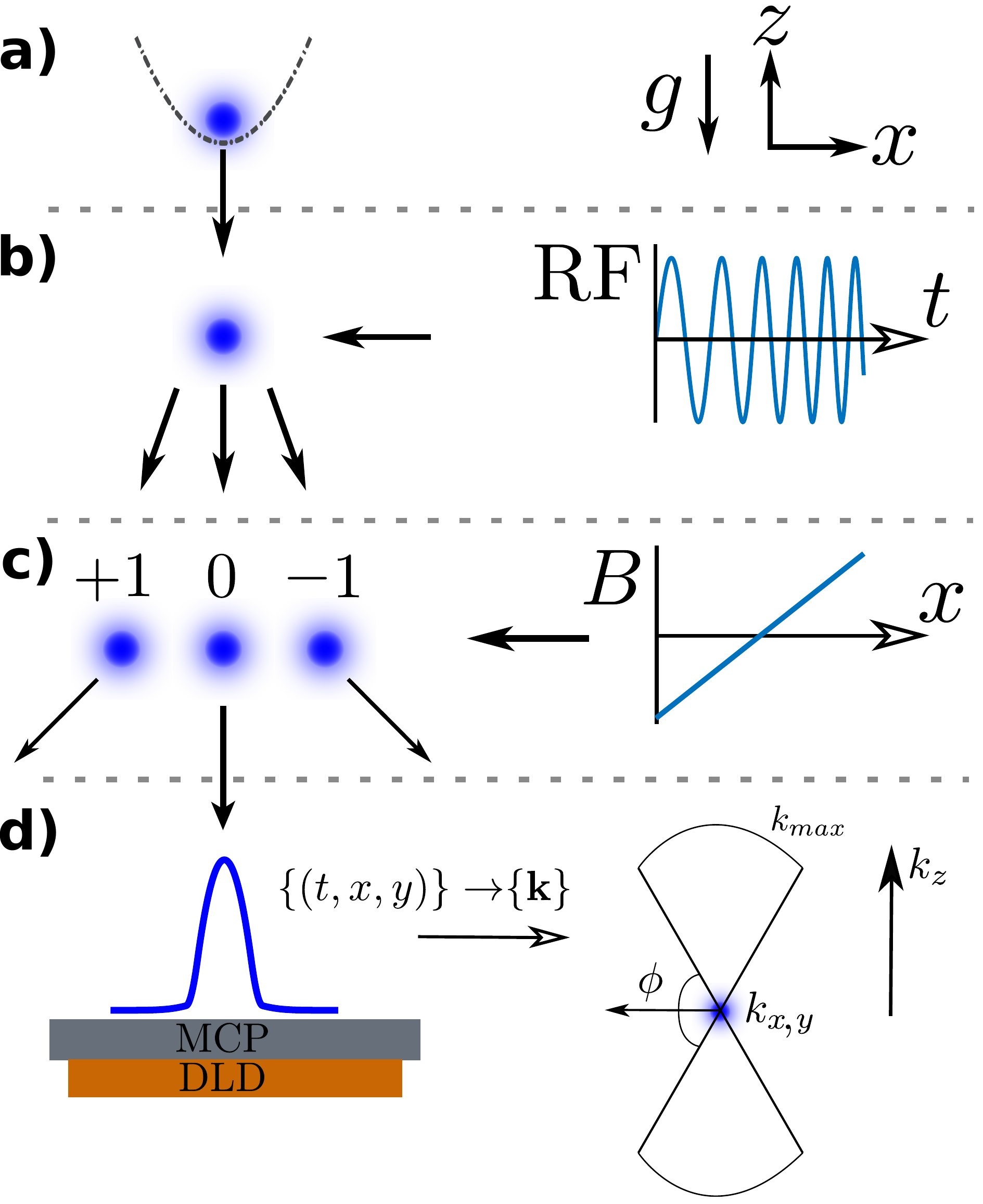}
	    \end{minipage}
	\end{figure}

	The experimental sequence, depicted schematically in Fig. \ref{fig:sequence}, began with BECs consisting of between $2\times 10^5$ and $5\times 10^5$ $^4$He atoms polarized in the $\metastable(m_J=1)$ state and cooled to $\sim$ 300 nK by forced evaporative cooling in a harmonic magnetic trap generated by field coils in a Bi-planar Quadrupole Ioffe configuration \cite{Dall07}. 
	After the trap was switched off,  about one quarter of the atoms are transferred to the magnetically insensitive $m_J=0$ state with a radio-frequency (RF) Landau-Zener sweep to avoid distortion by stray magnetic fields.
	The $m_J=\pm 1$ clouds were deflected outside the detector field of view by a Stern-Gerlach scheme implemented by switching on a magnetic field immediately after the RF pulse.
	The centre of mass of the cloud then impacted on the detector after a $\tau = 417$ ms time of flight following the trap switch-off. 
	The measurements just described were interleaved with calibration measurements to determine the shot-to-shot variation in atom number, trapping frequencies, magnetic state transfer efficiency, and noise contributions. I discuss these calibrations  in section \ref{sec:exp_details}.

\subsection{Analysis of dilute tails} 
\label{sec:analysis}

  \begin{figure}
       \includegraphics[width=\textwidth]{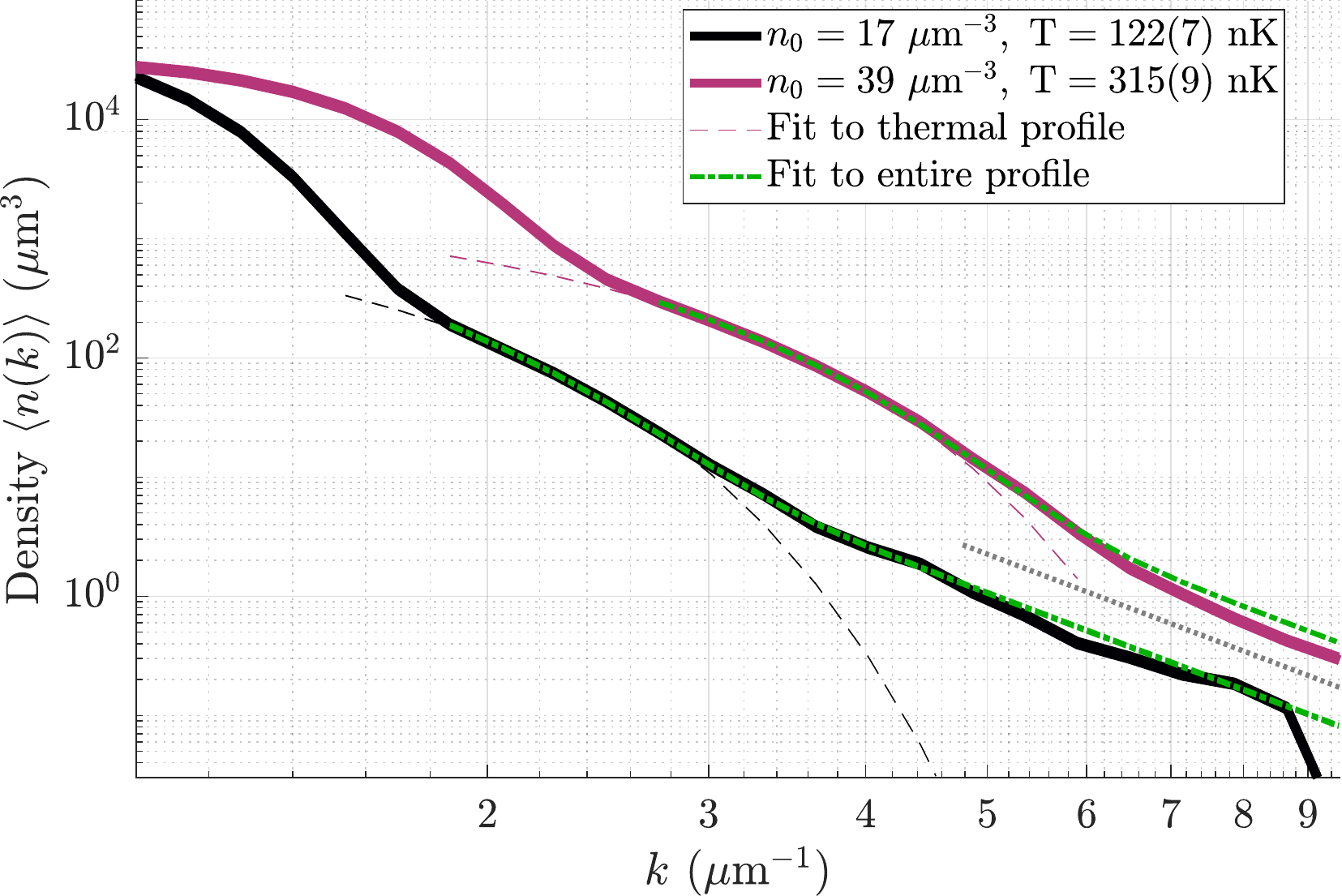}
	        \caption{The empirical density of particle momenta in the far field from two trap configurations (black and magenta). Three regions are shown: At low $k$ the parabolic distribution of the BEC dominates. For larger $k$, the thermal parts (fits shown by dashed lines) decay super-exponentially as $e^{-k^2}$. For even larger $k$, these give way to the depletion region. A combined fit of the form $n_T(k) + C_4/k^4$ (green dot-dash lines) yields temperatures consistent with the thermal fit and also an amplitude $C_4$ of the depleted tail. The grey dotted line is a guide to the eye showing a $k^{-4}$ decay.}
	        \label{fig:empirical_density}
        \end{figure}

	Fig. \ref{fig:empirical_density} shows the empirical density $n(k)$ for two data collection runs at the extreme values of $n_0$ used in the experiment\footnote{The spatial density of the atoms is given per cubic micron. The wavenumber $k$ has units $\micron^{-1}$, thus the particle density in $k$-space has units $(\micron^{-1})^{-3}=\micron^{3}$.}. The three regimes of the condensate, thermal depletion, and quantum depletion span over five orders of magnitude in density. The thermal part of the distribution is well fitted by the momentum distribution of an ideal Bose gas \cite{Dalfovo99}
	\begin{equation}
		\frac{n_T(k)}{{(2\pi)^3}} =\frac{N_T}{\zeta(3)} ~\left(\frac{\lambda_{dB}}{2\pi}\right)^3 g_{3/2}\left(\exp\left(-\frac{k^2 \lambda_{dB}^2}{4\pi}\right)\right)
		\label{eqn:th_fun}
	\end{equation}
	wherein the thermal de Broglie wavelength $\lambda_{dB} = \sqrt{2\pi\hbar^2/(m k_B T)}$ yields an estimate of the temperature $T$ which ranges from 100 to 320 nK in these experiments. Here, $g_{3/2}(\cdot)$ is the standard Bose integral, $\zeta(\cdot)$ is the Riemann zeta function, and $N_T$ is the number of atoms in the thermal component. 
	Note that for a non-interacting gas in the thermodynamic limit, the {number of thermal atoms} 
	is simply ${N_T^{\rm id}} = \zeta(3)(k_B T / \hbar\bar{\omega})^3{=\eta_T N}$, but for these condensates the critical temperature is reduced by $\approx20\%$ by interactions (see section \ref{sec:BEC_theory}).
	This, and the attendant twofold increase in the thermal fraction {$\eta_T$}  (relative to the non-interacting case), is accounted for by explicitly using $N_T$ as a fit parameter. 
	The lower bound of the $k$-interval for the thermal fits is easily identified by the transition point between the thermal and condensed region. The upper bound was chosen individually for each dataset to ensure simultaneous near-total coverage of the thermal region and minimal mean-squared error. Thus the fits generally exclude the condensate and the high-momentum region where the thermal fraction makes a negligible contribution. The fit parameters were used as initial guesses for the fits described in the next section.
	The same lower bounds were used when fitting the combined thermal- and power-law functions (next section), but the upper bound was fixed at $k=10~\micron^{-1}$ for all cases.

	The thermal population decays super-exponentially with $k$, and hence cannot account for the counts we observe beyond $k\gtrsim 6~\micron^{-1}$. 
	In the rest of this section we present evidence in support of the identification of these counts with the quantum depletion. 
	First, though, I argue that the usual approach of a least-squares regression with a density function is unsuitable for the purpose of inferring power-law-like behaviour of the dilute tails, motivating the alternative approach presented in section \ref{sec:regression}.

\subsubsection{Difficulties with analysis of power laws}	
\label{sec:pow_issues}

	A standard approach would be to proceed with a routine fit of the $k$-space histogram with an additional term of the form $C_\alpha/k^\alpha$ to estimate the parameters of the purported quantum-depleted tail.
	However, fitting histograms with power laws is prone to return biased estimates of parameters and to drastically under-report uncertainties \cite{Clauset09,Virkar14}, especially when data is available over less than a couple of decades of dynamic range\footnote{As the authors of \cite{Clauset09} note, ``In practice, we can rarely, if ever, be certain that an observed quantity is drawn from a power-law distribution. The most we can say is that our observations are consistent with the hypothesis that $x$ is drawn from [...] a power law"}.
	In this section I demonstrate some of the problems with a least-squares regression using these measurements as a case study.
	Hereafter, $C_\alpha$ refers to a fit parameter, as distinct from the total contact $\mathcal{C}$ computed using the sweep theorem (Eqn. (\ref{eqn:sweep_theorem})), and $C_\textrm{sim}$ as determined from  numerical simulations. All three have dimensions m${}^{-1}$.

    \begin{figure}
       \includegraphics[width=\textwidth]{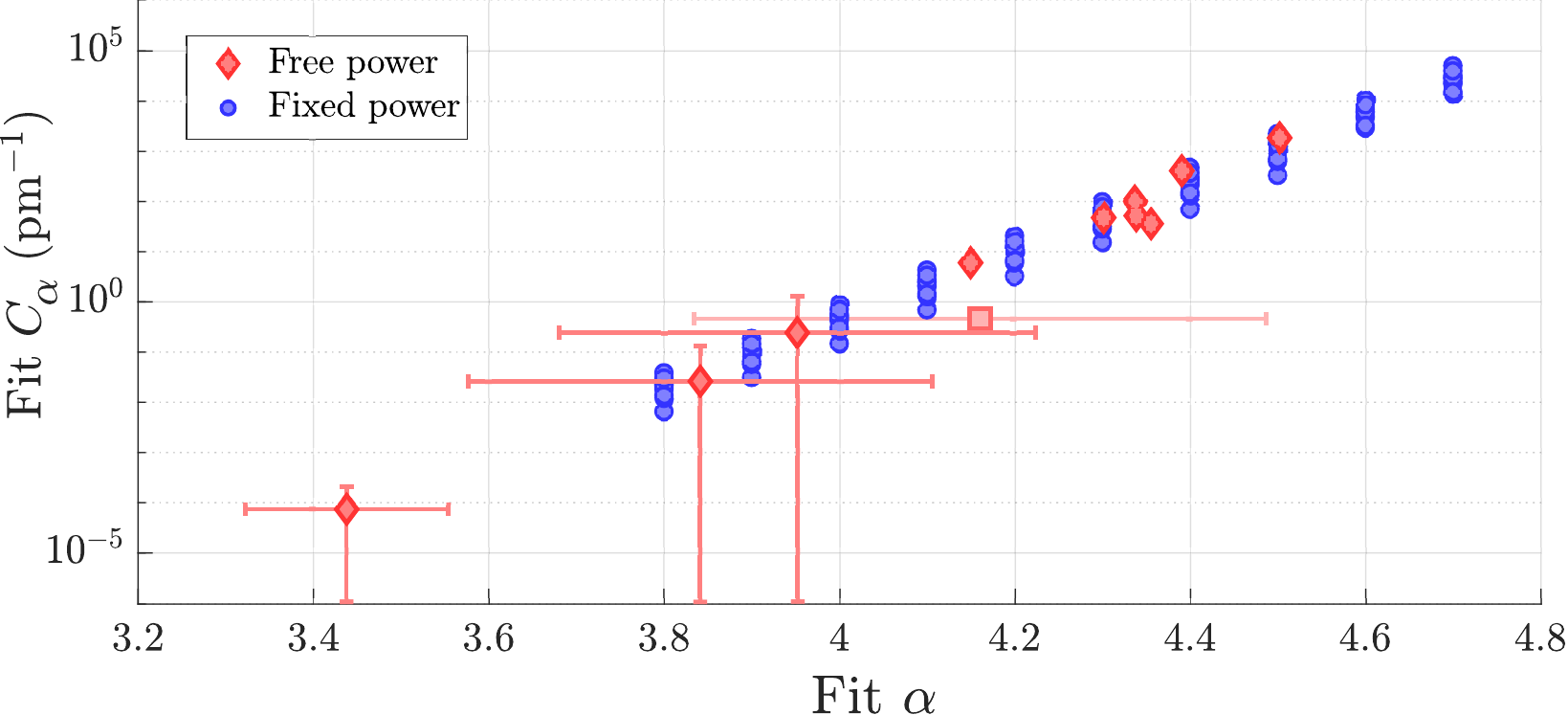}
	        \caption{Illustrating the large systematic errors in a fit to the density using a power law ansatz. If the fit to the density profiles (fig \ref{fig:empirical_density}) is replaced by $C_\alpha/k^\alpha$ and $\alpha$ used as a fit parameter it gives a wide variation in best-fit exponents and scale coefficients between each data set (red diamonds). The red square shows the mean fitted $\alpha$ and (geometric) mean $C_\alpha$ with standard deviation (in $\alpha$) shown as error bars. Blue circles show the amplitude coefficient $C_\alpha$ obtained from the same data sets with $\alpha$ fixed at values within the range of the free fit results. The choice of $\alpha$ strongly determines the coefficient $C_\alpha$, but the error bars (standard errors in fit parameters) are smaller than the markers in all cases. }
	        \label{fig:contact_determination_issues}
	\end{figure}

	If one augments the {thermal} fit function (Eqn. (\ref{eqn:th_fun})) with a power-law term and leaves $\alpha$ as a free parameter, the average exponent over all runs is 4.2(4). 
	For comparison, the prior work \cite{Chang16} reported power-law tails with an exponent 4.2(2).
	At first glance, one could simply determine the amplitude of the tails by fixing the exponent to 4 (being `within error' of the mean value), and thus find $C_{\alpha=4}$ which is, on average, 8(2) times greater than the coefficient predicted by Eqn. (\ref{eqn:pred_scaling}), and in general agreement with Ref. \cite{Chang16}.
	However, there are issues which undermine the utility of this otherwise standard approach.

	First, Fig. \ref{fig:contact_determination_issues} illustrates how the scale coefficient $C_\alpha$ depends exponentially on the choice of scaling exponent $\alpha$ when fitting to a fixed dataset.
	The red diamonds in Fig. \ref{fig:contact_determination_issues} (b) show the fit exponent and amplitude obtained from a fit (of the form ${n(k)} 
	= n_T(k) + C_\alpha/k^\alpha$) to each data set, each representing a different BEC density. 
	The variation in the fit exponent is small (mean 4.2, standard deviation 0.4) but the corresponding variation in fit amplitude spans over six orders of magnitude.
	Although the mean $\alpha$ from the fits is only half a standard deviation (one standard-error interval) different from 4, the corresponding difference in amplitude $C$ varies by a factor of order 100.
	Therefore the decision of which value to take for the exponent is extremely consequential. 
	There is no justification for picking the adjusted value (4) over the mean value, especially given the point of contention is whether or not the tails are the depletion, which \emph{could} justify fixing the exponent.
	
	Moreover, fitting with a fixed exponent returns an unrealistic estimate of the systematic error associated with the fitting procedure.
	The blue circles in Fig. \ref{fig:contact_determination_issues} show the amplitudes returned from fits to all datasets when $\alpha$ is constrained at some specific value within about one standard deviation of the mean $\alpha$. 
	In this case, the variation in $C_\alpha$ is much reduced, down to a factor of about 10 between the extremes.
	However, $C_\alpha$ varies over about six orders of magnitude as $\alpha$ varies within the range of uncertainties reported here and in the prior work \cite{Chang16}. 
	A linear fit reveals that $d \log_{10} C_\alpha/d\alpha \approx 6.8$.
	These fits scarcely differ in their goodness-of-fit criterion (the mean square error) and so offer no obvious way to reconcile the expected distribution with these divergent  conclusions.
	The underlying issue here is that there is not enough data to accurately constrain $\alpha$, but assuming some fixed $\alpha$ effectively discards the large covariance between these parameters.
	Furthermore, this problem is not reflected in the error estimates in the fitting routines: The error bars representing the uncertainty in parameters from the fixed-$\alpha$ fits are smaller than the markers used in Fig. \ref{fig:contact_determination_issues}. 
		
	It is easy to see that a well-intentioned choice of $\alpha$, which is not statistically different from the best-fit estimates, can lead to a conclusion which either agrees perfectly or disagrees catastrophically with the predictions of Eqn. (\ref{eqn:pred_scaling}).
	In particular if one assumes that the data conforms to a power law with $\alpha=4$ (i.e. that the data conforms to the Tan-Bogoliubov theory), one arrives at a contradiction (that the coefficient $C_4$ does not conform to the theory) but the inherent uncertainty in this approach precludes the possibility of such a definitive statement.

	A deceptively reassuring result could be found by multiplying the empirical density by $k^4$ in the hope of observing a flat region, and identifying this with a $k^{-4}$ tail (see, for example, \cite{Makotyn14,Chang16}). 
	The fitting procedure would be to appropriately scale the model of the thermal region (i.e. multiply Eqn. (\ref{eqn:th_fun}) by $k^4$) and add a constant term to fit the tail\footnote{Applying a fit of the form ${f_{\rm fit}(k)=}n_T(k)k^4+C$ to data scaled as ${f(k)} = k^4 n(k)$ minimizes the error term $\sum_i (k_{i}^{4}n(k_i)-k_{i}^4n_{fit}(k_i))$ which is equivalent to weighting the error terms {for the original $n(k)$ data} by $k^4$. 
	Moreover, the fit is highly sensitive to the choice of weighting function.
	If one constrains the exponent in the fit {of such an $f(k)$} at $\alpha=4$, then changing the {power of the} \emph{weighting function} of the squared errors in the fit from {$k^4$} by 0.1 {to $k^{4\pm0.1}$} changes the fitted {$C$} by a factor of ten.}.
	Unfortunately this offers no recourse from the issues described above and is not a definitive test for the presence of a power law or a way to obtain its parameters.
	Suppose the density actually decays as $n(k) \propto k^{-(\alpha+\delta)}$ with an exponent different by some $\delta$ from the expected value.
	Following the rescaling operation $n'(k) = k^\alpha n(k)$, one has a {function} 
	which would have the form $n'(k) \propto k^{-\delta}$. 
	Considering the range of $k$ examined in this (and the prior \cite{Chang16}) work, the scaled {function} 
	at the upper and lower end of the range would be predicted to differ by a factor $(k_\textrm{max}/k_\textrm{min})^\delta \approx 2^{0.1}\approx1.07$.
	This variation is dominated by the statistical fluctuations in the tails of the individual density profiles, and so would not be distinguishable by eye (or by fit) in a plot of the scaled density.
	
	The preceding discussion points to one of the primary challenges with power laws; the exponents are strongly entwined with the rate of occurrence of rare events, which by definition are subject to large statistical fluctuations and thus subvert even the most meticulous investigations.
	As I discuss in the next section, this underscores the challenge of identifying power-law behaviour in range-limited data.

	To summarize, these problems with fitting power laws are ubiquitous, and made more difficult by the small range of $k$ which are visible in the helium experiments.
	In general, estimating the exponent of a purported power law is difficult and requires data spanning several orders of magnitude in scale \cite{Goldstein04,Clauset09,Virkar14,Hanel17}, which are not present in either {of the} helium experiment{s to date}.
	The preferred statistical tools for analysing power law distributions are maximum likelihood estimators, as discussed in Refs. \cite{Clauset09,Virkar14}.
	However, in these particle detection experiments the limited sampling region and presence of spurious detection events (see below) mean that such estimators are not appropriate.

	In the next section I discuss an alternative approach which does not require -- and conversely cannot determine -- a precise value of the scaling exponent. 
	The approach discussed below permits a test of the Tan theory's validity in the far-field without assuming any specific properties of (and is clear about what can and cannot be said about) the actual form of the measured density distribution.
	
\subsubsection{Testing Tan's Tails}
\label{sec:regression}

	The prediction of the high-$k$ tail shape (via Eqn. (\ref{eqn:pred_scaling})) captures a great deal of information. 
	An important feature is that the amplitude of this tail is predicted to scale in proportion to the product $n_0 N_0$.
	One can integrate Eqn. (\ref{eqn:pred_scaling}) to predict the number of atoms in the trap whose wavevector has a modulus in the interval $k\in (k_\textrm{min}, k_\textrm{max})$. 

	\begin{equation}
		N_{k_\textrm{min},k_\textrm{max}} =\frac{\mathcal{C}}{2\pi^2}\left(\frac{1}{k_\textrm{min}}-\frac{1}{k_\textrm{max}}\right)
		\label{eqn:pred_num}
	\end{equation}
	For fixed $k_\textrm{min}$ and $k_\textrm{max}$, Eqn. (\ref{eqn:pred_num}) has the form {
	\begin{equation}
	N_{k_\textrm{min},k_\textrm{max}} = \Lambda N_0n_0
			\label{eqn:Lambda}
	\end{equation}
	}%
	(c.f. Eqn. \ref{eqn:pred_scaling} - note that the integral of $n(k)$ is most easily performed in spherical coordinates and requires the Jacobian $(2\pi)^{-3}d \kvec$ to ensure normalization.). 
	This form is directly testable by measuring the number of counts detected in the interval $(k_\textrm{min},k_\textrm{max})$ after producing a BEC of $N_0$ atoms with peak density $n_0$. 
	
	As above, $k_\textrm{min}$ was fixed at $6~\micron^{-1}$ which lies outside the thermal part for all the data sets (see Fig. \ref{fig:empirical_density}).
	However, the $k$-space field of view is restricted by the detector radius to $k\lesssim5\times 10^6$ m$^{-1}$ in the $(x,y)$ plane, which is only just sufficient to reach past the edge of the thermal region. 
	In the face of this tradeoff in the choice of $k_\textrm{max}$, I defined the bounds of the region of interest (ROI) by the minimum elevation angle $\phi_c=\pi/3$ rad above the $(x,y)$ plane and an upper bound of $k_\textrm{max} = 10~\micron^{-1}$.
	This amounts to an ROI consisting of two vertically oriented conical sections, each with half-angle $\pi/6$ from the $z$ axis, encompassing a total solid angle of $0.13\times 4\pi$ steradians. 
	The detector quantum efficiency of 0.08(2) and state-transfer efficiency of 25(2)\% combine  into the total efficiency $\epsilon\approx0.23(5)\%$.
	After discussing the results of this analysis, I will demonstrate their robustness against uncertainty in $\epsilon$ and the choices of $\phi_c$ and the $k$ bounds.

\begin{figure}
	\begin{center}
		\includegraphics[width=\columnwidth]{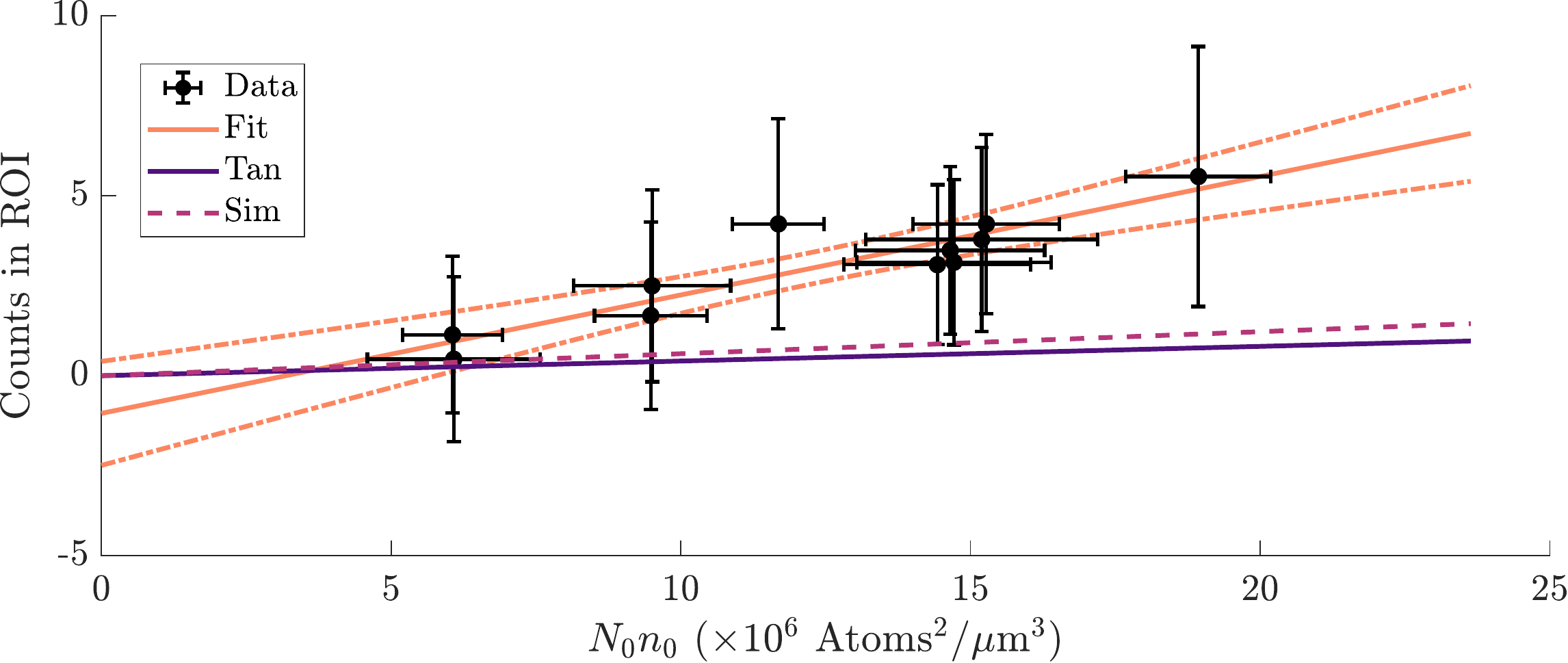}
			\caption{{Analysis of Tan's tails.} A linear fit shows that the product $N_0n_0$ is a good predictor of the number of counts within the region $(k_\textrm{min}=6~\micron^{-1},k_\textrm{max}~\micron^{-1})$, consistent with Eqn (\ref{eqn:pred_scaling}) 
			(solid orange line, dashed lines 95\% CI). The gradient {$\Lambda$} 
			{in Eqn. (\ref{eqn:Lambda})} can be predicted using Eqn (\ref{eqn:pred_scaling}) ({$\Lambda_{\rm pred}$} solid purple line) but this disagrees with the experiment by a factor of about 8. Our simulations (dashed line) show an increase in counts after release but by less than in the experiment. 
			}

		\label{fig:exp_results}
	\end{center}
	\end{figure}

	A linear fit of the form $\hat{N}_{k_\textrm{min},k_\textrm{max}} = \Lambda_\textrm{fit} n_0 N_0 + \beta$ yields {an intercept} 
	consistent with zero ($\beta$=-0.9,  95\% confidence interval (CI) = (-3.1, 1.2)) and a good correlation ($r^2\approx0.8$), providing evidence supporting the expected linear relationship, and is shown in Fig. \ref{fig:exp_results}. 
	The fit has $p=1\times10^{-3}$ which indicates it is extremely unlikely that the results could be obtained from a null model (i.e. one where the measured counts are independent of the independent variable $N_0 n_0$)
	The correlation coefficient between the 
	variables {$N_{k_{\rm min},k_{\rm max}}$} 
	and $N_0n_0\propto(N_0^7\bar{\omega}^6)^{1/5}$ is 0.9.
	This is consistent with the claim that the product $N_0n_0$ is a predictor of the depleted population, which is consistent with Eqn. (\ref{eqn:pred_scaling}).
	No other physically-motivated combination of the independent variables provides a better fit.
	For comparison, a linear fit proves that the atom number {$N$} itself is a poor predictor of the detected number ($r^2=0.05~,p=0.54$), as is the density {$n_0$} alone ($r^2=0.4~,p=0.04$).

\begin{figure}
	\begin{center}
		\includegraphics[width=\columnwidth]{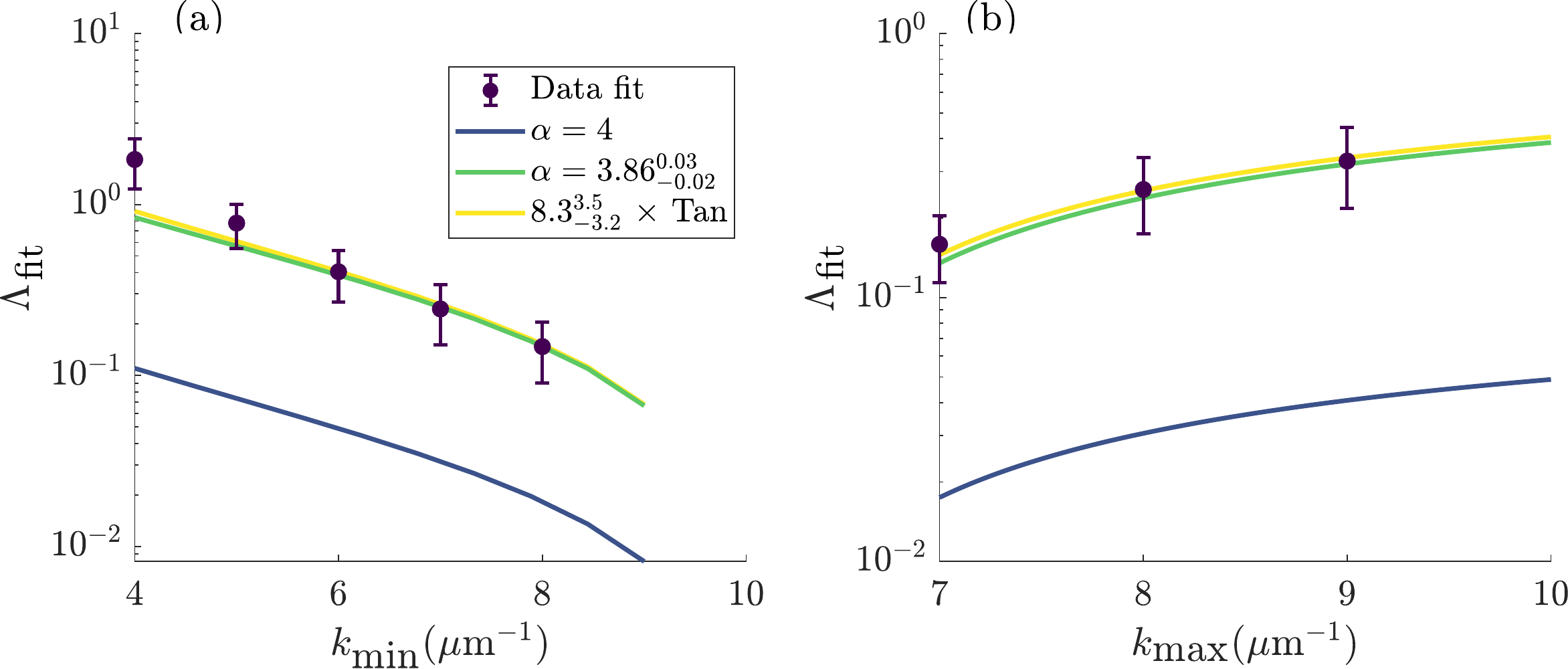}
			\caption{The experimental data {$\Lambda_{\rm fit}$} (points) show the $\Lambda_{\rm fit}$ to the entire data set with variable $k$ bounds ($k_\textrm{max}=10\micron^{-1}$ in (a) and $k_\textrm{min}=6\micron^{-1}$ in (b)).
		Predictions of $\Lambda$ based on Eqn (\ref{eqn:pred_scaling}) ({$\Lambda_{\rm pred}$}, blue), are shown along with the predictions {from Eqn. (\ref{eqn:Lambda}) using} a density function $n(k){=\mathcal{AC}/k^4}$ that {has an additional prefactor $\mathcal{A}=$8(3)} 
		(green) and one that has 
		a modified exponent of {$\alpha=3.86(2)$ via $n(k)=\mathcal{C}/k^{\alpha}$} (yellow). 
		The data does not distinguish between either of these hypotheses, and therefore provides limited information about {the prefactor $\mathcal{A}$ or exponent}
		$\alpha$.
		The error bars show the 95\% CI of the fit parameter $\Lambda_\textrm{fit}$, and parentheses enclose the uncertainty in the least-significant digit which is consistent with the 95\% CI.
			In (b), the deviation from the predictions at $k_\textrm{min}\lesssim6~\micron^{-1}$ is because the collection area starts to overlap with the thermal region.
			}
		\label{fig:k_indep}
	\end{center}
	\end{figure}

	The gradient $\Lambda_\textrm{fit}$ is of particular interest because it can be predicted using Eqn. (\ref{eqn:pred_num}).
	Given an ROI, one can calculate $\Lambda_\textrm{pred} = 32\epsilon a^2(k_{\textrm{min}}^{-1}-k_{\textrm{max}}^{-1})/7$.
	The predicted slope disagrees with the empirical fit by a factor of $\Lambda_\textrm{fit}/\Lambda_\textrm{pred}= 8.3$, 95\% CI $(5.5,11)$.
	Figure \ref{fig:k_indep} shows that this conclusion holds regardless of the choice of $k$ boundary values: No matter the choice of $k_\textrm{min}$ or $k_\textrm{max}$, the resulting fit parameter $\Lambda_\textrm{fit}$ is 8(3) times larger than the Tan-Bogoliubov prediction. 

	However, while this method can accurately predict the number of detected atoms, it does not distinguish whether or not the data follows a power-law distribution, let alone a power-law with a particular exponent. 
	Consider these simple alternatives: that the tail amplitude is simply larger than expected (i.e. $n(k)=A\mathcal{C}/k^4$), or that the tail decay is somehow slower (i.e. $n(k)=\mathcal{C}/k^{4+\delta})$, or that it comes from another distribution altogether.
	Specifically, the density profiles $n(k)=A\mathcal{C}/k^4$ with $A=8(3)$ and $n(k)=\mathcal{C}/k^{\alpha}$ with $\alpha=3.86(2)$ both predict the variation of $\Lambda_{\rm pred}$ with $k_\textrm{min}$ and $k_\textrm{max}$ with comparable accuracy, as shown in Fig. \ref{fig:k_indep}. 
	Without the ability to precisely determine the exponent $\alpha$, there is insufficient evidence to conclude which of these is correct.
	Indeed, the predicted $k^{-4}$ behaviour is only a strict constraint inasmuch as the tails are \emph{known} to be the quantum depletion escaping the condensate without perturbation - which is itself the claim under test and as argued below is not necessarily expected.

	We are then tasked with reconciling the nonlinear scaling of the detected counts, which is consistent with the quantum depletion, and the disagreement over the absolute number of detected counts.
	Ultimately, as I describe in the section \ref {sec:discussion}, the most prudent way through this dilemma is via alternative experimental designs.
	One conclusion remains robust: There are {almost} about ten times as many detections in the depletion region as one would expect based on the Tan theory {\emph{in situ}}. 
	Nonetheless, the particular nonlinear scaling of detected counts with the predictor $N_0n_0$ is concordant with the tails' originating in the quantum depletion.
	Furthermore, I describe below how the simulations indicate that the quantum depletion can survive the condensate expansion, presenting depleted tails in the far-field with modified amplitude. 
	We are therefore faced with the conclusion that tails are indeed a signature of quantum depletion, albeit subject to some effect during expansion into which the present analysis can see no further. 
	In order to find more insight into this observation, we performed numerical simulations of the time-dependent evolution of the expanding condensate, which I describe in the next section \ref{STAB}.
	In the next section I provide technical details of the experiments and rule out several systematic factors which could lead to this disagreement.

\subsection{Experimental details}
\label{sec:exp_details}
\subsubsection{Trap configuration}

	I prepared the BECs by evaporatively cooling helium in a harmonic magnetic trap with trap frequencies $\approx(45,425,425)$ Hz and a DC bias stabilized by our auxiliary field compensation coils \cite{Dall07,Dedman07}. For the tight trap, I increased the coil current after the cooling sequence to obtain trapping frequencies $\approx(71,902,895)$ Hz, ramping the current as a sigmoid step function to minimize in-trap oscillations. Note that the weak ($x$) axis of the trap is horizontal, with tight vertical confinement. The trap switched off with a $1/e$ time of $\approx38~\mu$s. The condensates then expanded for 2 ms before I transferred some of the initial $m_J=1$ condensate into the magnetically insensitive $m_J=0$ state via a Landau-Zener sweep to mitigate against distortion by stray magnetic fields during the free fall to the detector. The RF pulse was created by a function generator, amplified, and applied to the experiment chamber by a coiled antenna inserted into the BiQUIC coil housing. The pulse swept from 1.6-2.6 MHz over 1 ms and was centred on the resonance between the $m_J$ states. The determination of the transfer efficiencies $\eta_J$ for each of the $m_J$ states is discussed below. The sweep was $10^6$-fold wider than the RF width of the BEC which ensured uniform transfer at all momenta. Immediately after the RF sweep, the bias coils switched off and auxiliary push coils in the vertical (Z) and weak horizontal (X) axes are activated using a fast MOSFET switch to implement a Stern-Gerlach separation of the $m_J = -1,~0,$ and $+1$ pulses.

	The Roentdek DLD80 multichannel plate and delay-line detector stack, located 848mm below the trap, registers the arrival times and positions $(t_i,x_i,y_i)$ of each atom, indexed by $i$ \cite{Manning10}. 
	The velocity of each atom relative to the centre of mass of each cloud is calculated by $(v_x,v_y,v_z) = t_{i}^{-1}(x_i-\bar{x},y_i-\bar{y},\tfrac{1}{2}g_0(t_{cen}^2-t_{i}^{2}))$, where $g_0$ is the local gravitational acceleration, the overbar denotes the within-shot average and $t_{cen}$ is the time of flight of the centre of mass of the cloud. 
	The far-field momentum is thus obtained via $m\textbf{v} = \hbar\kvec$.
	The space and time resolution of the detector are 100 $\mu$m and 3 $\mu$s, respectively \cite{Henson18_BCR}, where the latter is comparable to a vertical resolution of $\approx 12~\mu$m.
	The detector efficiency of $8(2)\%$ was determined from analysis of the squeezing parameter of correlated atoms on the opposite sides of scattering halos \cite{Shin19,Shin20,Jaskula10}. 

	{The dominant uncertainty in the collection efficiency $\epsilon$ is the 25\% error in the detector quantum efficiency (QE), whereas the other factors (cutoff angle $\phi_c$ and transfer efficiency $\eta_0$) are more precisely known.
	Neither the uncertainty in the collection efficiency nor the choice of elevation angle cutoff $\phi_c$ have a significant effect on the findings. A change in quantum efficiency (QE) presents by changing the value of $\epsilon$ used in the prediction and also through the factor of $N_{0}^{7/5}$ used to compute the condensate density $n_0$. These effects partially cancel to produce a weak scaling of $\Lambda_\textrm{fit}/\Lambda_\textrm{pred}$ with respect to $\epsilon$.  Re-running the analysis using different QE yields fits that barely differ in the goodness-of-fit criterion and present comparable results for $\Lambda_\textrm{fit}$. The choice of collection area defined by the elevation angle cutoff $\phi_c$ has a weak effect on the result, but below statistical significance. 
	The relevant statistics are summarized in Table \ref{tab:choice_indep}.}

	\begin{table}
		\begin{minipage}{0.5\textwidth}
		\vspace{0cm}
		{\fontsize{11}{11}\selectfont
		\begin{tabular}{c c c c}
			\hline\hline

			QE & fit $r^2$ &  $\Lambda_\textrm{fit}$ & $\Lambda_\textrm{fit}/\Lambda_\textrm{pred}$\\      
			\hline
			0.05    &   0.83   &   0.2(0.1,0.3)  &  6.8(4.5,9.1)\\
			0.06    &   0.83   &   0.3(0.2,0.4)  &  7.4(4.9,9.8)\\
			0.07    &   0.83   &   0.3(0.2,0.4)  &  7.8(5.2,10.5)\\
			0.08    &   0.83   &   0.4(0.3,0.5)  &  8.3(5.5,11)\\
			0.09    &   0.83   &   0.5(0.3,0.6)  &  8.7(5.8,11.6)\\
			0.1     &   0.83   &   0.6(0.4,0.7)  &  9.0(6.0,12.1)\\
			0.11    &   0.83   &   0.6(0.4,0.8)  &  9.4(6.2,12.5)\\
			\hline
			$\phi_c$ & fit $r^2$ &  $\Lambda_\textrm{fit}$ & $\Lambda_\textrm{fit}/\Lambda_\textrm{pred}$\\
			\hline
			80$^\circ$    &   0.82   &   0.1(0.0,0.1) &  9.3(6.0,12.7)\\
			70$^\circ$    &   0.86   &   0.2(0.1,0.3) &  9.2(6.4,12)\\
			60$^\circ$    &   0.83   &   0.4(0.3,0.5) &  8.3(5.5,11)\\
			\hline\hline
		\end{tabular}
		}
		\end{minipage}
		\hfill
		\begin{minipage}{0.5\textwidth}
		\vspace{0cm}
		\caption{Sensitivity analysis for the systematic uncertainty in detector efficiency and choice of $\phi_c$. Terms in brackets are the upper and lower 95\% confidence intervals.}
		\label{tab:choice_indep}
		\end{minipage}
	\end{table}

\subsubsection{Peak density calibration}
\label{sec:n0_cal}

    The quantum depletion and contact are both predicted to depend solely on the condensed number and trapping frequencies via the condensate density, hence it is important to determine both quantities accurately. 
    The sole experimental parameters in the expression for the peak density (Eqn. (\ref{eqn:n0})) are the geometric trap frequency $\bar{\omega} = \left(\omega_x\cdot\omega_y\cdot\omega_z\right)^{1/3}$, and $N_0$, the number of atoms in the condensate. 
    The total atom number $N$ and trap frequency $\bar{\omega}$ can be determined simultaneously in a single shot using a pulsed atom laser, and the thermal fraction $\eta_T$ (section \ref{sec:th_spin}) thus determines the condensed number $N_0 = (1-\eta_T)N$. 

	The pulsed atom laser consists of a series of Fourier-broadened RF pulses centred on the minimum Zeeman splitting in the trap. 
	The pulse transfers atoms in the trap to the untrapped $m_J=0$ state with an approximately constant transfer rate across the cloud \cite{Manning10,Henson18_BCR}. 
	Approximately 2\% of the atoms are outcoupled per 100$~\mu$s pulse for $\approx$200 pulses, which eventually depletes the entire trap. 
	The atom laser thus prevents the detector from saturating and allows an accurate determination of the atom number, up to a factor of the quantum efficiency. 
	The trapping frequencies are determined by inducing centre-of-mass oscillations with a magnetic impulse, and finding the oscillation period from the atom laser pulses, detailed below.

	To illustrate how we obtain the trapping frequencies, consider an undamped harmonic oscillator in one dimension. The centre-of-mass momentum $p(t) = A \cos(2\pi f t))$ oscillates at a frequency $f$ Hz, and the centre of mass of a pulse outcoupled at time $t'$ lands on the detector at a position $x'(t+\tau) = p(t')\tau/m$, where $\tau$ is the time of flight of the centre of mass. Sampling the cloud motion with a sampling period T starting at initial time $t_0$ produces a series of pulses whose centres of mass land at $\{x_n = (A\tau/m) \cos(\omega(t_0+nT))\}$. The sampling frequency $f_s=1/T$ determines the \emph{Nyquist frequency} $f_N=f_s/2$, which is the maximum frequency that can be reconstructed unambiguously from such a sampling regime. When $f>f_N$ the signal manifests as a lower-frequency oscillation known as the \emph{alias} of the signal. The aliased frequency $f_a$ is

	\begin{equation}
	 f_a =
	  \begin{cases}
	   \frac{Z_N f_s}{2} - f & \text{if } Z_N \text{ even} \\
	   f - \frac{(Z_N-1)f_s}{2}       & \text{if } Z_N \text{ odd},
	  \end{cases}
	  \label{eqn:Z_N}
	\end{equation}
	
	where $Z_N = \lceil{f/f_N}\rceil$ is the Nyquist zone number, as described in numerous RF engineering references\footnote{Some good tutorials are presented \href{https://www.analog.com/media/en/training-seminars/tutorials/MT-002.pdf}{here} and \href{https://www.taborelec.com/Multi-Nyquist-Zones-Operation-Solution-Note}{here}. The trap frequency measurement technique specifically is presented in detail in \cite{Henson22_PAL}}. While it is not possible to determine $Z_N$ for a signal with an unknown (and not necessarily stationary) $f$ from samples taken at a fixed $f_s$, one can vary $f_s$ and determine both $Z_N$ and $f$.
	
	\begin{figure}
	\centering
		\includegraphics[width=0.7\textwidth]{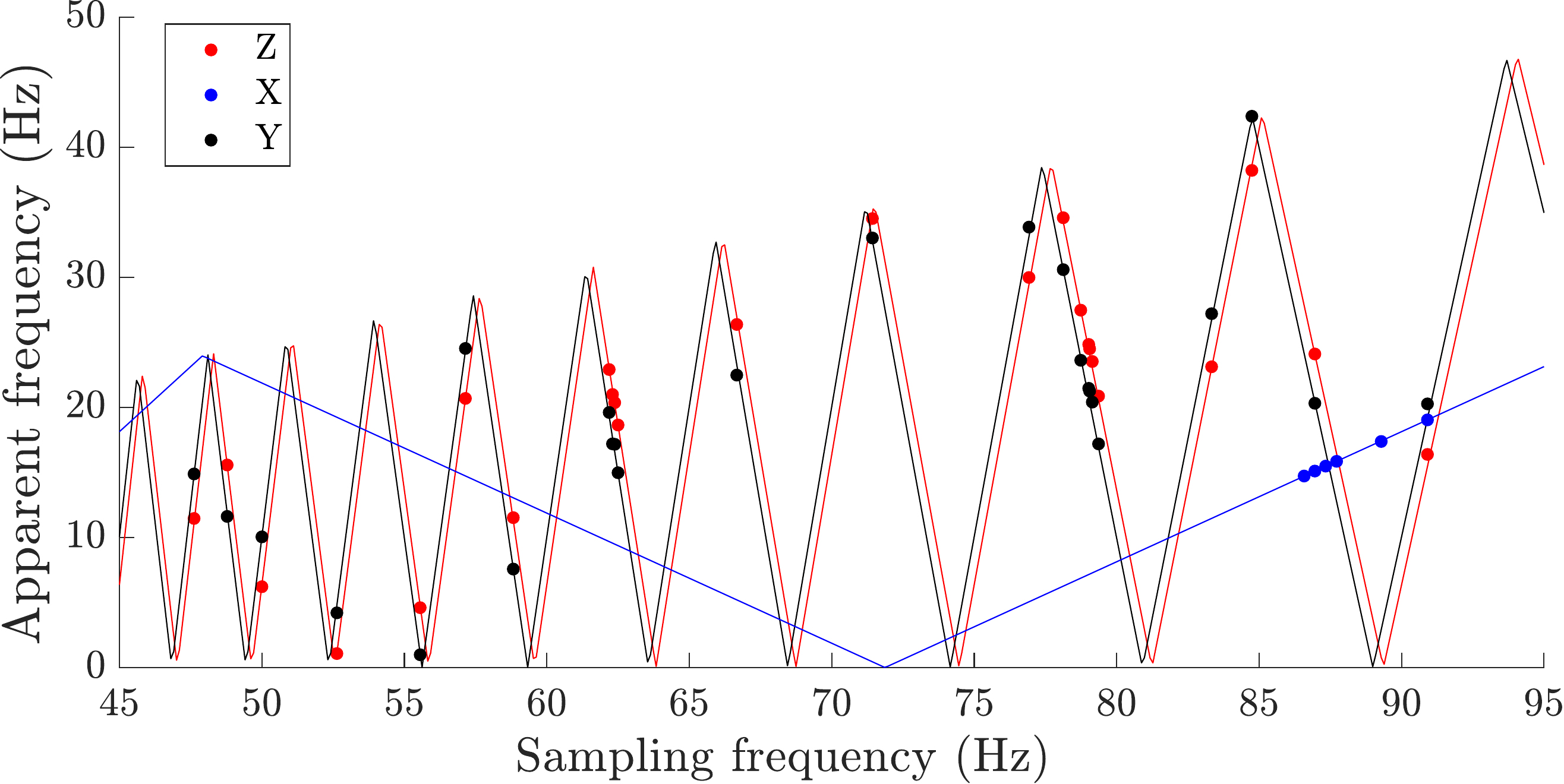}
			\caption{Aliasing of the trap frequency oscillations under different sampling regimes. The sawtooth function is a one-parameter fit (Eqn. \ref{eqn:Z_N}) which determines the underlying frequency of the centre-of-mass oscillations.}
			\label{fig:sawtooth_plot}
	\end{figure}	

	Figure \ref{fig:sawtooth_plot} shows the \emph{apparent} frequencies of oscillation in each axis as a function of sampling frequency, as measured using the pulsed atom laser method described above.
	The maximum sampling frequency is limited by the temporal width of the BEC pulse landing on the detector. 
	We fit the apparent oscillation frequency in each axis with the one-parameter model Eqn. \ref{eqn:Z_N} and thus determine the characterisic frequencies of the magnetic trap.

	In fact, the measurements of $\omega_y$ and $\omega_z$ in Fig. \ref{fig:sawtooth_plot} use much more data than is strictly necessary to obtain $f$ unambiguously, and is presented to illustrate the technique. 
	One can obtain faster estimates by varying $f_s$ slightly and computing the gradient $df_a/df_s$ which unambiguously fixes $Z_n$ (assuming one does not accidentally cross a zone boundary, see eqn \ref{eqn:Z_N}).
	Therefore by selecting a sampling frequency near the middle of a known Nyquist zone, one can then obtain the Nyquist zone number with just a few measurements, as done in the $X$ axis (blue in Fig. \ref{fig:sawtooth_plot}). 
	
	For example, if the oscillations in the \(y\)-axis are determined to be in the 5\textsuperscript{th} Nyquist zone when sampling at 125~Hz, the corresponding correction to obtain the underlying frequency from the aliased frequency would be
	\begin{equation}
	    f_{\text{real}}=3 f_{\text{sampling}} + f_{\text{aliased}}.
	\end{equation}
	where \(f_{\text{real}}\) is the underlying frequency, \(f_{\text{aliased}}\) is the apparent frequency obtained from the fit, and \(f_{\text{sampling}}\) is the sampling frequency. 

	One can monitor changes in the underlying frequency $f$ with single measurements, which is especially relevant in the experiment of Chapter \ref{chap:tuneout} (again, assuming the shift is not great enough to push the signal into a different Nyquist zone\footnote{In Chapter \ref{chap:tuneout}, the 125Hz sampling rate means the zones are 62.5 Hz wide. The sub-Hz stability of the magnetic trap and the small perturbation from the probe beam ensure that the net oscillation frequency signal remains within a single Nyquist zone over the entire data collection campaign. }).	
	The final uncertainty in the trap frequency determination is simply the fit error as the Nyquist zone number is known exactly.

\subsubsection{Determining spin transfer efficiency}
\label{sec:th_spin}

	\begin{figure}
	\begin{center}
		\includegraphics[width=\columnwidth]{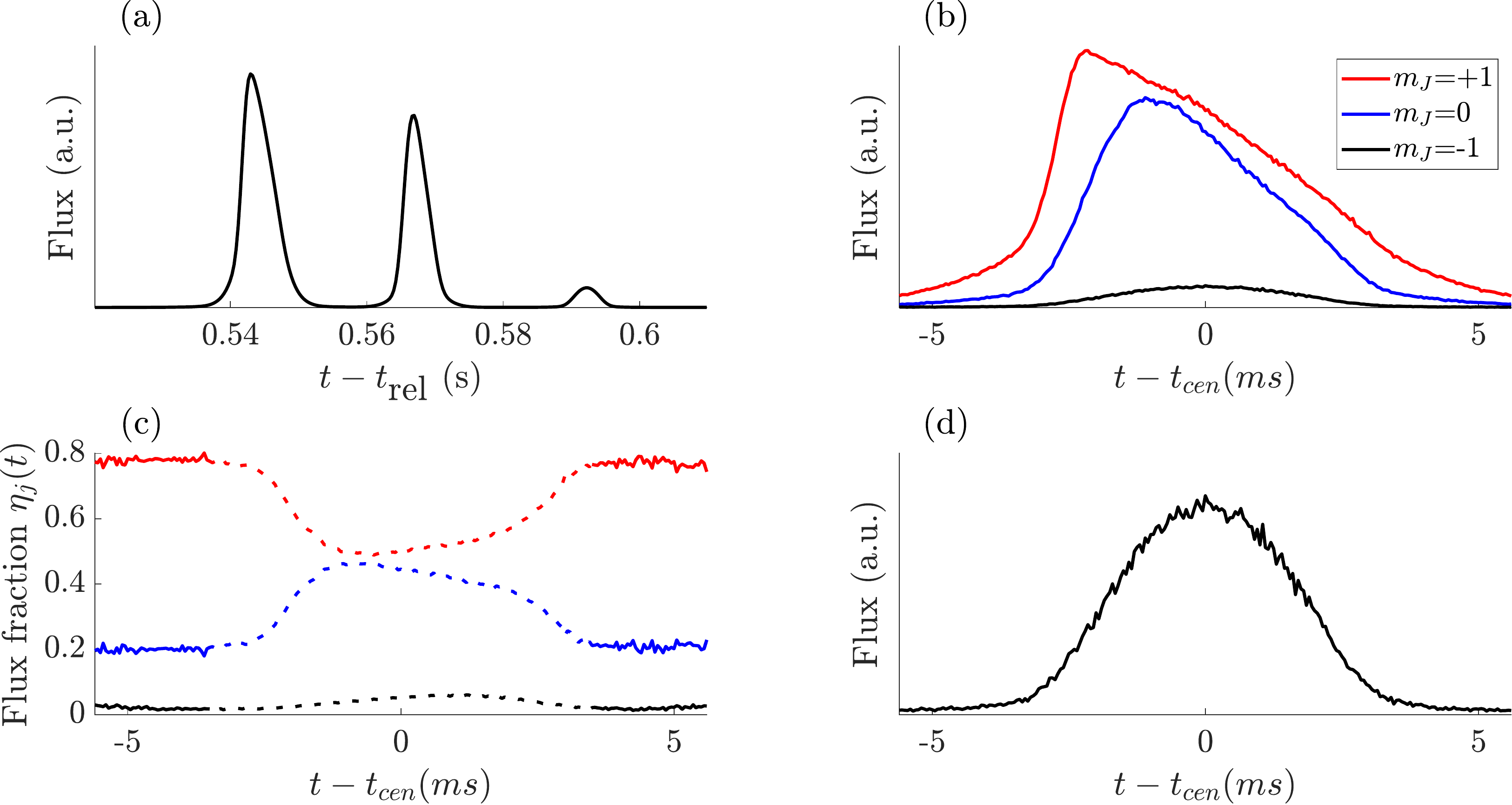}
		\caption{Determining the RF transfer efficiency. The time-of-flight profiles of each pulse are resolved (a) by applying a weak Stern-Gerlach pulse during the time of flight. The pulses are aligned with respect to their centre-of-mass (b) and used to determine the pointwise fraction ((c), dotted line). Detector saturation is evident in the peaks (dashed lines), but not in the thermal tails (solid lines), which are used to compute the transfer efficiency. Because of its lower flux, the $m_J=-1$ pulse does not show any clear evidence of saturation (d) and is used to determine the thermal fraction and hence $N_0$.}
		\label{fig:frac_cal}
	\end{center}
	\end{figure}
	To calibrate the transfer efficiencies, a weaker Stern-Gerlach is used, such that each $m_J$ cloud hits detector at different times and positions, as illustrated in Fig. \ref{fig:frac_cal}. 
	The efficiencies $\eta_J$ cannot be calculated by counting the atoms in each cloud because the detector saturates during the peak condensate flux.
	However, the thermal parts are unsaturated and thus are comparable by aligning each cloud along the time (Z) axis.
	We can then compute the pointwise fraction of the atomic flux $\phi(t)$ accounted for by each cloud, $\eta_j(t) = \phi_j(t)/\sum_j\phi_j(t)$, as depicted in Fig. \ref{fig:frac_cal} (a-c).
	The ratio of densities between the clouds is roughly constant in the thermal part (Fig. \ref{fig:frac_cal} (c)), indicating the absence of saturation effects in the thermal part and a spin transfer efficiency that is independent of $k$. 
	The fraction of the original cloud transferred into each $m_J$ state is determined by taking the average $\langle\eta_j(t)\rangle$ over the thermal tails. 
	These efficiencies are approximately 74\%, 24\%, and 2\% in all runs for the $m_J=+1$, 0, and -1 states, respectively.
	While the $m_J=0$ and $m_J=1$ clouds clearly saturate the detector, the small fraction ($\approx2\%$) of the atoms transferred to the $m_J=-1$ state does not (Fig. \ref{fig:frac_cal} (d)). 
	A bimodal fit to the condensed and thermal parts, plus constant background, yields the thermal fraction $\eta_T$ and condensed fraction $1-\eta_T$.

\subsubsection{Noise sources}
\label{sec:spinpop}

	In early tests of the measurement sequence I noticed a contamination of the signal by spurious counts. 
	I identified these as remnant counts from the $m_J=+1$ cloud as they were still visible when we ran an experimental sequence without the Landau-Zener transfer. 
	This contamination appeared in a particular region of the detector image, and as such can be corrected for by subtracting their contribution from the counts collected during measurement shots. 
	While the cause of the cross-contamination is unclear, the count density outside the region of interest is similar in both the shots with the RF pulse and those without. 
	We could thus hypothesize that the remnant counts are atoms transferred into the $m_J=0$ state by non-ideal behaviour of the Stern-Gerlach pulses or magnetic field switches. 
	Note that only about one in a million atoms from the $m_J=1$ cloud are present in this manner in a given shot.
	{Such counts constitute about 10(5)\% of the detection events in the ROI. 
	The calibration without the Landau-Zener transfer enables a correction for their contribution by subtracting the counts measured in the calibration runs from the total used in Sec \ref{sec:regression}.
	However, because it is not possible to distinguish \emph{which} individual atoms are spurious and which are genuine quantum-depleted particles, the preferred power-law analysis (the maximum-likelihood estimator) is not available. 
	Basic MLEs are built assuming one has a sample of data drawn from a single category; the probabilistic combination of two indistinguishable categories of events falls outside the scope of this framework.}

\section{Numerical simulations}
\label{STAB}

	The measurements described above are complemented by numerical simulations\footnote{ The simulations were run by my collaborator P. Deuar and their methods are detailed in \cite{Ross21}.} of the dynamics of the momentum distribution after the trap release using a Stochastic Time-Adaptive Bogoliubov (STAB) method in the positive-P framework  \cite{Deuar11,Kheruntsyan12}. 
	These show that the non-adiabatic release of the trap is responsible for survival of the depletion, and that the depleted particles acquire additional kinetic energy from the mean-field energy of the condensate during the subsequent adiabatic expansion. 
	These factors result in an amplification of the momentum tails relative to the in situ values {by a factor of up to about two}, and are not captured in the hydrodynamic approximation. 
	However, there remains a quantitative disagreement between simulation and the experiment {in that the experimental tails are even heavier by a large margin.}

	\begin{figure}
	        \includegraphics[width=\columnwidth]{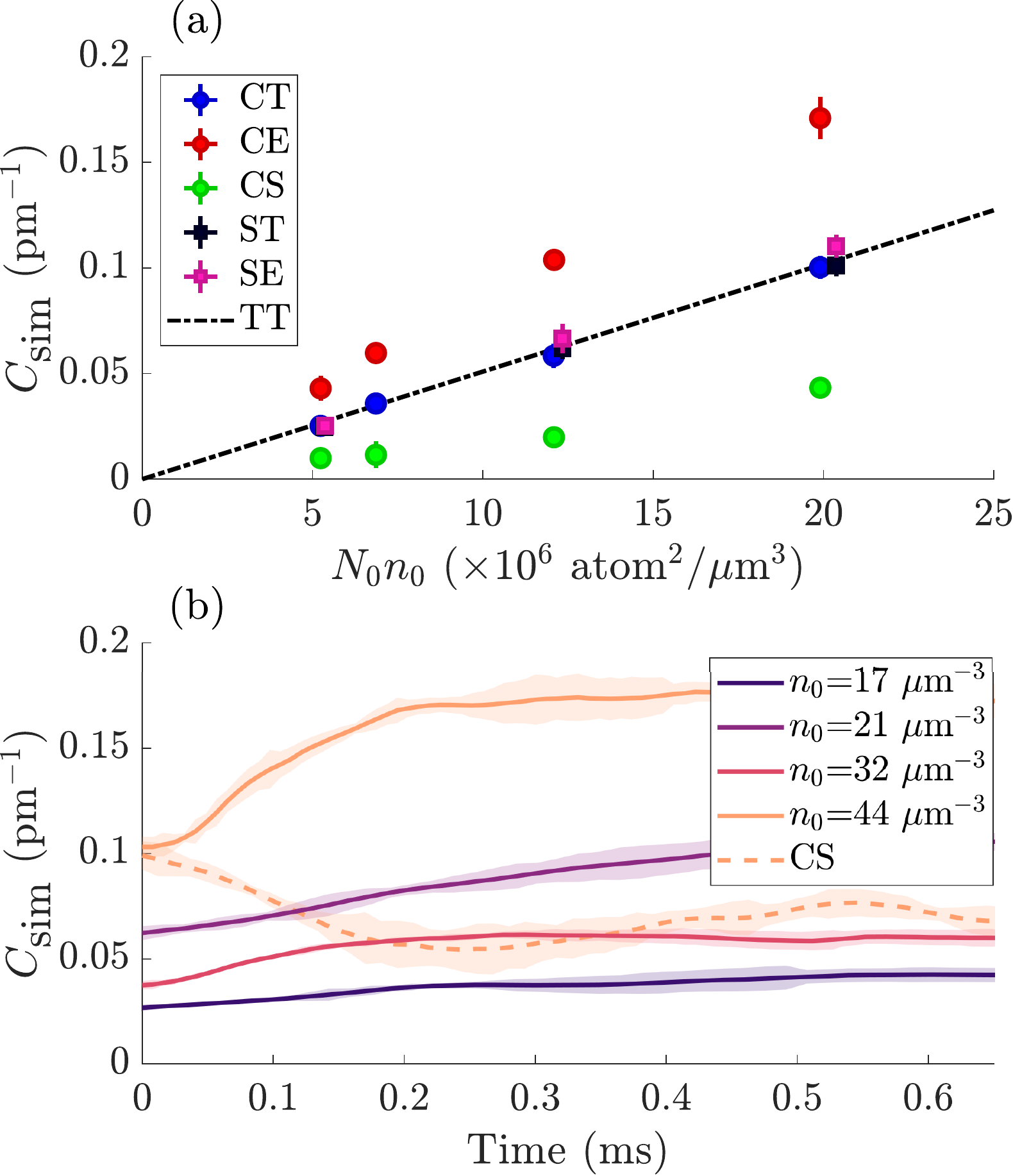}
	        \caption{Simulations of release from the trap. (a) Steady-state values of the simulated contact. Simulations of condensates released from a cigar-shaped trap (CT) are consistent with the Tan theory (TT) before release, and show an increase in contact after the trap release (CE). A slow relaxation of the transverse trapping frequencies (CS) shows a decrease in line with the predicted value of the lower density. Spherical traps (ST,SE) lack any directions of tight confinement, wherein a longer interaction time prevents the escape of depleted particles as seen in cigar traps. (b) the time-dependence of the contact stabilizes after a time on the order of $1/\omega_x$, several hundred $\mu$s. The expanded contact is consistently about 1.7 times the Tan theory. For comparison, the experimental control pulses are implemented after 2 ms of expansion. When the transverse trapping frequencies are slowly reduced by half (dotted line), the in-situ contact relaxes.}
	        \label{fig:sim_fig}
	\end{figure}
	
	The simulations started from a cigar-shaped trap with parameters matched to the experimental conditions. 
	The in-trap state before release from the trap at time $t=0$ (marked CT in Fig. \ref{fig:sim_fig} (a)) was consistent with the adiabatic sweep theorem. 
	Following expansion from the cigar trap, the simulated tail amplitude increased and stabilized within a few hundred microseconds (CE in Fig. \ref{fig:sim_fig} (a)), much sooner than the 2 ms delay between the trap release and application of the rf and Stern-Gerlach pulses. 
	Fig. \ref{fig:sim_fig}~(b) shows the time evolution of the tail amplitude {$C_{\rm sim}$ extracted from a $n(k)=C_{\rm sim}/k^4$ fit to the simulated density} 
	using {an ROI with the same cutoff elevation angle $\phi_c=\pi/3$ as in the experiment.}  
	In this configuration the steady-state value of the momentum tails was a factor of 1.64(9) above the predictions of Eqn. (\ref{eqn:pred_scaling}). 
	{An analysis of the occupation of the tails according to (\ref{eqn:pred_num}), gives very similar factors $\mathcal{A}_{\rm sim}$ for the increase in the strength of the tails (relative to \emph{in-situ} predictions) during evolution, as shown in Table~\ref{tab:Nminmax}.}

	\begin{table}

	\begin{minipage}{0.5\textwidth}
	\vspace{0pt}
	\setlength{\tabcolsep}{0.5pt}
	{\fontsize{8}{11}\selectfont
	\begin{tabular}{c|ccc|c|c}
	\hline\hline
	$n_oN$	&$k_{\rm min}$&$k_{\rm max}$ & $\phi_c$ & $N_{k_{\rm min},k_{\rm max}}$	&ratio $\mathcal{A}_{\rm sim}$ 	\\
	($10^6\mu$m${}^{-3}$)& \multicolumn{2}{c|}{($\mu$m${}^{-1}$)}& (deg) &final& 	final/in situ\\
	\hline
	\multicolumn{6}{l}{Rapid release from trap (CE)}\\
	5.237(16)   & 2.0 & 3.5 & 60 & 193(8)    & 1.38(5)\\
	12.09(6)    & 2.25 & 3.5 & 60 & 362(10)   & 1.43(4)\\
    6.86(3)     & 2.75 & 4.0 & 60 & 163(4)   & 1.45(4)\\
    19.90(8)    & 3.0 & 4.0 & \ \ 40 & \ \ 423(8)    & \ \ \ 1.37(3) \\
    19.90(8)    & 3.0 & 4.0 & \ \ 50 & \ \ 408(8)    & \ \ \ 1.56(3) \\
    19.90(8)    & 3.0 & 4.0 & 60 & 368(7)    & 1.80(4) \\
    19.90(8)    & 3.0 & 4.0 & \ \ 70 & \ \ 282(6)    & \ \ \ 2.00(4) \\
    19.90(8)    & 3.0 & 4.0 & \ \ 80 & \ \ 157(5)    & \ \ \ 2.19(6) \\
	\hline
	\multicolumn{6}{l}{Rapid release, spherical (SE)}\\
	5.35(2)     & 2.0 & 3.25 & 60 & 120(9)    & 0.92(6)\\
	12.23(6)    & 2.0 & 3.0 & 60 & 256(11)   & 1.05(4)\\
    20.38(10)   & 3.0 & 4.0 & 60 & 232(10)    & 1.10(5) \\
	\hline
	\multicolumn{6}{l}{Slow ramp down of trap (CS)}\\
	5.237(16)   & 2.0 & 3.5 & 60 & 55(10)   & 0.37(6)\\
	12.09(6)    & 2.0 & 3.5 & 60 & 103(21)   & 0.31(5)\\
	6.86(3)     & 2.5 & 4.0 & 60 & 50(8)   & 0.34(6)\\   
	19.90(0)    & 2.5 & 4.0 & 60 & 181(14)   & 0.54(4)\\
	\hline\hline
	\end{tabular}
	}
	\end{minipage}
	\hfill
	\begin{minipage}{0.5\textwidth}
	\vspace{0pt}
	\caption{
	Tail strength data in simulations, referenced by the value of $n_0N$. $N_{k_{\rm min},k_{\rm max}}$ is calculated as per (\ref{eqn:pred_num}). The ratio $\mathcal{A}_{\rm sim}$ of tail strength  in the expanded cloud compared to Tan theory predictions is obtained by dividing $N_{k_{\rm min},k_{\rm max}}$  at the final time by its value in situ at $t=0$. $k_{\rm max}$ is chosen to still contain all $4\pi$ steradians inside the square lattice, while  $k_{\rm min}$ to avoid overlap with the expanding condensate.
	}
	\label{tab:Nminmax}
	\end{minipage}
	\end{table}

	To understand the disagreement with earlier theory \cite{Qu16}, which predicted no depletion survival, we also investigated the effect of adiabatic expansion on the in-trap depletion by simulating a slow decrease of the transverse trapping frequencies by a factor of two (CS), and found that the in-trap contact {$C_{\rm sim}$ as well as the the tail strength $N_{k_{\rm min},k_{\rm max}}$ from (\ref{eqn:pred_num})} 
	decreased roughly as predicted by Eqn. (\ref{eqn:pred_scaling}) in these instances --- see the dashed line in Fig.~\ref{fig:sim_fig} (b) {and Table~\ref{tab:Nminmax}.} 
	We also found that the relative factors between Eqn. (\ref{eqn:pred_num}) and the simulated tails (i.e. $C_\textrm{sim}/\mathcal{C}$ {and $\mathcal{A}_{\rm sim}$) depend} on choice of cutoff angle $\phi_c$. 
	{We find that apparent tail strengths $C_\textrm{sim}$ {and $\mathcal{A}_\textrm{sim}$ are} larger for smaller collection regions that are more tightly concentrated about the 
	strong trapping axis, whereas larger collection angles (that include areas closer to the weak axis) produce lower apparent tail strengths. 
	{Data on $\mathcal{A}_{\rm sim}$ are shown in Table~\ref{tab:Nminmax} while 
	$C_\textrm{sim}$ was 1.6(1), 1.9(2), and 2.2(3) times $\mathcal{C}$ for $\phi_c$ values of 60, 70, and 80 degrees, respectively.}
	We note that the experimental results (Tab. \ref{tab:choice_indep}) could be said to display a similar trend, but the result is not statistically significant.
	Details regarding the implementation of the simulations can be found in \cite{Ross21}.

\section{Discussion}
\label{sec:discussion}

\subsection{Summary of results}

    One potential interpretation of these results is that the depleted atoms are accelerated by the non-uniform mean-field energy of the condensate during the expansion.
	In detail, after a quench into the free particle regime, the condensate expands hydrodynamically on timescales of $1/\omega$. 
	This is an adiabatic process for the low momentum depletion, whereby some depleted atoms are absorbed back into the condensate in agreement with \cite{Qu16}.
	However, the characteristic time for reabsorption is $\hbar/gn_0$, slow enough that quasiparticles in the particle branch of the Bogoliubov dispersion have sufficient velocity to escape the expanding cloud without being reabsorbed and thus transition to free atoms. 
	Quasiparticles whose wavenumber exceeds that of the speed of sound lie within the particle-like branch of the Bogoliubov spectrum, which for our considerations is on the order of $k\gtrsim2~\micron^{-1}$, well within the thermal velocities. Therefore most atoms with $k\gtrsim 6 \micron^{-1}$ would avoid absorption.
	
	Moreover, an atom inside the BEC experiences an effective force from the gradient of the mean-field potential $\textbf{F} = -4\pi\hbar^2 m^{-1}a \nabla  n(x)$. 
	This endows escaping depleted particles with a greater momentum, increasing the amplitude of the tails in the far-field, and has been observed for the thermal part of the cloud in other experiments \cite{Ozeri02}.
	Further, it is much easier for depletion atoms to escape and be accelerated along the tightly-confined axes of a cigar-shaped cloud because the distances $R_{TF}=\frac{1}{\omega}\sqrt{2gn_0/m}$ are reduced by $\bar{\omega}/\omega_{y,z}$, whereas the initial mean depletion velocities \textit{in situ} $v\sim \sqrt{2gn_0/m}$ are isotropic.
	Indeed, spherical clouds (SE) exhibit a much weaker effect than the elongated clouds (CE) owing to the longer escape time.
	This also presents as an increase in $C_\textrm{sim}$ {and $\mathcal{A}_{\rm sim}$ for} collection regions {with a higher $\phi_c$}. 
    These effects could lead to the persistence of populated high-$k$ modes in the far-field, as were detected in the experiment.
	On the other hand, the thermal population decays super-exponentially with $k$, and hence does not account for the atoms we observe beyond $k\gtrsim 6~\micron^{-1}$, even though it is subject to the same mean-field forcing as the depletion \cite{Ozeri02}.
	We can show this with a simple calculation, noting that the maximum energy that can be imparted is $\mu=g n_0$, where $n_0$ is the initial density in the centre of the cloud.
	For an atom with momentum $k=6~\micron^{-1}$, at the edge of the thermal region in the densest cloud we consider (44 $\micron^{-3}$), the additional energy $\mu$ imparts at most a momentum shift of order 0.7 $\micron^{-1}$, which is insufficient to account for the population detected as far out as $k=10~\micron^{-1}$.
	
	Although it is not possible to accurately determine the  exponent $\alpha$ and coefficient $\mathcal{C}$ of the power-law tails (using either the present method or standard curve-fitting), one conclusion remains robust: There are {almost} about ten times as many detections in the depletion region as one would expect based on the Tan theory {\emph{in situ}}. 
	The theory of the contact does not explain this excess in terms of the short-lived mixed-species condensate (where the $m_J=0,1$, and -1 clouds overlap after the Landau-Zener sweep). 
	The expression for the contact in a mixed-species bosonic gas \cite{Werner12_boson} can be combined with the energy of a condensed mixture \cite{PethickSmith} and the known values of the s-wave scattering lengths \cite{Vassen16} to show that the contact is maximized in a pure $m_J=1$ condensate.
	Any mixture of spin states is thus predicted to have a lower contact (and thus less-populated tails) than the initial condensate, in contrast to our observations.
	Futher, despite the ultracold clouds being realized at finite temperatures, the thermal population of quasiparticles does not account for the observed counts. 
	The thermal quasiparticles in the Bogoliubov picture simply map onto the thermal population of constituent particles of the same momentum\footnote{See, for example, \cite{PethickSmith} Chap. 8.3. or \cite{Vogels02}}. 
	{Therefore the phonon/particle changeover is not responsible for the inflections seen at high $k$ in Fig.~\ref{fig:contact_determination_issues}(a).} 
	Hence, given the doubly-exponential decay of thermally populated states, the large-$k$ tails are unambiguously \emph{not} thermal effects.

	The interpretation just described is supported by another observation within the simulations:
	During the expansion we observe a decrease in the total number of depleted particles (reabsorption) and a simultaneous increase of the large-k population (forcing). 
	This {also corroborates the above interpretation} 
	of the experimental findings, and results in tail amplitudes which are consistent with the quantum depletion multiplied by a constant factor. 
	
	In summary, we find that the number of atoms in the large-$k$ tails is predicted by the product $N_0n_0\propto(N_{0}^7\bar{\omega}^6)^{1/5}$, in line with Tan's theory of the contact (Eqns. \ref{eqn:pred_scaling},\ref{eqn:pred_num}). 
	However, the sensitivity of this relationship is significantly different than expected by a factor of order 8(3), which is not accounted for by any known systematic effects.
	As mentioned above and detailed in the supplementary materials, there are fundamental challenges in the analysis of heavy-tailed distributions which preclude any decisive evidence regarding the functional form of the momentum tails.
	We thus surmise that the tails are indeed a signature of quantum depletion (per the first point), albeit subject to some physical effect during the expansion or some nonequilibrium enhancement in the trapped state.
	We do note that the prior work \cite{Chang16} reported results which fall within the uncertainty range of our analysis.
	Indeed, Fig. 4 in \cite{Chang16} could be said to display the relevant scaling with respect to $N_0n_0$, but the reported values of the apparent contact should be considered alongside the caveats discussed in Section \ref{sec:pow_issues}.

\subsection{Further testing}

	The utility and ubiquity of far-field imaging (including in MCP-DLD setups using helium) stems from the relative simplicity of implementation and interpretation.
	A systematic deviation from the expected \emph{in-situ} amplitude of the high-$k$ tails could mean that such techniques are not the appropriate tool to study the details of the quantum-depleted momentum distribution.
	It is plausible that meaningful inferences about the in-trap physics could follow from  a better understanding of the mechanism underlying the anomaly.
	Toward this, it would be informative to determine whether the outstanding discrepancy originates in the trapped condensate or is due to some unknown non-equilibrium effect during expansion.
	This question invites complementary studies of the \emph{in situ} depletion in \mhe~BECs. 
	Such an investigation requires an \emph{in situ} probe of the contact, such as RF spectroscopy or Bragg spectroscopy.
	The latter may be the most fruitful of the two because of the difficulty of interpreting the results from the former, sketched here.
	
	The basic principle of RF contact spectroscopy is to apply a monochromatic RF probe which is detuned from the resonance between two spin states, coupling atoms in the initial spin state to an untrapped channel. 
	One then performs a differential measurement of the atom number and expects the signal strength to scale as $\omega_\textrm{RF}^{3/2}$ with the detuning from the RF resonance.
	The loss rate is also proportional to the difference of reciprocal scattering lengths $\Gamma\propto(1/a_\textrm{i,i}-1/a_\textrm{i,f})$ between pairs of atoms in initial-initial ($a_{i,i}$) and initial-final ($a_{i,f}$) spin states \cite{Braaten10,Wild12}. 
	For He$^*$ (spin 1) the scattering lengths $a_{1,1}$ and $a_{1,0}$ are identical \cite{Leo01}, rendering the preferred $m_J=1-m_J=0$ transition unusable. 
	On the other hand, $a_{1,-1} = 3/7 a_{1,1}$ \cite{Vassen16}, and the singlet transition can be driven without populating the $m_J=0$ state. 
	In principle this could produce a detectable flux of atoms to perform sensitive in-trap contact measurements, however, collisions in the $^1\Sigma_{g}^{+}$ channel have large Penning ionization rates which lead to significant trap losses \cite{Leo01}. 
	The ionization products would be detectable by in-vacuum channel electron multipliers but require theoretical work to disentangle from the spectroscopic signal. 
	Further, while other atomic species offer Feshbach resonances by which to tune the inter-species scattering length (and hence signal or ionization rate), \mhe~has no such feature. 
	While such a measurement is not \emph{prima facie} impossible, Bragg spectroscopy may yield more readily interpretable results.

\subsection{Conclusion}
	Taken together with the simulations, our findings show that the survival of the quantum depletion into the far-field is plausible, but not as a straightforward mapping into the far-field density.
	In a non-interacting ballistic expansion, the far-field density distribution would be a direct realization of the in-trap momentum distribution of the cloud. 
	This correspondence is known not to be completely faithful because the dispersal of the mean-field energy into kinetic energy, known as the release energy, imparts some acceleration to the atoms during the early stages of the expansion \cite{Ozeri02}. 
	This effect is responsible for the famous inversion of the cloud aspect ratio upon release from harmonic traps. 

	In conclusion, this work expands the growing suite of far-field investigations of quantum depletion \cite{Cayla20,Chang16}.
	The inherent challenges of analysing heavy-tailed distributions make definitive comparison of the decay exponent impossible.
	However, these findings present statistically robust evidence that the quantum depletion can, remarkably, survive the expansion and dilution of its original condensate. 
	Our simulations clarify how the depletion can be visible in the far-field momentum distribution here and in earlier experiments, and that the hydrodynamic approximation does not capture sufficient short-wavelength information to make detailed predictions about the high-momentum behaviour. 
	This leads to a  partial explanation for the deviation of the far-field distribution from both the in-situ and the hydrodynamic pictures. The major factors are the momentum-dependent reabsorption of Bogoliubov excitations and the dispersal of the {interaction energy} into kinetic energy. 
	Together, these result in a growth of the $k^{-4}$ tails of the momentum distribution during freefall. 
	However, a mystery remains: Why is there an excess of particles in the depletion region which is so much greater than accounted for by this picture? 
	This issue should be resolved if far-field observations are to be interpreted in terms of the in-trap physics of interest.

%% file: latex/24_conclusion.tex

\chapter{Conclusion}
\markboth{CONCLUSION}{}
\label{chap:conclusion}
\begin{flushright}
\singlespacing
\emph{``Good judgment depends mostly on experience, and\\ experience usually comes from poor judgment"} \\- Anonymous\footnote{The origin of this frequently misattributed aphorism is unknown \url{https://quoteinvestigator.com/2017/02/23/judgment/}}
\end{flushright}
\onehalfspacing
\vspace{1cm}

\section*{Summary of findings}
	{The} central element of this thesis was helium-4, cast as the workhorse for the experiments for the twin reasons of structural simplicity and unique affordance of three-dimensional single-particle momentum measurements.
	The dominant theme of this dissertation,  metrology, concerns the precision measurement of basic physical quantities to empirically ground present and future theories of physics and was the focus of chapters \ref{chap:transitions} and \ref{chap:tuneout}. 
	These works were motivated by the suitability of the (deceptively) simple helium atom to high-precision calculations.
	In the counterpoint, many-body physics, we turned attention outside the helium atom itself, towards the collective behaviour of interacting particles.
	This theme was elaborated in chapter \ref{chap:QD} (and appendix\ref{chap:lattice}) with particular emphasis on the scientific contributions made possible by `momentum microscopy' of quantum gases.

	Chapter \ref{chap:transitions} presented the first measurement of the spin-forbidden $2\triplet P_2\rightarrow 5\singlet D_2$ line in helium along with new measurements of transitions from the $2\triplet P_2$ state to the $5\triplet S_1$ and $5\triplet D_J~(J\in\{1,2,3\})$ states. 
	These measurements were made by disturbing an optical cooling process during the BEC production sequence, wherein the upper state of the 1083 nm cooling transition served as the lower state for the target transition. 
	When atoms were excited by the cooling light, they were able to absorb resonant probe light and thus degrade the phase-space density of the atomic sample by either absorbing the energy (heating the sample) or decaying to an untrapped state (reducing the number density).
	Subsequently, when the evaporative cooling process amplified the phase space density, the perturbation in the initial condition resulted in a decrease in the total number of atoms cooled to degeneracy.
	As far as I know, this is the first use of such a technique for atomic spectroscopy.
	Other transition measurements are usually made by probing the trapped ground state \cite{Thomas20,Rengelink18,Notermans14}, rather than excited states.
	The resulting determinations of transition frequencies improve upon previous measurements, where they exist \cite{Martin60}, by at least an order of magnitude in precision and are consistent with state-of-the art theoretical predictions.
	However, the precision of this measurement was limited to the level of a few MHz by the wavemeter we employed as the frequency reference.
	In comparison, the uncertainty of theoretical predictions for these transitions is dominated by the 700 kHz uncertainty in the lower state energy.
	The use of a frequency comb locked to a stable frequency reference would be the major instrumental requirement toward challenging the precision of current atomic structure theory calculations - additional considerations are discussed at the end of the chapter.

	The content of chapter \ref{chap:tuneout} was a variation on the theme of frequency measurement, focusing instead on the $\TO$ tune-out frequency.
	This measurement was underpinned by a novel technique for trap freqency measurement that yields a single-shot determination of the trapping frequencies with accuracy at the level of tens of ppm.
	This technique was employed to detect the shift in the net trapping frequency of a hybrid magnetic- and optical-dipole trap as a function of the probe laser frequency, and improved on the precision of the ANU group's previous measurement \cite{Henson15} by a factor of 15.
	A further advance over the prior work was made by characterizing the polarization-dependence of the tune-out, eliminating any dependence on parameters which were not precisely measurable and permitting an unambiguous comparison with theory.
	The experimental determination of $725,736,700(40_\mathrm{stat})(260_\mathrm{sys})$ MHz differs from the theoretical prediction of 725,736,053(9) MHz by 2.5$\sigma$, where the experimental accuracy is limited by the imprecise knowledge of the exact polarization of the probe laser light at the atoms.
	This measurement is the first tune-out measurement to resolve corrections for finite-wavelength retardation effects and yields the most precise constraint on the ratio of transition dipole matrix elements in any neutral atom to date.
	Furthermore, the highly accurate trap frequency measurements meant that our longest data collection run at a fixed polarization would be capable of detecting, with an SNR of 1, a differential Stark shift at the level of $10^{-35}$ J.

	In appendix \ref{chap:lattice} I recount the major milestones I achieved towards the construction of a new BEC machine with the objective of trapping \mhe~atoms in a 3D optical lattice.
	Over the course of work on this project, I built the vacuum system, the optomechanical and optoelectronic systems used to control and distribute laser light for cooling, trapping and imaging applications, and the imaging system (including hardware assembly and attendant software suite), all of which are still in use in the laboratory.
	This lab has since gone on to achieve condensation of $^4$He in a record time of only 3.3 seconds, with very large condensed populations on the order of a million atoms \cite{Abbas21}.
	Work towards the operation of the optical lattice is ongoing.

	Finally, the topic of chapter \ref{chap:QD} was the physics of the condensed state itself.
	This work concerned the density measurements of the ultra-dilute, large-momentum halo around the condensate and outside the thermal fraction.
	In this region of momentum space, we found evidence suggestive of the persistence of the quantum depletion through the expansion into the far-field, seemingly  consistent with Ref. \cite{Chang16}.
	Moreover, we looked in some detail at the shortcomings of least-squares regression against density histograms when attempting parameter estimation of power-law distributions, and provided an alternative analysis which circumvented this problem.
	The regression technique yields the robust conclusion that the large-momentum tails persist into the far-field, in contravention of theoretical expectations \cite{Qu16,Xu06}.
	These findings were complemented by simulations of the condensate expansion \cite{Ross21}. 
	Together, the experiment and simulations support the interpretation that the quantum-depleted particles are accelerated by the non-uniform mean-field potential energy during the condensate expansion.
	While this effect has been observed for thermal atoms \cite{Ozeri02}, we showed that the observed tail population could not have originated in the thermal part and indeed showed that the tail population exhibited the very same scaling with condensate size as one would expect of the depletion, based on Tan's theory of the contact.
	However, the population observed in the experiment was larger still than in the simulations, which could not be attributed to any identified systematic effects.

	Each chapter contains a short discussion of the near-term prospects for extending the work discussed therein.
	With all that behind us, let us take a moment to look farther forward by the light of the luminaries on whose shoulders we stand.


\section*{Closing remarks}
	
	Creating and studying quantum matter with the techniques of laser cooling and trapping has endowed the experimentalist with an ever-expanding arsenal of tools.
	Within this field the precise, quantitative, and functional characterization of individual and interacting atoms has led to the frontiers of space \cite{Aveline20} and time \cite{LeTargat13}.
	Increasingly specialized machines are driving the production of atom traps mounted on chip-scale systems \cite{AtomChips} which, packaged alongside supporting vacuum and optical systems in small form-factor housings produced with additive manufacturing methods \cite{Madkhaly21}, herald the approach of integrated `atomtronic' devices \cite{Amico21}.
	Future evolutions of ultracold-atom techniques may present themselves in such exotic manifestations as:
	
	\begin{itemize}
	 \item A sensing medium for gravitational wave astronomy \cite{Kolkowitz16}, dark matter detection \cite{Derevianko14}, spacetime geodesy and absolute gravimetry \cite{Bidel20}, multiparametric gradiometry \cite{Hardman16}, or ascertaining the non-classicality of gravity \cite{Howl21};
	 \item A gain medium for a gamma-ray laser with condensates of positronium  \cite{Wang14} or antihydrogen atoms produced by laser cooling \cite{Baker21};
	 \item A storage medium \cite{Heller20} for large entangled, topologically ordered \cite{Zhang19}, fault-tolerant quantum information \cite{Briegel99}, integrated within networks of diverse quantum substrates \cite{Maring17};
	 \item A computational medium in which to study the cosmology of the early universe \cite{Fischer04}, false-vacuum collapse \cite{Ng21}, the AdS/CFT correspondence \cite{Wei21}, and chromodynamics \cite{Zohar15}, or;
	 \item A working medium \cite{Niedenzu19} for powering and regulating engineered processes using quantum resources \cite{Chitambar19}.
	\end{itemize}

	\begin{figure*}
	\includegraphics[width=\textwidth]{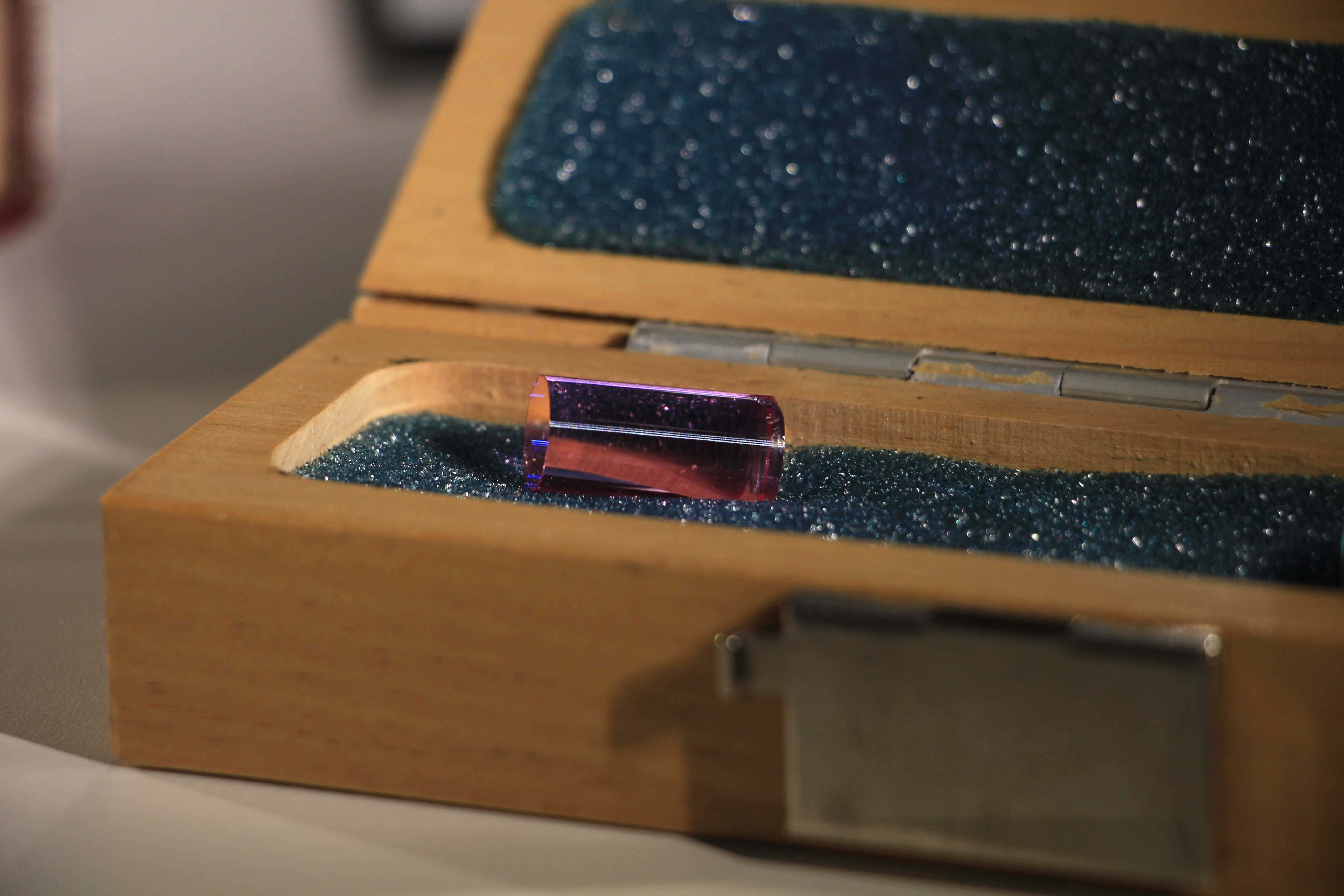}
	\caption*{The synthetic ruby crystal used by T. Maiman and colleagues in the first human-made laser. Taken by the author on a visit to the Max Planck Institute of Quantum Optics, Garching, Germany.}
	\end{figure*}

	\noindent  Just as James Watt could not comprehend how the development of resource extraction, mass production, and intercontinental transit would be spurred on by his steam engine, we too are surely blind to the world we {may} be building.
	It took two centuries for the largest steam turbine capacity to grow from the kilowatt to the gigawatt scale, but the intervening investment in industrial and information infrastructure tends to compress societal learning curves \cite{SmilEnergy}.
	Indeed, Claude Cohen-Tannoudji himself reflected that few expected Bose-Einstein condensation to follow so rapidly after the pioneering works of Chu, Phillips, and colleagues \cite{CTNote}.
	One readily doubts Theodore Maiman glimpsed the extent of modern scientific finesse in the first glint of laser light from his ruby crystal in 1960.
	Indeed, while `quantum technology' is a young term, its use has exploded since it was coined in the 1960s.
	One of the earliest uses of the term in print, in a US National Research Council report \cite{Isotopes69} on the needs and uses of separated stable isotopes in 1968, reads as a cautionary tale for the technological optimist:

	\bigskip
	\hfill
	\begin{minipage}{0.9\textwidth}
	\emph{``While it is dangerous to predict the future, the Committee strongly believes that our society is entering an era of nuclear and quantum technology, that will be different from that of the present which is based largely on nonquantum, non-nuclear science. The uses of separated isotopes in this new technology - beyond those envisioned and discussed above - will probably be multifold. Among the diverse uses that may beome important in the future are structural components of nuclear reactors and devices, industrial product labelling, measuring and controlling air pollution, and industrial process and product control."}
	\end{minipage}
	\bigskip

	\noindent The early decades of the 21st century {may} see the proliferation of cold-atom systems into wider application, following the path of the laser from R\&D to ubiquity.
	The children of 2121 {may} live in a world shaped as much by engineered quantum systems as today's society is by the repercussions of the first quantum revolution.
	They {may} look upon today's cold-atom machines as `primitive' in the same way that we look upon vaccum tubes and typewriters as quaint vestiges of a bygone era.
	We \emph{may} cross the precipice of the 21st century \cite{OrdPrecipice} and realize some of the innumerable possibilities bejeweling the vastness of Hilbert space.
	If we do, may we enjoy the burgeoning fruits of science with discerning wisdom and boundless generosity.
	After all - unlike the atoms in these experiments, we do not live in a vacuum.

\vfill

\section*{\begin{figure*}[!h]
\centering
\includegraphics[scale=2]{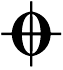}
\end{figure*}}


	{Albert} Michelson may have lived to rue his proclamation at the inauguration of the Ryserson Physics Laboratory that `the great principles had already been discovered, and that physics would henceforth be limited to finding truths in the sixth decimal place.'
	Nobody could repeat this mistake today as we can readily identify subjects awaiting clear insight, such as the nature of dark matter and dark energy, or the synthesis of quantum field theory and general relativity.
	At first glance, the dogged pursuit of the next digit of precision could appear a peculiar lens with which to bring such quandaries into focus, and one could demur the endeavour as searching for a revolution in the margins of Nature.
	But surely such an enterprise would find vindication in a discovery that the fundamental constants were contingent, in deciding whether or not gravity is fundamentally quantum, or in establishing control of dark matter?
	Albert Einstein, on whose shoulders this work squarely rests, noted that a well-formulated question contains its own answer. 
	Although we can't be sure from the side of the revolution, perhaps we are beginning to ask the right questions.
	
\vfill	

\begin{center}
\vfill
\singlespacing
\emph{``The key to discoveries is to look at those\\
 places where there is still a paradox.\\
It’s like the tip of an iceberg.\\
If there is a point of \\
dissatisfaction, \\
take a closer \\
look at it.'' \\-\\
\emph{Edral Arikan}}
 \end{center}




%% file: latex/A1_lattice_build.tex
\chapter{Towards an optical lattice trap for helium}
\markboth{\thechapter. TOWARDS A HELIUM LATTICE}{}
\label{chap:lattice}

	\begin{flushright}
	\singlespacing
	\emph{``It is not necessary to succeed in order to persevere.\\
	As long as there is a margin of hope, however narrow, \\
	we have no choice but to base all our actions on that margin"}\\
	- Leo Szilard\footnote{\emph{LIFE magazine} volume 51, issue 9, 1961 }\\
	\end{flushright}
	\onehalfspacing

	\blankfootnote{\noindent The contents of this chapter relate to the work published in \textbf{Rapid generation of metastable helium Bose-Einstein condensates} by A. H. Abbas, X. Meng, R. S. Patil, J. A. Ross, A. G. Truscott, S. S. Hodgman, \href{https://journals.aps.org/pra/abstract/10.1103/PhysRevA.103.053317}{Physical Review A} \textbf{103} (2021)}

This appendix presents some of the contributions I made to the re-establishment and upgrade of a decommissioned cold-atom system. The main components were a vacuum system extension, magneto-optical trap, an optical dipole trap loaded from an evaporatively-cooled magnetic trap, and an absorption-imaging system. Roughly halfway through my PhD, I changed course to focus on the laser spectroscopic measurments that were presented in the main body of this dissertation.

\section{The many-body problem}

	In his 1972 essay \emph{More Is Different} \cite{Anderson72},  Philip W. Anderson argued that the constructionist hypothesis, that one can reconstruct the universe starting from known physical laws,  `breaks down when confronted with the twin difficulties of scale and complexity.' 
	In this he argues that composing a system out of many well-understood pieces with well-understood interactions will eventually give rise to  phenomena in need of `new laws, concepts, and generalizations'.
	Familiar examples in physics include quasiparticles and hydrodynamics, and even atoms, which compress enormous amounts of microscopic information into an effective description based on observed emergent regularities.
	Anderson acknowledges the utility of reductionism by noting that emergent features are always understandable in terms of their microscopic composition.

	A major challenge in science, then, is traversing this bridge between `macro' and `micro'.
	Simulations (on paper or \emph{in silico}) contribute to the `upward-pointing' inquiry which aims to explain emergent features in terms of elementary concepts.
	The rapid rise of computing power in the latter half of the twentieth century provided a boon to scientists by permitting direct interrogation of previously intractable theoretical models.
	Simulating systems can now be done en masse, in parallel, and with much less human effort than iterations of manufacturing, transporting, and characterizing physical samples.
	The design-build-test-learn cycle thus contracts, and humankind grows more proficient at the design of compounds or materials with desirable properties such as their light-harvesting efficiency or efficacy at neutralizing pathogens.
	In the quantum-mechanical context this project is particularly challenging because analytical models at the microscale may be conceivable but are often analytically intractable.

	Further, the amount of information required to describe a quantum state, and the computational effort needed to forecast its evolution, quickly become so large as to exhaust even the largest supercomputers.
	It was in 1982 that Richard P.	Feynman laid out a proposition for the now-burgeoning field of \emph{quantum simulation} \cite{Feynman82}:

	\bigskip

	\begin{flushright}
	\singlespacing
	\emph{The first question is, What kind of computer are we going to use to simulate physics? Computer theory has been developed to a point where it realizes that it doesn't make any difference; when you get to a universal computer, it doesn't matter how it's manufactured, how it's actually made. Therefore my question is, Can physics be simulated by a universal computer? ... the physical world is quantum mechanical, and therefore the proper problem is the simulation of quantum physics ... But the full description of quantum mechanics for a large system with R particles  has too many variables, it cannot be simulated with a normal computer with a number of elements proportional to R ... And therefore, the problem is, how can we simulate the quantum mechanics? There are two ways that we can go about it. We can give up on our rule about what the computer was, we can say: Let the computer itself be built of quantum mechanical elements which obey quantum mechanical laws.  ... Can you do it with a new kind of computer--a quantum computer? (I'll come back to the other branch in a moment.) Now it turns out, as far as I can tell, that you can simulate this with a quantum system, with quantum computer elements. It's not a Turing machine, but a machine of a different kind.}
	\end{flushright}
	\bigskip

	The hope, then, is that one can engineer \emph{quantum} systems that behave like a certain other, potentially hypothetical, system of interest and obtain useful data from it more efficiently than via conventional computing, when the cost is measured in some combination of time, energy, or some other convenient currency.
	One flourishing branch of this field of inquiry is the use of {optical lattices} to synthesize artificial crystals, scaled up some 10,000 times and many orders of magnitude colder than conventional solids.
	The ultracold atoms in such systems then play the role of electrons bound in the ion lattice of solid systems.
	Ultracold metastable helium offers an extension of the questions one can ask of optical lattice simulators by giving access to single-particle momentum information in three dimensions.
	This affordance opens lines of inquiry not accessible via low-resolution momentum measurements or particle position measurements, as discussed in the sections below.

	In this chapter I recount efforts towards constructing an optical lattice trap for helium.
	I will describe the motivation for the project and the basic model of interacting bosons which can tunnel between nearest-neighbour sites on a lattice, the Bose-Hubbard model.
	Then I will detail the stages of construction that were achieved, and give a short account of the progress that has been accomplished by other students in the meantime, resulting in the publication \cite{Abbas21}.

\subsection{There's always a bigger Hilbert space}

	Given two quantum systems $A$ and $B$ with $d_A$ and $d_B$ degrees of freedom each, the natural way to represent the composite of the two systems is the tensor product $\mathcal{H}_C$ of their respective Hilbert spaces $\mathcal{H}_A$ and $\mathcal{H}_B$ \cite{Carcassi21}, whose basis can be written as
	\begin{equation}
	\{\ket{e_C}\} = \{\ket{e_A e_B}=\ket{e_A}\otimes\ket{e_C}\}
	\end{equation}
	for all pairs $(\ket{e_A},\ket{e_B})$ of the basis vectors of $\mathcal{H}_A$ and $\mathcal{H}_B$, respectively.
	Thus the dimension of the composite system $d_C = |\{\ket{e_C}\}| = d_A d_B.$ A composite system of $N$ two-level systems provides a simple illustration of the resource challenge in that the total dimension of the Hilbert space is $2^N$.
	A pure state can be represented as a vector with $2^N-1$ coefficients (the asymptotically trivial reduction by 1 owing to the normalization of the wavefunction), but a general mixed state must be encoded in a density matrix while tracking $\mathcal{O}(2^{2N})$  coefficients.
	The Hamiltonian, if similarly stored as a full matrix, requires the full $\prod_i d_i\times\prod_i d_i$ matrix to be stored in memory (albeit typically with real-valued entries).
	If the complex coefficients are to be represented by pairs of floating-point numbers with single precision (for 8 bytes = 64 bits for the real and imaginary parts of each coefficient), the memory of a consumer-grade laptop is exhausted by the state vector of 16 two-level systems; the density matrix and Hamiltonian exceed available memory for even smaller systems.

	Fortunately, it is not always necessary to store the full Hamiltonian in memory.
	A sparse representation often suffices as most systems of interest exhibit local coupling only, and so most coupling constants are zero.
	However, matrix diagonalization algorithms exploit tradeoffs between memory use and compute depth.
	For instance, fully general matrix diagonalization executes in $\mathcal{O}(D^{3})$ time, where $D$ is the matrix dimension.
	Some specialized algorithms can perform faster (as low as $\mathcal{O}(D^{2})$) but this is still unfavourable when dealing with exponentially-growing matrices.
	Thus storing sparse matrices may save on memory but slows down the diagonalization.

	In the case of the physical state, physical insight can lead to reformulations of the problem in (potentially many) fewer dimensions, such as by identification of a few relevant degrees of freedom, symmetries in the system wavefunction, or by finding conserved quantities.
	In this respect we are sometimes limited by our own ingenuity, but in general we are limited by the need to include higher-order correlations when many-body degrees of freedom cannot be integrated out in a mean-field or Hartree-type approach.
	In these situations, the memory requirements for the state vector can  be reduced for purposes of simulations by using inventive representations.
	A full review of numerical techniques is far beyond the scope of this thesis, but we will note that exact diagonalization \cite{Zhang10} is complemented by an expanding suite of compression techniques including matrix product states \cite{Schollwock11} employing the iterative density-matrix renormalization group  method \cite{Dechiara08}, the multiscale entanglement renormalization ansatz \cite{Evenbly15}, quantum Monte Carlo simulations \cite{BeccaQMC}, and emerging techniques using artificial neural network representations of many-body states \cite{Carleo17}.

	Quantum-native platforms offer a route by which to circumvent some of these issues; clearly Nature has no trouble evolving the state of some $\mathcal{O}(10^{23})$ nuclear spins in solid materials (let alone all their electronic states).
	If one encodes the physics of interest \emph{directly} into another quantum system, then the space (memory) requirements of simulating quantum systems is obviated.
	This does present another challenge, which is the means of initialising the state, implementing time evolution, and obtaining accurate readout.
	A broad pallette of methods exists for quantum simulation with digital quantum computers, including Trotterization, linear combinations of unitary operators, qubitization (these and other methods are compared in \cite{Low19}).

	Important milestones have been reached along the way in trapped ions \cite{Hempel18}, quantum dots \cite{Hensgens17}, microcavity polaritons \cite{Boulier20}, and superconducting circuits \cite{Wilkinson20}, 
	while cavity QED \cite{Zhu18}, and photonic circuits \cite{Arrazola21} are also promising platforms for quantum simulation.
	A comprehensive review of the conceptual background and platforms and problems of interest (as of 2014) can be found in \cite{Georgescu14}.
	As of the time of writing this dissertation, the largest digital quantum computers have around 50 qubits, and are not able to implement fault tolerant quantum computation at a physically meaningful scale.
	While early claims to quantum advantage have made it to press \cite{Arute19}, so far these results relate to algorithms which do not correspond to any class of familiar physical systems, or even any otherwise-useful calculation.
	It is worth noting that this observation may well be out of date in half a decade given the rapid progress in quantum engineering.

	An important counterpart to universal digital quantum simulation is \emph{analogue} simulation.
	The closest analogy in conventional `classical' computing is the application-specific integrated circuit (ASIC).
	These present-day systems that are optimized to perform specific tasks much more efficiently than a general microprocessor built within the Von Neumann architectural paradigm, but might not be Turing-complete.
	Similarly, analogue simulators are direct and dedicated simulations of a specified family of Hamiltonians for exploration of that system in particular.
	The focus of this chapter is one particular platform for quantum simulation: Ultracold neutral atoms trapped in an optical lattice. 
	While schemes have been proposed for universal quantum computing with neutral atoms in optical lattices \cite{Brennen99,Henriet20}, this potential is still some way off \cite{Markov00}.
	Meanwhile, analogue simulation in optical lattices is flourishing.
	For the second time, we will note that `any attempt to produce a comprehensive review is out of date by the time it is published' and, in the below, survey some pivotal results in order to clarify the specific contributions that metastable helium atoms could make to the suite of capabilities in optical lattices.

\section{Quantum simulation with optical lattices}

	The groundbreaking experiment in quantum simulation with optical lattices was the achievement of the Mott insulating state\footnote{A hallmark of the Mott insulator is the suppression of number fluctuations at each lattice site (i.e.
	\emph{number squeezing}), which was first observed a year prior in \cite{Orzel01}.
	However, the later result was the first to completely eliminate site occupancy fluctuations in 3D - see \cite{Morsch06} for discussion.} in a lattice filled with ultracold bosonic atoms \cite{Greiner01}, realizing the proposition by Dieter Jaksch \emph{et al.} from three years prior \cite{Jaksch98}.
	The subsequent few years saw an explosion of foundational research and technical development, summarized in the reviews \cite{Morsch06,Bloch08}. 
	The following decade heralded many major advances in quantum simulation \cite{Bloch12,Gross17}.
	Theoretical foundations of the myriad models realized in lattices, and their context in a more general condensed-matter setting, are discussed in Refs. \cite{LewensteinLattices, Lewenstein07}.

	The central principle of an optical lattice is the careful deployment of the optical dipole force \cite{Grimm00} to synthesize a potential energy landscape with a persistent periodic structure.
	In simpler lattice configurations (as considered in this chapter), a stable laser is reflected back on itself, thus creating a standing wave so atoms in this optical field are subject to a periodic potential along the axis of propagation of the laser.
	Lasers which are red-detuned relative to the nearest transition and have a Gaussian profile provide an overall, approximately harmonic, confinement, but this is typically negligible over the scale of the occupied lattice.
	
	Additional laser beams can break the radial symmetry and induce two- or three-dimensional periodic structures.
	The simplest of these is a square lattice, but different optical arrangements can be used to generate more complicated geometries such as the hexagonal honeycomb \cite{Jotzu14},  triangular \cite{Becker10}, and Kagome lattices \cite{Jo12}, and the recently-emerging quasicrystal lattices \cite{Viebahn19}.
	The translational symmetry can be intentionally broken by superposing the speckle pattern of another laser \cite{Pasienski10} or by collinear propagation of another laser whose wavelength is close to an irrational multiple of the main lattice \cite{Rispoli19}.
	More generally, arbitrary potentials can be generated in up to two dimensions using a digital mirror device (DMD) in the Fourier plane to synthesize user-defined intensity patterns in the plane of the lattice \cite{Gross17}.

	An ever-expanding suite of additional techniques allow the experimentalist to furnish their lattices with sophisticated control over the dynamics.
	One can realize paradigmatic condensed-matter models including the Bose \cite{Greiner01,Miranda15,Rispoli19,Sherson10,Preiss15a} and Fermi \cite{Bakr09,Cheuk15,Haller15,Chiu18} variants of Hubbard models, their disordered cousin the Aubry-Andre \cite{Rispoli19} model, and Ising models \cite{Simon11}.
	Within these models, characteristic phenomena like higher-order tunnelling \cite{Folling07} and particle-hole pair formation \cite{Endres11} have been observed in optical lattices.
	The controllability of these machines has permitted extensive exploration of the phase diagrams of these models.
	Through varying parameters such as the ratio of the tunnelling and interaction energies (the latter often achieved via Feshbach resonances \cite{Chin10}), optical lattice experiments can explore the phase diagram of a range of systems \cite{Greiner01,Eckardt05,Jordens08,Jo09,Haller10,Simon11,Baumann10,Leonard17,Landig16,SachdevQPT,Endres12,Anquez16,Clark16}.
	Whereas a classical phase transition is associated with a non-analytic free energy at some boundary in the space of state variables, a quantum phase transition occurs at non-analytic points of the ground state energy of a combination of Hamiltonians $H = H_0 + g H_1$, where $g$ is some real coefficient \cite{SachdevQPT}.
	If $H_0$ and $H_1$ commute, then the eigenstates are common but varying the coupling leads to a level-crossing at some critical value $g_c$, either side of which the ground state has markedly different character (more generally the system will exhibit an avoided crossing, but the definition holds in either case \cite{SachdevQPT}).
	
	Quantum phase transitions have been central objects of study with lattice simulators \cite{Greiner01,Eckardt05,Jordens08,Baumann10,Endres12,Haller10,Leonard17,Landig16,SachdevQPT}.
	Over the preceding decade there has been much activity investigating \emph{topological} phase transitions, which are associated with a nonlocal order parameter capturing the long-range entanglement structure inherent in different topological phases \cite{Goldman16,Nakagawa14}.
	Condensed matter models featuring this kind of transition in optical lattices include Hofstadter models\cite{Aidelsburger13,Tai17,Miyake13} and the Haldane model \cite{Jotzu14}.
	These models are synthesized by applying additional optical fields to create lattices with complex tunneling terms, imbuing particle motion with a path-dependent Aharonov-Bohm phase \cite{Aidelsburger11,Aidelsburger13,Miyake13}.
	This permits creation of systems wherein neutral atoms can be made to follow the equations of motion of charged particles under the influence of electromagnetic fields \cite{Aidelsburger13,Tai17,Endres11,Rispoli19,Jo09,Simon11,Miyake13,Folling07,Jotzu14}.
	The transition between regimes of time-evolution characterized by distinct scaling laws, called \emph{dynamical} phase transitions between, can also be produced in lattices \cite{Clark16}, including dynamical topological transitions \cite{Nakagawa14}.
	On the other end of the energy scale, efforts are ongoing to extend the reach of the cold-atom lab to simulations of high-energy phenomena like matter coupled to non-abelian gauge fields \cite{Zohar16,Schweizer19,Tagliacozzo13} and the Schwinger effect of particle-antiparticle pair production \cite{Pineiro19}.

\subsubsection{Frontier experiments with optical lattices}
	Another fascinating branch of inquiry led by optical lattice labs lies at the intersection of quantum information theory and the foundational questions regarding decoherence, the emergence of entropy in thermodynamics from unitary evolution, and the resulting familiar phenomena of classical states \cite{Osborne02,Osterloh02,Isakov11,Jiang12,Dalessio16,Goold16,Srednicki94,Amico08,Eisert15}.
	A modern picture entails that so-called eigenstate thermalization gives rise to classical thermodynamics even in unitary evolution \cite{Srednicki94,Dalessio16,Goold16}.
	The connection with quantum information is principally that sub-components of the system become entangled with one another as the system evolves, `scrambling' information about the inital configuration in non-local correlations.
	More specifically, although the global state remains pure (with nearly zero entropy) during isolated evolution, information exchange between subsystems entangles them.
	When performing local measurements, then, one discovers that the entropy of subsystems increases despite the whole system undergoing unitary evolution from an initially pure state.
	Sophisticated interferometric techniques have been implemented to measure the degree of entanglement between sites \cite{Brydges19,Daley12,Mouraalves04,Palmer05}.
	They then exhibit nonzero quantum \emph{mutual information} which is the difference between the entropy of the subsystems and the entropy of the entire system.
	The sub-additivity of the von Neumann entropy is a hallmark of the regime of quantum thermodynamics, whereas classical entropy is strictly additive.
	
	Optical lattice experiments are beginning to resolve relevant phenomena such as the growth of entanglement between subsystems coincident with the onset of thermalization \cite{Clos16,Kaufman16}.
	This work is enabled by several schemes for the measurement of entanglement in cold-atom systems \cite{Chiu18,Brydges19,Daley12,Mouraalves04}.
	The application of disordered and quasi-random lattices has permitted direct inquiry into wavefunction localization \cite{Anderson58,Dalessio16,Goold16,Srednicki94,Clos16,Kaufman16,Nandkishore15}, wherein the presence of disorder prevents the system from thermalizing and instead retains local information about its initial state over long timescales.
	As local order parameters are not obviously applicable to quantum hall states or spin liquids \cite{Isakov11}, which feature topological phases that are not locally distinguishable, non-local order parameters based on entanglement measures can distinguish between these collective states \cite{Isakov11,Jiang12}.
	Long-range entanglement also manifests in quantum phase transitions \cite{Osborne02} and shows scaling behaviour analogous to correlation functions at classical phase transitions \cite{Osterloh02}.
	Diatomic molecules also feature in lattices, contributing to cold chemistry \cite{Balakrishnan16} through studies of basic mechanisms like dipolar spin exchange \cite{Yan13}, enhanced interactions via Feshbach resonance \cite{Yang19} or microwave-dressed resonant interactions \cite{Yan20}.
	In fact, some schemes have been proposed for universal computation with dipolar molecules in optical lattices \cite{Yelin06, Micheli06}.
	
\subsection{Quantum state microscopy}

	A key enabler of the studies described above are the techniques of quantum gas microscopy \cite{Bakr09,Cheuk15,Endres11,Haller15,Miranda15,Parsons15,Rispoli19,Sherson10,Miranda17,Preiss15a} which include state preparation and measurements of spin and on-site particle parity (and recently even population up to $N=4$ \cite{Preiss15a}), each with single-site resolution.
	Direct access to microscopic information renders density correlations readily measurable \cite{Endres11,Rispoli19}.
	Correlation functions are central in quantum field theory and quantify the probability distribution of joint measurements of many-particle operators $\mathcal{O}_i(x_i)$ at some coordinates $x_i$, generally taking the form
	\begin{equation}
		G^{(N)}(\textbf{x}) = \langle \prod_{i=1}^{N}\mathcal{O}_i(x_i)\rangle = \textrm{Tr}\left(\prod_i\mathcal{O}_i(x_i) \rho\right).
	\end{equation}
	Such correlation functions characterize the many-body nature of interacting systems through the Wick decomposition \cite{Wick50}.
	This theoretical result guarantees that when pairs of degrees of freedom do not interact, their correlation functions for all $N>2$ are expressible in sums of products of correlation functions with $N\leq2$.
	Thus observing a factorization (or not) of many-particle correlation functions of order $M$ into lower-order correlations serves to indicate the absence (or presence) of relevant interactions of more than $M$ degrees of freedom \cite{Schweigler17}.
	One implication is that experimental results can constrain the number of required terms in controlled approximation methods like cluster expansion or basis truncation techniques \cite{Hodgman17}.
	Another is that collecting correlation functions can effectively provide an operational solution to the many-body problem in that all relevant quantities of interest (higher-order correlation functions) can be inferred \cite{Schweigler17,Hodgman17}.

	Going a step further and predicting products of arbitrary observables would require determination of the many-body density matrix $\rho$.
	The process of determining $\rho$ is also called quantum state tomography because the density matrix contains non-local information but is only amenable to local projective measurements.
	Some platforms for quantum computation (such as superconducting qubits and photonic systems) are furnished with the ability to perform arbitrary unitary operations. 
	This capacity, coupled with the fast cycle time of such systems, makes it possible to perform exhaustive tomography of the density matrix for modestly-sized systems.
	For an atomic system containing tens of thousands to millions of atoms, using machines which operate over several seconds rather than microseconds, such a task is clearly not feasible - even leaving aside the computational resource demands.
	Therefore, correlation functions between the degrees of freedom of interest remain important tools of inquiry into the many-body dynamics of ultracold gases.
	
	An immediate caveat is that one is limited in the extent to which one can infer correlations between degrees of freedom other than those one has measured to obtain the lower-order correlations. 
	This presents an obstacle to obtaining certain interesting quantities from analogue simulators equipped with quantum gas microscopes.
	For instance, in particle physics a common exercise is computing the expected momentum correlations between the product particles after collision in accelerators, which is not readily achievable using the gauge-field simulators described above.
	On the other hand, condensed-matter physicists tend to study correlation functions for \emph{collective} degrees of freedom, a pertinent example being Bogoliubov modes (or phonons in general) which are not generally detectable by means of spatial correlations, and manifest in the correlated momenta of many particles.

	One aspect which has been hitherto underrepresented in the cold atom toolbox is \emph{momentum microscopy} \cite{Ott16}.
	As noted in \cite{Bergschneider18}, `many important aspects of quantum states are not accessible by position-space imaging alone: properties such as long-range coherence, currents and phase fluctuations are related to off-diagonal order, or coherences, in the many-body states.'
	Momentum correlations provide access to information about these many-body coherences through Glauber's foundational work in quantum optics \cite{Glauber63,Naraschewski99}.
	Optical lattice experiments have, to date, typically employed coarse measurements of momentum-space information, for instance by absorption imaging in the far-field.
	This provides estimates of the momentum density, which has been used to verify the Bogoliubov picture of phonon formation \cite{Vogels02}, to verify coherence of a Mott-insulating state through noise correlations \cite{Folling05}, and to obtain the second-order momentum correlation function for a 1D Bose gas \cite{Fang16}, but does not resolve single-particle information or higher-order correlations.
	
	\noindent As discussed in the introduction, metastable helium has previously been used to probe the coherence of degenerate quantum gases by measuring high-order momentum correlations (see the introduction to chapter \ref{chap:apparatus} and references therein).
	In particular, data from the BiQUIC machine at ANU was used to show that the three-particle correlations in ultracold scattering halos can be expressed in terms of two-particle correlations \cite{Hodgman17}.	
	Some promising methods for momentum microscopy with other elements have also been developed. 
	Most recently, fluorescence imaging of $^6$Li with a high numerical aperture objective has been used to measure the position and, separately, momentum of up to three atoms  \cite{Bergschneider18}. 
	While this method is currently limited to small numbers of atoms in one dimension, another advance \cite{Bucker09} can image the far-field momentum of some tens of thousands of atoms.
	The latter method uses a single-photon sensitive camera to detect atoms as they fall through a 2D sheet of resonant light, thus projecting out the momentum information in the vertical direction. 
	While this could be employed in a stroboscopic manner to obtain 3D information, this could require many experimental cycles, unlike \mhe~experiments with an MCP-DLD stack which intrinsically provide three-dimensional momentum information.
	Therefore while the new optical methods will surely bear worthy fruit, \mhe~experiments currently stand as the most effective way to perform three-dimensional momentum microscopy.

\subsection{The Bose-Hubbard model}

	One of the most accessible many-body systems to realize with ultracold bosons in an optical lattice is the Bose-Hubbard model.
	The Hubbard model was originally introduced to study the transition metals in an attempt to understand their magnetic properties \cite{SachdevQPT}.
	The Bose-Hubbard (or `boson Hubbard') model replaces the spin-1/2 fermions with spinless bosons, which might represent Cooper pairs of electrons in a superconductor tunnelling between superconducting regions - or bosonic atoms hopping between sites in an optical lattice.
	The proposal to realize this model in optical lattices was put forward by Dieter Jaksch \emph{et al.} in 1998 \cite{Jaksch98}.
	The Hamiltonian of this model is 
	\begin{equation}
		\hat{H}_B = -J\sum_{\langle i j\rangle}\left(\cre{a}_i\anh{a}_j+\cre{a}_j\anh{a}_i\right) 
		+	\sum_{i} \frac{U}{2}\hat{n}_{i}(\hat{n}_{i}-1) 
		- \mu\sum_{i} \hat{n}_i
		\label{eqn:BH_ham}
	\end{equation}
	where $n_i = \cre{a}_i\anh{a}_i$ is the particle number operator, $\mu$ is the on-site chemical potential, $U$ is the interaction strength, and $J$ is the tunneling energy.
	The interaction energy $U$ and $J$ are generally evaluated numerically  based on the respective integral expressions.
	The interaction energy is
	\begin{equation}
		U = \frac{4\pi\hbar^2 a}{m}\int |w(\vec{x})|^4~d^3 x,
	\end{equation}
	where $w(\vec{x})$ is the Wannier function \cite{Wannier37,Marzari00} representing the wavefunction of a particle localized at a single site.
	By approximating the potential minimum as a harmonic oscillator with trapping frequencies $\omega_i = \sqrt{4 V_j}/\hbar$, where $V_j$ is the depth of the $j^{\textrm{th}}$ beam, one can arrive at the approximate expression 
	\begin{equation}
		U \approx \frac{\hbar \bar{\omega}a_s}{\bar{a}_{HO}\sqrt{2\pi}}
	\end{equation}
	where $a_0$ is the scattering length, $a_{HO}$ is the size of the ground state wavefunction, and the overbars indicate the geometric mean of the respective quantities	\cite{Jaksch98}.
	The tunneling rate is given in general by
	\begin{equation}
		J_{i,j} = \int w(\vec{x}_i) \left(-\frac{\hbar^2}{2m}\nabla^2 + V_0(x)\right)w(\vec{x}_j)~d^3 x,
	\end{equation}
	which in the deep-lattice limit is approximately \cite{Jaksch98}
	\begin{equation}
		J \approx \frac{4}{\sqrt{\pi}} E_r\Big(\frac{V_0}{E_r}\Big)^{3/4}e^{-2\sqrt{V_0/E_r}},
	\end{equation}
	where $E_r = \hbar^2k^2/2m$ is the recoil energy of a single photon from the lattice beams.
	The simplest model to consider is one where only nearest-neighbour hops are possible, but in a homogenous (infinite) system one has $J_{i,j} = J(|i-j|)$ and this is not an onerous addition.
	The second-order tunnelling between next-nearest neighbours is much smaller than the tunnelling between nearest neighbours, but nonetheless has been observed in quantum gas microscopes \cite{Folling07}.
	
	The optical lattice realization of this model relies on the optical dipole force \cite{Grimm00}, whereby two counterpropagating lasers (in each dimension) create a standing wave pattern\footnote{The upgrade at ANU includes a design to reflect the lattice beams back on themelves, similarly to the MOT trapping beams.}.
	If the laser has wavelength $\lambda$, then the electric field creates a lattice with twice the spatial frequency (half the wavelength) $k_L = 2k_\lambda = 4\pi/\lambda$.
	In a lattice formed by a collimated, retro-reflected Gaussian beam, the time-averaged electric field gives rise to an optical potential that can be approximated by
	\begin{equation}
		V_0(\vec{x}) = - V_{L} e^{-2\frac{{\vec{r_i}}^2}{{w_i}^2}} \sin^2(k x_i)
	\end{equation}
	where the trap depth is given by the optical dipole potential of a two-level system,
	\begin{equation}
		V_{L} = \frac{3\pi c^2}{2{\omega_0}^3}\frac{\Gamma}{\delta}\frac{2P}{\pi {w_0}^2},
	\end{equation}
	and the exponential envelope is inherited from the Gaussian profile of the laser.
	
	In the case of a focused laser, there will be an additional longitudinal envelope due to the divergence of the beam.
	In a 3D lattice created by mutually orthogonal beams of equal power, and considering only the sites closest to the centre of the trap, the potential can be approximated by 
	\begin{equation}
		V_0(\vec{x}) = -\sum_{i=1}^3\Big( V_{L} \sin^2(k x_i) + \frac{m}{2}(\omega_i x_i)^2\Big)
	\end{equation}
	where the Gaussian profile is predominantly governed by the first ($x^2$) term in its Taylor series with small-amplitude trapping frequencies $\omega_i$.	
	For very cold gases, only sites near the centre of the lattice will be occupied, and the potential can be well approximated by retaining only the first term and considering the lattice to be uniform.

	The Bose-Hubbard model exhibits a quantum phase transition between superfluid and Mott insulating states at the point $U/zJ \approx 5.8$ where $z=2d$ is the number of nearest neighbours \cite{Jaksch98}.
	For a system of N particles on M lattice sites (with $N = nM$), the  Mott-insulating ground state has the form of a product of Fock states
	\begin{equation}
		\ket{\Psi_{MI}} \propto \prod_{i=1}^M (\cre{a}_i)^n\ket{0_i},
	\end{equation}
	which exhibits vanishing number fluctuations between sites \cite{Greiner01}.	
	In the regime where tunnelling dominates (at $U\rightarrow 0$) the ground state takes the form
	\begin{equation}
		\ket{\Psi_{SF}} \propto \left(\sum \cre{a}_i\right)^N\ket{0_i}
	\end{equation}
	which is a product of (local) coherent states with Poissonian statistics \cite{Greiner01}.
	
	The momentum representation of the Bose-Hubbard model can be obtained by the Fourier transform \cite{Zhang10} of the bosonic creation operators,
	\begin{equation}
		\cre{b}_q = \frac{1}{\sqrt{M}}\sum_{j=1}^{M}e^{i(2\pi q j/M)}\cre{a}_j
	\end{equation}
	in terms of which the Hamiltonian is
	\begin{equation}
		\hat{H} = -2J\sum_{q=0}^{M-1}\cos\left(\frac{2\pi q}{M}\right)\cre{b}_q\anh{b}_q + \frac{U}{2M}\sum_{q_1,q_2=0}^{M-1}\sum_{q_3,q_4=0}^{M-1}\cre{b}_{q_1}\cre{b}_{q_2}\anh{b}_{q_3}\anh{b}_{q_4}\delta_{q_1+q_2,q_3+q_4}.
	\end{equation}
	The latter sum represents momentum-conserving collisions and commutes with the total momentum $K = \sum_{q=0}^{M-1}(2\pi q/M)n_q$.
	For a one-dimensional system with $N$ atoms and size $M$, the system has dimensionality $D = (N+M-1)!/(N!(M-1)!)$ which rapidly outstrips the storage space of consumer-grade computers.
	The problem can be made much more tractable by neglecting interactions, as Ramakumar \emph{et al.} did in an exact diagonalization study of non-interacting bosons (in a combined lattice and overall harmonic confinement) for up to $N=M=1000$ in one dimension and 2500 bosons in a $50\times50\times50$ lattice \cite{Ramakumar07}.
	The state-of-the-art quantum Monte Carlo calculations currently find good agreement with experiments \cite{Cayla18,Herce21}.

	Later in this course of study, an optical lattice was loaded with \mhe~by the group at Institut d'Optique in Paris \cite{Cayla18}, which therefter achieved the Mott insulator transition \cite{Carcy19} and showed the critical behaviour at the transition boundary conformed with the three-dimensional XY universality class \cite{Herce21}.
	There remain many interesting topics of study\footnote{Besides, it will be important to obtain independent reproduction of surprising discoveries.}, and potential future pathways are discussed at the end of this chapter.

\section{Infrastructure upgrade}

	In this section we deal with the (literal) nuts and bolts of assembling a cold-atom machine.
	The obvious route toward a helium-filled lattice would be to upgrade the BiQUIC machine.
	The biggest obstacle on this route is the limited optical access to the chamber.
	For one, the windows are relatively small and so the beams would need to be inserted parallel to existing beams.
	In principle one could install flipper mirrors and switch in the MOT beams for the loading sequence and insert the lattice beams along fixed, stable optics.
	However, if the laser generation table could be described as a `forest of optics', then the space surrounding the BiQUIC chamber is a veritable jungle.
	In short, it is certainly plausible that the machine \emph{could} host an upgrade, but it would decommission the machine for some time.
	It is much more convenient to instead blow the dust off another machine lying dormant in the laboratory one floor above level 0 \footnote{The BiQUIC lab is the sole room at the sub-basement level of the school.}.

	Before my arrival in the lab, Dr. Hodgman (the PI of the project) and some students had put in good early miles refurbishing the discharge source and pumping the existing chamber down to around $10^{-10}$ Torr.
	Some of the cooling and trapping beams were present but most of the AOM suite was yet to exist.
	The prior incarnation of the machine consisted of a source chamber and a Zeeman slower terminating at a MOT chamber, and a secondary science chamber.
	The pre-existing chamber only had ports that were small enough to clip the first-order diffraction peaks of the expanding lattice, and so we needed to overhaul the machine 
	with a new science chamber.
	This build would also include a detector chamber, additional vacuum maintenance and diagnostic tools, and an imaging system for the new magneto-optical, magnetic, and optical traps.
	Before we proceeded with the extension, we needed a cold source of atoms. 
	Hence, the first point of attention was rebooting the MOT.

\begin{figure}
	    \begin{minipage}{0.45\textwidth}
	    \vspace{0pt}
			\caption{Partial schematic of the control and distribution system for cooling light in the lattice machine.
	The fibre amplifier supplies light to each of several AOMs in cat's-eye configuration.
	Half-wave plates control the optical power supplied to each AOM via polarizing beamsplitters.
	The first diffracted mode is spatially selected by an aperture on the outgoing side.
	A quarter-wave plate ensures that the the polarization is rotated from V to H after the second pass and is transmitted through the beamsplitter, then to the beam-forming optics before insertion into the vacuum chamber.
	The parameters in the adjacent table were those used to achieve BEC \cite{Abbas21}.}
	    \label{fig:distribution_optics}
		\end{minipage}
	    \hfill
	    \vspace{5pt}
		\begin{minipage}{0.5\textwidth}
			\vspace{0pt}
			\includegraphics[width=\textwidth]{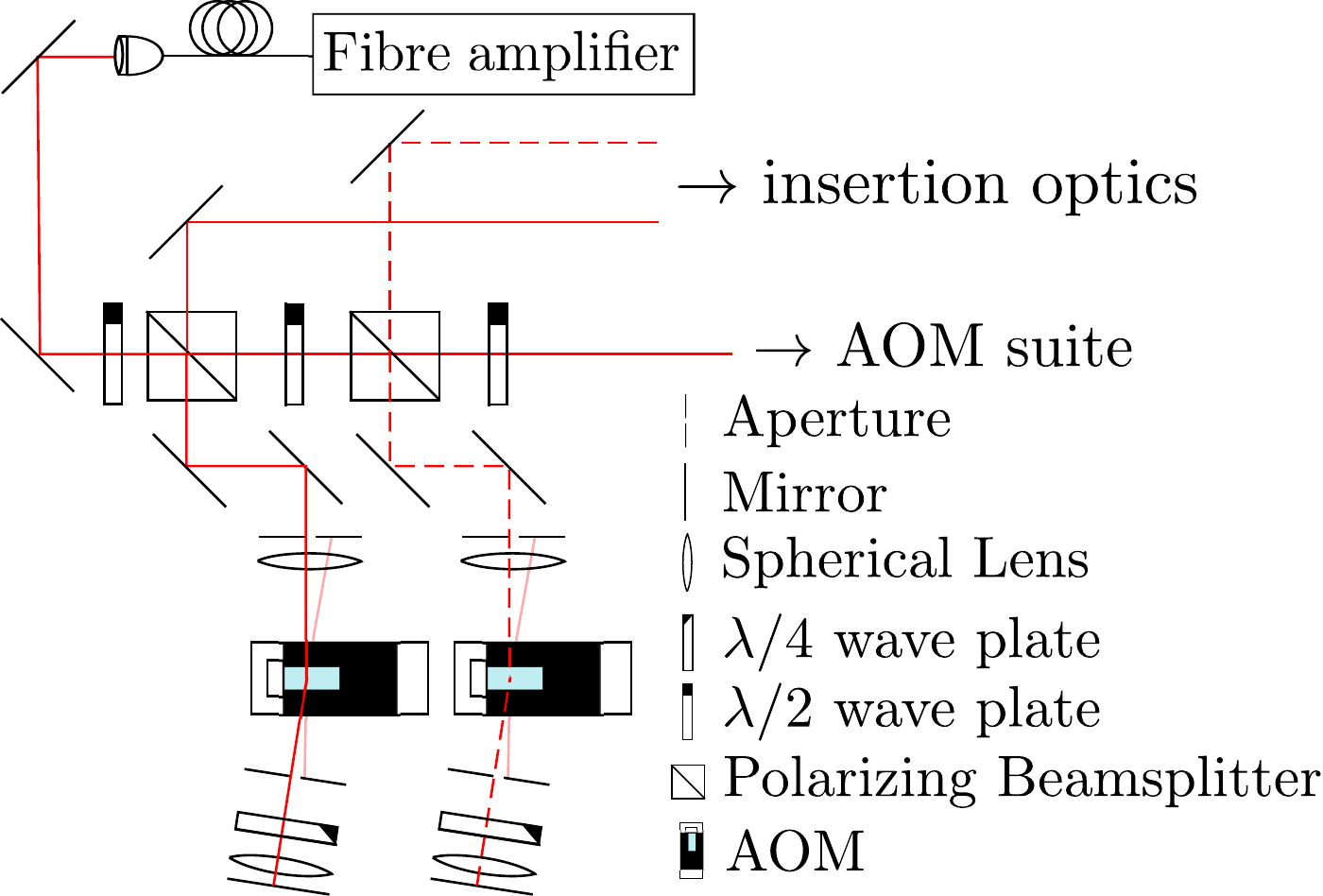}
		  {\fontsize{11}{13}
		  \begin{tabular}{l c c}
		  \hline\hline
		  Beam & Intensity & Detuning \\
		   & $(I_\textrm{sat})$ & ($\Gamma$) \\
		  \hline
		  Collimator &67& -5\\
		  Zeeman &92&-160 \\
		  MOT 1 &87 (Horz.)&-22 \\
		  		&110 (Vert.) & \\
		  Push &110&+3.3 \\
		  MOT 2 &37 (Horz.)& -33\\
		   &140,400 (Vert.)& \\
		  \hline
		  \end{tabular}}
	    \end{minipage}
	\end{figure}

	Achieving magneto-optical trapping required two preliminary steps: First, the cooling and trapping beams needed be produced from the amplified master laser ($\approx$3 W of continuous power tuned 253 MHz red of the 1083.331 nm transition, see previous chapter).
	Figure \ref{fig:distribution_optics} shows two AOMs in the double-pass configuration and illustrates how a single source beam can be split into multiple, approximately collimated, outgoing beams with controllable power and detuning.
	Lenses within the double-pass setup were used to increase the AOM efficiency and ensure the outgoing beam was collimated.
	The polarization and profile (hence also intensity) of the beams is fixed further downstream by a series of telescopes and wave-plates.
	The AOMs were driven in open-loop configuration by an amplified voltage-controlled RF oscillator, whose frequency and power were controlled via the LabView-based data acquisition and control system (DAQ).
	The LabView program was also used to control the current supplies driving the magnetic coils, relays or MOSFETs for switching off the various field generation coils, and a number of shutters on the table.
	The control system architecture is essentially the same as the one described in section \ref{sec:DAQ}, but some components of the 1550 nm laser system were controlled by other means, described below.
	
	We had to align the collimator and Zeeman slower beams before the MOT could be built.
	Naturally, having set everything up perfectly the first time, the MOT was nowhere to be found.
	After some further tweaking, we obtained a visual signature of the MOT on the camera, a 1-2 cm diameter smudge in the dark \footnote{We had to switch the room lights off to reduce scattered light in order to spot it.}.
	With some life breathed back into the old machine, it was time for a much larger operation.

\subsection{Vacuum build}

\begin{figure}
	  \begin{minipage}{0.55\textwidth} 
	  \vspace{0pt}
			\includegraphics[width=\textwidth]{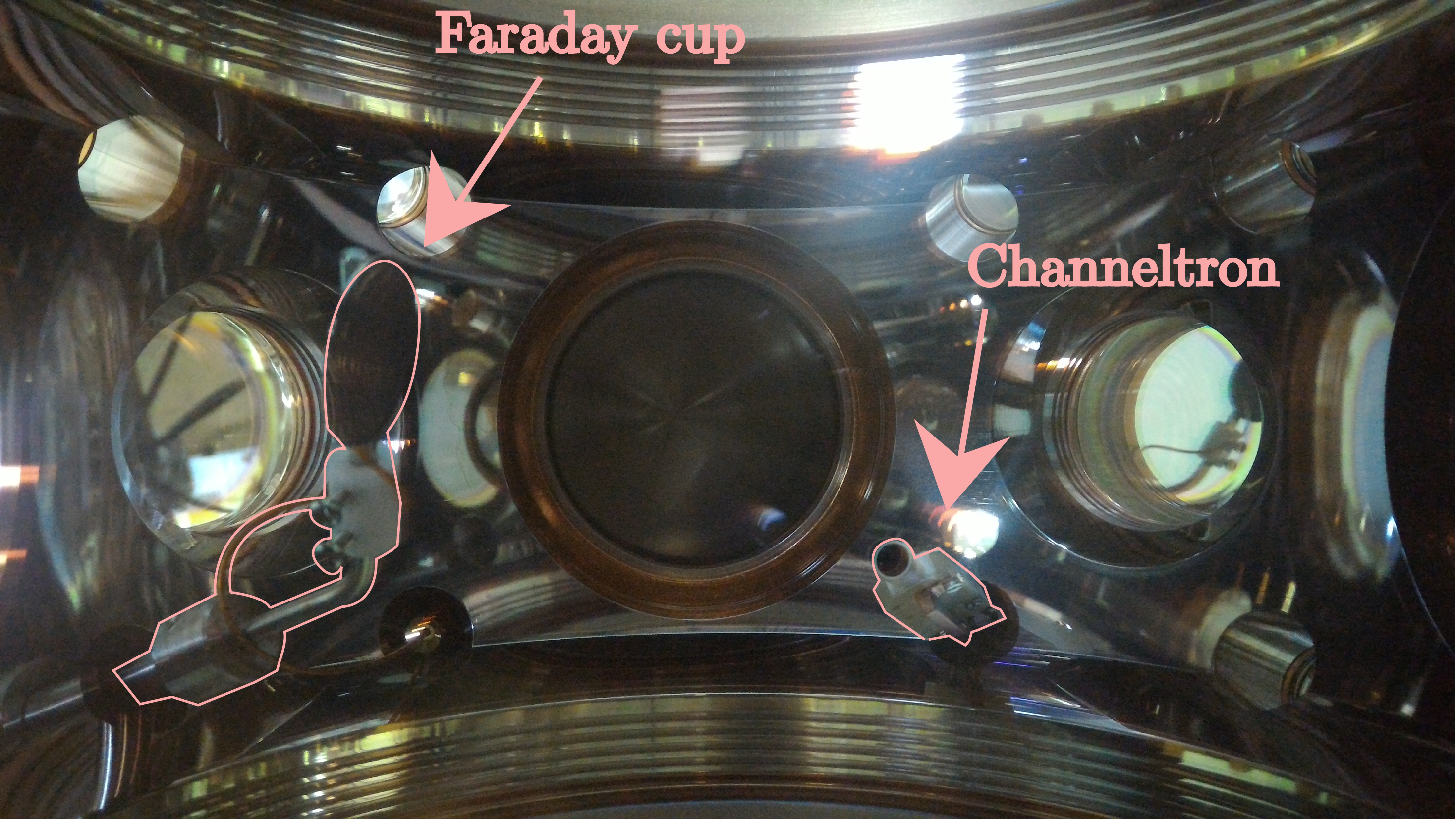}
		   {\begin{flushright}\caption{Initial build of the new science chamber.
	Right: The science chamber features a Faraday cup,  channeltron, and a NEG for vacuum maintenance.
	The camera was initially used to monitor the first MOT. Later uses of the camera are discussed in sections below.
	A 6" gate valve separates the chamber from a temporary assembly underneath the table.
	Top: Internal view of the chamber, taken the preceding day (after which the channeltron was moved to make room for the mounting plate, see below).}\label{fig:first_build}\end{flushright}}
	  \end{minipage}
	  \hfill
	  \begin{minipage}{0.45\textwidth}
	  \vspace{0pt}
		\includegraphics[width=\textwidth]{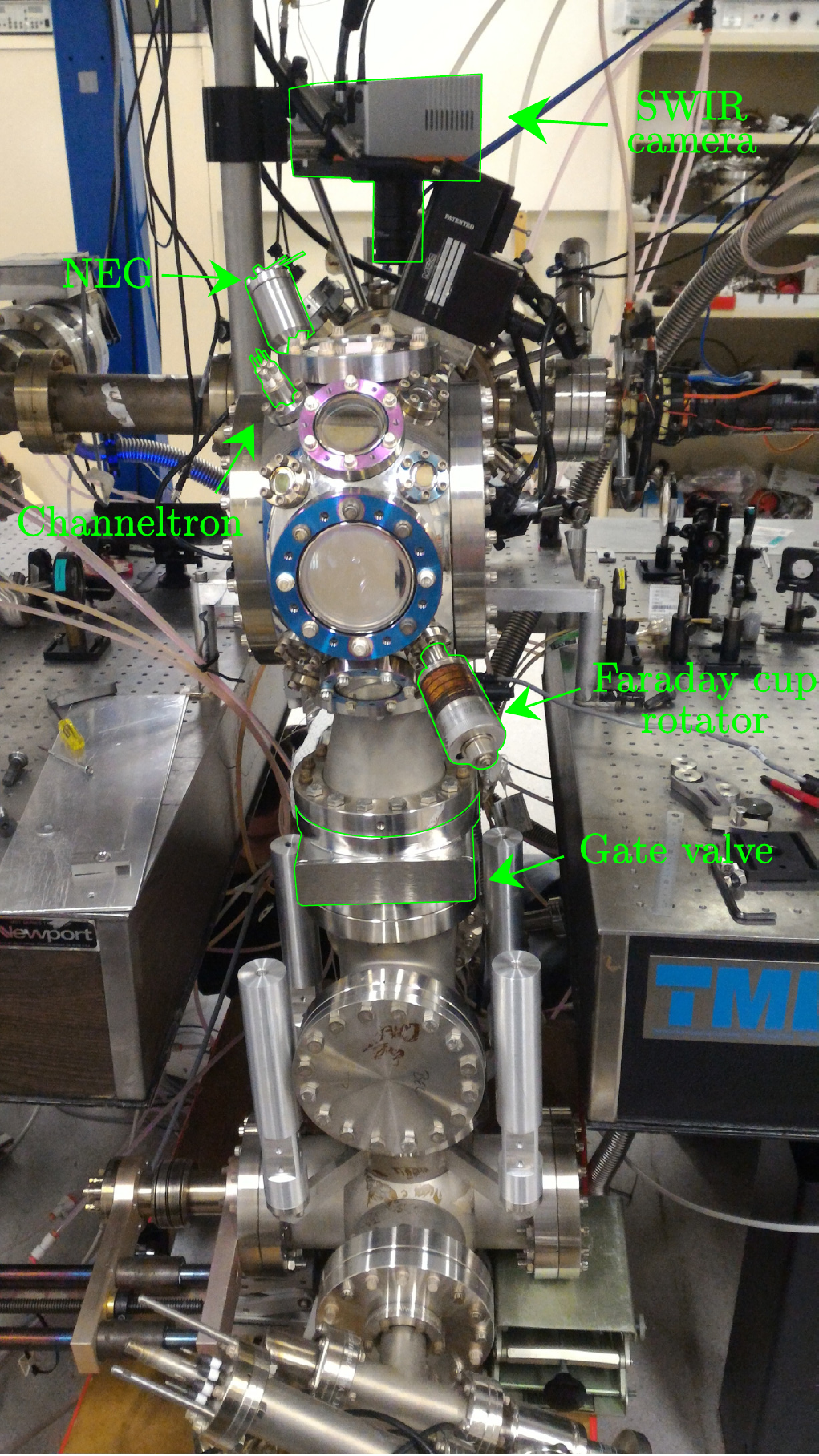}  
	  \end{minipage}
	\end{figure}

	The first stage of the build was to assemble the new science chamber.
	As we were waiting on delivery of the multichannel plates (MCPs) for the final detector, we assembled the second science chamber in order to get the low-velocity intense source (LVIS) operational in the interim.
	The first iteration of this build, prior to the addition of the MOT coils and optics, is shown in Fig. \ref{fig:first_build}.
	The large chamber (manufactured by Kimball physics) features two 8" windows on the flat sides through which the vertical MOT beams and absorption imaging are inserted, and 4.5" windows for insertion of the dipole beams on the top and side opposite the LVIS.
	Several $\frac{1}{2}$" ports are used as feed-throughs for in-vacuum instruments (namely a channel electron multiplier \emph{a.k.a.} channeltron for detecting ions produced by Penning ionization, a non-evaporative getter (NEG) to increase hydrogen adsorption and improve the vacuum, and a Faraday cup on a rotation stage to measure the flux of metastable atoms into the chamber via the LVIS) or covered with windows for optical instruments such as a photodiode to measure fluorescence of the trap.
	The lower assembly includes four-way and six-way crosses (6" flanges each).
	The four-way cross featured a turbomolecular pump (opposite the point of view) which was backed by roughing pump.
	The six-way cross (at the bottom of the stack) featured a residual gas analyser (RGA), whose purpose is described below.
	The entire assembly was put together while supported by four jack stands (one on each horizontal flange of the six-way cross) on a pallet jack.
	The pallet jack could then be wheeled into position and raised to height, and then carefully docked with the back of the gate valve on the MOT.
	The apparatus also featured an old discrete dynode electron multiplier (manufactured by ETP) that was used in early attempts to detect atoms dropped from a magnetic trap, but later stopped working and was thus of no use as a diagnostic for the optical dipole trap.

\begin{figure}
			\centering
		\includegraphics[width=\textwidth]{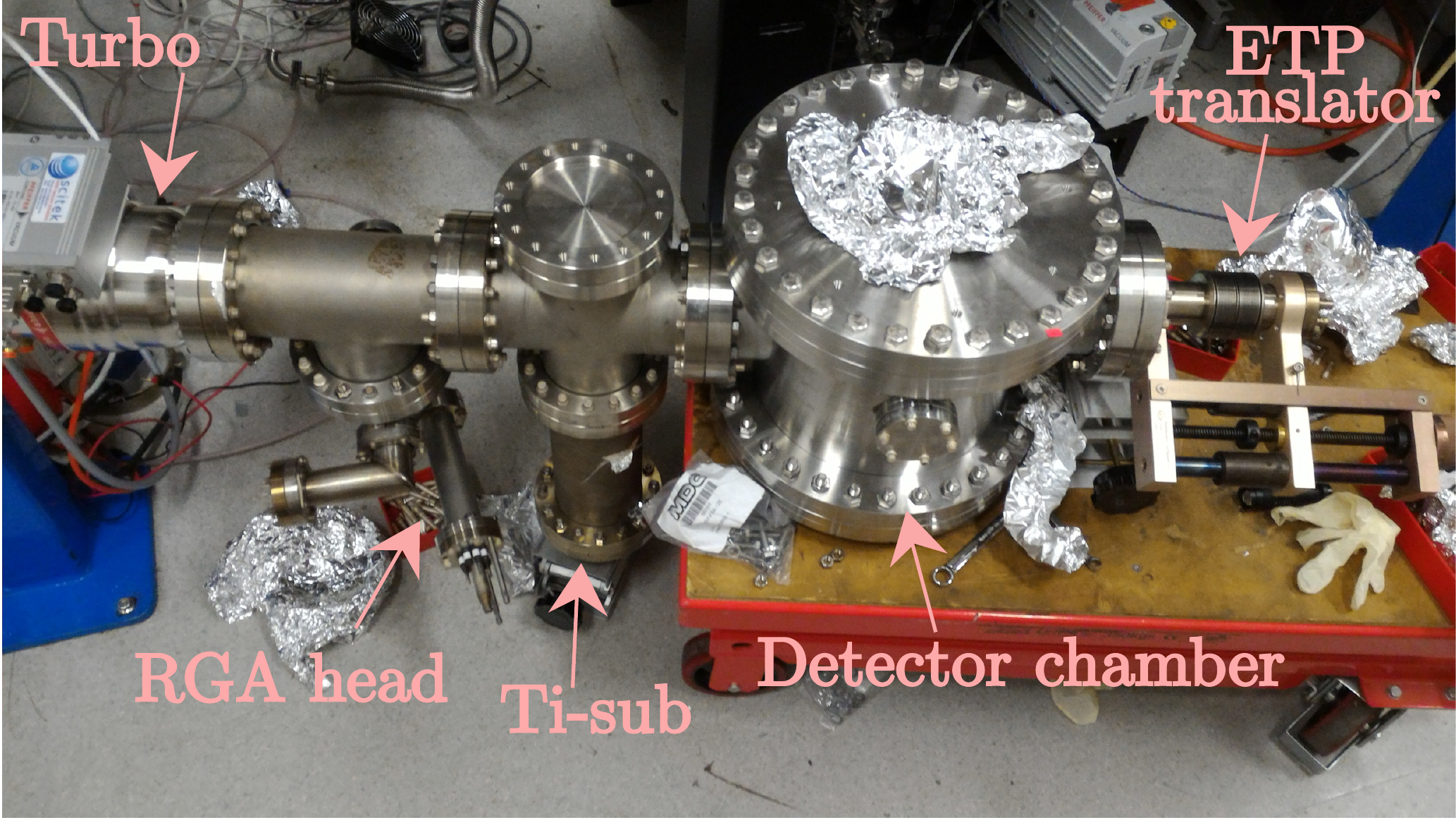} 
		\includegraphics[width=\textwidth]{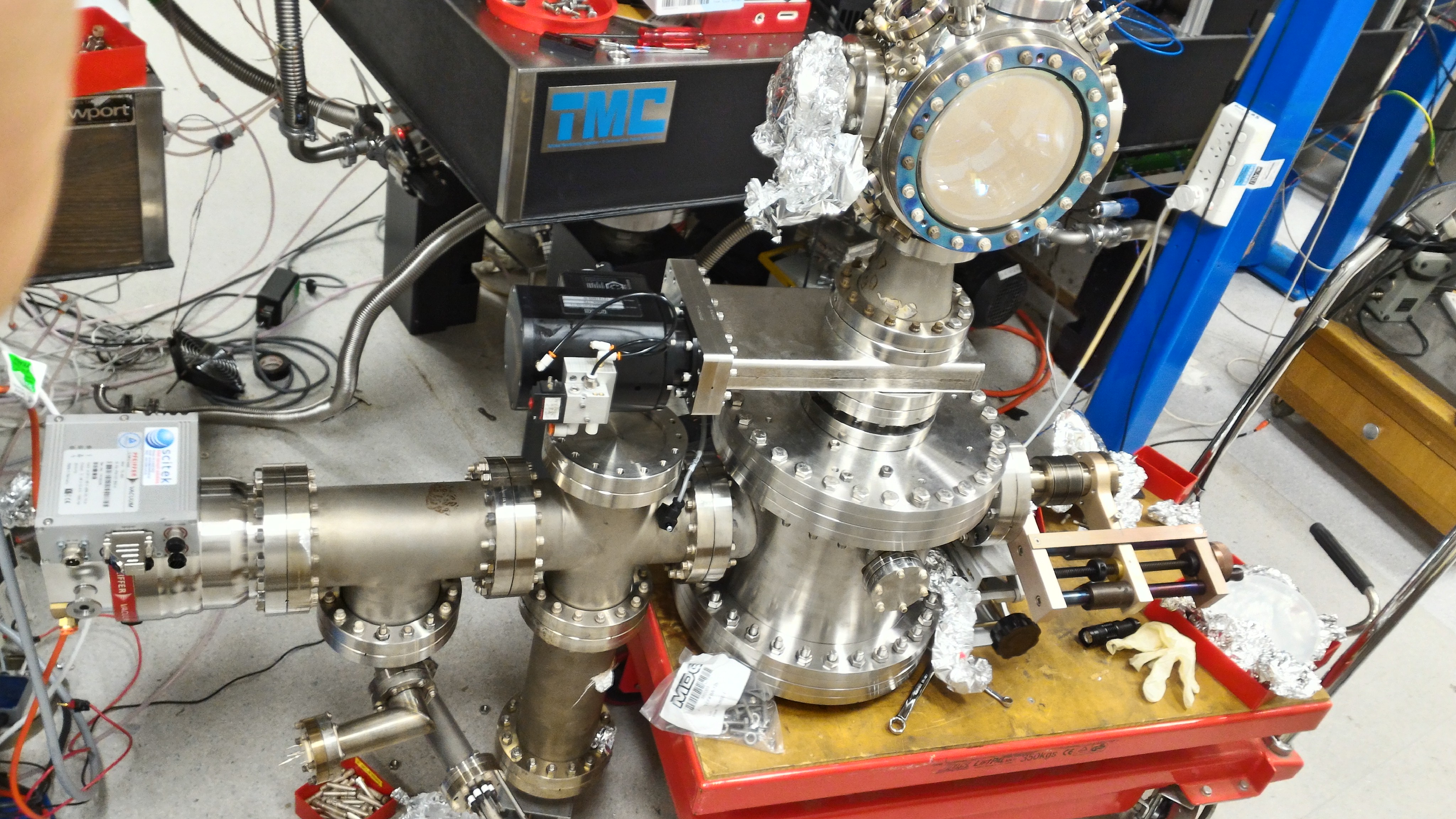} 
			\caption{Top: The leviathan-like underbelly of the science chamber resting after assembly on the mobile pallet jack and a smaller hand jack to support the TiSub.
		 Bottom: The science chamber was re-mounted with an intermediate gate valve and then wheeled into place.
		The assembly was stabilized on the blank flange at the bottom of the {24"}-diameter detector chamber, and by being \emph{really} heavy.}
	\label{fig:underbelly}
	\end{figure}

	The `push' beam was aligned by maximizing the flux from the first MOT into the science chamber as measured by the current on a Faraday cup.
	The sensor was positioned so that it could rotate into the helium beamline from its resting position out of any relevant optical paths.
	The current $I$ (in amps) gives an estimate of the order-of-magnitude flux $\phi_a$ of the atoms by the rule of thumb that $\phi_a\approx I/q$.
	This is an approximate method, but easily suffices for diagnostic purposes.
	Once we obtained a signal that the push beam was reading a current of order 200pA, indicating a flux on the order of 10$^9$ \mhe~atoms per second, we proceeded to assemble the rest of the vacuum system (shown in Fig. \ref{fig:underbelly}) and the MOT optics (detailed in section \ref{sec:new_optics}).
	The strategy was to assemble the MOT and absorption imaging systems to establish the magnetic and dipole traps, then use the ETP electron multiplier to optimize the dipole in the case we were still waiting on the MCP.
	The gate valve below the science chamber meant that we could close off the lower portion of the system and replace a single flange at the bottom of the 24" chamber when installing the MCP-DLD detector stack.
	Thus we would be able to perform the final upgrade without disturbing the MOT optics or breaking vacuum in the MOT chamber, which would be a significant advantage in that it would considerably simplify the necessary bake-out of the vacuum chamber, the subject of the next section.

	\subsubsection{Improving vacuum conditions}

		As mentioned in chapter \ref{chap:apparatus}, the viability of cold-atom experiments depends on having a good quality vacuum.
		Any background gas molecules will typically have thermal velocities of several hundred m/s, with kinetic energies many orders of magnitude larger than the depth of the trapping potentials.
		Thus an overabundance of background gas can dramatically reduce the lifetime of magnetic and dipole traps, rendering evaporative cooling to degeneracy impossible.
		In the case of helium, there is the added issue of the small atomic mass and possibility of Penning ionization (even in low-momentum collisions) which only exacerbate the issue.
		This lab does not maintain cleanroom conditions outside of the optics tables, so even though care was taken to cover components in new foil when they were not being handled, some environmental pollution would be inevitable.
		Water is a common contaminant, but handling errors can also be a problem, and dust or aerosols can accumulate invisibly on the steel surface.
		All of these contaminants become problems when pumping down to vacuum as the pressure rapidly drops below the vapor pressure of many chemicals that may be stable in atmospheric conditions.
		These surface contaminants may outgas slowly (or be comparatively large) and lead to persistent `virtual leaks' in the chamber.
		The standard remedy is to wrap the machine in highly resistive wires insulated with fibreglass, cover it with aluminium foil to keep heat in, and bake the entire apparatus at around 150 degrees celsius for several days.
		The initial evacuation of the chamber by connecting the turbomolecular pumps was enough to observe a magnetic trap, but would be insufficient to make further progress (see Fig. \ref{fig:lifetime}).

\begin{figure}
		\includegraphics[width=\textwidth]{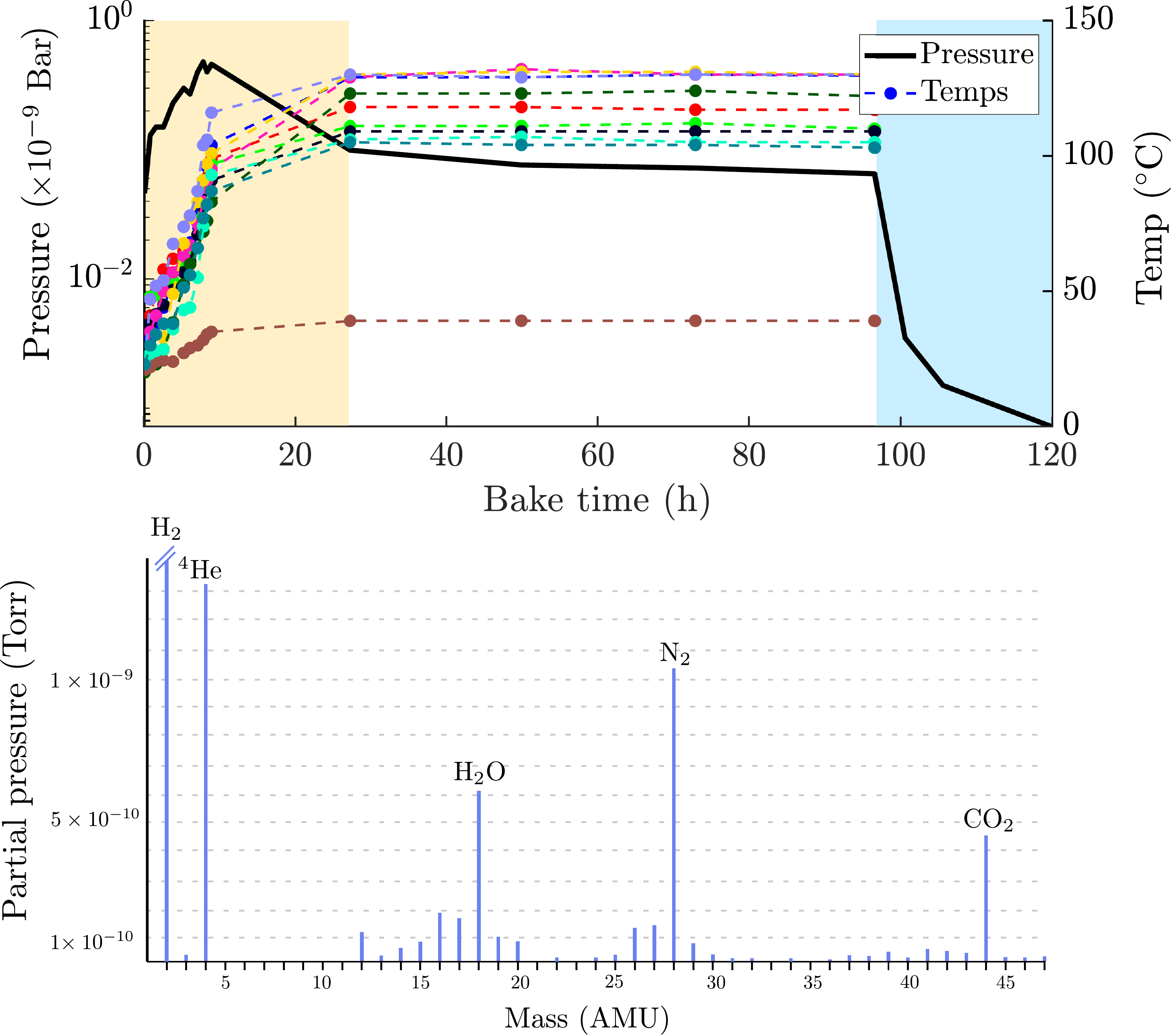}
		\caption{Top: The temperature measured by several thermocouples taped to the surface of the vacuum chamber (dots with dashed lines) and the pressure readout from the ion gauge on the science chamber during a bake.
		The heater tapes are supplied with current from variable voltage sources (variacs) which are turned up gradually to ensure even heating when starting the bake (orange region).
		After the tapes are switched off and the insulating foil opened, the steel quickly cools and the vacuum pressure stabilizes (blue region).
		The lower (brown) temperature curve was measured with a thermocouple on the flange connecting the 6" T-piece to the turbo (see Fig. \ref{fig:underbelly}) which was kept cool to avoid excessive heating of the turbo itself. 
		Lines in this figure are to guide the eye, individual measurements are denoted by the solid points.
		Bottom:
		Gas concentrations before the first bake as measured using the residual gas analyser (RGA).}
		\label{fig:bakeouts}
		\end{figure}

		The higher temperature during the bake increases the outgassing rate and can even liberate some contaminants from the surface, allowing them to be pumped out by the turbomolecular pumps.
		In this process one typically observes a sharp rise in pressure as the chamber heats and the contaminants vaporise, and then a decrease to a lower steady-state pressure as the contaminants cease outgassing and the remnants are evacuated.
		Once the pressure stabilizes, the heater tapes can be turned off, and as the apparatus cools the pressure then settles to a steady-state value.
		An example of this procedure is illustrated in Fig.	\ref{fig:bakeouts}.
		The chemical makeup of gaseous contaminants can be determined using a residual gas analyser (RGA) which are essentially compact mass spectrometers.
		Figure \ref{fig:bakeouts} shows the partial-pressure contributions of gaseous contaminants before the first bake.
		
		Even after baking, the pressure may not be low enough to permit long-lived traps.
		Hydrogen leaching from the surface of the steel chamber can be significant, and even permeate the steel on long enough timescales.
		One means of addressing this is with a titanium sublimation pump (TiSub).
		This is not a pump \emph{per se} but rather a titanium filament.
		A pulsed current of $\approx5$ A (duty cycle approximately 30 seconds on, 60 seconds off) heats the filament and causes titanium to sublimate off the filament and adsorb onto the interior surface of the steel vacuum chamber.
		Hydrogen adsorbs to the titanium surface but not to steel, thus creating a `virtual pump' and driving the pressure further down.

	Another pressure reduction can be obtained by `getters' which operate in a variety of ways (a TiSub is a kind of hydrogen getter).
	In our machine we used a non-evaporative getter (NEG) in the science chamber.
	The NEG has a highly porous active area which adsorbs hydrogen very strongly.
	This means that they tend to carry in a lot of contaminants and have little remaining active area when first pumping down.
	We cleared the surface of the NEG by passing up to 5 A of current through it.
	Again, the pressure spikes and eventually relaxes up to an hour later, when the current is switched off.
	As shown in Fig \ref{fig:bakeouts}, chamber bakes can improve the chamber pressure by up to three orders of magnitude.
	Afterwards, the TiSub can improve the pressure by a factor of five to ten, and the NEG by a factor of two to five.
		
	The payoff for this multi-stage procedure is the enormous reduction in background pressure that permits long-lived magnetic and (purely) optical traps.
	The lifetime can be quantified by any means that provides a reasonable proxy of the trapped population (the quantity of concern is the relative population at some time later), such as in Fig.	\ref{fig:lifetime}.
	Currently the machine operates at a background pressure below 3$\times10^{-11}$ mTorr (at which point the Ion gauge bottoms out).

\subsection{MOT and magnetic trap}
\label{sec:new_optics}

	\begin{figure}
		\includegraphics[width=\textwidth]{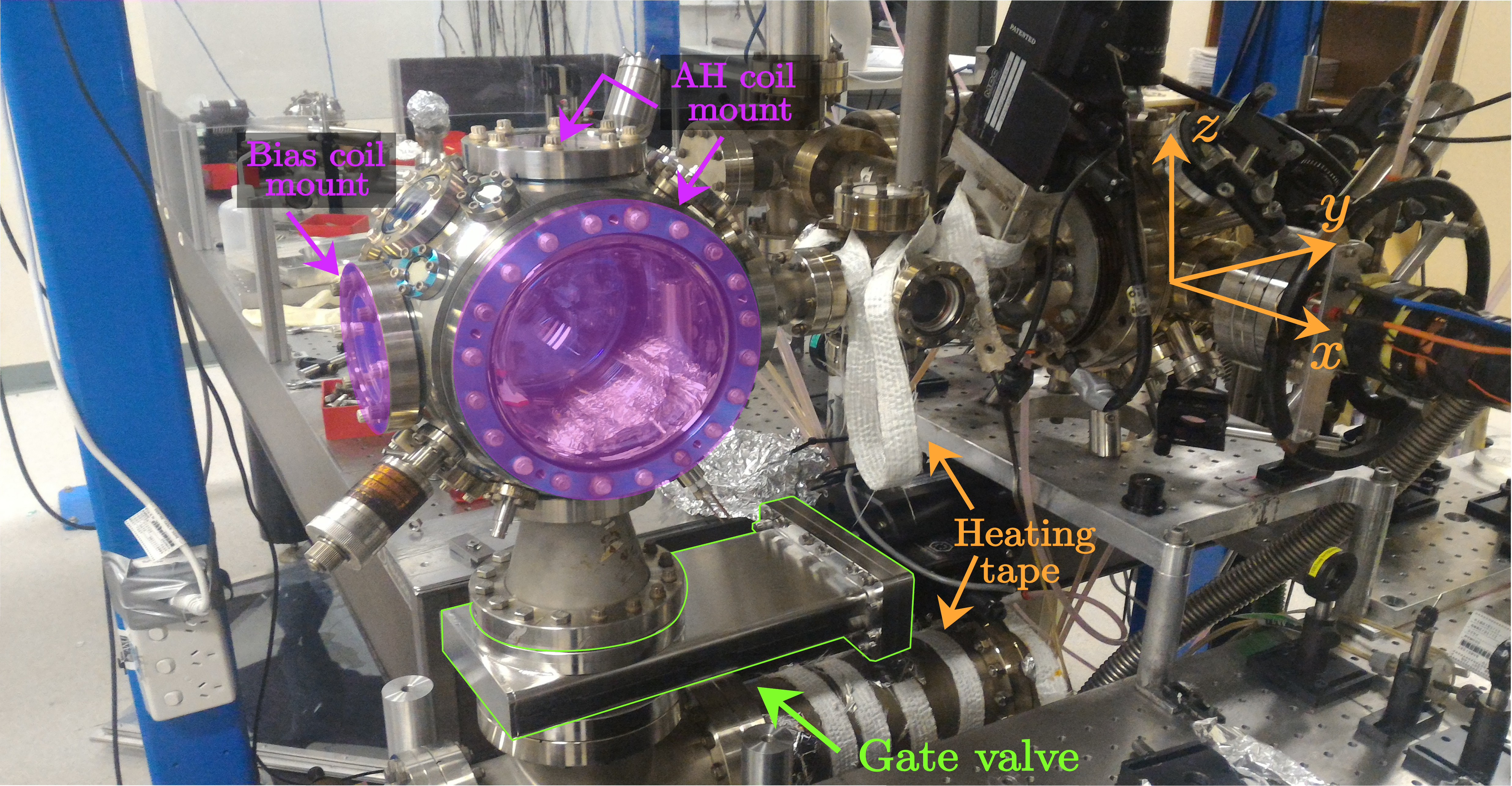} 
		\includegraphics[width=\textwidth]{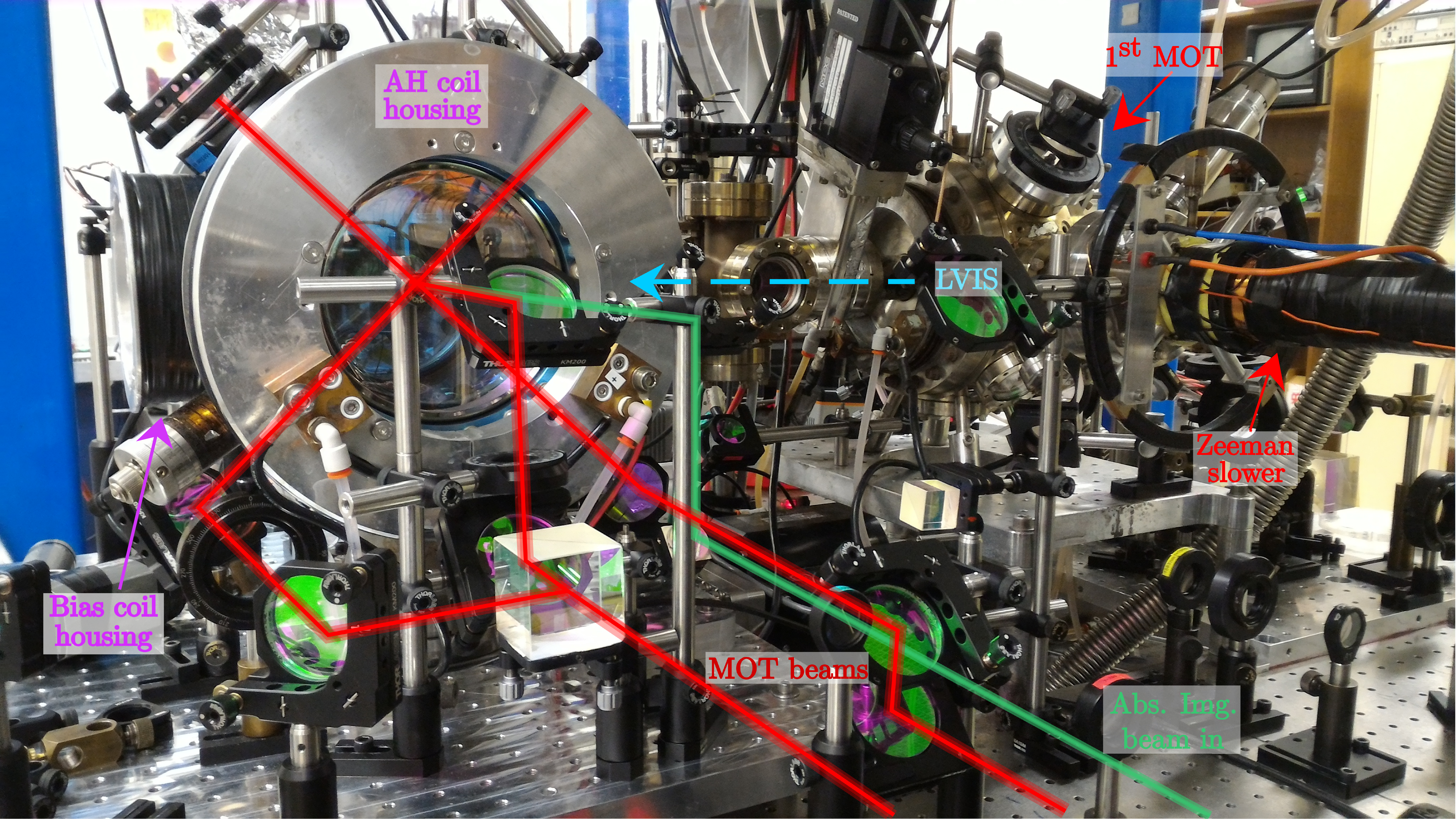} 
		\caption{The first iteration of the science chamber (in August 2016)  illustrating the coordinate axes used in the lab and the mounting surfaces for the magnetic coils (one of the Anti-Helmholtz (AH) coils was mounted on the back face of the large Kimball chamber). 
		The channeltron is visible on the lower right of the chamber (top), but was moved the next day to make room for the MOT mirror in a similar location in the later image.
		This configuration was used to test vacuum viability and align the LVIS.
		Once the new optic-mounting platform had ben constructed by technician Ross Tranter, we could install the MOT optics (bottom, mounting plate in lower-left quadrant).
		The MOT and imaging beam trajectories are marked, along with the direction of propagation of the LVIS.
		The Faraday cup and NEG are still installed in the lower picture, but the NEG is obscured by the chamber.}
		\label{fig:MOT_optics}
	\end{figure}
	With a good, clean, vacuum and ample optical bench space around the science chamber, the optical components could finally be installed.
	Figure \ref{fig:MOT_optics} shows the enclosing construction around the science chamber that supports a MOT, magnetic trap, and absorption imaging.
	First, the magnetic coils were mounted on the windows to create the anti-Helmoltz field (with axis of symmetry along $x$, parallel to the Zeeman slower) and a bias field directed along the $y$ axis, back along the LVIS.
	The diagonal MOT beams enter through the 2$\frac{3}{4}$" widows on the 45$^\circ$ ports.
	We used the channeltron as a sensor for the presence of a MOT during initial alignment.
	A poorly-aligned MOT (or one with unfavourably set beam parameters) would not have been easily visible by camera or photodiode, but the increase in Penning ionization rate would in principle be controllable by blocking a laser beam and hence provide a fast diagnostic.
	The MOT itself could only be constructed after installing a mounting plate around the base of the vacuum chamber, which was installed after baking the entire new assembly.

	\begin{figure}
		\includegraphics[width=\textwidth]{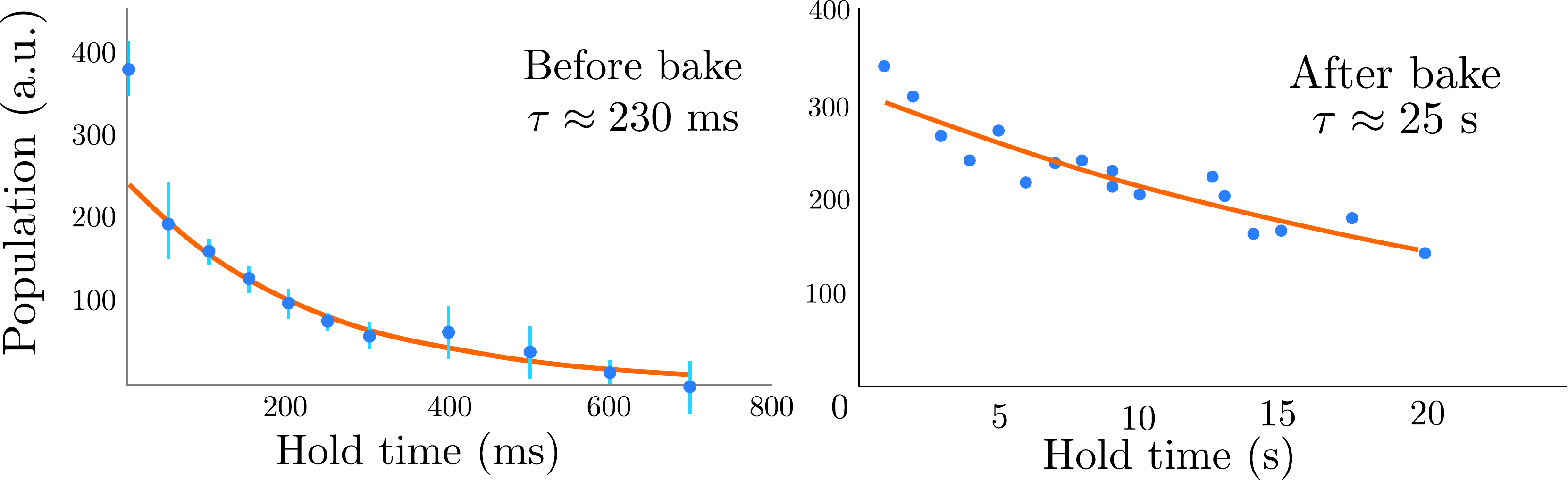}
		\caption{Magnetic trap lifetime measurements using before and after baking the new vacuum system. 
		The uncalibrated measurements only show the relative population, due to the measurement method described in section \ref{sec:abs_img}.
		Data points in the left plot are taken from averages of three shots, points on the right plot are single shots.
		The lines in both graphs are exponential fits.}
		\label{fig:lifetime}
	\end{figure}

\subsubsection{Absorption imaging}
\label{sec:abs_img}
		
	Alignment of the dipole trap was eventually made possible by absorption imaging.
	This system is preferred over the MCP-DLD for imaging MOTs and magnetic traps because 
	the clouds can be so wide on the detector that it is impossible to accurately determine the width, thus temperature, of the cloud.
	The analysis can be made more difficult by the potential saturation of the detector at large fluxes.
	A final, albeit lesser, concern is that MOTs and magnetic traps contain thousands of times as many atoms than remain after condensation and so they would shorten the detector's lifespan.

	\begin{figure}
	\includegraphics[width=\textwidth]{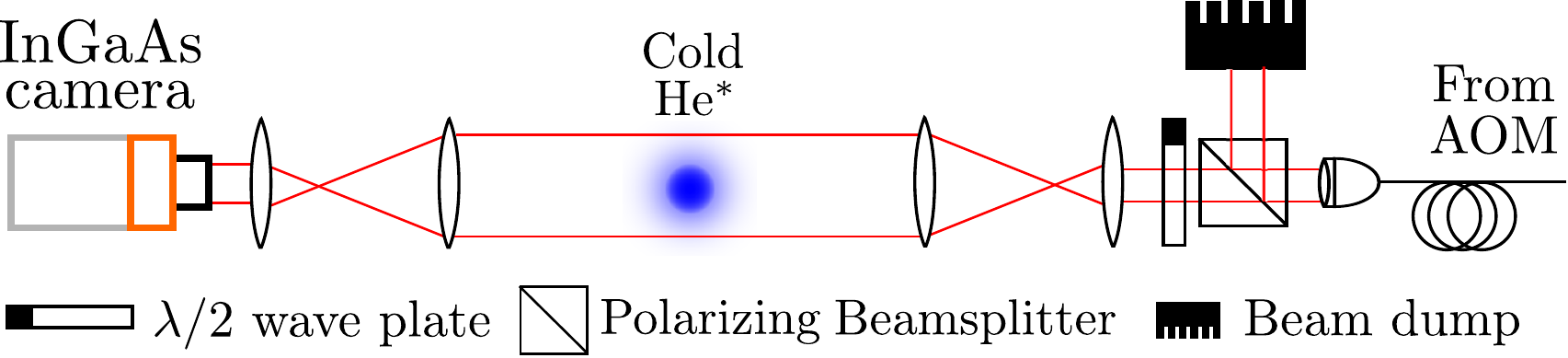}
	\begin{minipage}{0.44\textwidth}
	\vspace{0pt}
		\caption{ Above: Schematic of the absorption setup as described in the text.
		Adjacent: 
		Example image of the MOT.
		In practise the shot noise on the camera can be significant, and so for visual inspection a smoothing kernel is applied.
		Because of the issues in the text, it is nontrivial to make meaningful inferences about trap properties from in-trap images like these.}
		\label{fig:abs_img}
		\end{minipage}
		\hfill
		\begin{minipage}{0.54\textwidth}
	    \vspace{0pt}
	    \includegraphics[width=\textwidth]{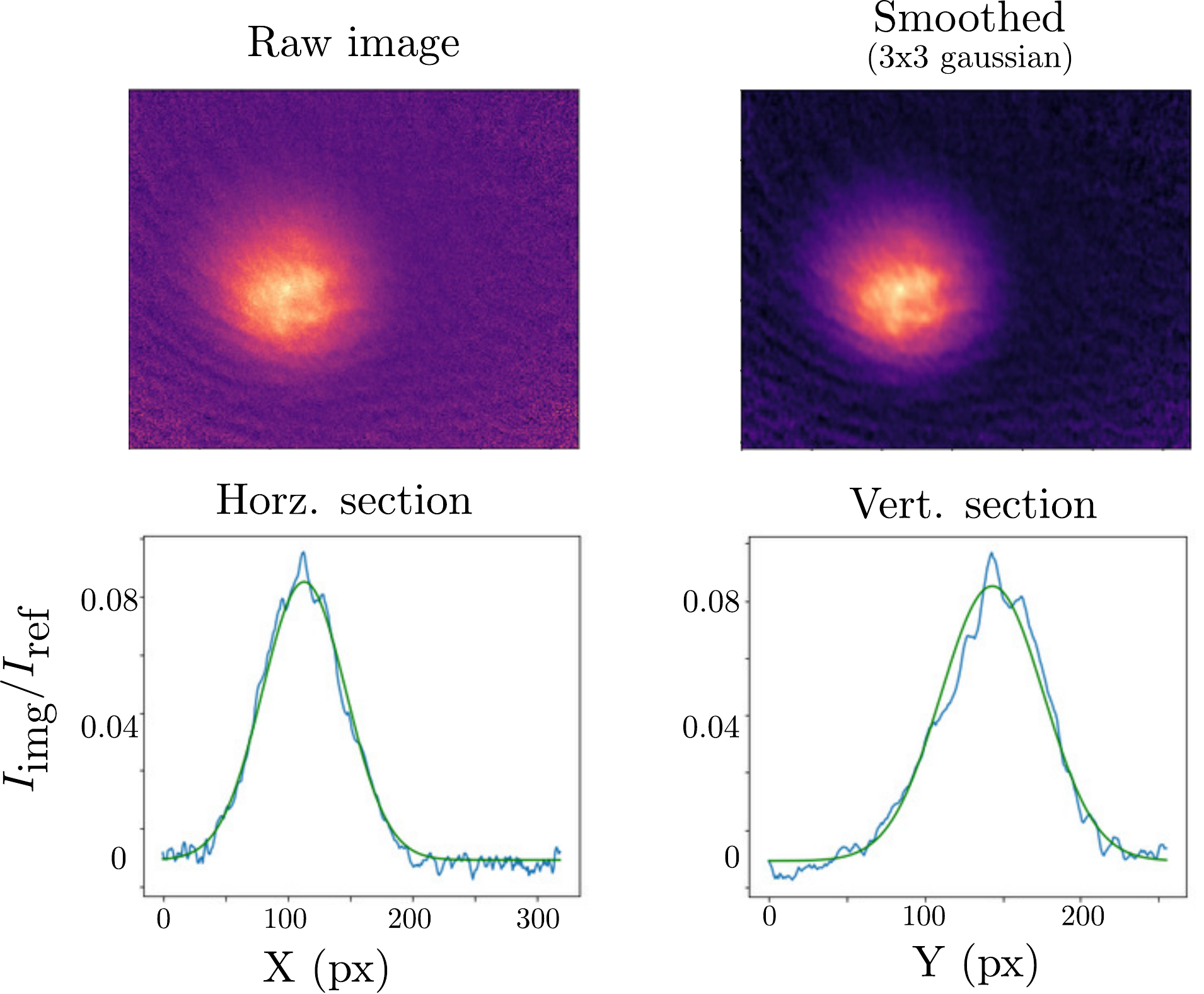}
	    \end{minipage}
	\end{figure}

	A simple description of absorption imaging is that one `takes a photo of the shadow of the atoms' by illuminating the sample with a collimated laser beam, and then projecting the beam onto a camera (Xenics Bobcat, 256x320 pixel InGaAs sensor, 20 $\mu$m pixel pitch).	
	Detailed discussions of the physical principles and implementations of this technique are presented in \cite{MakingProbingUnderstanding,TychkovThesis}.
	Here we present a simple picture, and an illustration in Fig. \ref{fig:abs_img}.
	A laser tuned close to resonance with the atomic sample is coupled via optic fibre to the absorption imaging setup.
	The light passes through a polarizing beam splitter and then a half-wave plate to fix the polarization, which is useful when imaging in the presence of a bias field\footnote{And indeed the earth's field, which was later found to be strong enough to suppress Penning ionization sufficiently to achieve condensation \cite{Abbas21}.}.
	The beam is then magnified by a 4:1 telescope to a $\approx1$ cm collimated waist and directed at the atomic sample through the vacuum windows.
	The electric field on the exit side of the cloud will have been attenuated by a factor $Ae^{i\phi}$, where the attenuation factor $A=\exp(-\frac{\tilde{n}\sigma_0}{2}\frac{1}{1+\delta^2})$ depends on the column density  $\tilde{n} = \int n(x) dx$ (integrated along the beam axis), the detuning in (half-linewidths) $\delta=2(\omega-\omega_0)/\Gamma$, and the absorption cross-section which is $3\sigma_0\lambda^2/2\pi$ in the two-level approximation \cite{MakingProbingUnderstanding}.
	The phase $\phi$ also depends on the atomic density and detuning from resonance, and is in principle useful for techniques like phase contrast imaging, but is not presently used in this apparatus.

	The camera records an image during the application of the laser light and a second image of identical exposure time after a {$\approx$10 ms delay}\footnote{We found it helpful to ensure that the AOM switched off in the meantime as well, and turned back on for the same cycle time for the second measurement.
	The AOMs can exhibit slow rise-times to their steady state pointing and efficiency due to dissipation of heat into the crystal from the RF drive.
	This was measurable as differences in the light intensity between subsequent images if the AOM was left on between exposures.}.
	By this point the atoms have scattered many photons and left the laser path.
	The second image is used as a reference to compute the integrated absorption at each pixel.
	A light-free image can be used as a `darkfield' to compensate for the camera's inherent noise profile.
	The ratio of (calibrated) pixel intensities in the presence of atoms $I_\textrm{abs}$ to the reference image without atoms $I_\textrm{ref}$ then provides a measure of the squared transmission coefficient,
	\begin{equation}
	A^2(x,y)=\frac{I_\textrm{abs}-I_\textrm{dark}}{I_\textrm{ref}-I_\textrm{dark}}.
	\end{equation}
	The 2D absorption profile can be fitted for purposes of further quantitative analysis. 
	Imaging in-trap is certainly possible and useful for optimizing performance, but it introduces the complication of a spatially-varying magnetic field.
	This means that the transmission coefficient varies due to the Zeeman shift around the trap.
	However, the initial imaging setup had too large a magnification to image the far-field density distribution which is necessary for thermometry.
	This design was shaped by the objective to load atoms into the dipole trap by using absorption imaging as the diagnostic, hence the magnification was chosen such that the MOT would almost fill the frame of the camera sensor.
	Therefore thermometry was not performed using this technique. 
	This was a shortcoming of the first design which was later remedied by upgrading the magnetic coils (see below), thus producing a more tightly-confining field and thus a denser trap.
	For detection and analysis of clouds released from a dipole trap, the lab currently uses the pulses of current across the MCP plates to infer the time-of-flight of atomic detection events \cite{Abbas21}.
	Calibration of the DLD for reconstruction of spatial information is currently underway.

	We could still perform (approximate) population measurements by fitting a Gaussian profile to the images.
	Unfortunately this was not adequately sensitive to measure the lifetime of the magnetic trap.
	After a short decay time it became impossible to get a good fit or even to see the cloud by eye in the absorption images, and thus the long-time behaviour was inaccessible.
	Instead, we released atoms by turning off the magnetic trap and used the number of detection events from the electron multiplier (manufactured by ETP) as a proxy for the population. 
	The current pulses resulting from atoms landing on electron multiplier were passed through a constant fraction discriminator, amplifier, and pulse rate counter which passed into the LabView software.
	The lifetime could then be obtained by an exponential fit to the measured populations over time.
	Measurements of the magnetic trap lifetime before and after the chamber bake are shown in Fig. \ref{fig:lifetime}.

\subsection{Dipole trap}
\begin{figure}
		\centering
		\includegraphics[width=\textwidth]{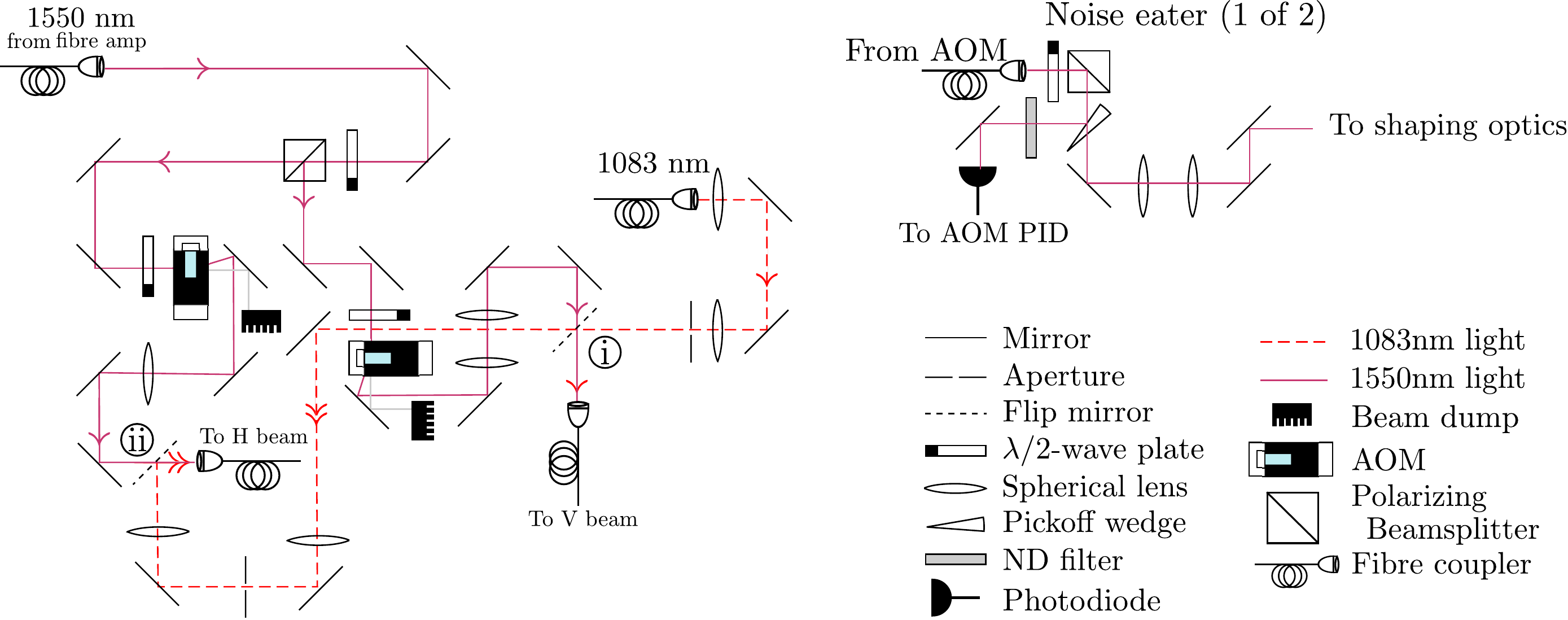}
		\caption{Schematic of the 1550 nm optics used to generate and control the dipole trap beams in the lattice machine.
		The AOMs are used for amplitude control and fast switching, and to run the dipole beams at slightly different frequencies to avoid mutual interference.
		They are driven by dedicated PID controls set in configurable sequences with the arbitrary-waveform generators.
		The beams then focus into an optical fibre which transports the beam to the shaping optics (example on top right).
		Initial alignment of the dipole beams is achieved one beam at a time by inserting one of the removable mirrors (marked (i) and (ii)) and coupling resonant 1083nm light into the fibre.}
		\label{fig:dipole_optics}
	\end{figure}

	While the main chamber was unworkable during bakeouts, construction could continue on other components of the machine, including the control and distribution systems for the optical dipole beams.
	These beams are generated by a 1550 nm diode laser with a linewidth of $\approx100$ kHz that seeds a 30 W fibre amplifier.
	The high-power beam is split into two AOM arms in a similar manner to the cooling optics described above.
	This configuration is illustrated  in Fig.	\ref{fig:dipole_optics}.
	
	To mitigate against misalignment of the dipole beams, which would eventually need to be overlapped in the focal region within less than 100 $\mu$m, we coupled light from the generation optics to the shaping optics via high-power NKT Photonics photonic crystal fibres to ensure the downstream optics are independent of the alignment upstream.
	Each AOM was driven at a fixed frequency by an amplified voltage-controlled oscillator, but the drive power was set in closed-loop configuration (in contrast to the cooling optics) to mitigate against heating the trap through intensity fluctuations.
	The voltage from a post-fibre photodiode was returned to a PID controller whose set point was determined by an arbitrary waveform generator pre-loaded with user defined waveforms and triggered by the main DAQ system.
	After losses in AOM transmission efficiency and fibre coupling losses, the maximum achievable (total) efficiency of each arm was about 30\% (defined as the ratio of power delivered into the chamber over the input power to the respective AOM).
	While I was working in the lab, the beam powers were roughly equal, but in the interim the dipole system has been redesigned, as discussed in \cite{Abbas21}.

	It is generally desirable to operate far-red-detuned dipole traps at very high intensities to create deeper traps (recall the ground-state shift scales with the intensity $I$, c.f. chapter \ref{chap:theory}).
	Deeper optical dipole traps can contain more atoms at a given temperature by virtue of trapping over a larger range of particle momenta, and so greater laser intensity is generally a good thing.
	It is true that spontaneous off-resonant scattering rates increase (approximately linearly) with intensity, but this can be remedied by detuning further (the scattering rate falls off with $\Delta^2$).
	Therefore one can obtain better loading efficiencies by using brighter beams.
	The purpose of the shaping optics are thus to shape the profile of the beam such that it reaches a tight focus.
	In order to achieve this, given the non-collimated laser profile emitted from the fibres, we first had to determine the profile of the beam, and then determine an arrangement of lenses to shape the beam to a tight focus at the desired trap location.
	The procedure we used to construct the shaping and insertion optics are discussed here.
	We aligned the beam output from the fibre parallel to a 1m-long rail which would be used to mount the lenses.
	We measured the profiles (at low power) by taking images with an InGaAs camera and fitting the output images with 2D gaussian profiles.
	A constrained two-dimensional Gaussian fit yields the waists $\sigma_x$ and $\sigma_y$ along the frame axes.

\begin{figure}
	\includegraphics[width=\textwidth]{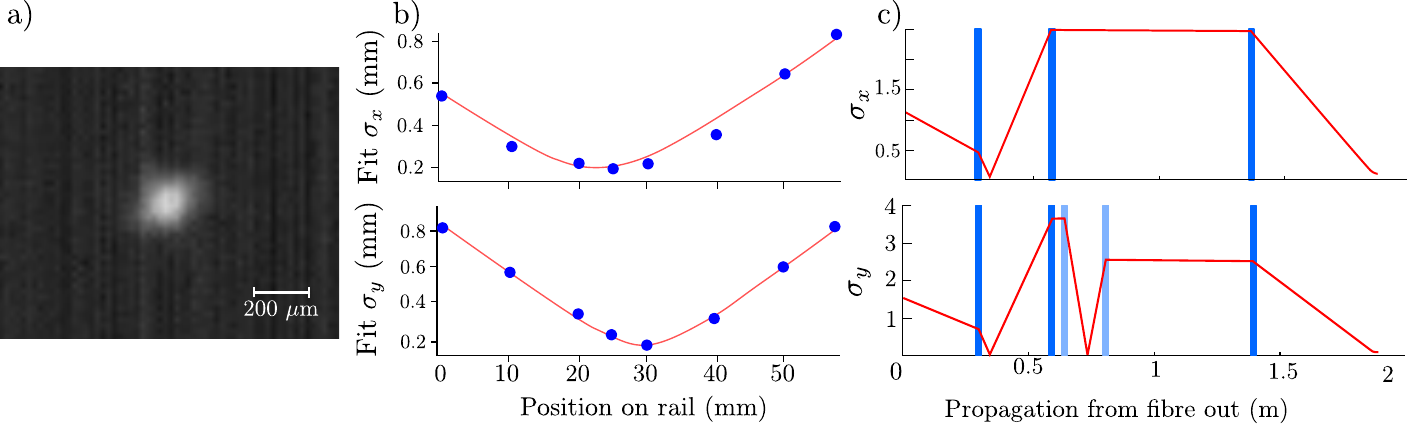}
	\caption{a) Image taken from the InGaAs camera near the focus of a dipole beam.
	b) Beam waists along the rail as obtained from two-dimensional Gaussian fits to the images.
	c) Forward-propagation simulation of the beam waists which was designed to reshape the profiles.
	Cylindrical lenses only act on one of the axes, and are shown in a lighter blue.
	This iterative process produced the image on the left, with waist sizes 130 $\mu$m and 124 $\mu$m.}
	\label{fig:profiling}
	\end{figure}

	A collection of measurements of the waists at positions $z_i$ along the rail can be fitted by the expression $w(z) = w(0)\sqrt{1 + (z/z_R)^2}$ for the waist of a Gaussian beam.
	In the preceding expression, $w(z)$ is the beam waist at distance $z$ from the focus and $z_R=\pi w(0)^2n/\lambda$ is the Rayleigh range (in terms of the laser wavelength $\lambda$ and refractive index $n$).
	We can then calculate the complex beam parameter $q(z) = z + i z_R$ and propagate it backward to determine the spot size and radius of curvature at the beginning of the rail.
	We used these values as input to a home-coded optical simulator which predicts the beam waist at a position $z$ after the beam propagates through a user-defined set of lenses and total optical path length.
	An example of such a simulated beam profile is shown in Fig. \ref{fig:profiling}.
	In reality some manual adjustment is necessary.
	The focus was finessed by deflecting the beam along a path of equal length to the distance to the desired trap centre and focusing the beam with respect to a camera at that position.

\subsubsection{Dipole trap}
\label{sec:dipole_trap}

	The dipole beams were first aligned to overlap with the magnetic trap by piping resonant light (at 1083 nm) through the dipole beam delivery fibres instead of 1550 nm.
	To do this, we redirected a few mW of light from the absorption imaging optics using a beamsplitter, passed it through an attenuator, and inserted it into the dipole optics via removable mirrors as illustrated in Fig.	\ref{fig:dipole_optics}.
	The endgame strategy was to achieve BEC by evaporative cooling in an optical dipole trap composed of two beams intersecting at right angles.
	Given total efficiencies of about 30\% through the dipole optics, we could deliver about 5W of power through each beam.
	With each beam focused to a waist of order 100 $\mu$m, the peak intensity of each beam would be $I_0 = 2P/\pi w_x w_y \approx 3.5\times10^9$ mW cm$^{-2}$, and combine to give a dipole trap about 70 $\mu$K deep.
	This highlights the need for very cold magnetic traps (and tightly-focused dipole beams) to ensure many atoms have sufficiently low energy to remain in the dipole trap.
	One also desires a tightly-confined magnetic trap to ensure good spatial overlap of the magnetic trap with the brightest parts of the dipole beam.
	Furthermore, the rate of elastic collisions between trapped atoms increases with tighter traps, which permits more efficient evaporative cooling \cite{Ketterle96}, underscoring the need for tight traps.
	Several iterations of magnetic field configurations and RF ramps were tried during the build but no significant improvement in loading efficiency (based on absorption-imaging measurements) was obtained.
	Due to the aforementioned issues with time-of-flight imaging given the small field of view, the challenges of in-trap imaging, and the low loading efficiency of the dipole trap, precise thermometry was not yet possible.  
	In the end, a key step forward was to install in-vacuum coils (at no small expense of effort) after I had left the lab.
	Details about the present trapping methods are presented in \cite{Abbas21}.
	This includes several re-assemblies of the optics and a change in the dipole trap configuration (namely switching to a second horizontal beam rather than a vertical beam as the upper flange of the Kimball chamber now hosts the coil feed-throughs including their cooling water).

	\begin{figure}
		\begin{minipage}{0.43\textwidth}
		\vspace{0cm}
		\caption{The $y$-axis dipole passes through a final focusing lens (lower middle) and is lifted by sturdy optics mounted to a heavy post.
	The upper mirror dials can be remotely controlled, using piezo-driven stepper motors, for precision alignment.
	The camera and refocusing optics for the absorption imaging system are just outside the left of the frame.}
		\label{fig:lifetime}
		\end{minipage}
		\hfill
		\begin{minipage}{0.55\textwidth}
		\vspace{0cm}
		\includegraphics[width=\textwidth]{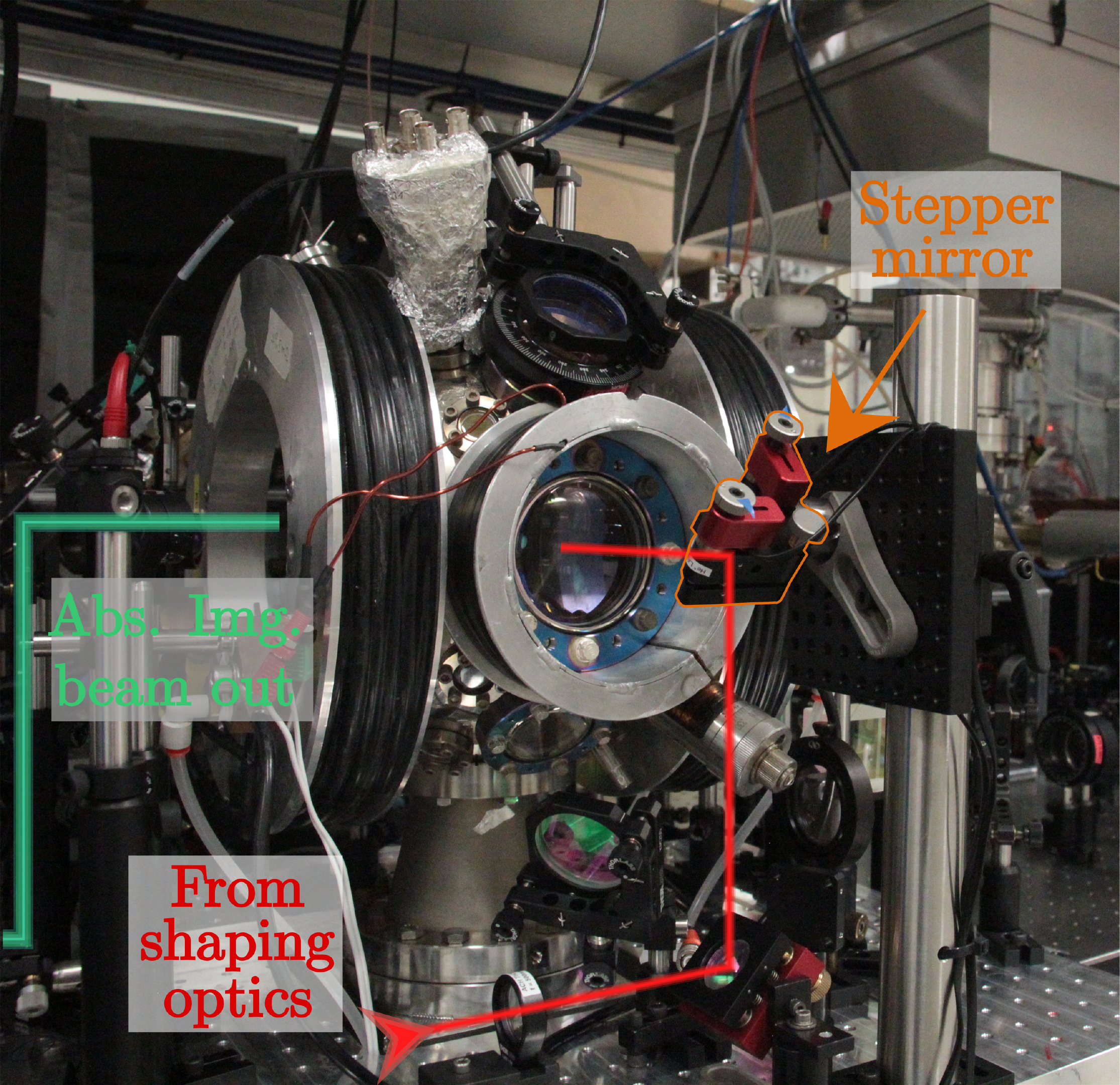} 
		\end{minipage}
	\end{figure}

\begin{figure}
	\centering
	\includegraphics[width=\textwidth]{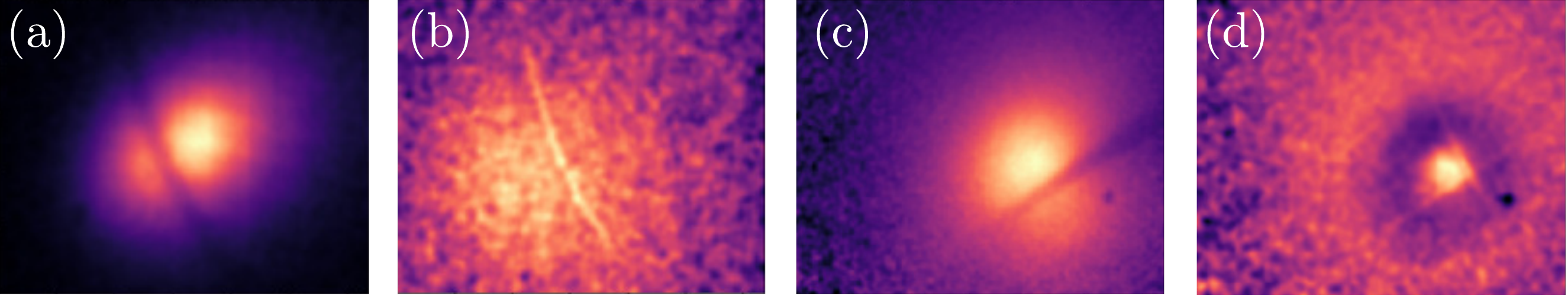}
		\caption{Stages of dipole beam alignment.
	The first alignment was achieved by passing resonant light through the cloud (a) and triggering the image acquisition (for 10 $\mu$s) at the same time as the resonant pulse.
	After switching to 1550 nm light, the beam deflected slightly due to chromatic aberration in the optics (b) and could then be resolved in averages of 8-10 images. 
	The beam appears to bend in these images because the imaging beam was not aligned with the optical axis of the lenses, which was discovered and corrected using this technique.
	Some time later the vertical dipole was aligned using the same method (c) and so two dipoles could be loaded simultaneously. 
	Finally, an image of the combined magnetic and dipole traps could be taken ((d), average of 10 shots). To obtain this image, a Gaussian fit to the central cloud was subtracted from the image, with a residual halo and bright spot (c.f. the aforementioned issues with in-trap imaging).}
	\label{fig:dipole_align}
	\end{figure}

	However, the basic operation of the dipole is unchanged, including the alignment procedure we discovered, and so is documented here.
	As discussed above, the alignment procedure used resonant light, directed through the dipole fibre into the trap.
	Coarse alignment could be found by supplying light near resonance through the dipole optics after loading a magnetic trap, and optimizing for the destruction of the cloud in response to the beam.
	Alignment could be finessed by reducing the intensity until the cloud recovers, then re-adjusting the beam to hit the centre of the trap and destroy it again.
	The cycle could be repeated until very low intensities were able to destroy the trap.
	Unfortunately we eventually found that there were multiple such optima.
	We determined that these were due to reflections of the beam from the inside of the chamber and scattering back across the trap.
	Afterwards, we acquired the image coincident with the resonant beam (rather than afterward) and found that when the beam intensity was reduced to a mere 4 $\mu$W, the beam appeared to bore a hole through the cloud, as shown in Fig.	\ref{fig:dipole_align}.
	
	After this initial alignment several attempts were made to optimize the dipole loading by adjusting the final lens position, mirror orientation, and experimental control sequences.	
	Matters were made more difficult by several concurrent issues.
	For one, we were \emph{still} waiting for parts for the MCP-DLD detector mount, the ETP electron multiplier had since failed, and the channeltron showed only modest changes in the ionization rate as a function of beam positioning.	
	This left absorption imaging as the sole means of ascertaining the loading efficiency.
	Unfortunately the signal-to-noise was poor and the dipole would only resolve after $\approx 8$ shots.
	At the time the experimental cycle was on the order of 15 seconds, and the long acquistion times were a major obstacle.
	It would eventually be found that the magnetic trap was simply not tight enough.
	The magnetic field gradient would later be improved by upgrade to in-vacuum coils during the tenure of the students who took up the mantle.

\section{Progress and outlook}

	After my departure, the lab sat motionless for some months as I had been the sole graduate student on the project.
	Later, the \mhe~lattice team was renewed by two PhD students (A. H.	Abbas and X. Meng) and a Masters student (R. S. Patil).
	The team has since achieved a BEC production time of as low as 3.3 seconds, nearly a factor of 2 faster than the prior art \cite{Bouton15} and a factor of 8 faster than the BiQUIC machine.
	The subsequent publication \cite{Abbas21} contains details about the new coils, additional infrastructure and cooling techniques, and characterization techniques.
	The large condensates, containing something on the order of a million atoms, are also competitive with the state-of-the art.
	I take my hat off to the students who picked up the trail where I had left off, and accomplished a goal I'd struggled towards for two years.

	Ahead, the team is faced with the challenge of aligning and optimizing three pairs of lattice beams.
	Once this is accomplished, there is still much ground to cover in exploring the Bose-Hubbard model.
	For instance, the time evolution of the system following a quench or sudden change in dimensionality would be a natural starting point.
	The onset of coherence could be probed via many-particle mometum correlations, extending the work in \cite{Carcy19} into the dynamical regime.	
	Beyond the Bose-Hubbard model, there lies a fork in the road.
	One path leads to the Fermi-Hubbard model by way of further infrastructure upgrades to achieve degenerate $^3$\mhe~in the same machine.
	The expertise gained through the ongoing upgrade of the BiQUIC machine to include $^3$He circulation would be instrumental in this mission.
	The other path is towards the study of the disordered Bose-Hubbard model (also known as the Aubry-Andr\'{e} model), which has been studied in quantum gas microscopes \cite{Rispoli19} but not, so far, with a momentum microscope.
	Theoretical studies predict that momentum-space localization occures in disordered lattices generated by lasers with incommensurate wavelengths \cite{Larcher11}, which is an aspect of localization which has received little attention.
	\vfill

	\begingroup
	\begin{flushright}
	
	\singlespacing{\fontsize{12}{12}\selectfont\emph{
	``It might be noted here, for the benefit of those interested in exact solutions, \\
	that there is an alternative formulation of the many-body problem:\\
	How many bodies are required before we have a problem? \\
	G.	E.	Brown points out that this can be answered by a look at history.	\\
	In eighteenth-century Newtonian mechanics,
	the three-body problem was insoluble.\\
	With the birth of general relativity and quantum electrodynamics in 1930,\\ 
	the two- and one-body problems became insoluble.
	And within modern\\
	quantum field theory, the problem of zero bodies (vacuum) is insoluble.\\
	So, if we are out after exact solutions, no bodies at all is already too many!"}\\
	- Richard Mattuck\footnote{R.
	Mattuck, \emph{A Guide to Feynman Diagrams in the Many-Body Problem}, Dover Books on Physics (1992)}
	}
	\end{flushright}
	\onehalfspacing
	\endgroup
	\vspace{1cm}